\documentclass[a4paper]{panl}
\usepackage[symbol]{footmisc}   
\usepackage{cite}
\usepackage{wrapfig}
\usepackage{graphicx}
\usepackage{amssymb}
\usepackage{amsfonts}
\usepackage{amsmath}
\usepackage{longtable}
\usepackage{rotating}
\usepackage{lscape}
\usepackage{epsfig}
\usepackage{multirow}
\usepackage{subfigure}
\usepackage[utf8]{inputenc} 
\usepackage{isotope}
\usepackage{tabularx}
\usepackage{dcolumn}

\newcommand\ba{\begin{eqnarray}}
\newcommand\ea{\end{eqnarray}}
\newcommand\nn{\nonumber}
\newcommand{\be}{\begin{equation}}
\newcommand{\ee}{\end{equation}}
\originalTeX
\def\bfg #1{{\mbox{\boldmath $#1$}}}
\begin{document}

\issuearea{ Prepared  for Physics of Elementary Particles and Atomic Nuclei. Theory}

\renewcommand{\thefootnote}{\fnsymbol{footnote}}     

\title{Possible  studies  at  the first stage of the NICA collider operation
 with polarized and unpolarized proton and deuteron beams}

\maketitle

\authors{V.V.\,Abramov$^{\,1}$, A.\,Aleshko$^{\,2}$,
 V.A.\, Baskov$^{\,3}$,
E.\,Boos$^{\,2}$,}
\authors{V.\,Bunichev$^{\,2}$, 
O.D.\, Dalkarov$^{\,3}$, R.\,El-Kholy$^{\,4}$,A.\,Galoyan$^{\,5}$, A.V.\, Guskov$^{\,6}$,}
\authors{
V.T.\, Kim$^{\,7,8}$,  E.\, Kokoulina$^{\,5,9}$, I.A.\,Koop$^{\,10,11,12}$, 
B.F.\,Kostenko$^{\,13}$,}
\authors{A.D.\, Kovalenko$^{\,5}$, V.P. Ladygin$^{\,5}$,A.\,B.\,Larionov$^{\,14,15\,}$,
A.I.\, L'vov$^{\,3}$, A.I.\,Milstein$^{\,10,11}$,}
\authors{V.A.\, Nikitin$^{\,5}$, N.\,N.\,Nikolaev$^{\,16,26\,}$, A.\,S.\,Popov$^{\,10}$, \,V.V.\, Polyanskiy$^{\,3}$,
 } 
\authors{J.-M. Richard$^{\,17}$, S.\,G.\,Salnikov$^{\,10}$,
A.A.\,Shavrin$^{\,18,}$, 
P.Yu.\,Shatunov$^{\,10,11}$,}
\authors{Yu.M.\,Shatunov$^{\,10,11}$, O.V.\,Selyugin$^{14}$,
M.\, Strikman$^{\,19,}$, \,E.~Tomasi-Gustafsson$^{\,20}$,}
	\authors{V.V.\, Uzhinsky$^{\,13,}$,
Yu.N.\,Uzikov$^{\,6,21,22,\,}$\footnote{E-mail: uzikov@jinr.ru; Corresponding author},
Qian Wang$^{\,23}$,  
Qiang Zhao$^{\,24,25}$, A.V.\, Zelenov$^{\,7}$}

\vspace{3mm}
\from{$^{1}$\,NRC ``Kurchatov Institute'' - IHEP,
Protvino 142281, Moscow region, Russia }
\from{$^{2}$\,Skobeltsyn Institute of Nuclear Physics, MSU, Moscow, 119991 Russia}
\from{$^{3}$ P.N.\, Lebedev Physical Institute,Leninsky prospect 53, 119991 Moscow, Russia }
\from{$^{4}$\,Astronomy Department, Faculty of Science, Cairo University, Giza, Egypt, 12613}
\from{$^{5}$\,Veksler and Baldin Laboratory of High Energy Physics, Joint Institute for Nuclear Research,%
	Dubna, Moscow region, 141980 Russia}
\from{$^{6}$\,Joint Institute for Nuclear Researches, DLNP, Dubna, Moscow reg. 141980 Russia}	
\vspace{-3mm}
\from{$^{7}$\,Petersburg Nuclear Physics Institute NRC KI, Gatchina, Russia}
\vspace{-3mm}
\from{$^{8}$\,St. Petersburg Polytechnic University, St. Peterburg, Russia}
\from{$^{9}$ Sukhoi State Technical University of Gomel,
Prospect Octiabria, 48,
246746 Gomel, Belarus}
\from{$^{10}$ Budker Institute of Nuclear Physics of SB RAS, 630090 Novosibirsk, Russia}
\from{$^{11}$ Novosibirsk State University, 630090 Novosibirsk, Russia}	
\from{$^{12}$ Novosibirsk State Technical University,630092 Novosibirsk, Russia}
\from{$^{13}$\,Laboratory of Information Technologies, Joint Institute for Nuclear Research,
	Dubna, Moscow region, 141980 Russia}
\vspace{-3mm}
\from{$^{14}$\,Joint Institute for Nuclear Researches, BLTP, Dubna, Moscow reg. 141980 Russia}
\vspace{-3mm}
\from{$^{15}$\,Institut f\"ur Theoretische Physik, Justus-Liebig-Universit\"at, 35392 Giessen, Germany}
\from{$^{16}$ L.D. Landau Institute for Theoretical Physics, 142432 Chernogolovka, Russia}
\from{$^{17}$\,Universit\'e de Lyon, Institut de Physique des 2 Infinis de Lyon, UCBL--IN2P3-CNRS,
4, rue Enrico Fermi, Villeurbanne, France}
\vspace{-3mm}
\from{$^{18}$\,St. Petersburg State University, St. Peterburg, Russia}
\vspace{-3mm}
\from{$^{19}$  Pensilvania State University,\, 104 Davey Laboratory
 University Park PA 16802 USA}
\vspace{-3mm}
\from{$^{20}$\,DPhN, IRFU, CEA, Universit\'e Paris-Saclay, 91191 Gif-sur-Yvette Cedex, France}
\vspace{-3mm}
\from{$^{21}$\,Dubna State University, Dubna, Moscow reg. 141980 Russia }
\vspace{-3mm}
\from{$^{22}$\, Department of Physics, M.V. Lomonosov State University, Moscow, 119991 Russia }
\from{$^{23}$\,Guangdong Provincial Key Laboratory of Nuclear Science, Institute of Quantum Matter,
South China Normal University, Guangzhou 510006, P.R. China}
\vspace{-3mm}
\from{$^{24}$\,Institute of High Energy Physics, Chinese Academy of Sciences, Beijing 100049, P.R. 
China}
\vspace{-3mm}
\from{$^{25}$\,University of Chinese Academy of Sciences, Beijing 100049, P.R. China}
\vspace{-3mm}
\from{$^{26}$ \, Moscow Institute of Physics and Technology (National Research University),
141701 Dolgoprudny, Russia}


\begin{abstract}
\vspace{0.2cm}

    Nuclotron based Ion Collider fAcility (NICA)  project is in progress   at the
  Joint Institute for Nuclear Research and  will  start experiments with heavy ions.  
  In the context of the NICA Hadronic Physics programme  double polarized 
  $pp$-, $dd$- and $pd$- collisions even  at lower  energies of $\sqrt{s_{NN}}=3.4-10$ GeV,
  which will be accessible already at the initial stage of experiments,are  essential tools
  for precise understanding the spin dependence of the nucleon-nucleon strong interactions, 
  in both elastic and deep-inelastic regimes. A special interest is interaction in few baryon
  systems  at double strangeness, charm and beauty thresholds.For instance, polarized large-angle 
  elastic $pp$ and $pn$ scattering near the charm threshold  allows one to get an access  to
  properties of possible exotic multiquark states  and their relation to the states recently 
  observed at LHCb.Large angle scattering of protons and deuterons on the deuteron contains 
  unique information on the short-range structure of the deuteron, its non-nucleonic degrees 
  of freedom and also on color transparency phenomenon. Furthermore,  double polarized proton-deuteron scattering offer a possibility to test the Standard Model through the search for time-invariance 
  (or CP- invariance under CPT symmetry) violation and parity-violation in single-polarized 
  scattering. This paper contains   suggestions for experiments with usage  of  the  Spin Physics
  Detector (SPD) and discusses perspectives of the first stage of the SPD Programme. This  includes experiments  with  non-polarized beams too as well as collisions like
   $^{12}$С-$^{12}$С and $^{40}$Сa-$^{40}$Ca.
  
\end{abstract}
\renewcommand{\thefootnote}{\arabic{footnote}}      
\setcounter{footnote}{0}

\eject

\counterwithout{footnote}{page}
\tableofcontents
\vfill
\counterwithin{footnote}{page}
\renewcommand{\thefootnote}{\arabic{footnote}}
\setcounter{footnote}{0}



\noindent
\eject

\section*{\bf Tests of QCD basics in the transition region} 

The Standard Model (SM)  of fundamental interactions  formulated five decades  ago
as a local gauge invariant  theory based on the $SU(2)_L\times U(1)_Y\times SU(3)_c$ 
spontaneously broken  symmetry, was perfectly  confirmed  by experiments in electroweak sector.
The only part of this model, Quantum Chromodynamics (QCD), connected with the colored  $SU(3)_c$ symmetry and  considered  as a basis of strong interactions  between quarks and gluons 
is still  under experimental verification. 

At  low energies, below the  GeV region the strong interaction is described in terms
of baryons exchanging mesons  in accordance with the
chiral effective field  theory, which is based on spontaneously broken chiral symmetry of the QCD  Lagrangian \cite{Epelbaum:2018ogq}.
Recent progress in our understanding of properties of the light nuclei and nuclear reactions
 achieved within this approach is outlined in Refs.
 \cite{Epelbaum:2019kcf,Carlson:2014vla}.
 At  much  higher energies and high transferred 4-momenta, perturbative Quantum Chromodynamics (pQCD) characterizes
 the strong force in terms of quark and gluons carrying  color charge, and  obeying to parton distribution functions (PDF) of hadrons and nuclei.
       Although these two pictures are  well determined in their  respective energy  scales, the transition between them is not well identified.
       Whereas the goal  of the Many Purposes Detector (MPD)  NICA project is to search    for phase transition of the baryon matter at high temperature and high density into the quark gluon plasma in heavy-ions collision,     and  on this way to study  properties of the early Universe, the main aim of the
       Spin Physics Detector 
       (SPD) project  \cite{Abazov:2021hku} 
       at its first stage with lower energies is  quite different and, in particular, is just connected with a search for the transition region   from  hadron to quark-gluon degrees of freedom in theoretical  describing of collisions of free nucleons or  lightest nuclei.
  QCD predicts that hadrons produced in exclusive processes at sufficiently high 4-momentum transfer will experience diminished final (initial) state interactions. This QCD prediction  named as color transparency (CT) \cite{Mueller}, \cite{Brodsky}  may help to identify  the transition between  these two alternative descriptions of strong forces after  the  onset of CT will be observed.  Another signal for the transition region in
  structure of the lightest nuclei is related to  onset of the predicted by pQCD dimensional scaling  in reactions with  these nuclei.
   A clear  indication for transition to quark degrees of freedom  in strong interactions would give a formation of  multiquark states, like dibaryon resonances observed in sector of  light quarks \cite{Clement:2016vnl}. 
 Production of heavy quarks in few-nucleon systems  can be related  to  formation of exotic type of 
 resonances, as ``octoquarks' $uuds\bar s uud$,   $uudc\bar c uud$ \cite{Brodsky:1987xw}
and  the  behaviour of double  spin correlation
$A_{NN}$  of $pp$ elastic   scattering measured  near    the charm threshold at large angles \cite{Court:1986dh} supports this assumption.
On the other hand, it is important  to understand  how this observation is related  to recently observed at LHCb pentaquark states $uudc\bar c$ \cite{Aaij:2015tga}. The  SPD NICA has all  possibilities  for study   these and  other issues of  QCD.
 Furthermore, polarization phenomena provide an unique possibility to  search for physics beyond the SM   by making test of fundamental discrete symmetries of the SM  related to the space (P), time (T) and charge (C) inversion. One of these options is connected  with double polarized proton-deuteron scattering providing  a search for T-invariance (or CP-invariance under CPT-symmetry) violation. 
 
 Experiments with unpolarized colliding beams are also  of importance in study of
 reactions at  heavy quark thresholds and  in search for both
 color transparency and scaling onset  or  multiquark (dibaryon) states.
 


\section{\bf {The SPD  setup and experimental conditions} \protect\footnote{This section is written by   A.V.\, Guskov (E-mail: alexey.guskov@cern.ch)
 and  A.D.\, Kovalenko (E-mail:kovalen@dubna.ru)}}
 
The SPD experimental setup  is being designed as a universal $4\pi$ detector with advanced tracking and particle identification capabilities based on modern technologies that can operate with polarized proton and deuteron beams at a collision energy up to 27 GeV and a luminosity up to $10^{32}$ cm$^{-2}$ s$^{-1}$ (proton collisions). Details of the SPD experimental setup  are described in its  Conceptual Design Report 
\cite{Abazov:2021hku}.
The silicon vertex detector will provide resolution for the vertex position on the level of below 100 $\mu$m needed for reconstruction of primary and secondary vertices. The straw tube-based tracking system  placed within a solenoidal magnetic field of up to 1 T at the detector axis should provide the transverse momentum resolution $\sigma_{p_T}/p_T\approx 2\%$ for a particle momentum of 1 GeV/$c$. The time-of-flight system with a time resolution of about 60 ps will provide $3\sigma$ $\pi/K$ and $K/p$ separation of up to about 1.2 GeV/$c$ and 2.2 GeV/$c$, respectively. Possible use of the aerogel-based Cherenkov detector could extend this range. Detection of photons will be provided by the sampling electromagnetic calorimeter with the energy resolution $\sim 5\%/\sqrt{E}$. To minimize multiple scattering and photon conversion effects for photons, the detector material will be kept to a minimum throughout the internal part of the detector. The muon (range) system is planned for muon identification. It can also act as a rough hadron calorimeter. The pair of beam-beam counters and zero-degree calorimeters will be responsible for the local polarimetry and luminosity control.  To minimize possible systematic effects, SPD will be equipped with a triggerless DAQ system. 

It is assumed that  up to
30\% of the collider running time will be devoted to polarized deuteron and proton experiments
from the beginning of  the collider commissioning. Thus,  some  polarized $pp$-, $dd$- and even $pd$- collisions at energy range  of $\sqrt{s_{NN}} = 3.4 \div 10$ GeV, could be possible already at the initial stage of the collider operation. The most accessible
is polarized deuteron beam from the Nuclotron in the energy range of $1 \div 4$ GeV/u.
Average 
luminosity of $dd$ - collisions is estimated to $8\times 10^{27}\div  2.5\times 10^{31}\,
cm^{-2} s^ {-1}$. Stable direction of the polarization vector is vertical.  A single and double polarized collisions are possible.  Transverse polarization of deuteron beam can be obtained at the specific energy point $\sim 5.6$ GeV corresponding to the spin integer resonance. 
  The adequate intensity of polarized proton beam from the Nuclotron
  ($\geq 10^{10}$ part./pulse) will be reached after commissioning of the new light ion injector LILAC scheduled to 2025-2026 and the spin control system have been designed for the collider.
  The existing proton injected chain put limit to the beam intensity due to very low output linac energy  (5 MeV). Thus, only experiments on the beam storage and acceleration are planning 
  for the commissioning phase.
Realization of pd - mode is more complicated because HILAC and LILAC both injection chains 
should be  involved in the process.   Moreover, only single polarized collision mode is available, namely: unpolarized deuteron with polarized proton. The peak luminosity in symmetric dp – mode, corresponding to equal momentum of the colliding particles per nucleon, can reach of 
$2\times 10^{31}\, cm^{-2} s^{ -1}$ at stored intensity of $6\times 10^{11}$
particles per each collider ring. 
Light ion collision studies at the SPD are possible also. The luminosity level can be scaled from that was specified for gold-gold collisions:  $1\times 10^{27}\, cm^{-2} s^{-1}$ at 
$\sqrt{s_{NN}}  = 11$ GeV.

\section{\bf {Elastic $pN$, $pd$ and $dd$  scattering} \protect\footnote{This section is written by   Yu.N. Uzikov; E-mail:uzikov@jinr.ru}  }
\label{Uzikov}

\begin{abstract}
\vspace{0.2cm}
 The spin-dependent Glauber theory is applied  to calculate spin observables of $pd$ elastic scattering at
 $3$-$50$ GeV/c using $pp$ amplitudes available in the literature
and parametrized
within the Regge formalism. The calculated vector $A_y^p$, $A_y^d$ 
and tensor $A_{xx}$, $A_{yy}$ analyzing powers  
 and the spin-correlation coefficients $C_{y,y}$, $C_{x,x}$,
$C_{yy,y}$, $C_{xx,y}$ can be measured at 
SPD NICA and, thus, will provide a test of the used $pN$ amplitudes.
Quasi-elastic scattering $pd\to \{pp\}_sn$ with formation of spin-singlet $pp(^1S_0)$ pair at zero
 scattering angle is  of special interest. The $dd$ elastic scattering 
is briefly outlined.
 The double polarized $pp$  and $pn$ elastic scattering at large c.m.s.  scattering
  angle  $\theta_{cm}=90^\circ$ is 
considered in the threshold of the
  charm production.
\end{abstract}
\vspace*{6pt}

PACS: 25.40.Cm,\,13.75.Cs,\,13.88.+e

\subsection{{Spin amplitudes of $pN$ elastic  scattering }} 
Nucleon-nucleon elastic scattering contains fundamental information on the dynamics of the $NN$ interaction and  constitutes a basic process in physics of atomic nuclei and  hadrons.
%
A systematic reconstruction of spin amplitudes of $pp$ and $pn$ elastic scattering
from $pN$ scattering  data is provided by the SAID partial-wave analysis \cite{Arndt:2007qn} 
and covers laboratory energies up to $3$ GeV ($p_{lab}\approx 3.8$~GeV/c) for $pp$ and $1.2$ GeV ($p_{lab}\approx 1.9$~GeV/c)
for $pn$ scattering.
At higher energies there is only incomplete experimental information on $pp$ scattering, whereas data for the $pn$ system are very scarce. 
In the literature there are several models and corresponding
parametrizations for $pN$ amplitudes. Some of them are obtained 
 in the eikonal approach  for the lab momentum $6$ GeV/c \cite{Sawamoto:1979cb} and for LHC energies \cite{Selyugin:2009ic} and recently in 
\cite{Selyugin:2015vao}
(see Sect. \ref{Selyugin}).
At moderate transferred momenta $-t$  and  large   invariant mass $s$  the Regge model is expected to be  valid to describe elastic $pN$ scattering. 
In  literature there are some parametrizations for $pN$ amplitudes, 
obtained    within the Regge 
phenomenology for values of $s$ above $6$ GeV$^2$ ($p_{lab} \ge 2.2$~GeV/c)
\cite{Ford:2012dg} and 
 for $p_{lab}=3$-$50$ GeV/c (corresponding to $2.77 < \sqrt{s} < 10$ GeV) 
 \cite{Sibirtsev:2009cz}.

 Assuming Lorentz-invariance and parity conservation, the elastic  $NN$ scattering is described
 by eight independent helicity amplitudes $\phi_i$ ($i=1,\dots 8$) determined in
 \cite{Bystricky:1982bk,Bystricky:1976jr}. Under  time-reversal invariance,
 one has ($\phi_5=\phi_8$, $\phi_6=\phi_7$) six independent
 amplitudes, and for identical nucleons $pp$ and $nn$ the number of independent helicity amplitudes is equal
 to five ($\phi_5=-\phi_6$, $\phi_7=-\phi_8$).
Full information about the 
spin dependent $pN$ amplitudes can be obtained, in principle, from a complete 
polarization experiment, which, however, requires to measure twelve (ten)
independent observables at a given collision energy for $pn$ ($pp$ or $nn$) 
and, thus, constitutes a 
too complicated experimental task.
Another possible way to check existing parametrizations
in addition to  direct measurement of spin observables of $pN$ elastic 
scattering is to study spin effects in proton-deuteron ($pd$) and neutron-deuteron ($nd$) 
elastic and quasi-elastic scattering. The polarized  $pd$-elastic scattering is 
discussed  below  using the Glauber diffraction theory.

At large $-t$  corresponding to  large  scattering angles in the c.m.s. pN system
($\theta_{cm}\approx90^\circ$), where the Regge model cannot be applied, very interesting features  were observed in the double spin 
 asymmetry $A_{NN}$ in the elastic pp scattering  at  laboratory momenta $p_{lab}= 5-10$
 GeV/c.
 Commonly accepted explanation of those features is absent in literature. In section 
 \ref{Uz-ANN} we give 
a short review of existing models based on usage  of  the  pQCD amplitudes and non-perturbative exotic
 multiquark resonances contribution.

\subsection{Polarized $pd$ elastic diffraction  scattering within
the Glauber model}

 As was noted  above, a possible way to check existing parametrizations of $pN$ elastic
 amplitudes is to study spin effects in proton-deuteron ($pd$) and deuteron-deuteron ($dd$) 
elastic and quasi-elastic scattering. At high energies and small four-momentum transfer $t$, $pd$ scattering can be described 
by the Glauber diffraction theory of multistep scattering, which involves as input on-shell $pN$ elastic scattering amplitudes. 
Applications of this theory with spin-dependent effects 
included \cite{Platonova:2016xjq} indicate a good agreement with 
the $pd$ scattering data at energies about $1$~GeV if the SAID data on $pN$ scattering amplitudes are used as input 
of the calculations \cite{Temerbayev:2015foa, Mchedlishvili:2018uur,Platonova:2019yzf}.
\begin{figure}[t]
\begin{center}
 \includegraphics[width=125mm]{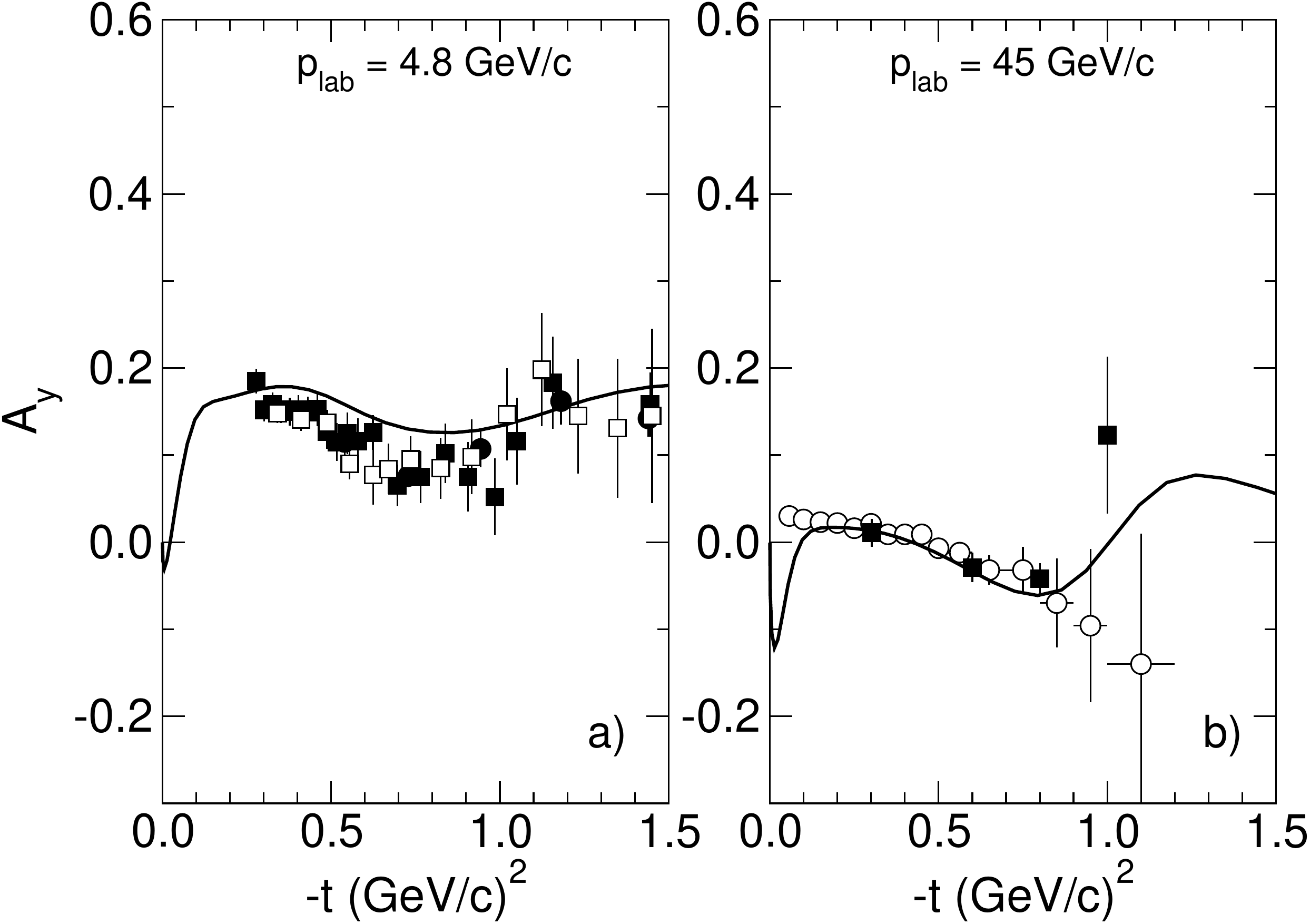}
\vspace{-3mm}
\caption{Analyzing power for $pp$ elastic scattering as a function of
the four-momentum transfer $-t$ at $4.8$ GeV/c (left) and $45$ GeV/c (right).
The results of calculations  \cite{Uz-Jh-TemerB}
based on  the Regge model parameterizations from
\cite{Sibirtsev:2009cz} are shown by the solid line
(see details in Ref. \cite{Uz-Jh-TemerB}).
Left: Data are taken from Refs.~\cite{Parry:1973fj} 
(filled squares: $4.4$ GeV/c;
open squares: $5.15$ GeV/c), and \cite{Abshire:1974ed} (circles).
Right: Data are taken from Refs. \cite{Corcoran:1980ew} (squares)
and \cite{Gaidot:1976kj} (circles).
}
\end{center}
\labelf{fig-uz01}
\vspace{-5mm}
\end{figure}

 The spin-dependent Glauber theory 
\cite{Platonova:2016xjq,Temerbayev:2015foa} is applied recently 
\cite{Uz-Jh-TemerB}
to calculate spin observables of $pd$ 
elastic scattering at $3$-$50$ GeV/c utilizing the $pp$ elastic scattering amplitudes $f_{pp}$ 
established and parametrized in Ref.~\cite{Sibirtsev:2009cz} within  
the Regge formalism.
The Regge approach allows one to construct $pn$ 
(and $\bar p N$) amplitudes together with the $pp$ amplitudes. 
This feature allows one to perform a test of broad  set of $pN$
amplitudes and applicability of the Regge model itself to $pN$ elastic scattering.
However, in view of the scarce experimental information about the spin-dependent $pn$ amplitudes and taking into account 
that the spin-independent parts of the $pp$ and $pn$ amplitudes at high energies are approximately the same, it was  assumed  in 
\cite{Uz-Jh-TemerB} as  a first approximation, 
 that $f_{pn}=f_{pp}$.   
The amplitudes of $pN$ elastic scattering are written as \cite{Platonova:2016xjq}
\begin{eqnarray}
\label{pnamp}
M_N({\bf p}, {\bf q};\bfg \sigma, {\bfg \sigma}_N)=
 A_N+C_N\bfg \sigma \hat n +C_N^\prime\bfg \sigma_N \hat n +
B_N(\bfg \sigma \hat {\bf k}) (\bfg \sigma_N \hat {\bf k})+\\ \nonumber
+ (G_N+H_N)(\bfg \sigma \hat {\bf q}) (\bfg \sigma_N \hat {\bf q})
+(G_N-H_N)(\bfg \sigma \hat {\bf n}) (\bfg \sigma_N \hat {\bf n}), 
\end{eqnarray}
where the complex numbers $A_N$, $C_N$, $C_N^\prime$, $B_N$, $G_N$, $H_N$ were fixed from 
the amplitudes of the SAID analysis \cite{Arndt:2007qn} and parametrized by a sum of Gaussians.
For the double scattering term in $pd$ scattering the unit vectors $\hat {\bf k}$, $\hat {\bf q}$,
$\hat {\bf n}$ are defined separately for each individual $NN$ collision.
Numerical values for the parameters of the Gaussians
are obtained by 
fitting to the helicity amplitudes from Ref.~\cite{Sibirtsev:2009cz}. Those for $p_{lab}=45$ GeV/c  
are given
in Ref. 
\cite{Uz-Jh-TemerB}.
The differential cross section of $pp$ elastic scattering and the vector analyzing power 
$A_y$ are reproduced with these parameterizations on the same level of accuracy as in 
Ref.~\cite{Sibirtsev:2009cz}, in the interval of transferred four momentum 
$-t< 1.5$ (GeV/c)$^2$. An example of calculations of $A_y$  at $p_{lab}= 4.8$ GeV/c and 
45 GeV/c is shown in Fig.~\ref{fig-uz01}.

 The spin observables $A_y$, $A_{ij}$, and $C_{ij,k}$ considered in the 
 work 
 \cite{Uz-Jh-TemerB}
 are defined in the
notation of Ref.~\cite{Ohlsen:1972zz} as following
\begin{eqnarray}
 \label{observs}
A_y^d=Tr M S_yM^+/TrMM^+ , A_y^p=TrM \sigma_yM^+/TrMM^+\\ \nonumber 
A_{yy}=TrM {\cal P}_{yy} M^+/TrMM^+,A_{xx}=TrM {\cal P}_{yy} M^+/TrMM^+ \\ \nonumber
  C_{y,y}=TrM{ S}_y\sigma_yM^+/TrMM^+,C_{x,x}=TrM{ S}_y\sigma_yM^+/TrMM^+,  \\ \nonumber 
C_{xx,y}=TrM{\cal P}_{xx}\sigma_yM^+/TrMM^+,  C_{yy,y}=TrM{\cal P}_{yy}\sigma_yM^+/TrMM^+, \nonumber 
\end{eqnarray}
where ${\cal P}_{ij}=\frac{3}{2}(S_iS_j+S_jS_i)-2\delta_{ij}$ and $S_j$ ($j=x,y,z$) are Cartesian 
components of the spin operator for the system with $S=1$,
 the transition operator $M$ depends on the momentum  of the initial ($\bf p$) and final (${\bf p}'$)
proton and contains the Pauli spin matrices
 ${\bfg \sigma}=(\sigma_x, \ \sigma_y, \ \sigma_z)$.
  We use the Madison reference frame with the axis OZ$||$${\bf p}$, OY$||$$ [{\bf p}\times {\bf p}']$
 and  OX choosen in such a way to provide a right-handed coordinate system.
 
\begin{figure}[t]
\begin{center}
\includegraphics[width=125mm]{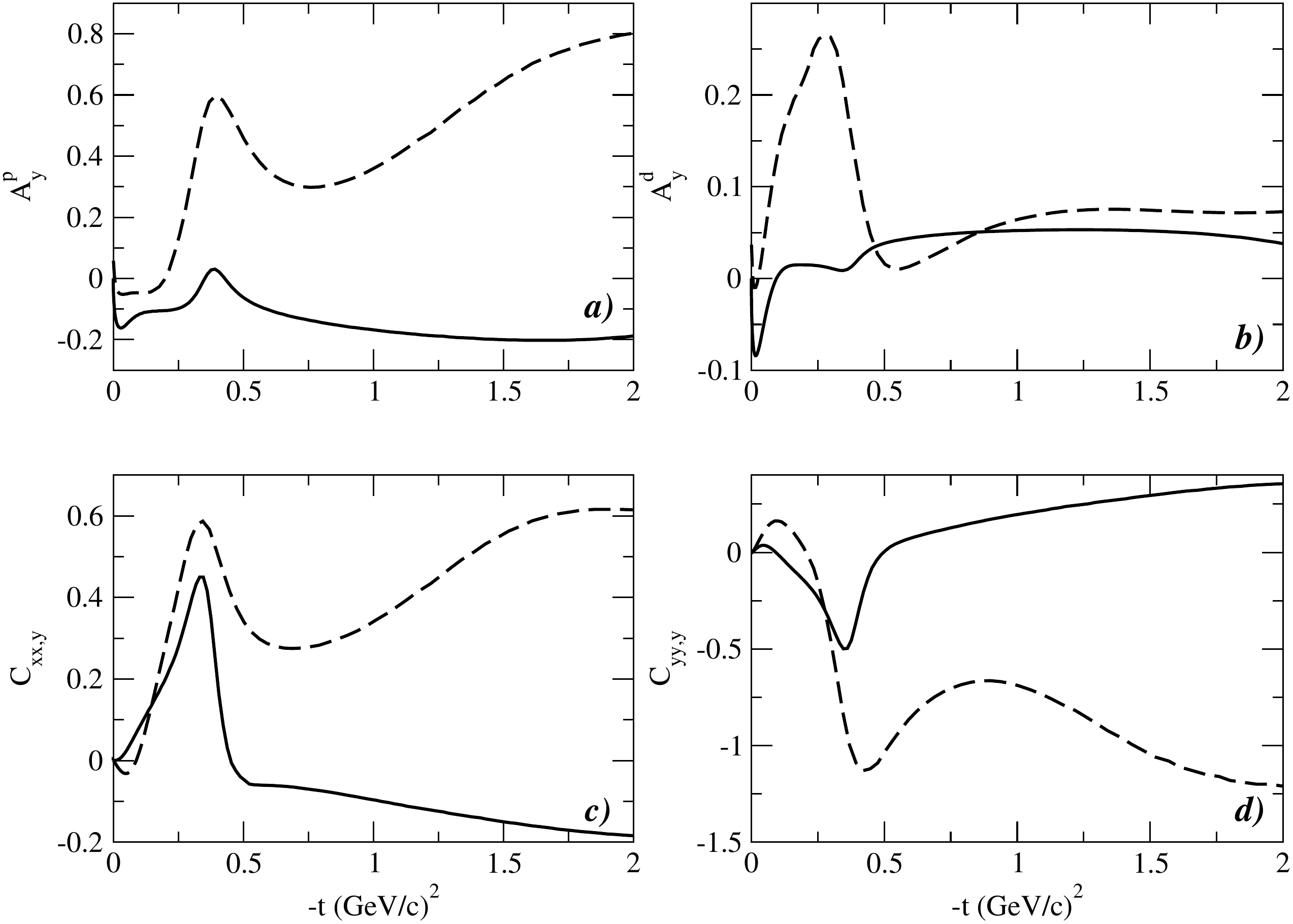}
\vspace{-2mm}
\caption{Results for spin-dependent $pd$ observables. 
Predictions from Ref. \cite{Uz-Jh-TemerB} 
for $p_{lab} = 4.8$ GeV/c are shown by dashed lines
while those at $45$ GeV/c correspond to the solid lines. 
}
\end{center}
\labelf{fig-uz02}
\vspace{-5mm}
\end{figure}
 
The unpolarized differential cross section, vector ($A_y^p$, $A_y^d$)
and tensor ($A_{xx}$, $A_{yy}$) analyzing powers and some 
spin correlation parameters
($C_{x,x}$, $C_{y,y}$, $C_{xx,y}$, $C_{yy,y}$)
\footnote{ We use here  notations of Ref. \cite{Ohlsen:1972zz}}
of $pd$ elastic scattering were calculated 
 at $p_l=4.85$ GeV/c and 45 GeV/c at $0<-t< 2$ GeV$^2$ 
 using $pN$ amplitudes  from  \cite{Sibirtsev:2009cz}. 
 The results obtained  for $A_y^p$, $A_y^d$, $C_{xx,y}$ and $C_{yy,y}$
 are shown in Fig. \ref{fig-uz02}.
 As shown in Ref. \cite{Uz-Jh-TemerB}
 available data on $pd$-elastic differential cross section in forward hemisphere are well described by this model.
 Most sensitive to the spin-dependent $pN$ amplitudes are vector analyzing
 powers $A_y$ and spin correlation parameters $C_{x,x}$ and $C_{y,y}$.
 So, even measurement of the ratio $A_y^d/A_y^p$ at low $t$ gives valuable information on the transverse spin-spin term
 in NN-amplitudes \cite{Uzikov:2019hzx}.
 In contrast, the tenzor analyzing powers $A_{xx}$ and $A_{xx}$ are very
 weakly sensitive to those amplitudes  and weakly changed with increasing energy. 
The calculated in \cite{Uz-Jh-TemerB} polarization observables
can be measured at 
SPD NICA
that will provide a  test of the used $pN$ amplitudes. 
The corresponding differential cross section  is rather  large  in the considered region $p_{lab}=3-50$ GeV/c   and   $|t|=0-2$ GeV$^2$ being   
 $d\sigma/dt >0.1$ mb/GeV$^2$.    Expected counting rate 
 $N$
 at $p_{lab}=50$ GeV/c  (${q_{pp}^{cm}}= 5$ GeV/c)  for the luminosity $L= 5\times 10^{30} cm^{-2}s^{-1}$ and
 for the solid angle $\Delta\Omega=0.03$ is $N \geq 10^2 s^{-1}$.

The $pN$ helicity amplitudes   $\phi_5$ and $\phi_1 +\phi_3$, which can be tested in the above described procedure  
are necessary   in search of time-reversal 
invariance effects in double-polarized $pd$ scattering \cite{Uzikov:2015aua,Uzikov:2016lsc}.
  Data
  of the spin-correlation parameters of $pp$  elastic scattering  being analyzed
  in the framework of the eikonal model
  \cite{Selyugin:2009ic}
   will allow one to obtain    space structure of the
  spin-dependent hadron forces \cite{Selyugin:2010xv}.

\subsection{{ Quasielastic pd-scattering $p+d \to\{pp\}(^1S_0)+n$}}

  Spin structure of the amplitude of 
  the reaction of quasielastic $pd$ scattering  with  formation of the $pp$ pair
  at small excitation energy $\leq 3$ MeV  
  \begin{equation}
     \label{uz1}
     p+d \to\{pp\}(^1S_0)+n
   \end{equation}
   is of special interest.
   In this
   reaction  the  final $pp$ pair  is  in  the $^1S_0$ state of the internal
  motion, therefore the number of  of independent transition  matrix elements
  is diminished to six instead of  twelve for the elastic $pd$- scattering.
  Since the angular momentum of the $pp(^1S_0)$ pair is zero, 
  in collinear kinematics the transition matrix element of this reaction is completely described by two independent
   amplitudes 
   ${\cal A}$ and  ${\cal B}$ as following
      \begin{equation}
\label{collinear}
{\cal F}={\cal A}({ \bf e}\cdot{\bf k})({\bfg \sigma}\cdot{\bf k}) +
{\cal B} {\bf e}\cdot {\bfg \sigma},
\end {equation}
   where ${\bf k}$  is unit vector  directed along the beam, ${\bf e}$  is  the deuteron 
    polarization vector and ${\bfg \sigma}$  is    the Pauli matrix.
    The modules of these amplitudes and  cosine of the relative  phase  $\varphi_{AB}$
      can be determined   by measurement of  unpolarized cross section of the reaction
     $d\sigma_0$ and  tenzor analyzing powers $T_{20}=A_{zz}/\sqrt{2}$ and $A_{yy}$.
     In order to measure  the sine of the relative phase  $\varphi_{AB}$
    one   has to measure only the sign  of  the spin-correlation coefficient
    $C_{xz,y}$.
   
    Within the approximation of the $pn$- single scattering the theoretical analysis of this reaction becomes more simple. In this case the ${\cal A}$
    and ${\cal B}$  amplitudes of the reaction (\ref{uz1}) 
    are expressed via the  spin amplitudes  of the charge
    exchange reaction
     \begin{equation}
     \label{uz2}
     p+n\to n+p.          
     \end{equation}
     The transition  matrix element of  reaction (\ref{uz2}) at zero  scattering angle
      can be written as
      \begin{equation}
\label{pncollwb}
f_{12}^{collin}=\alpha +\beta ({\bfg \sigma}_1\cdot{\bfg \sigma}_2)+
 (\varepsilon-\beta)({\bfg \sigma}_1\cdot {\bf k})
({\bfg \sigma}_2\cdot {\bf k}),
   \end{equation}   
   where ${\bfg \sigma}_1$  (${\bfg \sigma}_2$) the Pauli matrix acting on the spin state
   of the first (second) nucleon.
   
   

We can show  that measurement of $d\sigma_0$  and  $T_{20}$ provides 
  the modules of $|\varepsilon|$  and $|\beta|$ whereas the cosine of the relative phase 
  ( or $Re\varepsilon\beta^*$)   is determined  by the spin correlation parameters
  $C_{x,x}=C_{y,y}$. In order to measure the sine  of  this phase
  ($Im\beta \varepsilon^*$)
  one has to measure  the sign  of $C_{xz,y} (=-C_{yz,x})$.
 Therefore, measurement of $d\sigma_0$, $T_{20}$, $C_{y,y}$ and the sign of $C_{xz,y}$
  at zero scattering angle  completely determines the  spin amplitudes
  $\varepsilon$ and $\beta$.

\subsection{ {Elastic  $dd$ scattering}}

Spin observables of the  $dd$- elastic scattering  in forward hemisphere   also can be used to  test   spin-dependent
amplitudes of $pN$ elastic scattering  since the Glauber model can be used for description of these observables. 
 Unpolarized  differential  cross section of the $dd$- elastic scattering  in forward hemisphere  measured
  at energies $\sqrt{s}=53-63$ GeV \cite{Goggi:1978ae}
  was well described by  the modified Glauber theory  including Gribov inelastic  corrections.  At lower energies corresponding to the SPD NICA region, one may expect that inelastic corrections are not important, that can be checked by direct calculation of unpolarized  cross section and subsequent comparison with
  the data. In this calculations  the above considered  spin dependent amplitudes of the $pd$ elastic scattering 
  \cite{Uz-Jh-TemerB} can be used as input for the  Glauber calculations of the $dd$ scattering.
  
      At large scattering angles  $\theta_{cm}\sim 90^\circ$  
      the  $pd\to pd$ and $dd\to dd $  processes are  sensitive to the short-range (six-quark) structure of the deuteron. Therefore, measurement of any observables
      of these processes at large  $\theta_{cm}$  will be  important to search for  non-nucleonic degrees of freedom of the deuteron.

\subsection{{Double polarized large angle  $pN$ elastic scattering}}
\label{Uz-ANN}

The $pp$ and $pn$ elastic scattering at high energy $\sqrt{s}=5-7$ GeV  and large transferred momentum $-t= 5-10$ GeV$^2$ is powered by
 short-range properties  of NN-interaction corresponding to   small separation between nucleons 
 $r_{NN}\sim \hbar/\sqrt{-t}\leq 0.1$ fm.
There are three following  aspects of QCD dynamics in  these processes.
 ({\it i}) First,   the differential cross section $d\sigma^{pp}/dt({s},\theta_{cm})$ at fixed angle $\theta_{cm}\sim 90^\circ$ on the whole
   follows to the  pQCD  constituent counting rules $d\sigma^{pp}/dt({s},\theta_{cm})\sim s^{-10}$ \cite{Allaby:1968zz,Akerlof:1967zz,Perl:1969pg,Stone:1977jh}. However, a clear  deviation from this  prediction
    in form of oscillations with increasing energy is observed in the region
     $s=10\div 40$ GeV$^2$ \cite{Allaby:1968zz,Akerlof:1967zz,Perl:1969pg,Stone:1977jh}. The irregularity in the energy dependence is on the level of $\sim 50 \%$ in the region, where magnitude of the elastic pp- cross section falls down by 8 orders of magnitude.
 ({\it ii}) Second, anomalous polarization asymmetries were observed in hard pN-scattering at
 $p_{lab}=11.75$ GeV/c 
 \cite{Crabb:1978km,Crosbie:1980tb,Court:1986dh}.
 Elastic $pp$-cross section with spins of protons parallel and normal   
    to the scattering plane  is almost four time larger than the cross section 
    with antiparallel spins.
    The challenge is that in order to generate  such large
    polarization effect, one  needs to have  large contribution  from  double spin-flip  helicity amplitude $\phi_2$ or negligible contribution from helicity conserving
    $\phi_1$ amplitude.
   However, in pQCD, in contrast,  $\phi_2$ is the most suppressed and the $\phi_1$ is largest \cite{Sargsian:2014bwa}. 
   Predicted within the pQCD  (quark-interchange model) double spin asymmetry $A_{NN}$  does  not depend on energy \cite{Brodsky:1979nc},
    \cite{Farrar:1978by}, whereas the
     measured asymmetry demonstrates ''oscillating`` energy dependence. ({\it iii}) The third QCD aspect of hard NN scattering is related to the Color Transparency phenomenon (CT), that is a  reduction of the absorption in the nuclear medium of hard produced hadrons, both mesons and  baryons\cite{Mueller}, \cite{Brodsky}. Being in point like configurations, which are dictated by mechanism of high momentum transfer, the initial and final hadrons  participating in hard process  have  small color dipole  momenta and, therefore, small interaction cross section with nuclear medium. These expectations resulted in   huge  theoretical and experimental activities in  90's. While the CT effect is observed for the hard production of the $q\bar q$ systems, the similar effect for $qqq$  is elusive. The data \cite{Mardor:1998zf,Aclander:2004zm} on the reaction $p+A\to pp+X$ on the $^{12}$C  and $^{27}$Al   show  again an ''oscillatory`` effect, i.e.  the transparency increases
     with increasing momentum up  to $p_{lb}= 9$ GeV/c, and  then decreases  below the  Glauber calculation predictions at 14 GeV/c.
      An attempt to connect  all three above aspects together into one approach was undertaken in Ref. \cite{Sargsian:2014bwa}. However, recent measurement of the cross section
       of the reaction $^{12}$C(e,ep)X at $Q^2=8-14 \, (GeV/c)^2$ 
        \cite{Bhetuwal:2020jes}
         shows   no CT effect and this fact  raises new  questions to the  analysis made  in  \cite{Sargsian:2014bwa}.
         On the other hand, according to \cite{Brodsky:1987xw},   the observed 
         large variations in  spin correlations of $pp$-elastic scattering are consistent  with formation in the s-channel  of ``octoquark''resonances $uuds\bar s uud$  and    $uudc\bar c uud$ near the strangeness and  charm production thresholds, respectively. The  variations with increasing energy are explained as a result of interference of the pQCD background amplitude  with  nonperturbative resonant amplitudes.
         Furthermore, the model \cite{Brodsky:1987xw} provides a description of the oscillations in the unpolarized differential $pp$- elastic cross section.One should mentioned, however, 
         that  another explanation of the oscillation effect in the  $d\sigma^{pp}/dt({s},\theta_{cm})$
         was  suggested  in  Ref. \cite{Ralston:1982pa}.
         
         The considered questions about new types of charm-based resonances
         \cite{Brodsky:2016tew}
         became especially interesting after observation enhancement effects  in the decay $\Lambda_b^0\to J/\Psi p K^-$ interpreted as pentaquark states
         $uudc\bar c$ \cite{Aaij:2015tga} (see also Ref. \cite{Brodsky:2016tew}).         
       More insight into this issue can be get from the data on large angle   $pn$ elastic scattering.
     Different spin-isospin  structure of the   transition matrix elements
     for the  near threshold  $J/\Psi$ production in $pn$ and $pp$ collisions \cite{Rekalo:2002ck} means
    that  spin observables in $pn$- elastic scattering
     can give a valuable   independent information on the considered dynamics.       
      Data on  these observables  are  almost absent in the considered  energy region. 
       A task    to get such  data  in the energy interval $\sqrt{s_{NN}}\approxeq 
       3 \div 5$ GeV
       from the  ${\vec p}{\vec d} \to pnp$ and 
       ${\vec d}{\vec d} \to pnpn$ reactions is accessible for the SPD NCA.
 
 \subsection{{Summary} }

To conclusion, nucleon-nucleon elastic scattering is a basic process in the physics of atomic 
nuclei and the interaction of hadrons with nuclei.
Existing models and 
corresponding parametrizations of $pp$ amplitudes in the region of small 
transferred momenta can be effectively tested by a measurement
of spin observables for $pd$ and $dd$ elastic scattering and a subsequent comparison of 
the results with corresponding Glauber calculations. The spin observables
of $pd$ elastic scattering studied and evaluated in \cite{Uz-Jh-TemerB}
are found to be not too small and, thus, could be measured at the future 
SPD NICA facility. 
As extension of this study, the quasi-elastic processes  with formation of the spin-singlet
final $NN$-pair at small excitation energy $<3$ MeV in the $^1S_0$ state of internal motion,
$pd\to n\{pp\}_s$ and 
$pd\to p\{pn\}_s$  can be also investigated.

\section{\bf Studying periphery of the nucleon in diffractive $pp$ scattering
\protect\footnote{This section is written by V.A. Baskov,\,O.D.\, Dalkarov,
   A.I.\, L'vov (E-mail: {lvov@x4u.lebedev.ru}) and V.V.\, Polyanskiy}}

\begin{abstract} 

   Motivation is outlined for a precise study of high-energy
diffractive scattering of protons at $|t|\lesssim 1$~GeV$^2$ in the
experiment SPD.
   Small oscillations in the $t$-dependence of the
differential cross section at low and medium $t$, observed in
earlier experiments at Protvino, ISR, Fermilab and now also at
LHC, are probably related with the proton's structure at impact parameters
exceeding the size of the proton's quark core and thus indicate
involvement of meson periphery of the nucleon to diffractive
scattering. The experiment SPD can provide new precise data on small-angle
elastic $pp$-scattering for exploring this phenomenon.

\end{abstract}

\vspace*{6pt}

\vspace{1em}
   1. Scattering of high-energy hadrons at low $t$ is usually
described by a simple phenomenological dependence
$d\sigma/dt = Ae^{Bt}$
(not applicable in the Coulomb region,
$|t|\lesssim 0.01$~GeV$^2$, and at $|t|\gtrsim 0.4$~GeV$^2$).
   In the impact parameter representation, such a dependence
corresponds to a gaussian profile function $\Gamma(b) \sim
\exp(-b^2/2B)$ with the average transverse size $\langle
b^2\rangle^{1/2} = B^{1/2} \sim 0.6~$fm when $B\sim
10$~GeV$^{-2}$.
   This size corresponds well to the quark core size of the
nucleon, $r_q\sim 0.4-0.5$~fm, where the bulk of the nucleon mass
(and the energy and momentum) is concentrated.

   On the other hand, part of the nucleon components is clearly
located at larger distances, pion cloud being the most evident
example. The first evidence of the pion cloud effect in the
diffractive scattering, including rapid variation of the
effective slope $B$ at $|t|\sim 0.1$~GeV$^2 \approx 4m_\pi^2$,
have been found in ISR measurements (a comprehensive review
of the ISR data can be found in \cite{Amaldi:1979kd}).

   First explanations of this, presumably pion cloud effect,
were provided by Anselm and Gribov \cite{Anselm:1972ir} (see also 
\cite{Khoze:2014nia, Jenkovszky:2017efs}).
   Soon then a dedicated experiment was conducted in Protvino
\cite{Antipov:1976eja} in order to test the ISR results. However,
beyond confirming findings of ISR, 
one more oscillation in the differential cross section
at $|t|\sim 0.5$~GeV$^2$ was found. Being located at higher $t$,
it might be related with somewhat heavier mesons around the proton
(but not as heavy as vector mesons that are too heavy).

Actually, S.P. Denisov {\it et al.} recently suggested to continue exploring
$pp$ elastic scattering in that kinematical region at Protvino
\cite{Denisov:2016rir}, and the current proposal of doing similar
experiment at SPD was directly motivated by the Denisov's ideas.

\begin{figure}[!ht]
\centering
\includegraphics[width=0.5\textwidth]{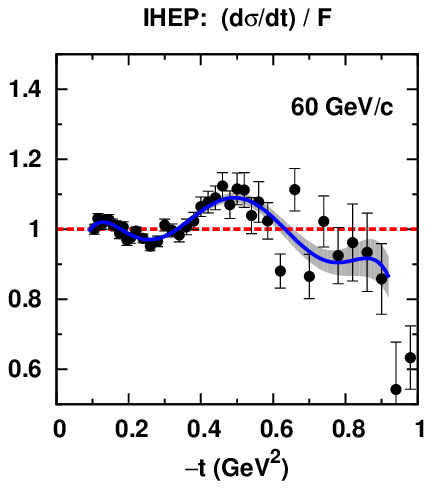}\hspace*{-1em}
\includegraphics[width=0.5\textwidth]{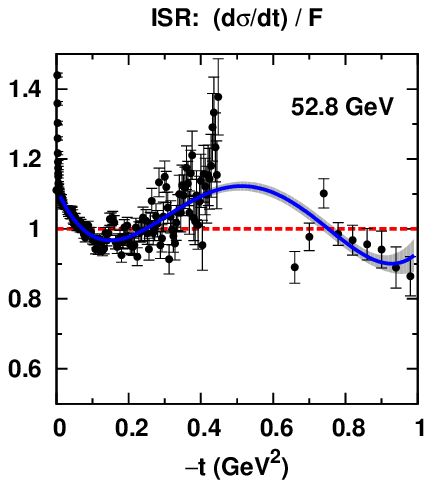}\hspace*{-1em}
\par
\includegraphics[width=0.5\textwidth]{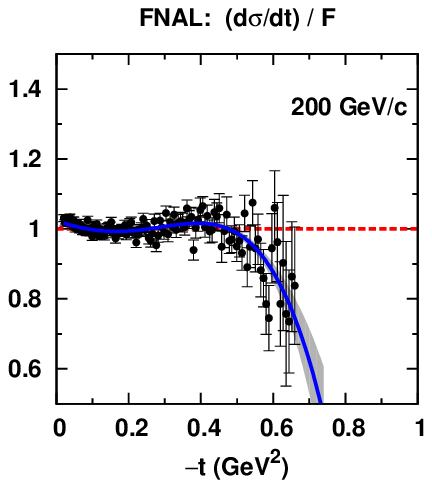}\hspace*{-1em}
\vspace{-3mm}
\caption{Deviations of the $pp$ differential cross section from
smooth dependences $F(t)=Ae^{Bt+Ct^2}$. Data are from Protvino, ISR and FNAL, see in the text.
Solid line is a polynomial smoothing of shown ratios.
Strips show statistical errors in the polynomial.}
\label{lvov-fig}
\vspace{-3mm}
\end{figure}

\vspace{1em}
2. It is essential that the Protvino experiment is not the only work
indicating an oscillation at $|t|\sim 0.5~\rm GeV^2$
in the fine structure of the $pp$ diffraction cone.
   In Fig.~\ref{lvov-fig} the most precise
data of three experiments --- from Protvino \cite{Antipov:1976eja}
(at the proton beam momentum $p=60$~GeV$/c$),
ISR \cite{Amaldi:1979kd} (at the total energy $\sqrt s=52.8$~GeV),
and Fermilab \cite{Schiz:1979rh} (at $p=200$~GeV$/c$),
see also a comprehensive compilation and parameterization
of world data in \cite{Belousov:2020rzj}  --- are compared to
exponential form $F(t) = Ae^{Bt+Ct^2}$ beyond
the Coulomb region of tiny $|t|$ and the region of
$|t|\lesssim 0.1$~GeV$^2$ where effects of the pion cloud contribute.
   One has to notice that ISR and FNAL data do not fully 
cover the region of $t$ with suspected oscillations and do not have sufficient accuracy there.
Therefore, further experimental studies in that region are well justified.

\vspace{1em}

\vspace{1em}
   In principle. information on the smoothed ratios $R(t)=(d\sigma/dt)/F(t)$
could be used in order to estimate the $pp$ scattering amplitude
$f(s,t)$ and to find then, through a Fourier-Bessel transformation,
a profile function $\Gamma(b)$ of the impact parameter $b$ \cite{Dremin:2012ke}.
A peak in $f(s,t)$ at $|t|\sim 0.5~\rm GeV^2$
corresponds to a peak in the profile function $\Gamma(b)$ at large distances
$b\sim 7.0/\sqrt{|t|} \sim 2$~fm (here 7.0 is the second maximum of the Bessel function $J_0(x)$),
However, a straightforward calculation of $\Gamma(b)$ in this way
does not give reliable results in the region where $\Gamma(b)$ becomes very small
and sensitive to assumed phase of the amplitude used, its spin structure,
behavior at higher $|t|$, etc.
Actually, more sophisticated and indirect approaches are to be used in order to analyze
data on the described oscillation --- see, for example,
\cite{Selyugin:2015pha, Prochazka:2016wno, Broniowski:2018xbg}.

\vspace{1em}
3.   In order to cover the region of interest, $|t| \sim 0.1-0.8$~GeV$^2$,
the experimental setup must detect protons (in coincidence)
scattered at angles $\theta \sim 3-10^\circ$
what needs detectors placed at distances $R\sim 4-15$~cm from the beam.
   Accuracy of determination of the momentum transfer squared $t$
in individual events of elastic $pp$-scattering must be better than
$\Delta t \sim 0.01{-}0.02$~GeV$^2$, and this can be achieved with planned
tracker endcap detectors and with angular spread of colliding protons
determined by beam emittance and beta-function at IP.

   Additional measurements of $d\sigma/dt$ and/or polarization observables
at higher $t$ are also desirable; they do not require high accuracy in
determination of $t$ \cite{Sharov:2017dkm}.

   Vertex detector, tracker system and software for track reconstruction in SPD 
are sufficient for identification and recording $pp$ elastic events at energies
$\sqrt s \lesssim 15$~GeV. For higher energies and smaller angles, when scattered protons
fly very close to the beam pipe, installing fast detectors
close to the pipe, with the time resolution
$\Delta T\lesssim 50$~ps,
for determination of hitting times of forward-flying protons (perhaps, using the so-called
PID system) would make it possible to study the discussed anomaly at the highest SPD energies.



\section {\bf Hadron structure and spin effects in elastic hadron scattering at NICA energies
\protect\footnote{This section is written by O.V. Selyugin; E-mail: selugin@theor.jinr.ru}}
\label{Selyugin}

\vskip 5mm
\begin{abstract}
 The spin effects in the elastic proton-proton scattering are analysed at NICA energies.
  It is shown the importance the investigation of the region of the diffraction minimum
  in the differential cross sections. Some possible estimation of spin effects are given for
  the different NICA energies in the framework of the new high energy generelazed structure
  (HEGS) model.
  \end{abstract}
  \vskip 4mm
  PACS:
      {13.40.Gp}, 
      {14.20.Dh}, 
      {12.38.Lg}
  \vskip 6mm

  One of the most important tasks of modern physics is the research into the basic properties of
  hadron interaction.
	The dynamics of strong interactions  finds its most
  complete representation in elastic scattering. 
  It is just this process that allows the verification of the
results obtained from the main principles of quantum field theory:
the concept of the scattering amplitude as a unified analytic function
of its kinematic variables connecting different reaction channels
were introduced in the dispersion theory by N.N. Bogoliubov\cite{Bogolyubov:1983gp}.
 Now many questions of  hadron interactions are connected with
 modern problems of  astrophysics such as unitarity and the optical theorem \cite{Goodhew:2020hob},
 and problems of baryon-antibaryon symmetry and CP-invariance violation \cite{Uzikov:2015aua}
 The main domain of  elastic scattering is  small angles.
  Only in this region of interactions can we measure the basic properties that
  define the hadron structure. 
  Their values
  are connected, on the one hand, with the large-scale structure of hadrons and,
  on the other hand, with the first principles that lead to the
  theorems on the behavior of scattering amplitudes at asymptotic
  energies \cite{Martin:1969ina,Roy:1972xa}.

  Modern studies
   of elastic scattering of high energy protons lead to several unexpected results reviewed, e.g., in \cite{Dremin:2016zwt,Dremin:2017qfo}.
    Spin amplitudes of the elastic $NN$ scattering constitute a spin picture 
     of the nucleon.
   Without knowledge of the spin $NN$-amplitudes  it is not possible to understand spin observable
   of  nucleon scattering off nuclei.
       In the modern picture, the structure of hadrons is determined by Generalized Distribution functions (GPDs),
       which include the corresponding parton distributions (PDFs). The sum rule \cite{Ji:1996nm}
       allow to obtain the elastic form factor (electromagnetic and gravitomagnetic)
       through the first and second integration moments of GPDs. It leads to remarkable properties of GPDs -
       some corresponding to inelastic  and elastic scattering of hadrons.
 Now some different
models examining the nonperturbative instanton contribution lead
to sufficiently large spin effects at superhigh energies \cite{Anselmino:1992wi,Dorokhov:1993nw} 
The research of such spin effects
will be a crucial stone for different models and will help us
to understand the interaction and structure of particles, especially at large
distances.
There are large programs of researching spin effects
at different  accelerators. Especially, we should like to note
the programs at  NICA, 
where the polarization of both the collider beams will be constructed.
So it is very important to obtain reliable predictions for the spin
asymmetries at these energies. In this paper, we extend the model
predictions to spin asymmetries in the NICA energy domain.

   The NICA SPD detector  bounded a very small momentum transfer.
   If in the first steps the angles start from $16$ mrad, then the minimum
   momentum transfer, which can be measured is more than $-0.01$ GeV$^2$. Hence it is needed to exclude the
   Coulomb-nuclear interference region, where the real part of the spin-non-flip amplitude can be determined,
  We should  move our researches on
   the region of the diffraction minimum, where the imaginary part of the spin-non-flip amplitude
   changes its sign. Note that in some models the absence of the second diffraction minimum
   is explained by the contribution in the differential cross section of the spin-flip amplitude \cite{Edneral:1979zk}
   The interference of the hadronic
 and electromagnetic amplitudes may give an important contribution not
 only at very small transfer momenta but also in the range of the
 diffraction minimum \cite{Selyugin:1999dn}.  However, for that one should know the
 phase of the interference of the Columbic and hadronic amplitude at
 sufficiently large transfer momenta too.

   Using the existing model of nucleon elastic scattering
   at high energies $\sqrt{s}>9$ GeV - 14 TeV \cite{Selyugin:2012np,Selyugin:2015pha}, which involves
   minimum of free parameters, we are going to develop its extended version aimed  to describe all available
   data on cross sections and spin-correlation parameters  at  lower energies down  to the  SPD NICA region.
   The model will be based on the usage  of known information on generalized
    parton distributions in the nucleon, electro-magnetic and gravitomagnetic  form factors of the nucleon
    taking into account  analyticity and unitarity requirements and providing compatibility with the
    high energy limit,  where the pomeron exchange dominates.

\subsection{HEGS model and spin effects in the dip region of momentum transfer}

  The differential cross sections of nucleon-nucleon elastic scattering  can be written as a sum of different
  helicity  amplitudes:
 \begin{eqnarray}
   \frac{d \sigma}{dt} = \frac{2 \pi}{s^2} ( |\Phi_{1}|^2 + |\Phi_{2}|^2+
   |\Phi_{3}|^2 + |\Phi_{4}|^2+4 |\Phi_{5}|^2.
\end{eqnarray}

\begin{eqnarray}
 A_{N} \frac{d \sigma}{dt} = -\frac{4 \pi}{s^2}
   [ Im (\Phi_{1}(s,t) + \Phi_{2}(s,t)+ \Phi_{3}(s,t) - \Phi_{4})(s,t) \Phi^{*}_{5}(s,t)]
\end{eqnarray}
and
\begin{eqnarray}
  A_{NN} \frac{d \sigma}{dt} = \frac{4 \pi}{s^2}
   [ Re (\Phi_{1}(s,t) \Phi^{*}_{2}(s,t) - \Phi_{3}(s,t) \Phi^{*}_{4})(s,t) + |\Phi_{5}(s,t)|^2]
\end{eqnarray}

  The HEGS model \cite{Selyugin:2012np,Selyugin:2015pha} takes into account all five spiral electromagnetic amplitudes.
   The electromagnetic amplitude can be calculated in the framework of QED.
    In the high energy approximation, it can be  obtained \cite{Buttimore:1978ry}
  for the spin-non-flip amplitudes:
  \begin{eqnarray}
  F^{em}_{1}(t) = \alpha f_{1}^{2}(t) \frac{s-2 m^2}{t}; \ \ \  F^{em}_3(t) = F^{em}_1;
  \end{eqnarray}
   and for the spin-flip amplitudes:
   with the  electromagnetic and hadronic interactions included, every amplitude $\phi_{i}(s,t)$
  can be described as
\begin{eqnarray}
  \phi_{i}(s,t) =
  F^{em}_{i} \exp{(i \alpha \varphi (s,t))} + F^{h}_{i}(s,t) ,
\end{eqnarray}
where
 $  \varphi(s,t) =  \varphi_{C}(t) - \varphi_{Ch}(s,t)$, and
   $ \varphi_{C}(t) $ will be calculated in the second Born approximation
 in order  to allow the evaluation   of   the Coulomb-hadron interference term $\varphi_{Ch}(s,t)$.
   The  quantity $\varphi(s,t)$
 has been calculated at large momentum transfer including
  the region of the diffaction minimum 
  \cite{Selyugin:1999dn,Selyugin:1996vs,Selyugin:1999ve} and references therein.

 Let us  define the hadronic spin-non-flip amplitudes as
\begin{eqnarray}
  F^{h}_{\rm nf}(s,t)
   &=& \left[\Phi_{1}(s,t) + \Phi_{3}(s,t)\right]/2; \label{non-flip}
 \end{eqnarray}
  The model is based on the idea that at high energies a hadron interaction in the non-perturbative regime
      is determined by the reggenized-gluon exchange. The cross-even part of this amplitude can have two non-perturbative parts, possible standard pomeron - $(P_{2np})$ and cross-even part of the 3-non-perturbative gluons ($P_{3np}$).
      The interaction of these two objects is proportional to two different form factors of the hadron.
      This is the main assumption of the model.
      The second important assumption is that we choose the slope of the second term
       four times smaller than the slope of the first term, by  analogy with the two pomeron cuts.
      Both terms have the same intercept.

      The form factors are determined by the Generalized parton distributions of the hadron (GPDs).
      The first form factor corresponding to the first momentum of GPDs is the standard electromagnetic
      form factor - $G(t)$. The second form factor is determined by the second momentum of GPDs -$A(t)$.
      The parameters and $t$-dependence of  GPDs are determined by the standard parton distribution
      functions, so by  experimental data on  deep inelastic scattering and by  experimental data
      for the electromagnetic form factors (see \cite{Selyugin:2009ic}). The calculations of the form factors were
      carried out in \cite{Selyugin:2014sca}.
The final elastic  hadron scattering amplitude is obtained after unitarization of the  Born term.
     At large $t$  our model calculations are extended up to $-t=15 $ GeV$^2$.
  We added a small contribution of the energy  independent  part
  of the spin flip amplitude in the form similar to the proposed    in \cite{Galynskii:2013esa}
  and analyzed in \cite{Selyugin:2015vao}.
  \begin{eqnarray}
  F_{sf}(s,t) \ =  h_{sf} q^3 F_{1}^{2}(t) e^{-B_{sf} q^{2}}.
  \end{eqnarray}
    The energy dependent part of the spin-flip amplitude is related to the main amplitude
    but with an additional kinematic factor and the main slope taken twice more,
    conformity with the paper \cite{Predazzi:2001sh,Cudell:2004ev}.     The form factors incoming in the
    spin-flip amplitude are determined by the GPD functions $H(s,t,x)$ and    $E(s,t,x)$,
    which include the corresponding PDF distributions.
  The model is very simple from the viewpoint of the number of fitting parameters and functions.
  There are no any artificial functions or any cuts which bound the separate
  parts of the amplitude by some region of momentum transfer.

 Now we shell restrict our discussion to the analysis of  $A_N$ as there are some experimental data
 in the region of NICA energies.
   In the standard pictures the spin-flip and double spin-flip amplitudes
    correspond to the spin-orbit $(LS)$ and spin-spin $(SS)$ coupling terms.
  The contribution to
  $A_N$ from the hadron double spin-flip amplitudes
   already at $p_L = 6 \ $GeV/c is of the second order
  compared to the contribution from the spin-flip amplitude.
   So with the usual high energy approximation for the helicity amplitudes
   at  small transfer momenta we suppose that
   $\Phi_{1}=\Phi_{3}$ and we can neglect the contributions of the hadron parts
   of $\Phi_2-\Phi_4$.
   Note that if $\Phi_{1}, \Phi_3, \Phi_5$ have the same phases, their interference contribution
   to $A_N$ will be zero, though the size of the hadron spin-flip amplitude can be large.
   Hence, if this phase has  different $s$ and $t$ dependencies, the contribution from the hadron
   spin-flip amplitude to $A_N$ can be zero at $s_i, \ t_i$ and non-zero at other $s_j, \ t_j$.
e experimental data ($\sum \chi^2/n_{dof} =1.24$).

\begin{figure}
\begin{center}
  \includegraphics[width=0.4\textwidth]{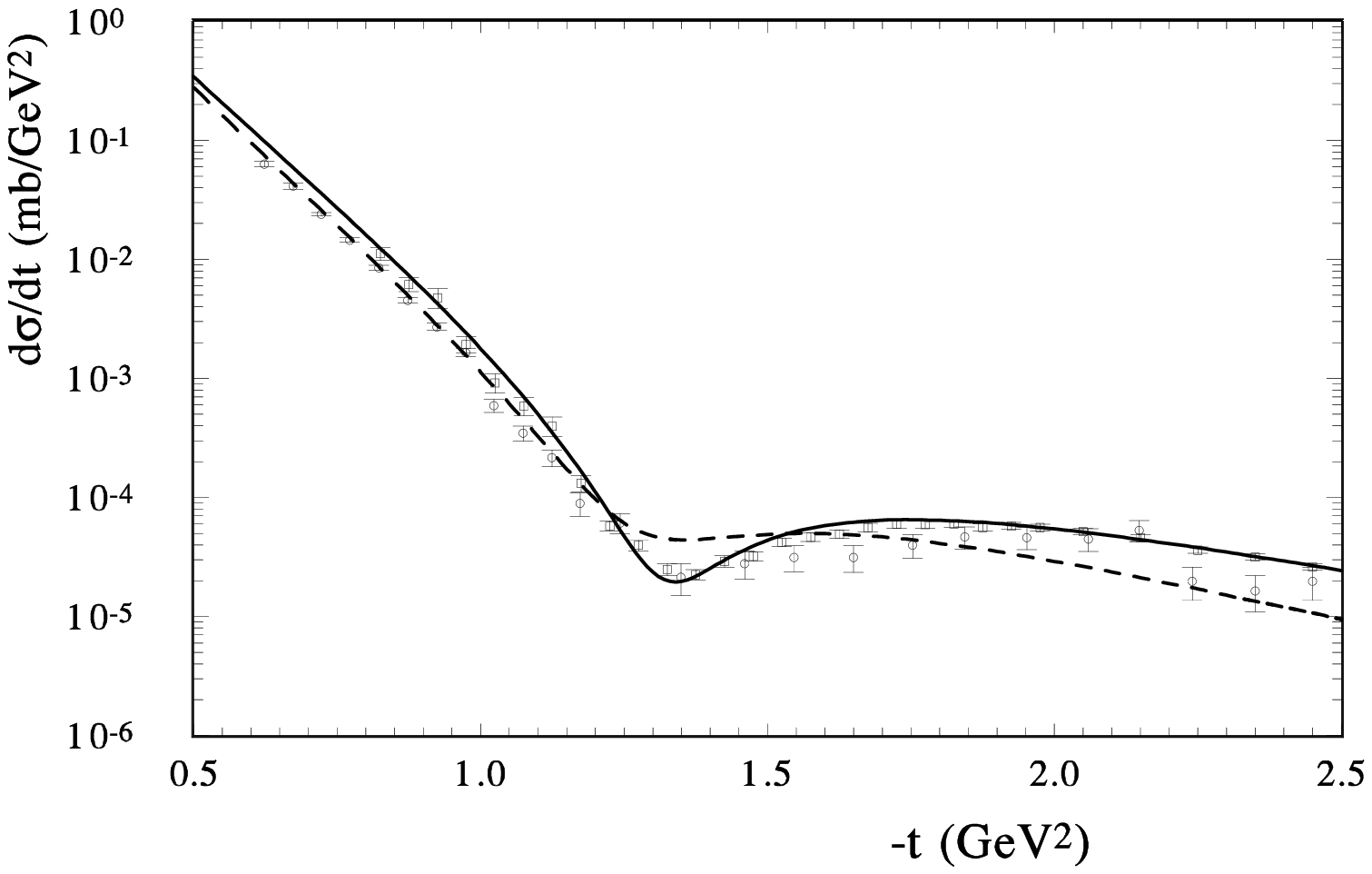}
 \includegraphics[width=0.4\textwidth]{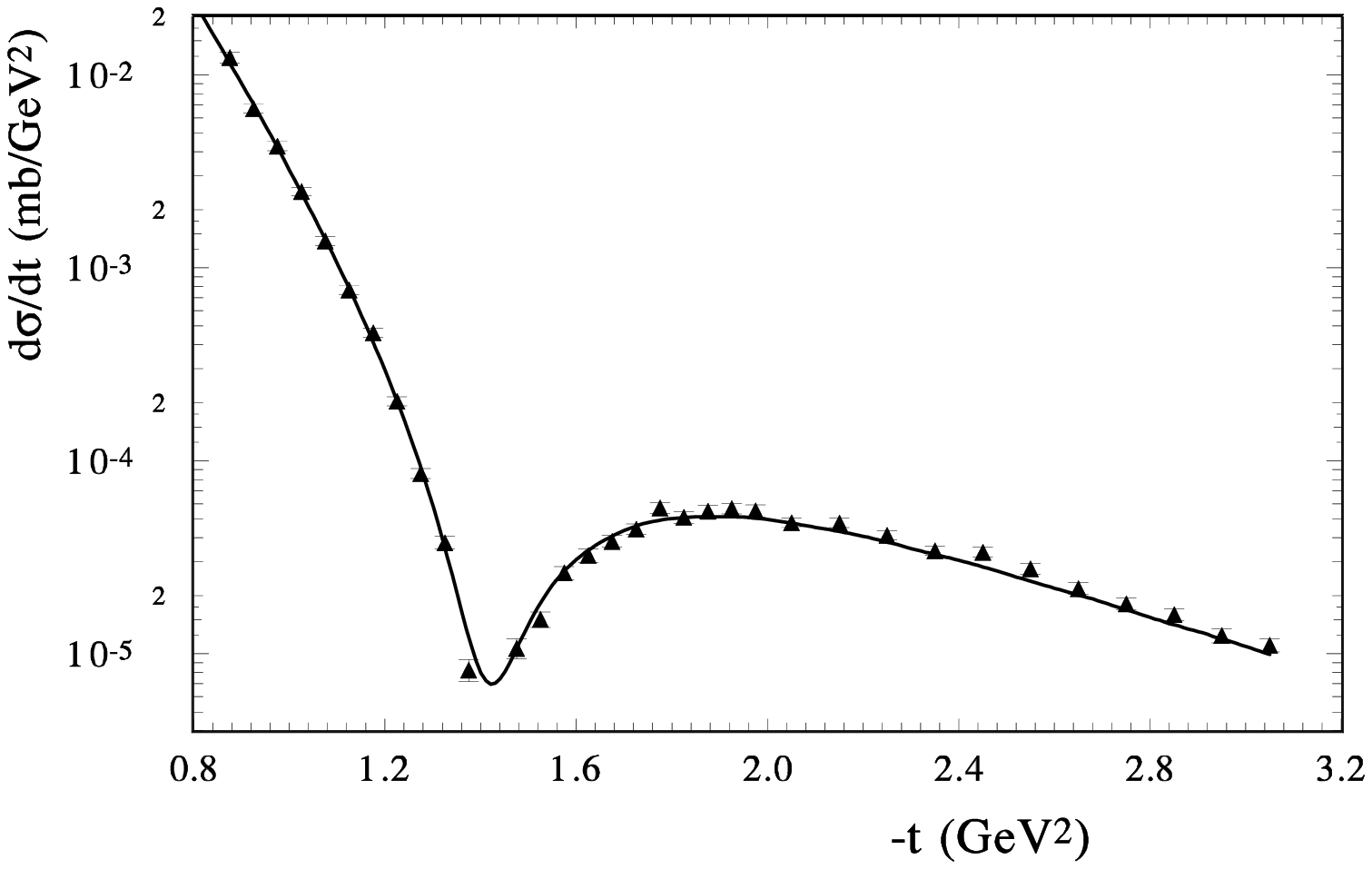}
\end{center}
\caption{ The model calculation of the diffraction minimum in $d\sigma/dt$ of  $pp $ scattering
   [left] at  $  \sqrt{s}=30.4  $~GeV;
 [right] for $pp$ and $p\bar{p}$  at $\sqrt{s}=52.8$ GeV \cite{Whalley:1991mk} 
  scattering.
 }
\end{figure}
%

\begin{figure}
\begin{center}
\includegraphics[width=0.6\textwidth] {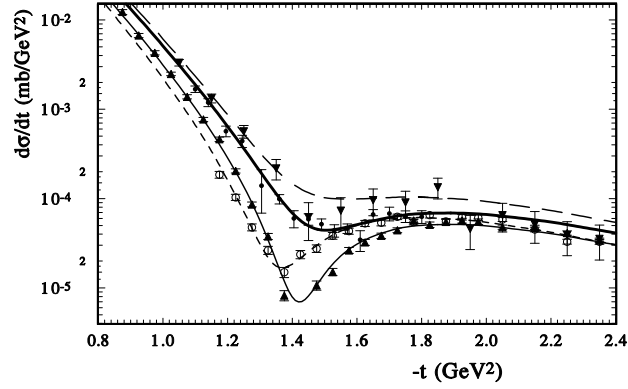}
\end{center}
\vspace{1.cm}
\caption{ The model calculation of the diffraction minimum in $d\sigma/dt$ of  $pp $
   at  $  \sqrt{s}=13.4; 16.8;  30.4; 44.7;  $~GeV;
 (lines, respectively,  long dash;  solid; thin-solid, and short - dash ); 
and experimental data \cite{Whalley:1991mk}, respectively,  - the triangle down,  the circles (solid),
  triangle up, and  circles    ) 
 }
\end{figure}

 Now let us examine the form of the differential cross section in the region of the momentum transfer
     where the diffractive properties of
     elastic scattering appear most strongly - it is the region of the diffraction dip.
 The form and the energy dependence of the diffraction minimum  are very sensitive
   to different parts of the scattering amplitude. The change of the sign of the imaginary part
   of the scattering amplitude determines the position of the minimum and its movement
    with changing  energy.
   The contributions of the real part of the spin-non-flip scattering amplitude and
   the square of the spin-flip amplitude  determine the size and the energy dependence of the dip.
   Hence, this depends    heavily on the odderon contribution.
   The spin-flip amplitude gives the contribution to the differential cross  sections additively.
   So the measurement of the form and energy dependence of the diffraction minimum
   with high precision is an important task for future experiments.

The HEGS model reproduces   $d\sigma/dt$ at very small and large $t$ and provides a qualitative description of the dip region
at $-t \approx 1.6 $~GeV$^2 $, for $\sqrt{s}=10 $~GeV and  $-t \approx 0.45 $~GeV$^2 $   for $\sqrt{s}=13 $~TeV. 
 Note that it gives a good description for the proton-proton and proton-antiproton elastic scattering
  or $\sqrt{s}=53 $~GeV and for $\sqrt{s}=62.1 $~GeV (Fig.2a).

  The dependence of the position of the diffraction minimum on $t$ is determined in most part by the growth of the total cross sections  and the slope of the imaginary part of the scattering amplitude.
  Figures 2b and 3  show this a dependence obtained in the HEGS model at different energies.

\begin{figure}
\begin{center}
\includegraphics[width=0.4\textwidth] {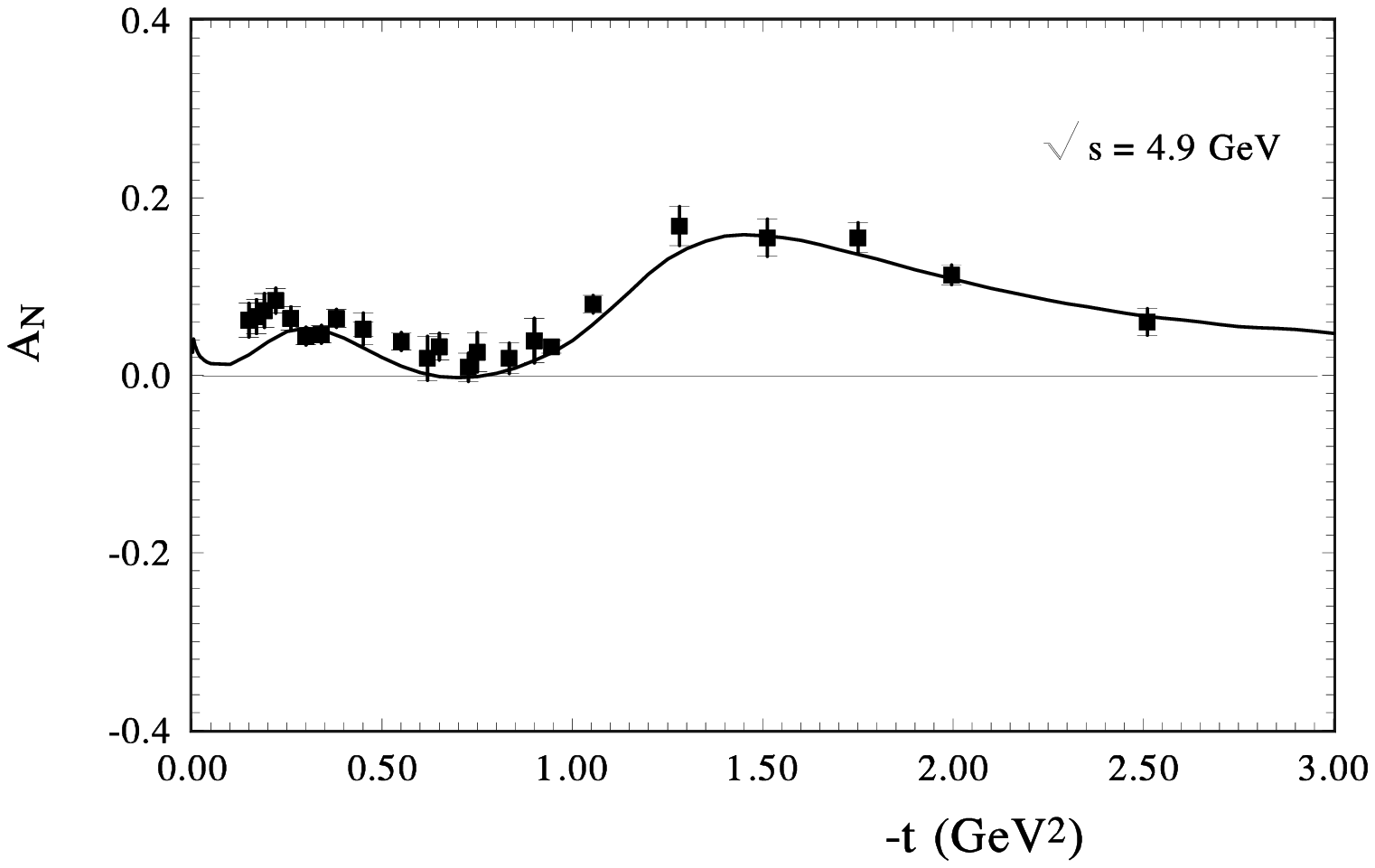}
\includegraphics[width=0.4\textwidth] {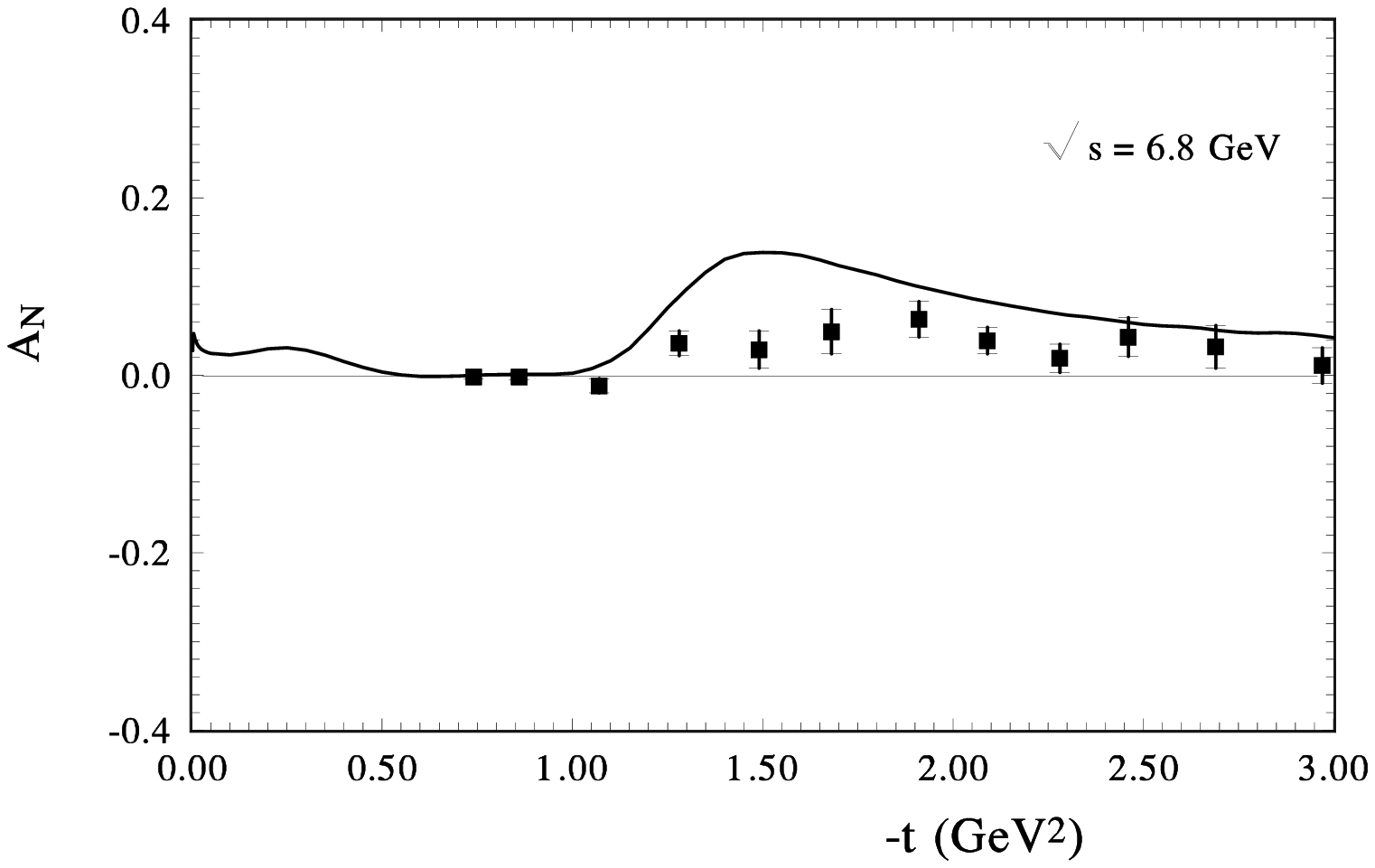}
\end{center}
\caption{The analyzing power $A_N$ of pp - scattering
      calculated:
  a) at $\sqrt{s} = 4.9 \ $GeV  (the experimental data \cite{Kramer:1977pf}),
   and
  b) at $\sqrt{s} = 6.8 \ $GeV    (points - the existence experimental data \cite{Antille:1981tf} ).
   }
\label{fig:1}       
\end{figure}

\begin{figure}
\begin{center}
\includegraphics[width=0.4\textwidth] {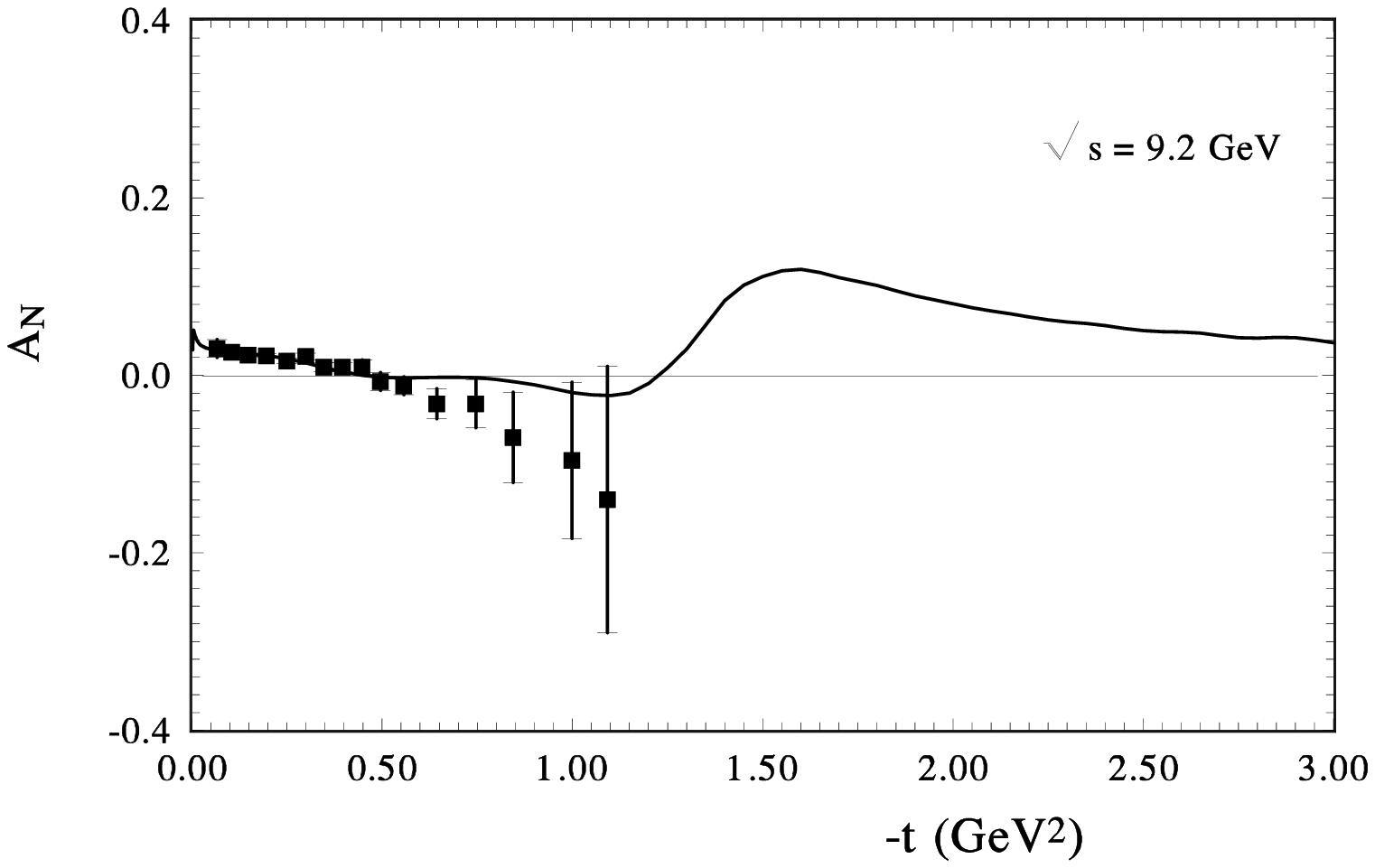}
\includegraphics[width=0.4\textwidth] {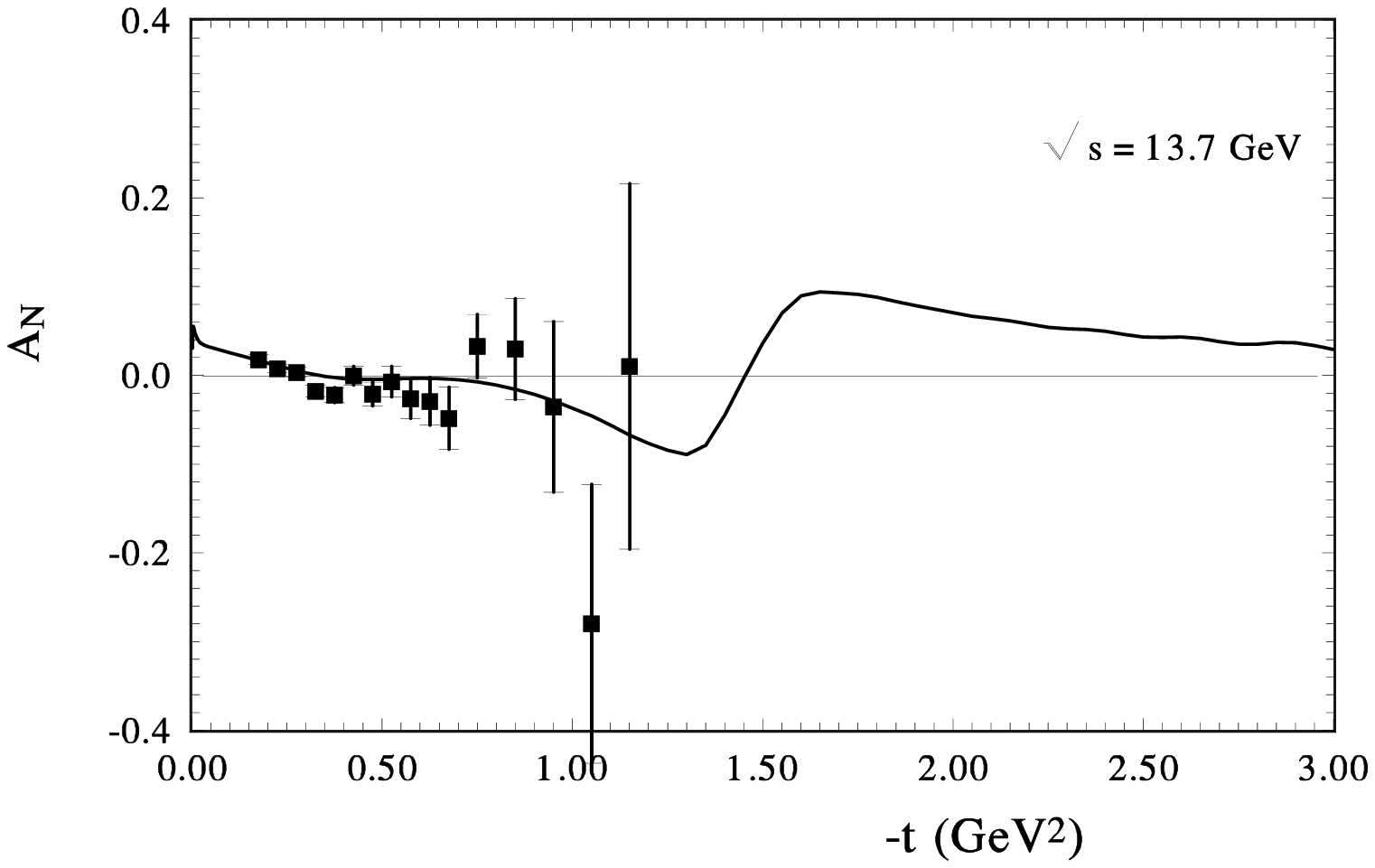}
\end{center}
\caption{The analyzing power $A_N$ of pp - scattering
      calculated:
   a) at $\sqrt{s} = 9.2 \ $GeV, and  (the experimental data \cite{Gaidot:1976kj}), and
   b)    at $\sqrt{s}= 13.7 \ $GeV (points - 
       the experimental data \cite{Kline:1980cs}).
          }
\label{fig:1a}       
\end{figure}

     In Fig.3, the description of the diffraction minimum in our model is shown for NICA energies.
     The HEGS model reproduces  sufficiently well  the energy dependence and the form of the diffraction dip.
     In this energy region the diffraction minimum reaches the sharpest dip at  $\sqrt{s}=30 $~GeV  near
     the final NICA energy.
     Note that at this energy the value of $\rho(s,t=0)$ also changes its sign in the proton-proton
     scattering.

The calculated analyzing power  at $p_L = 6 \ GeV/c$ is shown in
  Fig.4a.
  One can see that a good description 
   of experimental data on the analyzing power
  can be reached only with
   one  hadron-spin flip amplitude.

    The experimental data at $p_L = 11.75 \ $GeV/c  seriously
 differ from those at $p_L = 6 \ $GeV/c but our calculations
 reproduce $A_{N}$ sufficiently well (Fig.4b ).
 It is shown that our energy dependence of the spin-flip amplitudes
  was chosen correctly and we may hope that  further we will obtain
  correct values of the analyzing power and other spin correlation parameters.

  From Fig.4 we can see that in the region
   $|t| \approx  0.2  \div 1 \ $ GeV$^2$ the contributions from the hadron spin-flip amplitudes
  are most important.
  At last, Fig.5a shows our calculations at $p_L = 200 \
  GeV/c$.

  At this energy, the contributions of the phenomenological energy independent part
  of the spin-flip amplitude is  compared with the energy dependent part.
  The spin effect is sufficiently large and has a specifical form,
   which is determined by the form of the differential
  cross section in the diffraction dip domain.

\begin{figure}
\begin{center}
\includegraphics[width=0.4\textwidth] {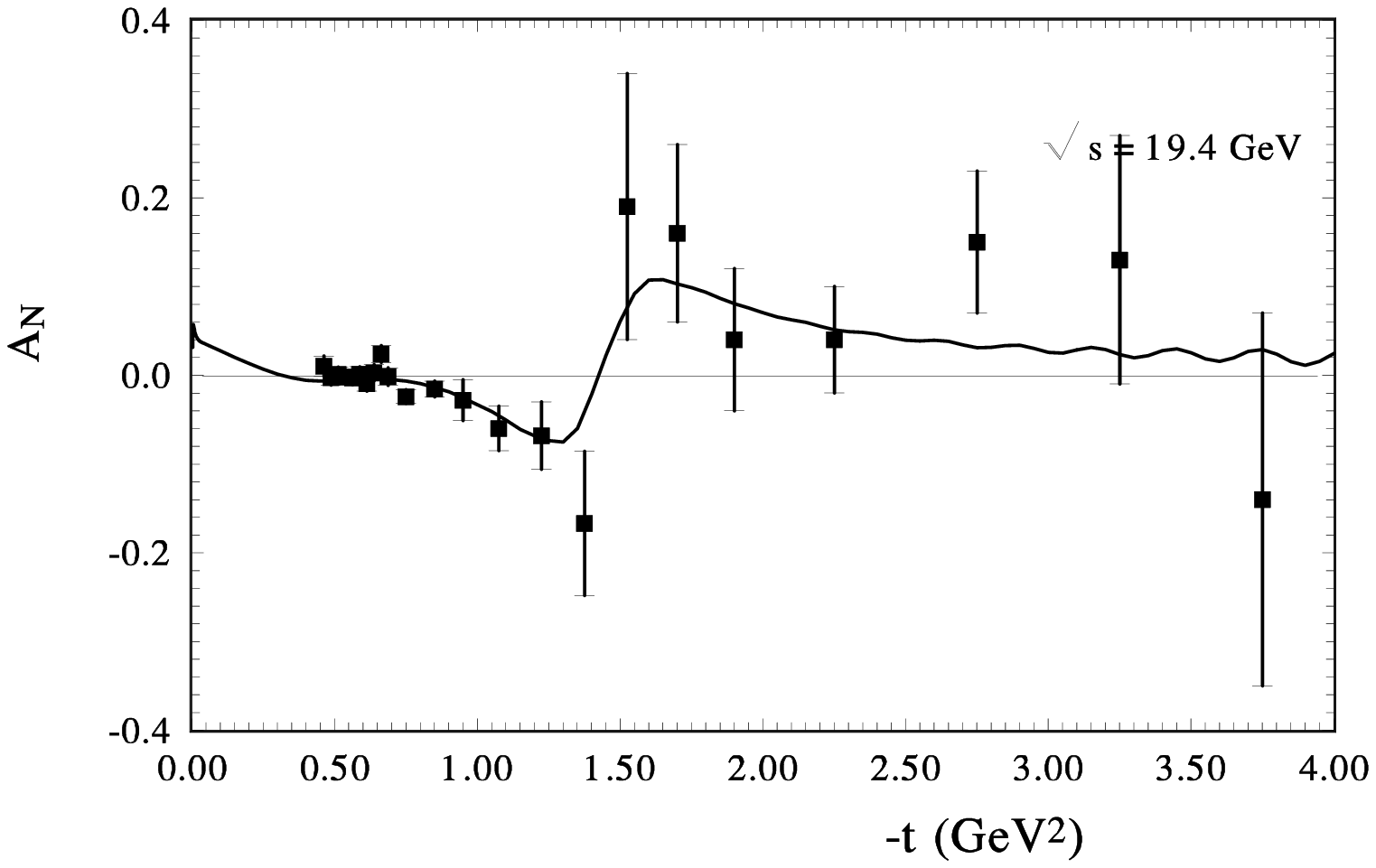}
\includegraphics[width=0.4\textwidth] {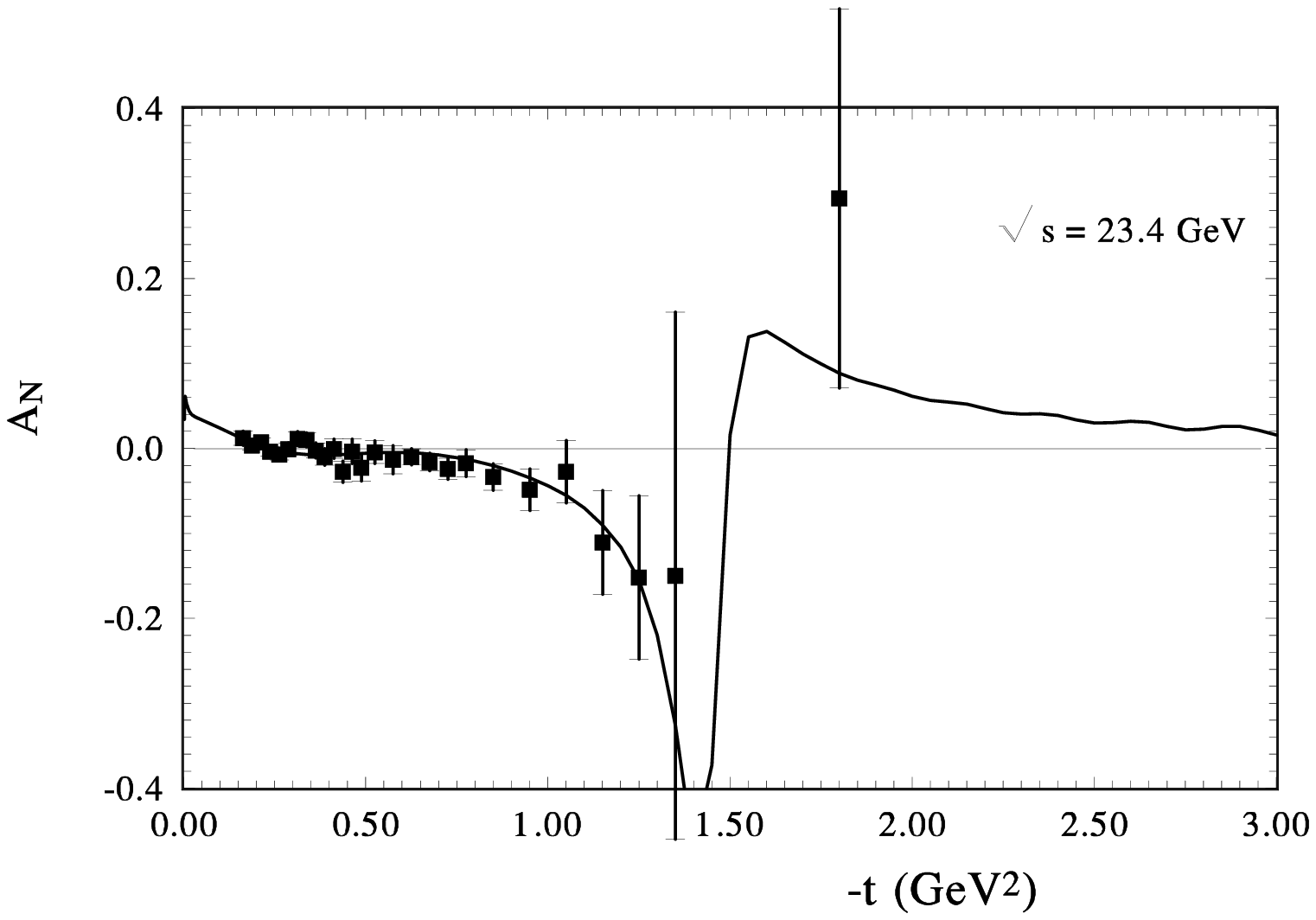}
\end{center}
\caption{The analyzing power $A_N$ of pp - scattering
      calculated:
   a) at $\sqrt{s} = 19.4 \ $GeV  (the experimental data \protect\cite{Fidecaro:1981dk}),
   and
   b)    at $\sqrt{s}= 23.4 \ $GeV
       (points - the  experimental data \protect\cite{Kline:1980cs})
          }
\label{fig:1b}      
\end{figure}

\subsection{Conclusions}

     The Generelized parton distributions (GPDs) make it possible to better understand the thin hadron
     structure and  to obtain the hadron structure in the space frame (impact parameter representations).
     It is tightly connected with the hadron elastic hadron form factors.
    The research into the  form and energy dependence of the diffraction minimum of the differential cross sections
    of elastic hadron-hadron scattering at different energies
    will give  valuable information about the structure of the hadron scattering amplitude
     and  hence  the hadron structure and the  dynamics of strong interactions.
    The diffraction minimum corresponds to the change of the sign of the imaginary part of the spin-non-flip hadronic scattering amplitude
     and is created under a strong impact of the unitarization procedure.
    Its dip depends on the contributions of the real part of the spin-non-flip amplitude and the
    whole contribution of the spin-flip  scattering amplitude.
    In the framework of HEGS model, we show a deep connection between elastic and inealastic cross sections,
    which are tightly connected with the hadron structure at small and large distances.

    The HEGS model reproduces well the form and the energy dependence of the diffraction dip  of the proton-proton and proton antiproton
    elastic scattering \cite{Selyugin:2016lls}. The predictions of the model in most part reproduce the form of the differential cross section at $\sqrt{s}=13 $ TeV.
    It means that the energy dependence of the scattering amplitude  determined in the HEGS model
     and unitarization procedure in the form of the standard eikonal
     representation satisfies the experimental data in the huge energy region (from $\sqrt{s}=9 $ GeV up to $\sqrt{s}=13 $ TeV.
     It should be noted that the real part of the scattering amplitude,
      on which the form and energy dependence of the diffraction dip heavily depend, is
     determined in the framework of the HEGS model only by the complex $\bar{s}$, and hence
      it is tightly connected with the imaginary part of the scattering amplitude
      and satisfies the analyticity and the dispersion relations.
            Quantitatively, for different thin structures of the scattering amplitude,
         a wider  analysis is needed.
          This concerns the fixed intercept taken from the deep inelastic processes and the fixed Regge slope $\alpha^{\prime}$,
           as well  as the form of the spin-flip amplitude.
           Such an analysis requires  a wider range of  experimental data, including the polarization data
        of $A_N(s,t)$, $A_{NN}(s,t)$, $A_{LL}(s,t)$, $A_{SL}(s,t)$.
                        The obtained information about the sizes and energy dependence of the
                        spin-flip and double-flip amplitudes will make it possible
    to better understand 
     the results of  famous experiments carried out by A. Krish at the ZGS  to obtain the
          spin-dependent differential cross sections 
          \cite{OFallon:1977fkf,Lin:1978xu} and
          the spin correlation parameter $A_{NN}$   
          and   at the AGS \cite{Crabb:1990as} to obtain  the spin correlation parameter $A_{N}$
           showing the significant spin  effects at large momentum transfer.


\section{\bf {Single-spin physics
\protect\footnote{This section is written by V. Abramov; \,E-mail: Victor.Abramov@ihep.ru}}}

\begin{abstract}
Physics of single-spin processes for the SPD NICA project is proposed. This includes transverse 
single-spin asymmetry ($A_N$) and hyperon  polarization ($P_N$) measurements in various types of collisions, including \it{p}\rm{+}\it{p}\rm, \it{d}\rm{+}\it{d}\rm,  \rm{C+C} and \rm{Ca+Ca}. The polarized 
$p$- and $d$-beams in the NICA collider 
can be used to study $ A_N $ for more than several dozen reactions at different energies in the 
$3.4 < \sqrt{s} < 27$ GeV range. 
A number of interesting phenomena have been predicted, such as the oscillation for $A_{N}(x_{\rm{F}})$ and $P_{N}(x_{\rm{F}})$, the resonance dependence on the energy $\sqrt{s}$ for  $A_N$ and $P_N$, and the threshold dependence of $A_{N}$ on the c.m. production angle for some reactions. The role of quark composition of particles involved in the reaction is discussed.



\end{abstract}
\vspace*{6pt}

\noindent 
PACS: 12.38.$-$t; 12.38.Qk; 13.60.$-$r; 13.75.$-$n; 13.85.$-$t; 13.85.Ni; 13.87.Fh; 
%
\label{sec:intro} 
\subsection*{Introduction}
All previous experience in the development of spin physics testifies to its fundamental importance for understanding the laws of the micro-world, including for the construction of the theory of strong interactions. It should be noted that large values of transverse single-spin asymmetries ($A_N$) and hyperon polarizations ($P_N$) in a wide energy range have not yet received an unambiguous and convincing explanation within the framework of the theory of strong interactions - quantum chromodynamics (QCD), which is one of the components of the Standard Model. The experimental data accumulated to date point to a very interesting phenomenology in the field of transverse single-spin phenomena, including the nontrivial dependence of the spin observables $A_N$ and $P_N$ on the collision energy ($\sqrt{s}$), the Feynman variable ($x_{\rm{F}}$), the transverse momentum ($p_T$), the atomic weights of the colliding particles ($A_1$ and $A_2$), the multiplicity of charged particles ($N_{ch}$) in the event and the centrality of collisions. It is equally important to measure $A_N$ and $P_N$ for as many reactions as possible in order to understand how spin effects depend on the quark composition and other quantum characteristics of the particles involved in the reaction. Data on dozens of reactions have been accumulated, but the accuracy of measurements and the limited kinematic region in most experiments do not yet allow unambiguous conclusions to be drawn about the origin of polarization phenomena and even about their dependence on various variables. The purpose of this proposal is to significantly expand the amount of polarization data available for analysis and to improve their accuracy. This will help advance the creation of adequate models of polarization phenomena and their discrimination when compared with the entire dataset.

 Planned measurements at the SPD facility in the energy range for a pair of colliding nucleons from 3.4 to 27 GeV in the reaction c.m. frame are very important for the systematic and detailed study of polarization phenomena and the study of their dependence on various variables. Analysis of the available data within the framework of the chromomagnetic polarization of quarks (CPQ) model \cite{Abramov:2008zz} shows that the  unambiguous determination of the model parameters is possible only if there are measurements with several (three or more) values for each of the listed above variables. It should be noted that the maximum energy of the accelerator in Dubna is high enough to register particles with large transverse momenta in the range 
$p_T$ = 1 - 4 GeV/c, for which the polarization effects are significant and the quark degrees of freedom are already manifested. The identification of particles in this energy range is much easier than at large accelerators, and this is an important condition for the systematic study of polarization phenomena in a large number of reactions.

The conditions for making measurements at the SPD facility at the first stage of the NICA collider can be found in 
\cite{Meshkov:2019utc}. Maximum energy in the c.m. of two colliding nucleons will be 27 GeV for \it{p}\rm{+}\it{p} \rm collisions and 
14 GeV for \it{d}\rm{+}\it{d}\rm, \rm{C+C} and
 \rm{Ca+Ca} collisions. Vector polarization will be 50\% for protons and 75\% for deuterons.
\begin{table}[htb] 
\caption{Inclusive reactions for which the spin-spin asymmetry $A_N$ was measured.}
\bigskip
\begin{tabular}{|c|c|c|c|c|c|}
\hline
%
$\rm N^{\underline{o}}$ &  Reaction &  $\rm N^{\underline{o}}$ &  Reaction  & $\rm N^{\underline{o}}$ &  Reaction    \\
\hline
 1 & $p^{\uparrow} p(A) \rightarrow \pi^{+} X$   & 
10 & $p^{\uparrow} p(A) \rightarrow J/\psi X$      &
19 & $\bar{p}d^{\uparrow} \rightarrow \pi^{0} X$     \\
 2 & $p^{\uparrow} p(A) \rightarrow \pi^{-} X$   &
11 & $p^{\uparrow}p(A) \rightarrow \eta X$      &
20 & $\pi^{+}p^{\uparrow} \rightarrow \pi^{+} X$      \\
 3 & $p^{\uparrow}p \rightarrow \pi^{0} X$     &
12 & $d^{\uparrow} p(A) \rightarrow \pi^{+} X$     &
21 & $\pi^{-}p^{\uparrow} \rightarrow \pi^{-} X$ \\
4 & $p^{\uparrow} p(A) \rightarrow K^{+} X$    & 
13 & $d^{\uparrow} p(A) \rightarrow \pi^{-} X$      &
22 & $\pi^{-}p^{\uparrow} \rightarrow \pi^{0} X$ \\
 5 & $p^{\uparrow} p(A) \rightarrow K^{-} X$     & 
14 & $p^{\uparrow}p \rightarrow \Lambda X$        & 
23 & $\pi^{-}d^{\uparrow} \rightarrow \pi^{0} X$  \\
 6 & $p^{\uparrow} p \rightarrow K^{0}_{S} X$     &
15 & $\bar{p}^{\uparrow}p \rightarrow \pi^{+} X$ &  
24 &  $K^{-}d^{\uparrow} \rightarrow \pi^{0} X$     \\
 7 & $p^{\uparrow} p(A) \rightarrow n X$       & 
16 & $\bar{p}^{\uparrow}p \rightarrow \pi^{-} X$  &
25 & $K^{-}p^{\uparrow} \rightarrow \pi^{0} X$    \\
 8 & $p^{\uparrow} p(A) \rightarrow p X$ &  
17 & $\bar{p}^{\uparrow}p \rightarrow \pi^{0} X$  &
26 & $\pi^{-}p^{\uparrow} \rightarrow \eta X$     \\
 9 & $p^{\uparrow}p(A) \rightarrow \bar{p} X$    &
18 & $\bar{p}^{\uparrow}p \rightarrow \eta X$   & 
27 &  $\bar{p}p^{\uparrow} \rightarrow \pi^{0} X$  \\
\hline
\end{tabular}
\label{MeasAN}
\end{table} 

Table \ref{MeasAN} presents 27 inclusive reactions for which there are already data on the single-spin asymmetry of hadrons 
\cite{Abramov:2008zz,Abramov:2000jg}. The first 14 reactions from the Table \ref{MeasAN} can potentially be studied at the NICA collider using the SPD facility. A list of other possible 28 reactions is shown in Table \ref{NotMeasAN} and includes various particles and resonances.
The initial state can be any with a polarized beam: $p^{\uparrow}p$, 
$p^{\uparrow}d$, $d^{\uparrow}p$, $d^{\uparrow}d$.   Their detailed study will reveal the dependence of $A_N$ on kinematic and other variables, including the quark composition of the particles involved, their spin, isospin, and atomic weight.

\begin{table}[htb] 
\caption{Inclusive reactions to be studied at the SPD for which $A_N$  has not yet been measured. The reaction is $p^{\uparrow}p \rightarrow h + X$. Final decay mode of the detected particle $h$ is indicated only.}
\bigskip
\begin{tabular}{|c|c|c|c|c|c|}
\hline
%
$\rm N^{\underline{o}}$ & Decay mode &  $\rm N^{\underline{o}}$ &  Decay mode  & $\rm N^{\underline{o}}$ & Decay mode  \\
\hline
 1 &  $K^{0}_{L} \rightarrow \pi^{+} \pi^{-} \pi^{0}$   &
10 & $\phi \rightarrow K^{+} K^{-}$      & 
19 & $\bar{\Xi}^{0} \rightarrow \bar{\Lambda} \pi^{0}$      \\
 2 & $\eta' \rightarrow \pi+ \pi- \eta $     & 
11 & $ \rho^{0}(770) \rightarrow \pi^{+} \pi^{-}$      &
20 & $\Sigma^{0} \rightarrow \Lambda \gamma$     \\  
 3 & $ a_{0}(980) \rightarrow  \eta \pi^{0}$    &
12 & $ \rho^{+}(770) \rightarrow \pi^{+} \pi^{0}$    &
21 & $\bar{\Sigma}^{0} \rightarrow \bar{\Lambda} \gamma$ \\
4  & $K^{0*}(892)\rightarrow K^{+}\pi^{-} $     & 
13 & $ \rho^{-}(770) \rightarrow \pi^{-} \pi^{0}$      &
22 & $\Delta^{++} \rightarrow p \pi^{+}$    \\
 5 & $K^{0*}(892)\rightarrow K^{-}\pi^{+} $     & 
14 & $ \rho^{0}(770) \rightarrow \mu^{+} \mu^{-}$     & 
23 & $\Delta^{+} \rightarrow p \pi^{0}  $      \\
 6 & $K^{+*}(892)\rightarrow K^{+}\pi^{0} $     &
15 & $\bar{\Lambda} \rightarrow \bar{p} \pi^{+}$ &  
24 & $\Delta^{0} \rightarrow p \pi^{-}$       \\
 7 & $K^{-*}(892)\rightarrow K^{-}\pi^{0} $       & 
16 & $\Xi^{-} \rightarrow \Lambda \pi^{-}$     &
25 & $\Delta^{-} \rightarrow n \pi^{-}$     \\
 8 & $\omega(782) \rightarrow \pi^{+} \pi^{-} \pi^{0} $  &   
17 & $\Xi^{0} \rightarrow \Lambda \pi^{0}$     &
26 & $\bar{\Delta}^{--} \rightarrow \bar{p} \pi^{-}$    \\
 9 & $\omega(782) \rightarrow \gamma \pi^{0} $ &
18 & $\bar{\Xi}^{+} \rightarrow \bar{\Lambda} \pi^{+}$     &
27 & $\bar{\Delta}^{0} \rightarrow \bar{p} \pi^{+}$  \\
\hline
\end{tabular}
\label{NotMeasAN}
\end{table} 
Data on the transverse polarization of hyperons and antihyperons are no less interesting. The list of reactions available to date, for which their polarization $P_N$ was measured, is presented in Table \ref{MeasPN} and includes 32 reactions \cite{Abramov:2008zz,Abramov:2001rh}.  The first 14 reactions can potentially be studied at the SPD setup. This list can be supplemented with reactions such as  $p p \rightarrow \Sigma^{0\uparrow}(1385) X$,  $p p \rightarrow \bar{\Sigma}^{0\uparrow} X$, $p p \rightarrow \Lambda^{\uparrow}(1405) X$, 
$p p \rightarrow \Lambda^{\uparrow}(1520) X$. The initial state can be any with a polarized or unpolarized beam: $p^{\uparrow}p$, $p^{\uparrow}d$, $d^{\uparrow}p$, $d^{\uparrow}d$ and  $AA$.  
\begin{table}[htb] 

\caption{Inclusive reactions for which the polarization ($P_N$) of hyperons was measured.}
\bigskip
\begin{tabular}{|c|c|c|c|c|c|}
\hline
%
$\rm N^{\underline{o}}$ &  Reaction &  $\rm N^{\underline{o}}$ &  Reaction  & $\rm N^{\underline{o}}$ &  Reaction  \\
\hline
 1 & $p p(A) \rightarrow \Lambda^{\uparrow} X$   & 
12 & $A_{1}A_{2} \rightarrow \Lambda^{\uparrow} X$     &
23 & $\pi^{-} A \rightarrow \Xi^{-\uparrow} X$   \\
 2 & $p p(A) \rightarrow \Xi^{-\uparrow} X$     & 
13 & $A_{1}A_{2} \rightarrow \Lambda^{\uparrow(G)} X$    &
24 & $\pi^{-} A \rightarrow \bar{\Xi}^{+\uparrow} X$  \\
 3 & $p p(A) \rightarrow \Xi^{0\uparrow} X$     &
14 & $A_{1}A_{2} \rightarrow \bar{\Lambda}^{\uparrow(G)} X$      &
25 & $\pi^{-} p \rightarrow \Lambda^{\uparrow} X$   \\
4  & $p p(A) \rightarrow \Sigma^{+\uparrow} X$      & 
15 & $\Sigma^{-} A \rightarrow \Sigma^{+\uparrow} X$          & 
26 & $\pi^{-} p \rightarrow \bar{\Lambda}^{\uparrow} X$     \\
 5 & $p p(A) \rightarrow \Sigma^{0\uparrow} X$     & 
16 & $\Sigma^{-} A \rightarrow \Xi^{-\uparrow} X$       & 
27 & $\pi^{+} p \rightarrow \Lambda^{\uparrow} X$       \\
 6 & $p p(A) \rightarrow \Sigma^{-\uparrow} X$    &
17 & $\Sigma^{-} A \rightarrow \Lambda^{\uparrow} X$ & 
28 & $K^{-} A \rightarrow \Xi^{-\uparrow} X$     \\
 7 & $p p(A) \rightarrow \Omega^{-\uparrow} X$      & 
18 & $\Sigma^{-} A \rightarrow \bar{\Lambda}^{\uparrow} X$     &
29 & $\bar{p} A \rightarrow \bar{\Lambda}^{\uparrow} X$     \\ 
 8 & $p p(A) \rightarrow \bar{\Lambda}^{\uparrow} X$        & 
19 & $K^{-} p \rightarrow \Lambda^{\uparrow} X$      &
30 & $e^{+} e^{-} \rightarrow \Lambda^{\uparrow} X$      \\
 9 & $p p(A) \rightarrow \bar{\Xi}^{+\uparrow} X$ &  
20 & $K^{-} p \rightarrow \bar{\Lambda}^{\uparrow} X$    &
31 & $\nu_{\mu} A  \rightarrow \Lambda^{\uparrow} X$  \\
10 & $p p(A) \rightarrow \bar{\Xi}^{0\uparrow} X$    &
21 & $K^{+} p \rightarrow \Lambda^{\uparrow} X$      &
32 & $e^{+} A \rightarrow \bar{\Lambda}^{\uparrow} X$  \\
11 & $p p(A) \rightarrow \bar{\Sigma}^{-\uparrow} X$       &
22 & $K^{+} p \rightarrow \bar{\Lambda}^{\uparrow} X$     & 
33 &  -  \\
\hline
\end{tabular}
\label{MeasPN}
\end{table} 

It is important to note that for hyperons it is possible to simultaneously measure both the transverse polarization $P_N$ and the single-spin asymmetry $A_N$.  Comparing $A_N$ and $P_N$ for a specific reaction with the predictions of various models will bring us closer to revealing the mechanism of the origin of polarization phenomena at high energies and will shed light on the physics of strong interactions in the confinement region.

A systematic study of polarization data assumes the presence of a model that describes, within a single mechanism, a large number of reactions depending on the variables listed above. An example of such a model is the model of chromomagnetic polarization of quarks (CPQ) \cite{Abramov:2008zz}. 

References to most of publications devoted to polarization experiment data can be found in 
\cite{Abramov:2008zz,Abramov:2000jg,Abramov:2001rh}  and other in papers  listed in the cited literature.
The following sections describe in more detail the model of chromomagnetic polarization of quarks and consider examples of existing data and calculations of $A_N$ and $P_N$ for various reactions that can potentially be studied using the SPD setup at the NICA collider in Dubna.

\label{sec:model}
\subsection{Model of chromomagnetic polarization of quarks}

The phenomenological model of chromomagnetic polarization  of quark is based on the following basic assumptions \cite{Abramov:2008zz}:

1) As a result of collisions of hadrons, a pair of quarks with a large transferred transverse momentum $p_{T}$ is scattered. Further, the scattered (test) quark with large $p_{T}$ moves in the effective chromomagnetic field  $\rm\bf{B}^{\it{a}}$ and experiences the action of the Stern-Gerlach force proportional to the product of the field gradient components and the corresponding components of the quark chromomagnetic moment. The direction of the Stern-Gerlach force and the additional transverse momentum received by the test quark in the effective chromomagnetic field depend on the projections of the quark spin onto the quantization axis. Subsequently, the polarized quark from the incident polarized proton recombines with other quarks to form the observed hadron. The angular distribution of such hadrons has an azimuthal dependence, i.e., a single-spin asymmetry arises. If unpolarized hadrons collide, then the action of the Stern-Gerlach force imparts an additional transverse momentum directed to the left or to the right, depending on the direction of the projection of the quark spin up or down, when the quark moves, for example, to the left. Thus, when scattering to the left, a quark has predominantly one polarization sign, and when scattering to the right, the opposite. The hyperons formed from these quarks acquire transverse polarization relative to the scattering plane.

2) The effective chromomagnetic field  $\rm\bf{B}^{\it{a}}$   is created by spectator quarks, that is, all quarks that will not be included in the recorded hadron. Spectator quarks are moving in the c.m. in the direction 
of the colliding hadrons and create for a short time a circular transverse chromomagnetic field. The sign of the circular chromomagnetic field to the left and to the right of the collision axis is opposite, but the field gradient does not change its direction, which ensures a nonzero polarization effect due to the action of the Stern-Gerlach force. The predominant direction of polarization of quarks in a chromomagnetic field arises, hence the name of the model.

3) When taking into account the interaction of a test quark with the field created by a moving spectator quark, it is necessary to take into account the color factor for the corresponding pair of quarks (spectator and test quarks). An analysis of the data showed that the quark-antiquark pair interacts predominantly in the color-singlet state with the color factor $C_{F}$ = 4/3, and the quark-quark or antiquark-antiquark pair interacts in the color-triplet state with 
$C_{F}$ = 2/3. For a hydrogen-like potential, the wave function of two quarks or a quark and an antiquark at zero coordinate is proportional to 
$|\psi(0)| \propto (C_{F}\alpha_{S})^{3/2}$ 
\cite{Baranov:1995rc}, which leads to the ratio of contributions from {\it{qq}}  and $q$$\bar{q}$ interactions to an effective field of the order
\begin{equation}\label{eq:lambda}  							
\lambda \approx -|\psi_{qq}(0)|^{2}/|\psi_{q\bar{q}}(0)|^{2}= 
-1/8 = -0.125. 	\end{equation}  
The minus sign in (\ref{eq:lambda})  takes into account the opposite sign of the field created by a moving spectator quark and a moving spectator antiquark. Experimentally, the value of the global parameter, obtained as a result of the global fit of the polarization data, turned out to be 
$\lambda = -0.1363 \pm 0.0003$. 

  If the  spectator  quark is a product of target fragmentation and moves in the c.m. in the opposite direction, then its contribution to the effective field will be additionally suppressed by the factor $-\tau$, where $\tau = 0.0267 \pm 0.0012$ is another important global parameter of the CPQ model. This suppression of the contribution of quarks from the target is due to the fact that the chromomagnetic field they create is in a different region of space-time and, therefore, has almost no effect on the test quarks moving forward. 

4) The presence of an effective chromomagnetic field should lead to the precession of the test quark spin when it moves in the field. Analysis of the data showed that the effective field length and the corresponding precession angle are proportional to 
$x_{A} = (x_{R} + x_{\rm{F}})/2$ and  $x_{B} = (x_{R} - x_{\rm{F}})/2$
in the fragmentation region of the incident particle A and target B, respectively.   As a result, this leads to oscillations of the dependence of $A_N$ and $P_N$ on the kinematic variables $x_A$ and $x_B$, and hence on $x_{\rm{F}}$ and $p_T$. These oscillations are the main feature of the CPQ model and should manifest themselves in the case of strong fields, when the precession angles reach values of the order of $\pi$ or more.

\begin{figure}[ht]
\includegraphics[width=18.0pc]{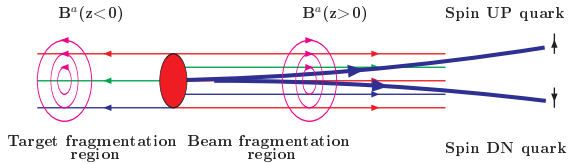}
\caption{\label{SG}
The mechanism of origin of single-spin polarization phenomena.
}
\end{figure}
The mechanism of origin of single-spin polarization phenomena is shown schematically
in fig.~\ref{SG}. The interaction of colliding particles $A$ and $B$ is considered in the c.m. of pair of colliding nucleons.

Observables $A_N$ and $P_{N}$ are described by equations (\ref{eq:PN}) and (\ref{eq:Gfun}):  
\begin{equation}\label{eq:PN}
P_{N} = C(\sqrt{s}) F(p_{T},A)[G(\phi_{A}) - \sigma G(\phi_{B})],
\end{equation}
\begin{equation}\label{eq:Gfun}
G(\phi) = (1-\cos\phi)/\phi + \epsilon \cdot \phi , 
\end{equation}   					
where function (\ref{eq:Gfun}) takes into account the action of the Stern-Gerlach forces and the precession of the quark spin, and
where $\epsilon = -0.00497 \pm 0.00009$ is the global parameter of the CPQ model, $\sigma$ is the local parameter. 
The integral angles of precession of the quark spin are
\begin{equation}\label{eq:phi}
\phi_{A} =  \omega^{0}_{A}y_{A}, \hskip 1.5cm  \phi_{B} =  \omega^{0}_{B}y_{B},  
\end{equation}   					
in the fragmentation region of colliding particles $A$ and $B$, respectively. The oscillation frequency  $\omega^{0}_{A(B)}$ is described by the equation
\begin{equation}\label{eq:omega}  							
\omega^{0}_{A(B)} =  g_{s}\alpha_{s}\nu_{A(B)}m_{r}(g^{a}_{Q}-2)/M_{Q},  							
\end{equation}   					
where $\alpha_{s} = g_{s}^{2}/(4\pi)$ is the current strong interaction constant, $g_s$ is the color charge, $M_Q$ is the mass of the constituent quark $Q$, $g^{a}_{Q}$ is the the Lande gyromagnetic color factor of a quark, $m_{r} = 0.2942 \pm 0.0072$ GeV is the global parameter that can be considered as the ratio of the maximum longitudinal extension of the chromomagnetic field to the square of its radius.

The total contribution of spectator quarks (with weights $\lambda$ and $-\tau$) to $\nu_{A(B)}$ in the fragmentation region of colliding particles $A$ and $B$, respectively, is calculated using quark diagrams and the quark counting rules \cite{Abramov:2008zz}.

\begin{figure}[b!]
  \centering
  \begin{tabular}{cc}
     \includegraphics[width=55mm]{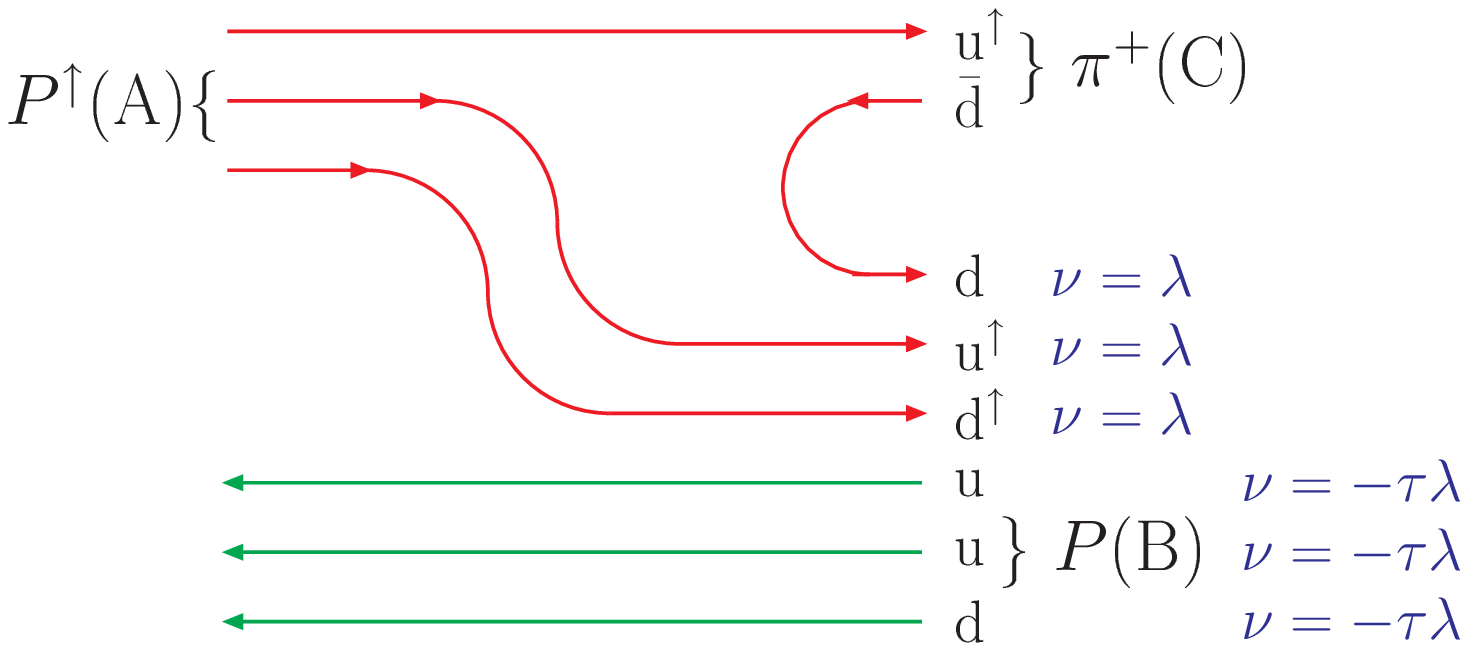} &
    \includegraphics[width=55mm]{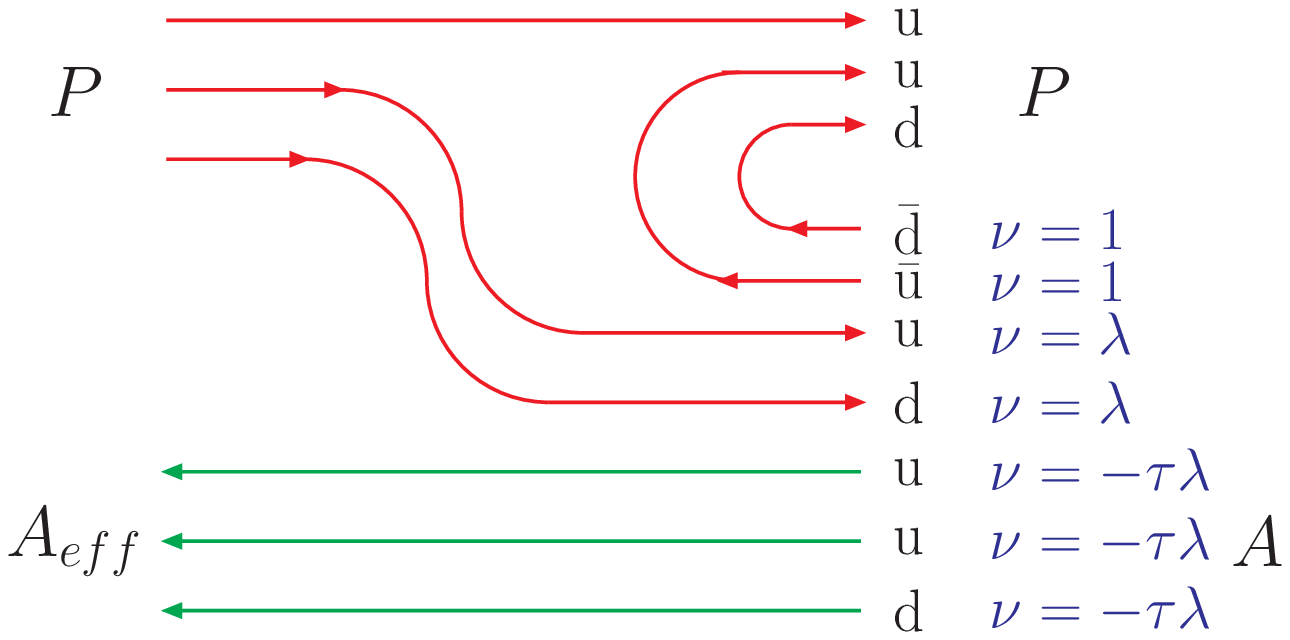} \\
    \textbf{(a)} & \textbf{(b)}
  \end{tabular}
  \caption{%
  Quark flux diagrams for the reaction 
 $p^{\uparrow}$+$p$$\rightarrow$$\pi^{+}$+$X$  \textbf{(a)}
 and $p^{\uparrow}$+$p(A)$$\rightarrow$$p$+$X$ \textbf{(b)}.
  }
  \label{re06re13diag}
\end{figure}

Quark diagrams for the reactions 
$p^{\uparrow}$+$p$$\rightarrow$$\pi^{+}$+$X$ and
$p^{\uparrow}$+$p(A)$$\rightarrow$$p$+$X$
are shown in fig.~\ref{re06re13diag}a and \ref{re06re13diag}b, respectively. 
When nucleus is the target, as in the case of fig. \ref{re06re13diag}b,
the number of target spectator quarks is equal to $3A_{\rm{eff}} \propto A^{1/3}$, where $A$ is an atomic weight, since all target quarks hit by the incident proton contribute to the spectator quarks \cite{Abramov:2008zz}. Below we assume $A_{\rm{eff}} = A = 1$.
%
%

In the approximation of moderate energies ($\sqrt{s} < 70$ GeV), we obtain $\nu_{A}$ for the reaction 
$p^{\uparrow}$+$p$$\rightarrow$$\pi^{+}$+$X$: 
\begin{equation}\label{eq:nuRE6}  							
\nu_{A} = \nu_{B} = 3\lambda - 3\tau\lambda A_{\rm eff} = -0.398, 
\end{equation}
and for the reaction 
$p^{\uparrow}$+$p(A)$$\rightarrow$$p$+$X$:
\begin{equation}\label{eq:nuRE13}
\nu_{A} = \nu_{B} = 2+2\lambda - 3\tau\lambda A_{\rm eff} = 1.738. 
\end{equation}
To calculate $\nu_{A}$ we have to add up all the contributions ($\nu$) of the spectator quarks shown to the right of the quark diagram. The $\nu_A$ value for the reaction 
$p^{\uparrow}$+$p$$\rightarrow$$\pi^{+}$+$X$ 
is much less than 1 in absolute value. Consequently, the oscillation frequency 
$\omega^{0}_{A(B)}$ is also low, and the $A_{N}(x_{\rm{F}})$ dependence is close to linear. For the reaction 
$p^{\uparrow}$+$p(A)$$\rightarrow$$p$+$X$,
the value of $\nu_A$ is significantly greater than unity in absolute value, and for it, as we will see below, a nonmonotonic oscillating dependence $A_{N}(x_{\rm{F}})$  is indeed observed.

Kinematic variables
\begin{equation}\label{eq:yA}
y_{A} = x_{A}  - (E_{0}/\sqrt{s} + f_{0} )[1 + \cos\theta_{cm} ] + a_{0}[1 - \cos\theta_{cm} ],    				
\end{equation}
\begin{equation}\label{eq:yB}
y_{B} = x_{B}  - (E_{0}/\sqrt{s} + f_{0} )[1 - \cos\theta_{cm} ] + a_{0}[1 + \cos\theta_{cm} ],    				
\end{equation}
are expressed in terms of the scaling variables $x_A$ and $x_B$, the reaction energy $\sqrt{s}$, the emission angle $\theta_{cm}$ in c.m. and three local parameters $E_0$, $a_0$ and $f_0$. Function
\begin{equation}\label{eq:C}
C(\sqrt{s}) = v_{0} / [(1 - E_{R}/\sqrt{s})^{2} + \delta_{R}^{2}]^{1/2}, 
\end{equation}
takes into account the dependence of the rate of precession of the spin of a quark on its energy $E_Q$ in c.m. and the effect of attraction ($E_{R}> 0$) or repulsion ($E_{R} <0$) between the test quark and spectator quarks. 
The $E_R$ sign is determined by the factor $-g_{S}\nu_{A}$, where $g_S$ is the color charge of the test quark (positive for a quark and negative for an antiquark).  An example of a reaction with $E_{R}> 0$ is 
$p + p \rightarrow \bar{\Lambda} + X$, and reaction with $E_{R} <0$ is
$p + p \rightarrow \Lambda + X$. The global data fit confirms the $E_R$ sign rule for most of 85 investigated reactions (96,5\%).

The coefficient $v_0$ determines the value of $A_N$ and $P_N$ and is calculated as follows:
\begin{equation}\label{eq:v0}
 v_{0} = -D_{r}g^{a}_{Q}P_{Q}/2(g^{a}_{Q}-2), 
\end{equation}
where $D_r$ is a local dimensionless parameter of order 0.8, which is the ratio of the spectrum slope in $p_T$ to the transverse radius of the effective field, $P_{Q}$ is the polarization of the $Q$ quark in a polarized proton (+1 for $u$-quark and -1 for $d$-quark), $g^{a}_{Q}$ is the Lande gyromagnetic factor for the $Q$-type quark, which is a global parameter.  The $A_{N}$ or $P_N$ sign for most reactions at small $\phi_{A}$ is the product of three factors: $-g_{S}\nu_{A}P_{Q}$. When calculating the polarization of hyperons, we set $P_{Q}=1$.

The color form factor $F(p_{T},A)$ suppresses $A_N$ and $P_N$ at low $p_T$, when the colored quarks inside the hadron are not visible due to the uncertainty relation:
\begin{equation}\label{eq:FpT}
F(p_{T},A) = {1 - \exp[-(p_{T}/p^{0}_{T})^{2.5}]}(1 -\alpha_{A}\ln{A}), 
\end{equation}
where $p^{0}_{T}$ is a local parameter, and the other parameter $\alpha_{A}$ is zero for most of reactions. 

The dependence of a number of parameters on the atomic masses $A_1$ and $A_2$ turned out to be universal for most of the reactions in Tables \ref{MeasAN} and \ref{MeasPN} \cite{Abramov:2008zz,Abramov:2014sr}. Further development of the CPQ model is reflected in papers
 \cite{Abramov:2014sr,Abramov:2007zz,Abramov:2007sr,Abramov:2009tm,Abramov:2011zz,Abramov:2011sr,Abramov:2014bha,Abramov:2016lks,Abramov:2018vlp,Abramov:2020tne}. 

%
%
%
\label{sec:asymmetry}
\subsection{Single-spin hadron asymmetry}
%
\begin{figure}[b!]
  \centering
  \begin{tabular}{cc}
     \includegraphics[width=55mm]{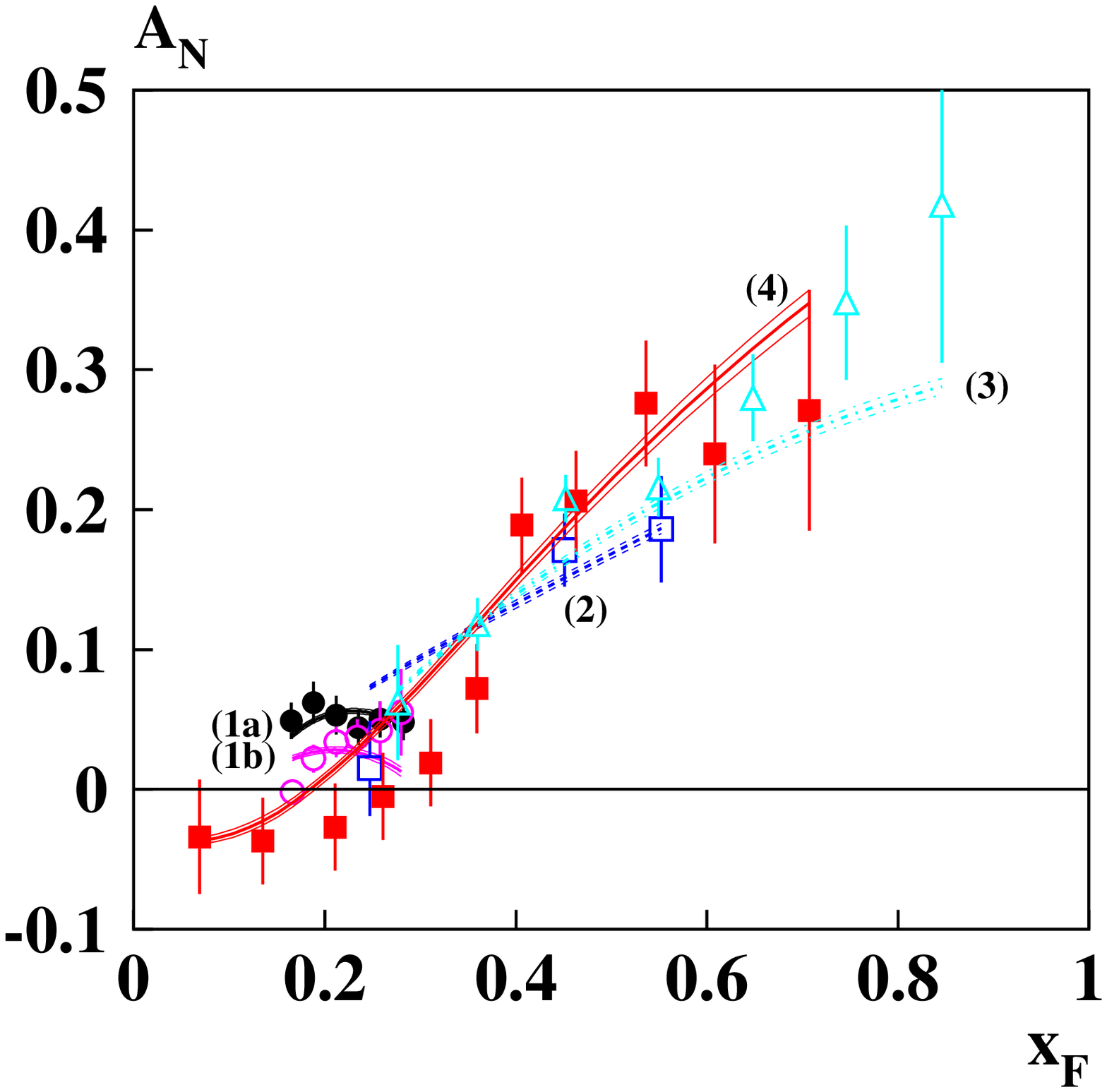} &
    \includegraphics[width=55mm]{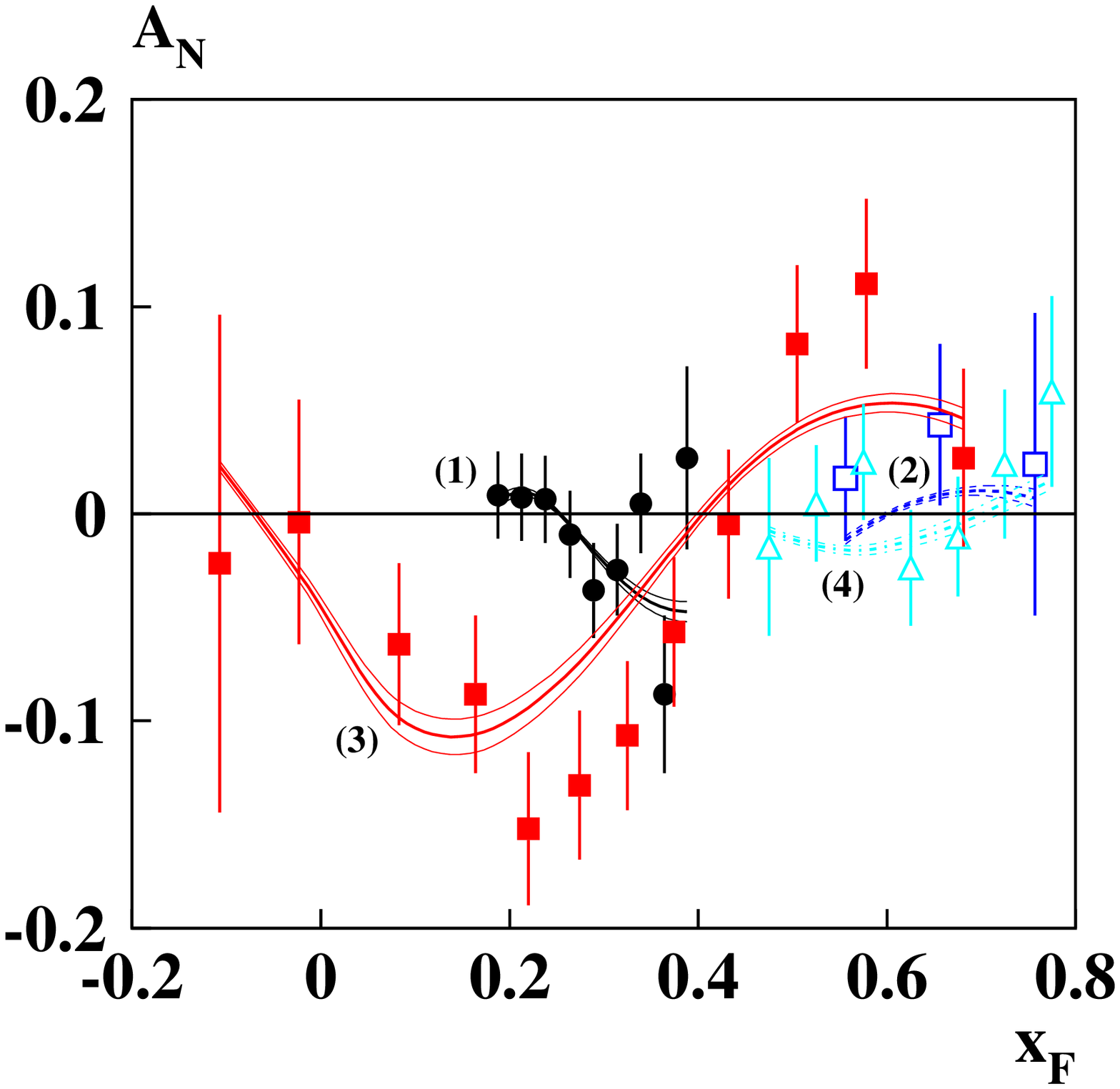} \\
    \textbf{(a)} & \textbf{(b)}
  \end{tabular}
  \caption{%
  $A_{N}(x_{\rm{F}})$ for the reaction $p^{\uparrow} + p(A) \rightarrow  \pi^{+} + X$  \textbf{(a)}  and
 $p^{\uparrow} + p(A) \rightarrow  p + X$ \cite{Abramov:2016lks}  \textbf{(b)} \cite{Abramov:2016lks}. 
  }
  \label{re06re13AN}
\end{figure}

The most numerous experiments to measure the single-spin asymmetry were carried out for the reactions of the production of charged and neutral 
$\pi$-mesons in $p^{\uparrow}p$ and $p^{\uparrow}A$ collisions. These data are included in the general database of polarization phenomena, which  contains 3608 experimental points for 85 different inclusive reactions, in which the polarization of one of the particles is known or measured in the initial or final state \cite{Abramov:2008zz,Abramov:2016lks}. A global fit was performed for the entire dataset using the CPQ model.

Data on $A_N$ for the reaction
$p^{\uparrow} + p(A) \rightarrow  \pi^{+} + X$  at different energies are shown in fig.~\ref{re06re13AN}a from~\cite{Abramov:2016lks}, where they are compared with the results of calculations using the CPQ model.  
As seen from fig.~\ref{re06re13AN}a, $A_{N}(x_{\rm{F}})$ dependence for the reactions 
 $p^{\uparrow} + p(A) \rightarrow  \pi^{+} + X$  at moderately high energies $\sqrt{s} <70$ GeV, is almost linear, which agrees with the predictions of the CPQ model. This is due to the insignificant value of the parameter
$\nu_{A} = \nu_{B} = 3\lambda - 3\tau\lambda = -0.398$, which follows from the quark diagram shown in fig.~\ref{re06re13diag}a. The positive sign of 
$A_{N}(x_{\rm{F}})$ for the reaction 
$p^{\uparrow} + p(A) \rightarrow  \pi^{+} + X$  
is explained by the dominant contribution of the positively polarized test
$u$-quark from a polarized proton.

\begin{figure}[b!]
  \centering
  \begin{tabular}{cc}
        \includegraphics[width=55mm]{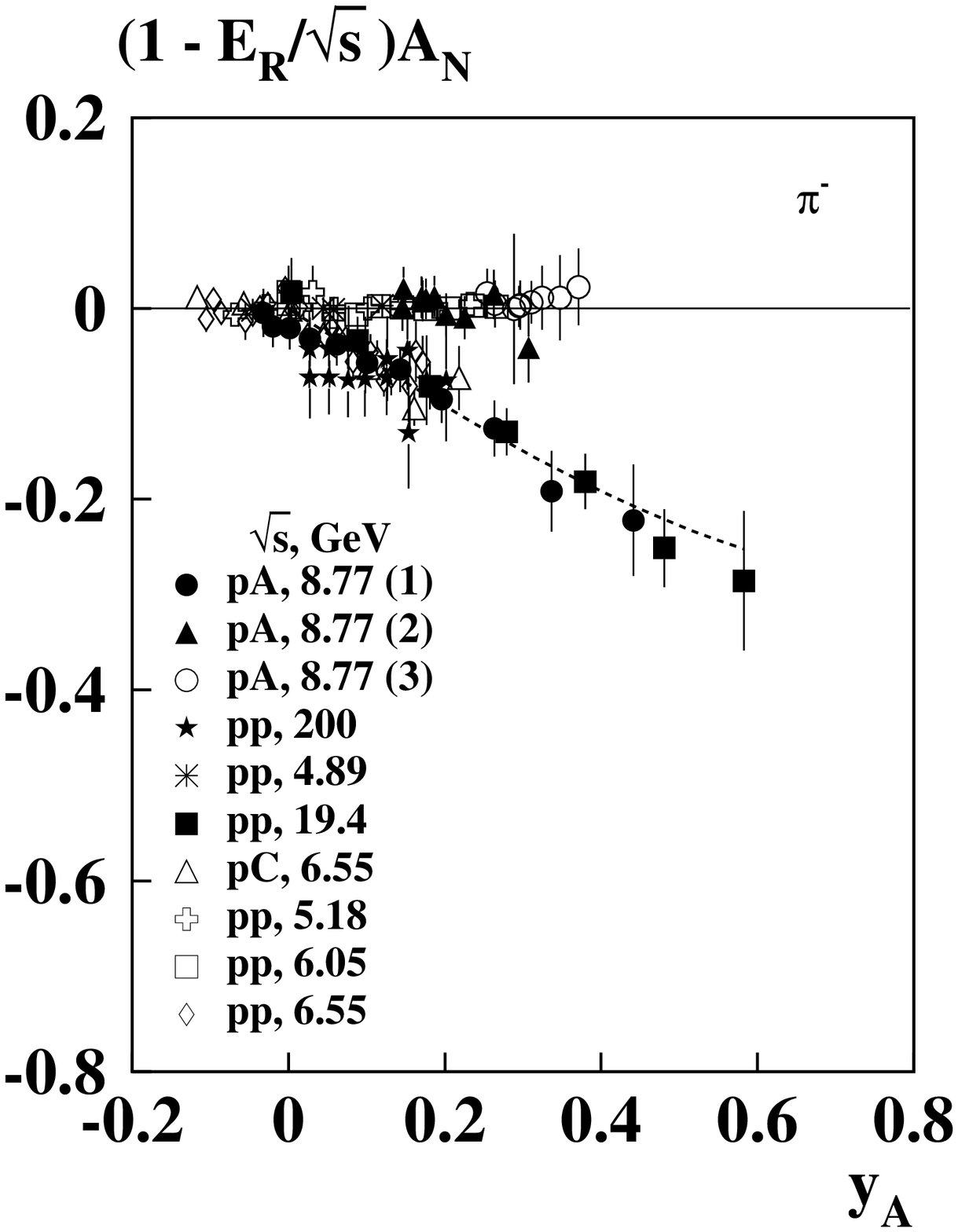} &
    \includegraphics[width=55mm]{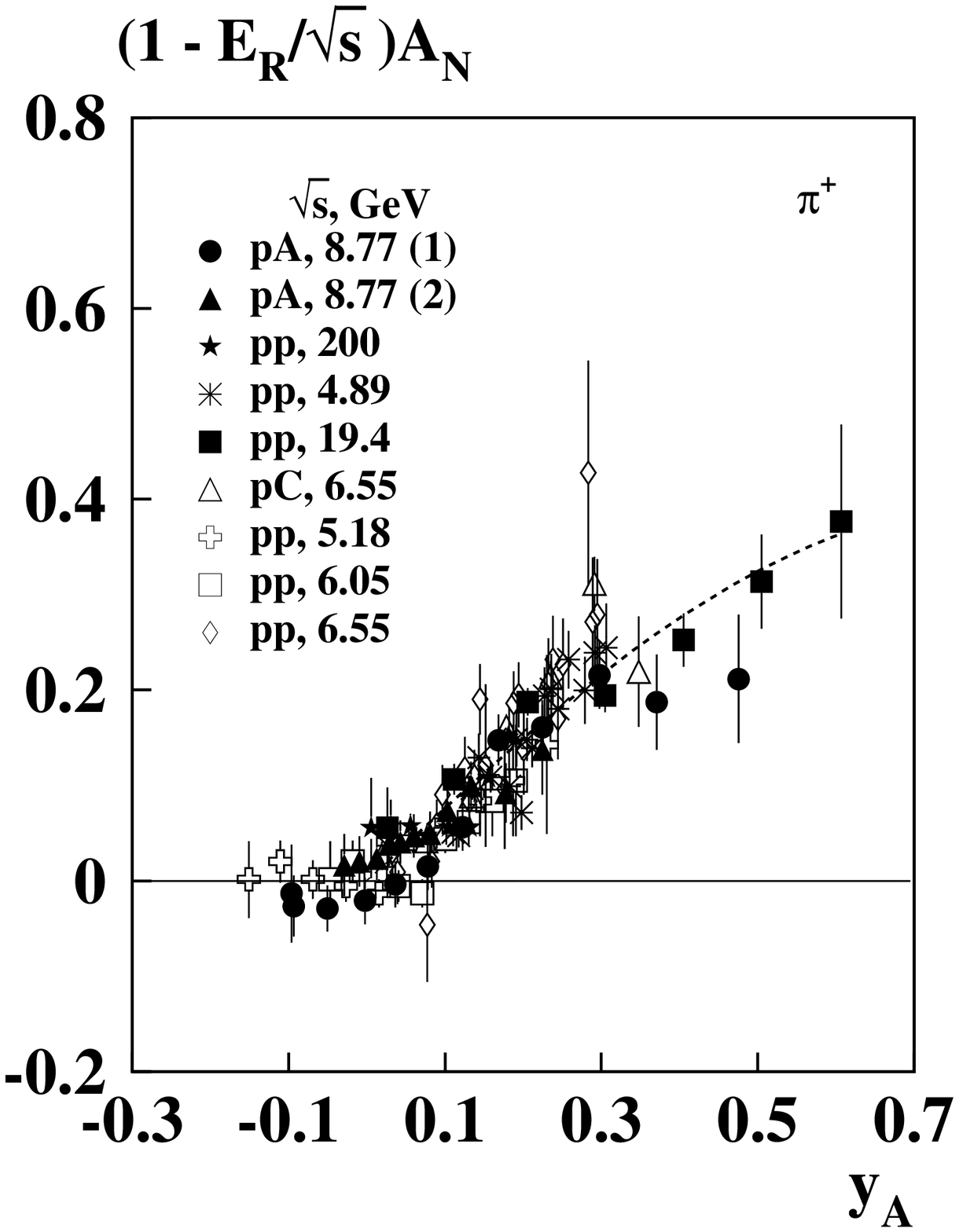} \\
    \textbf{(a)} & \textbf{(b)}
  \end{tabular}
  \caption{%
 Dependence of the value $(1-E_{R}/\sqrt{s})A_{N}$ on $y_{A}$, where
$E_{R} = 4.98 \pm 0.29$ GeV for the reaction 
$p^{\uparrow} p(A) \rightarrow  \pi^{-} X$   \textbf{(a)}  and
$E_{R} = 1.92 \pm 0.30$ GeV, for the reaction 
 $p^{\uparrow}  p(A) \rightarrow  \pi^{+} X$  \textbf{(b)}  
 \cite{Abramov:2007zz}.
  }
  \label{ThreshPI}
\end{figure}

A very unexpected and interesting feature of the reaction 
$p^{\uparrow} + p(A) \rightarrow  \pi^{-} + X$
 turned out to be the threshold dependence of $A_{N}(y_{A})$ on the angle of production $\theta_{cm}$ in c.m. In fig.~\ref{ThreshPI}a from \cite{Abramov:2007zz} is shown the dependence of the quantity 
$(1 - E_{R}/\sqrt{s})A_{N}$ on $y_A$, where $E_{R} = 4.98 \pm 0.29$ GeV. It turned out that this quantity is described by the universal function of 
$y_{A}$ if $\theta_{cm} < 74^{o}$, 
and is equal to zero if $\theta_{cm} > 74^{o}$. 
In fig.~\ref{ThreshPI}a, two clearly distinct branches are visible, into which the experimental points are grouped.

Within the framework of the CPQ model, the threshold effect for
 $A_{N}(y_{A})$ can be qualitatively explained by the greater mass of the constituent $d$-quark as compared to the mass of the $u$-quark.

In fig.~\ref{ThreshPI}b from~\cite{Abramov:2007zz} is shown the dependence of the quantity $(1-E_{R}/\sqrt{s})A_{N}$ on $y_{A}$ for the reaction 
$p^{\uparrow} + p(A) \rightarrow  \pi^{+} + X$,
 where $E_{R} = 1.92 \pm 0.30$ GeV. Most of the light test $u$-quarks flying into the front hemisphere will be from a polarized proton, which means that the asymmetry $A_{N} > 0$ for $\pi^{+}$ mesons \cite{Abramov:2007zz}. All data in fig.~\ref{ThreshPI}b 
 are located on the same branch, for a wide range of 
energies $\sqrt{s}$ and production angles in c.m.

The positive value $E_{R} = 4.98 \pm 0.29$ GeV for the reaction
$p^{\uparrow} + p(A) \rightarrow  \pi^{-} + X$, found in the framework of the CPQ model, is a manifestation of the effect of "attraction" of test quarks and spectator quarks.
According to formula (\ref{eq:C}), $A_N$ reaches its maximum value at energy $\sqrt{s} \approx E_{R}$  \cite{Abramov:2008zz,Abramov:2007zz}. Investigation of the effect of "attraction" of test quarks for various reactions is one of the objectives of this proposal and involves scanning in energy $\sqrt{s}$ near $E_{R}$. This phenomenon is observed not only for single-spin asymmetry, but also for the  polarization of hyperons, in those reactions for which $E_{R}$ is positive and amounts to several GeV \cite{Abramov:2008zz}.

Finding of scaling (independence of $(1 - E_{R}/\sqrt{s})A_{N}$ from energy $\sqrt{s}$) in the variable $y_{A}$ was one of the stages in the process of creating the CPQ model \cite{Abramov:2008zz,Abramov:2000jg,Abramov:2007zz}. Investigation of the scaling for polarization  observables $A_{N}$ and $P_{N}$ is of independent interest and can be one of the tasks for the SPD setup. In the framework of the CPQ model, scaling in polarization phenomena is due to the occurrence of processes at the quark level, in the limit of high energies and large transverse momentum \cite{Abramov:2008zz,Abramov:2000jg,Abramov:2001rh,Abramov:2007zz}.

Data and calculations of $A_{N}(x_{\rm{F}})$ for the reaction
$p^{\uparrow} + p(A) \rightarrow  p + X$,
  taken from  \cite{Abramov:2016lks}, are shown in 
  fig.~\ref{re06re13AN}b.  
  The data of the FODS-2 experiment \cite{Abramov:2007zzc}, measured in a wide range in the variable $x_{\rm{F}}$, at an energy of 
	$\sqrt{s} = 8.77$ GeV (solid squares in fig.~\ref{re06re13AN}b and curve (3)), demonstrate a nonmonotonic oscillatory dependence 	of $A_{N}(x_{\rm{F}})$.
	This is a consequence of the large value of the parameter $\nu_{A}$ and the significant precession angle of the quark spin in the chromomagnetic field. The quantity 
  $\nu_{A} = \nu_{B} = [2 + 2\lambda - 3\tau\lambda] = 1.738$ 
  is large enough, which follows from the quark diagram shown in 
  fig.~\ref{re06re13diag}b. In the energy region of the NICA collider, a  negative asymmetry $A_{N}(x_{\rm{F}})$ of about 10\% is expected near $x_{\rm{F}} = 0.2$ (fig. \ref{re06re13AN}b, curves (3) for $\sqrt{s} = 8.77$ GeV).

Another new and interesting direction in the study of polarization phenomena is associated with the dependence of $A_{N}$ and $P_{N}$ on the multiplicity of charged particles ($N_{ch}$) in an event. The first results in this region were obtained for reactions 
$p^{\uparrow} + p \rightarrow  \pi^{\pm} + X$
 in the BRAHMS experiment at an energy $\sqrt{s} = 200$ GeV \cite{Lee:2009ck}. The single-spin asymmetry $A_N$ increases in absolute value, if we select events with $N_{ch}$ above the average, and decreases, if we select events with $N_{ch}$ below the average. These data, together with the calculations, are disscussed in \cite{Abramov:2011sr}.
In the CPQ model, events with a multiplicity above the mean correspond to quark diagrams with additional quark-antiquark pairs compared to the minimum required number.  This effect, which  can manifest itself for both $A_N$ and $P_N$, can be studied at the SPD facility.

\label{sec:hyperons}
\subsection{Transverse polarization of hyperons}

Hyperons have the remarkable property that their decay in the weak interaction makes it possible to determine the transverse polarization to the scattering plane ($P_N$) - the only one possible in strong interactions, due to the conservation of parity in them. Therefore, the polarization of hyperons can be studied in collisions of practically any particles. In the case of the first phase of the SPD NICA project, we are interested in collisions $pp$, $pd$, $dd$, C+C and Ca+Ca. 
 The available data are discussed in detail in  \cite{Abramov:2001rh}.

 Quark diagrams for the production of 
 $\Xi^{-}$ hyperons in pp-collisions can be found  in~\cite{Abramov:2020tne}.  The effective number of spectator quarks for the  reaction   $p + p \rightarrow  \Xi^{-\uparrow} + X$ is
  $\nu_{A} = \nu_{B} = 2 + 2\lambda - 3\tau\lambda \approx 1.7383$.
  Similar calculations for the  reaction
  $p + p \rightarrow  \Lambda^{\uparrow} + X$ give
   $\nu_{A} = \nu_{B} = 1 + \lambda - 3\tau\lambda \approx 0.8746$. 
   Therefore, a nonmonotonic   
	dependence $P_{N}(x_{\rm{F}})$ can be expected in the case of the reaction 
    $p + p \rightarrow  \Xi^{-\uparrow} + X$.
\begin{figure}[hb!]
  \centering
  \begin{tabular}{cc}
     \includegraphics[width=55mm]{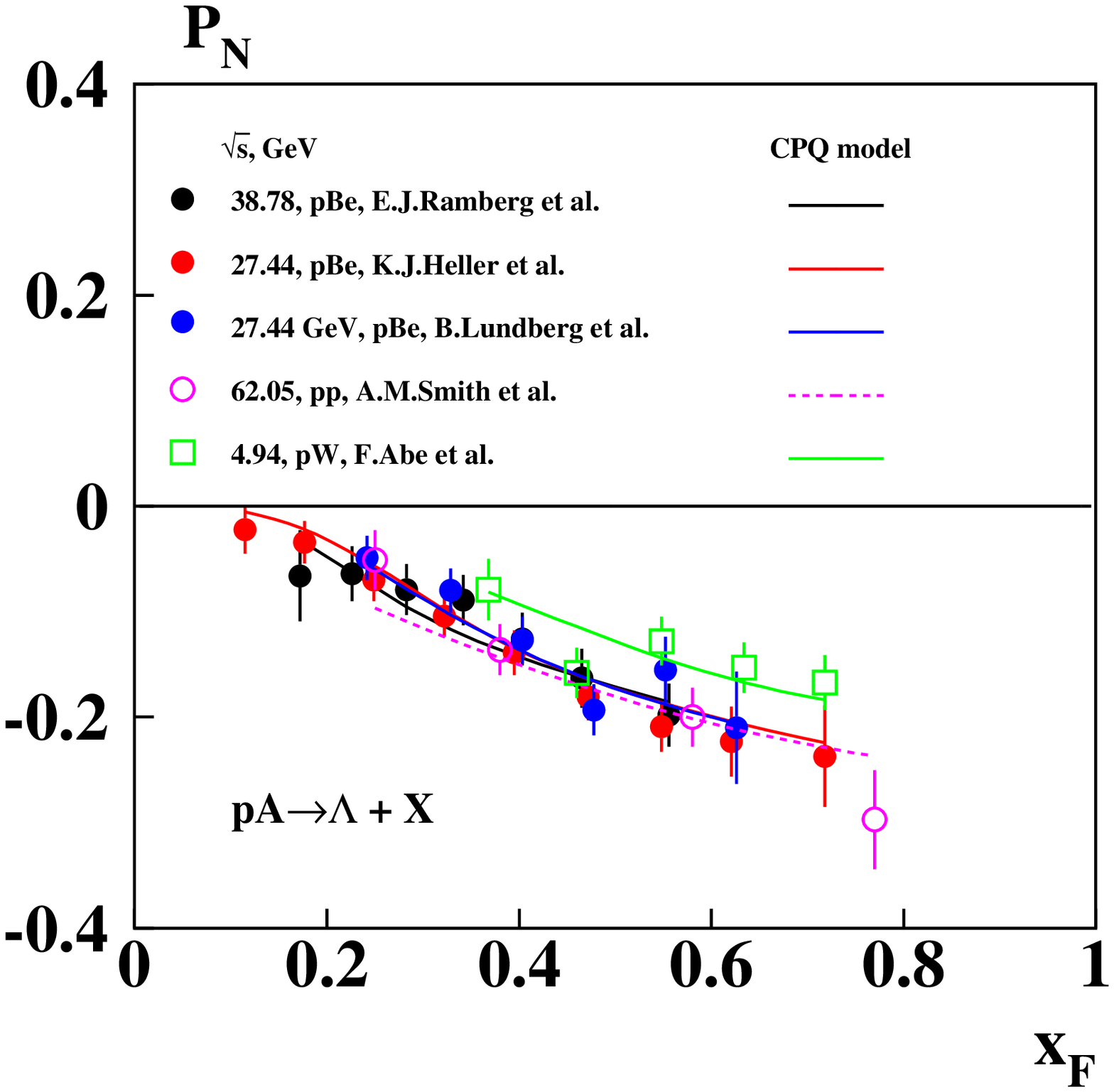} &
    \includegraphics[width=55mm]{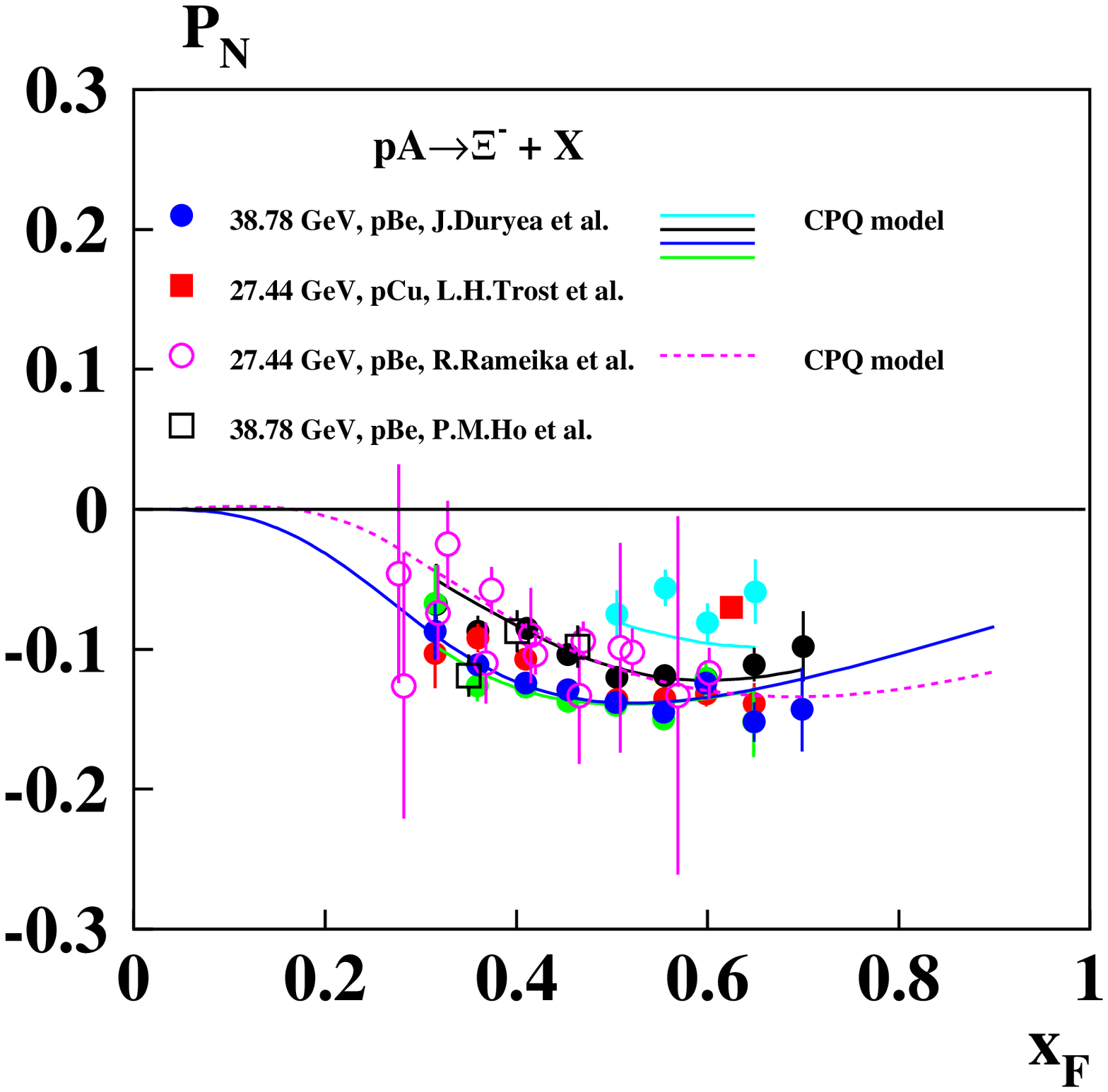} \\
    \textbf{(a)} & \textbf{(b)}
  \end{tabular}
  \caption{%
  $P_{N}(x_{\rm{F}})$ data and the CPQ model calculations   for the reaction
$p + p(A) \rightarrow  \Lambda^{\uparrow} + X$ \textbf{(a)} and 
$p + p(A) \rightarrow  \Xi^{-\uparrow} + X$ \textbf{(b)},
  taken from  \cite{Abramov:2020tne}.
  }
  \label{PNLamXim}
\end{figure}
		
		The $P_{N}(x_{\rm{F}})$ data for the reaction 
$p + p(A) \rightarrow  \Lambda^{\uparrow} + X$
 are shown in fig.~\ref{PNLamXim}a, 
and the data for the reaction 
$p + p(A) \rightarrow  \Xi^{-\uparrow} + X$
 are shown in fig.~\ref{PNLamXim}b, together with the CPQ model predictions
  \cite{Abramov:2020tne}.

As seen from fig.~\ref{PNLamXim}b, the $P_{N}(x_{\rm{F}})$ dependence for cascade hyperons is nonlinear function and $P_{N}(x_{\rm{F}})$ reaches its maximum absolute value at $x_{\rm{F}}$ in the range 0.5 - 0.6, in agreement with the calculations by the CPQ model. For the reaction $p + p(A) \rightarrow  \Lambda^{\uparrow} + X$,
 a close to linear dependence is observed, 
 since the parameter 
 $\nu_{A} = \nu_{B}  \approx 0.8746$ in this case is approximately two times smaller. The  maximum of the absolute value of polarization for the reaction $p + p(A) \rightarrow  \Lambda^{\uparrow} + X$
  is approximately twice that for $p + p(A) \rightarrow  \Xi^{-\uparrow} + X$,
   and continues to increase with increasing $x_{\rm{F}}$ up to 0.75, in agreement with the calculations in the framework of the CPQ model.

 Detailed calculations of 
$P_{N}(x_{\rm{F}})$ for reactions $p + A \rightarrow \Xi^{-} + X$ and 
$p + A \rightarrow \Xi^{0} + X$ can be found in~\cite{Abramov:2020tne},
    which also covers the energy range, available at the NICA collider.

\begin{figure}[hb!]
  \centering
  \begin{tabular}{cc}
      \includegraphics[width=55mm]{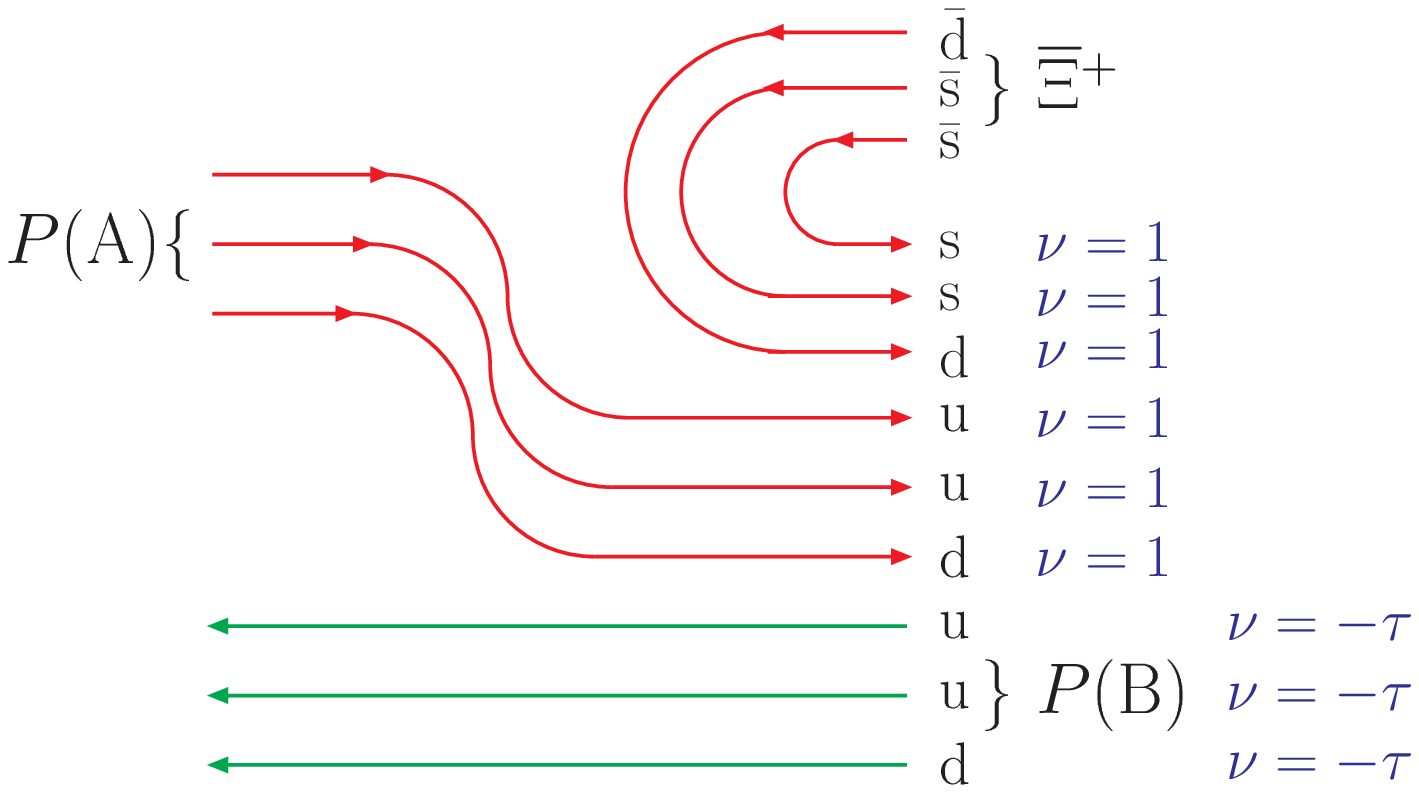} &
    \includegraphics[width=55mm]{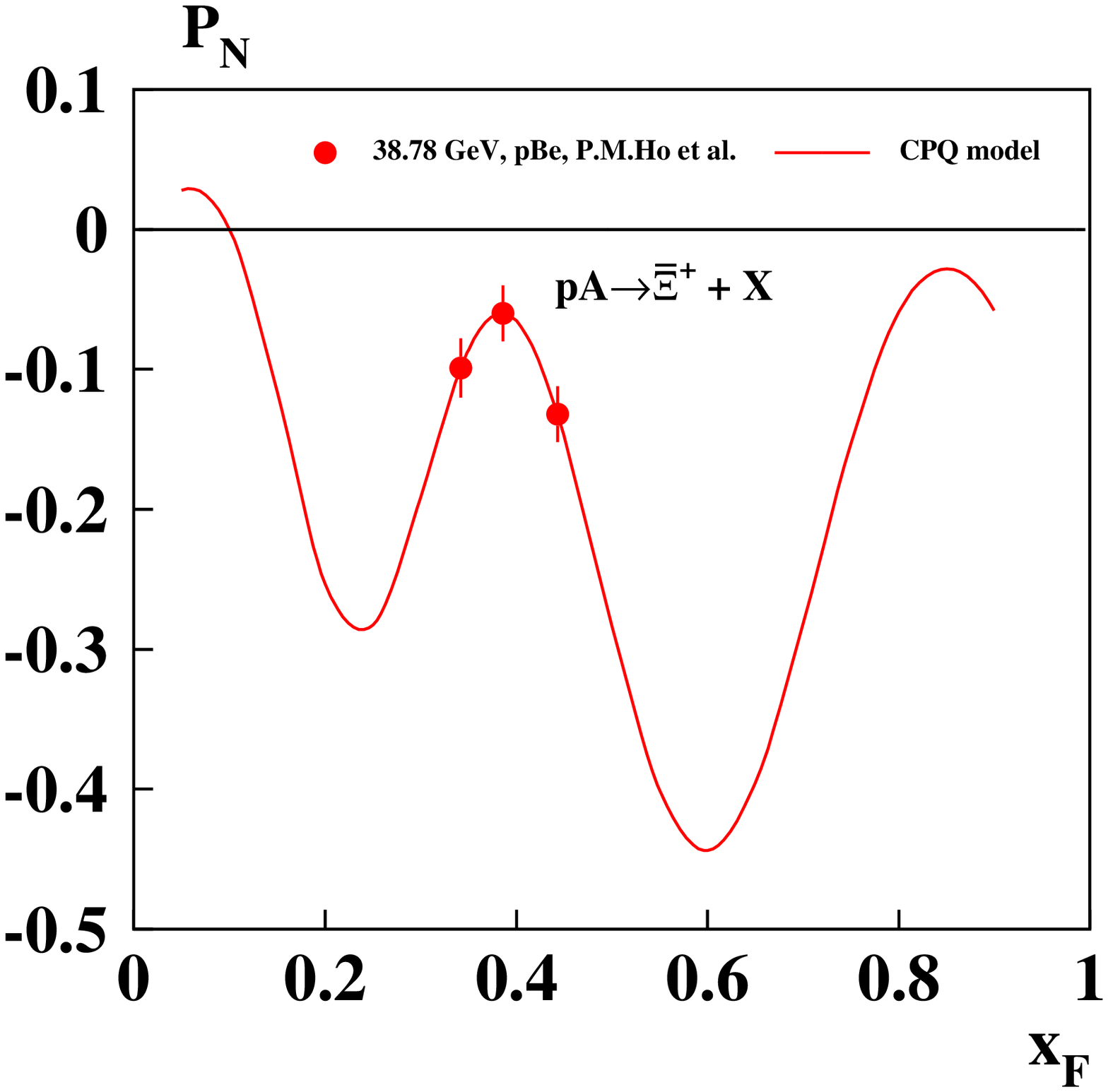} \\
    \textbf{(a)} & \textbf{(b)}
  \end{tabular}
  \caption{%
 Quark flow diagram \textbf{(a)} and
 $P_{N}(x_{\rm{F}})$  data  \cite{Ho:1990dd} \textbf{(b)}
 for the reaction 
$p + \rm{Be} \rightarrow \bar\Xi^{+} + X$, taken from   
\cite{Abramov:2020tne}. 
  }
  \label{DiagPNXiBar}
\end{figure}

The highest oscillation frequency $P_{N}(x_{\rm{F}})$ is expected, according to calculations by the CPQ model, in the reactions of antibaryon production in baryon collisions. This is due to the large number of spectator quarks from projectile (there are 6 of them, see fig.~\ref{DiagPNXiBar}a) accompanying the production of three antiquarks, which make up an antibaryon. There is a very limited set of data on the polarization 
$P_{N}(x_{\rm{F}})$ of antihyperons produced in nucleon-nucleon collisions. 
In fig.~\ref{DiagPNXiBar}a is \ shown quark diagram for the reaction $p + A \rightarrow \bar\Xi^{+} + X$.
 The weighted number of spectator quarks for both reactions is $\nu_{A} = \nu_{B} = 6 - 3\tau A_{eff} \approx 5.92$. This leads to a high oscillation frequency $P_{N}(x_{\rm{F}})$ according to (\ref{eq:omega}), so that in the range
$0 <x_{\rm{F}} <1$, several complete cycles can be observed.
%

%
In fig.~\ref{DiagPNXiBar}b are shown  the data for the reaction 
 $p + A \rightarrow \bar\Xi^{+} + X$  \cite{Ho:1990dd}. 
 There are also shown there the calculations
 of $P_{N}(x_{\rm{F}})$ according to the CPQ model~\cite{Abramov:2020tne}. Although the available data agree with the calculations of $P_{N}(x_{\rm{F}})$  using the CPQ model, the number of experimental points are clearly insufficient to prove the phenomenon of  $P_{N}(x_{\rm{F}})$ oscillations. New additional data are required in the range  $0 <x_{\rm{F}} <1$ to observe several cycles of
$P_{N}(x_{\rm{F}})$ oscillations.
Examples of $P_{N}(x_{\rm{F}})$ calculations for the reactions
  $p + A \rightarrow \bar\Xi^{+} + X$ and 
  $p + A \rightarrow \bar\Xi^{0} + X$
  can be found in~\cite{Abramov:2020tne}.

The effect of "attraction" in the polarization of antihyperons should manifest itself most clearly in the reaction 
$p + A \rightarrow \bar\Lambda + X$ \cite{Abramov:2014bha}. 
 The dependence of $P_{N}$ on the energy 
$\sqrt{s}$ of the resonance type is expected, 
with a maximum at $\sqrt{s} = E_{R} = 6.98$ GeV. 
This behavior $P_{N}(\sqrt{s})$ is
 based on a single non-zero $P_{N}$ report for reaction
$ p + A \rightarrow \bar\Lambda + X $, observed in experiment E766 
at $\sqrt{s} =7.31$ GeV \cite{Felix:1997ck}. 
It is very important to repeat such measurements that are within the energy range available at the NICA collider in
\it{p}\rm{+}\it{p}\rm, \it{d}\rm{+}\it{d}\rm,  \rm{C+C} and \rm{Ca+Ca} collisions. 
 The width of the "resonant" peak  is small, since the precession of only one test $\bar{s}$-quark is important in this case \cite{Abramov:2014bha}. In case of the reaction 
$p + A \rightarrow \bar\Xi^{+} + X$, 
 there are two $\bar{s}$-quarks and one $\bar{d}$-quark with different precession frequencies, which broadens the "resonant" peak.

Investigation of the dependences $P_{N}(\sqrt{s})$ of the "resonant" type and $P_{N}(x_{\rm{F}})$ of the "oscillating" type for the reaction
$p + A \rightarrow \bar\Lambda + X$
  is a very interesting problem affecting many aspects of strong interactions, such as color forces between quarks, precession of quark spin in a chromomagnetic field, quark counting rules for spectator quarks creating the field, anomalous chromomagnetic moment of quarks, the role of constituent (dressed) quarks in hadron interaction and  formation and quark confinement phenomenon.

 An example of possible studies of $P_{N}$ in collisions of ions 
can be found in~\cite{Abramov:2009tm}. 
It is shown that the higher is the atomic weight of the ions, the higher is the frequency of the oscillations, since the effective chromomagnetic field is increased by the quarks, coming from colliding ions. 
	
The only available data for the $A + A \rightarrow \Lambda X$ reaction in heavy ion collisions, where $P_{N}$ was measured, were used as input to the CPQ model.
    The data were obtained in a fixed-target experiment, where 
		$\Lambda$ was produced in Au+Au collisions at c.m. energy
		$\sqrt{s} = 4.86$ GeV \cite{Bellwied:2002rg}.

Already at the first stage of the SPD NICA project, it is possible to start studing the transverse polarization of hyperons and antihyperons in ion collisions. We also note the possibility of simultaneous measurement of the so-called global polarization with respect to the reaction plane. In this case the rotation of hadron or quark matter after collision of two nuclei leads to the hyperon polarization with respect to the reaction plane, determined by an impact parameter.

\label{sec:conclusion}
%
To conclusion, the study of single-spin polarization phenomena in the SPD NICA project makes it possible to reveal the regularities in the behavior of the single-spin asymmetry of hadrons and the transverse polarization of hyperons and antihyperons. Such studies are possible due to the $4\pi$ geometry of the SPD facility, a developed identification system, a fairly wide range of available energies, the presence of beams of polarized protons and deuterons, as well as ion beams.
Among the most interesting tasks on this topic are the following:

1) Measurement of $A_{N}$ and $P_{N}$ at several energies $\sqrt{s}$ in a wide range for $x_F$ and $p_T$, in order to separate the dependences on these three kinematic variables. The form of these dependences reflects the mechanism of the origin of polarization phenomena. These measurements should be carried out for as many reactions as possible, which is important for studying the dependence of $A_N$ and $P_N$ on the type of particles participating in the reaction. In general, this study will significantly expand the database available for theoretical analysis and discrimination of theoretical models.

2) Investigation of the scaling phenomenon for $A_N$ and $P_N$ and corrections to it, reflecting the peculiarities of the mechanism of the origin of polarization phenomena. 

3)Investigation of the threshold phenomena for $A_N$, including the measurement of the threshold angle of hadron production in the c.m. on which 
$A_N$ becomes null. 

4) Investigation of the phenomenon of $A_N$ and $P_N$ oscillations and the relationship between the oscillation frequency and the number of spectator quarks and the type of hadrons participating in the reaction. Particularly interesting in this respect are antihyperons and cascade hyperons, as well as secondary protons and neutrons, for which the oscillation frequency reaches a significant value, which facilitates its measurement. High oscillation frequency is expected also in heavy ion collisions. 

5) Investigation of the phenomenon of "resonance" dependence of $A_N$ and $P_N$ on energy $\sqrt{s}$. Disclosure of the mechanism of this phenomenon.

6) Study of the dependence of $A_N$ and $P_N$ on the atomic weights of the particles involved in collisions. This will allow not only to link the data obtained with different nuclei, but also to use the nuclei as tools for investigating the mechanism underlying the polarization phenomena. Research using ion collisions will provide a new insight into the phenomena previously studied in hadron-hadron collisions.  Until now
there is only one experiment in which the transverse polarization of a hyperon was measured in heavy ion collisions. Global polarization with respect to the reaction plane can be measured in addition to the 
$P_N$, which is measured with respect to the production plane. 

7) Additional possibilities for studying the mechanism of polarization phenomena are provided by the use of such variables as the multiplicity of charged particles in an event, as well as the centrality of collisions and the impact parameter in the case of collisions of nuclei.

The data obtained in the proposed studies will significantly expand the general world database on polarization measurements and become the basis for their systematic theoretical analysis, within the framework of a unified approach. One of the models that makes it possible to carry out a systematic global analysis of polarization data is the model of chromomagnetic polarization of quarks, which makes it possible to analyze various reactions in a wide range of kinematic and other variables that determine the experimental conditions. Global analysis of the entire dataset is suggested.

\medskip




\section{\bf Vector light and charmed meson production 
\protect\footnote{ This section is written by \,E.~Tomasi-Gustafsson;\,{E-mail: egle.tomasi@cea.fr}.}}

\begin{abstract}


In the context of NICA-SPD project, the motivation of the study of vector meson,  charm production (hidden) $ p+p\to p+p+V$, $V=\rho$,$\phi$, $J/\Psi$ and  open $N+N\to \Lambda_C(\Sigma_C) +\bar D +N$  is recalled. Backward vector meson production, that should be background free in a collider, can possibly be measured and be used also as an alternative method of producing neutron beams. Simple estimations of cross sections are presented on the basis of  existing literature. When possible, model independent statements on polarization effects are highlighted. 
\end{abstract}
\vspace*{6pt}

\noindent

PACS: 13.85.-t; 14.40.-n; 14.40.Lb

\subsection*{Introduction}

Among the wide possibilities opened by the future availability of the beams from the NICA collider and the operation of the large acceptance  SPD detector, we focus here on two issues: charm production (hidden and open) and backward vector meson production. The study of such channels will take full advantage of the possibility of accelerating polarized $p$, $d$ beams (as well as heavier ions) in a kinematical region where data are scarce on cross sections and polarization effects are mostly unmeasured. New, precise data will be extremely useful for the understanding of the mechanism of charm creation and of hadronic matter dynamics in the non-perturbative region of QCD. In general, threshold meson production channels in $NN$-collisions,  $p+p\to p+p + \omega(\phi)$, $p+p\to \Lambda(\Sigma^0)+K^+ +p $, and $p+p\to p+p + \eta(\eta ')$,  give deeper insight in the reaction mechanisms as it is shown by the experimental programs at different proton accelerators as SATURNE and COSY.

In this respect, $J/\psi$ production has a specific interest: the production and the propagation of charm in ion-ion collisions have been considered as one of the most promising probe of quark-gluon plasma (QGP) \cite{Matsui:1986dk}, but in order to state the evidence of a clear signal, it is necessary to analyze in detail all possible mechanisms for $J/\psi$ production in ion-ion collisions, and also all other processes which are responsible for the dissociation of the produced $J/\psi$ meson. The  study of charmonium (hidden strangeness) and $D$ $(D^*)$ mesons (open charm) are equally important.

\subsection{Charm production}
\label{sec:charm}
The elementary $pp$ cross section are collected and illustrated in Ref. 
\cite{Zyla:2020zbs}.

In the energy region that can be investigated with the NICA-SPD facility,  $3.4 \le \sqrt{s}$[GeV]$ \le 27$ \cite{Uzikov} for $pp$ collisions,  the total $pp$ cross section is relatively constant around 40 mb, whereas the  elastic cross section decreases due to the opening of different inelastic channels as the energy increases  \footnote{Fundaments of elastic $pp$ scattering up to LHC energies have been recently reviewed in Ref. \protect\cite{Pancheri:2016yel}  and references therein.}. The order of magnitude of the inelastic cross section can therefore be sizable, reaching 30 mb at the highest energies. Among these inelastic cross sections, the channels  $p+p\to p+p+ J/\Psi$ and 
$ p+p\to p+ \Lambda_C (\Sigma_C) + D $ open around $\sqrt{s_{Thr}}\sim 5$ GeV and they are expected to grow up to several $\mu b$ in the considered energy range.  

The  production mechanisms for  charmonium (hidden strangeness) and $D$ $(D^*)$ mesons (open charm) in nucleon-nucleon collision are not yet understood. The question is  how charm quarks - that are not preexisting in the nucleon as valence quarks - are formed and how they hadronize.  To interpret the production and the propagation of charm in heavy ion collision as a  probe of quark-gluon plasma (QGP), it is necessary to have a solid theoretical background based on the understanding of elementary processes. 

Experimental data and theoretical studies of $J/\psi$ production  in different processes and of its decays exist: for a review, see \cite{Vogt:1999cu} and for a most recent data collection \cite{Maltoni:2006yp}. As a result of high statistics and high resolution experiments a large information on the properties of the $J/\psi$ meson have been collected, on the production processes and on its numerous decays. From a theoretical point of view, the interpretation of the data, in particular in confinement regime, is very controversial. As an example, the $c-$quark mass is too large, if compared to predictions from chiral symmetry, but for theories based on expansion of heavy quark mass (Heavy Quark Effective Theory), this mass is too small \cite{Isgur:1989vq}.

In the threshold region, the final particles are produced in $S$-state and the spin structure of the matrix element is largely simplified. Simple considerations indicate that this region is quite wide: the effective proton size, which is responsible for charm creation, has to be  small, $r_c\simeq 1/m_c \simeq 0.13$ fm, where $m_c$ is the $c$-quark mass, selecting small impact parameters  \cite{Brodsky:2000zc}. The $S$-wave picture can therefore be applied for $q\le m_c$, where $q$ is the norm of the $J/\psi-$ three-momentum in the reaction center of mass (CMS).  The momenta of the produced particles are small, but the mechanisms for the production of charmed quarks must involve large scales. In Ref. \cite{Rekalo:2002ck}, the near-threshold $J/\psi-$ production in nucleon-nucleon collisions was analyzed in the framework of a general model independent formalism, which can be applied to any reaction $N+N\to N+N+V^0$, where $V^0=\omega$, $\phi$, or $J/\psi$.  Such reactions show large isotopic effects: a large difference for $pp$- and $pn$-collisions, which is due to the different spin structure of the corresponding matrix elements.  

In Ref. \cite{Rekalo:2002ck} an estimation of $J/\Psi$ production was suggested from the comparison of  the cross sections for the $\phi$ and $J/\psi $ production in $pp$ collisions. The same approach, namely $\pi$ exchange in $N+N\to N+N+V^0$ and $\rho$ exchange for the sub process $\pi+N\to N+V^0$, with $V^0=\phi$ or $J/\psi $ was considered. 
For the same value of the energy excess, $Q=\sqrt{s}-2m-m_V$, taking into account the different phase space volumes, coupling constants  for the decay $V\to \pi\rho$,  monopole-like phenomenological form factor for the vertex $\pi^*\rho^*V$, with virtual $\pi$ and $\rho$,
one finds the following simple parametrization for the cross section, holding in the near threshold region only:
\be
\sigma [nb]=0.097 (Q [\mbox{GeV}])^2.
\label{eq:JPSI}
\ee
In Ref.  \cite {Craigie:1978bp} a parametrization of exponential  form 
\be 
\sigma [nb]= a e^{-bM_{J/\Psi}/\sqrt(s)};
\label{eq:Craigie}
\ee
was suggested. The values $a$= 1000 [nb], and $b$=16.7 GeV reproduce well the experimental data over threshold.

\begin{figure}
  \begin{center}
\includegraphics[width=0.98\columnwidth]{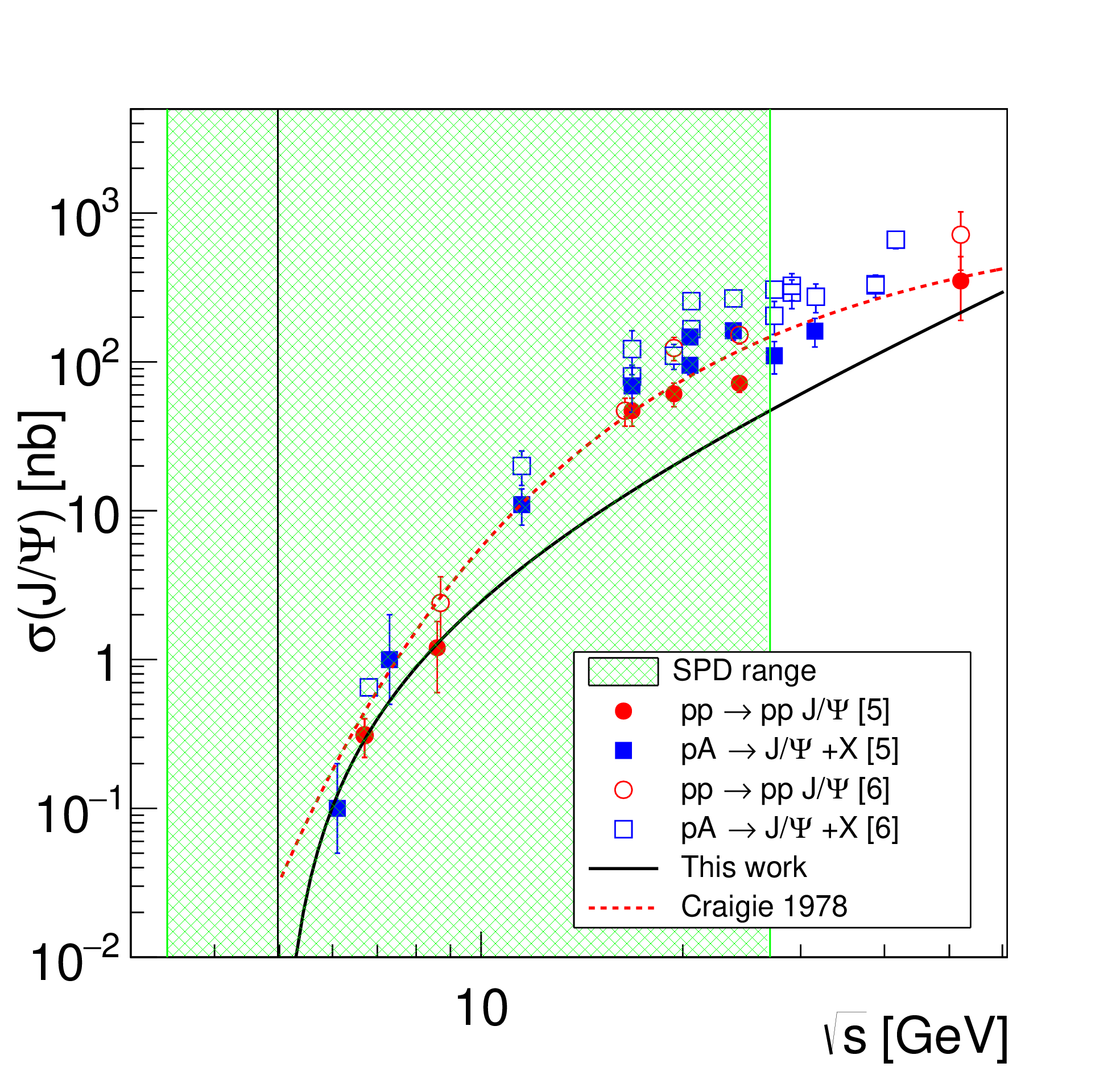}
  \caption{Experimental data on $J\psi$ production in $pp$ (red circles) and $pA$ (blue squares) reactions,  from the recollections in Refs. \cite{Vogt:1999cu} (filled symbols) and \cite{Maltoni:2006yp} (open symbols). The solid line is the calculation from Ref. \protect\cite{Rekalo:2002ck}. The plot is drawn from the $J/\psi$ production threshold (black line). The green filled region represents the range that can be investigated with NICA-SPD.}
 \label{Fig:jpsi}
    \end{center}
\end{figure}

The threshold for this reaction is $E_{th}$=12.24 GeV which corresponds to $\sqrt{s}=2m+m_{J/\psi}\simeq$ 4.97 GeV. In Fig. \ref{Fig:jpsi} the data for $p+p \to J/\psi + p+p$ (red circles) and $p+A \to J/\psi + X$ (blue squares) are plotted from the recollections in Refs. \cite{Vogt:1999cu} (filled symbols) and \cite{Maltoni:2006yp} (open symbols). Different symbols differentiate $J/\psi$ production in $pp$ or (extrapolated from) $pA$ collisions. The data, mostly collected at CERN, are reconstructed from the measurement using models and/or assumptions, and the compiled total cross section for $J/\Psi$ production may differ up to a factor of two. For example, the original reference for the measurement from Protvino at $\sqrt{s}=11.5 GeV$ \cite{ANTIPOV1976309} gives $\sigma (pp\to (J/Psi\to \mu+\mu^-)+X) =9.5\pm 2.5$ nb, whereas the same experimental point is referenced as  $\sigma =11\pm 3$ nb, in Ref. \cite{Vogt:1999cu} and $\sigma =20\pm 5.2$ nb, in Ref. \cite{Maltoni:2006yp}.  The cross section from Ref. \cite{Rekalo:2002ck} is also plotted in Fig. \ref{Fig:jpsi}  (solid line). 

Taking the value of luminosity ${\cal L} = 10^{30}$ cm$^{-2}$ s$^{-1}$, one expects 3 counts/hour  for such a process with a cross section of the order of 1 nb. This number is not corrected for the detector efficiency and reconstruction with identification, for example, in a missing mass. The reconstruction of $J/\Psi$ through its decay into a lepton pair, that is the preferred mode, requires two additional orders of magnitude as the branching ratio is $(\simeq 5.9\pm  0.5)10^{-2}$. 

Note also that in the framework of the considered model, one can find a large isotopic effect, due to the different spin structure of the matrix element at threshold: 
$$\displaystyle\frac{\sigma(np\to np J/\psi)}{\sigma(pp\to pp J/\psi)}=5,$$
which would require a correction of the experimental data on $pA$ reaction, where equal $np$ and $pp$ cross sections are usually assumed for the extraction of the elementary cross section in $pp$ collisions.
From Ref.  \cite{Rekalo:2002ck} one also learns that  only one polarization observable, the  $J/\psi$-polarization, is identical for  $pp$ and $pn$ collisions: the  $J/\psi$ meson is transversely polarized - even in collisions of unpolarized nucleons. The experimental determination of the ratio of the total cross sections for $np$ and $pp$ collisions gives important information for the identification of the reaction mechanism. 

The possibility of presence of intrinsic charm as a higher order component of the development of the Fock expansion for the proton state  
has been discussed in Ref. \cite{Vogt:1992ki}. Near threshold, all partons must transfer their energy to the charm quarks, within a time $t\sim 1/m_c$, thus selecting short range correlations between the valence quarks. Most interesting is the deuteron case, where all six quarks must be involved coherently, giving access to the hidden color part of the deuteron wave function.

\subsection {Open charm production}
\label{Section:opencharm} 
\begin{figure}
  \begin{center}
    \includegraphics[width=0.98\columnwidth]{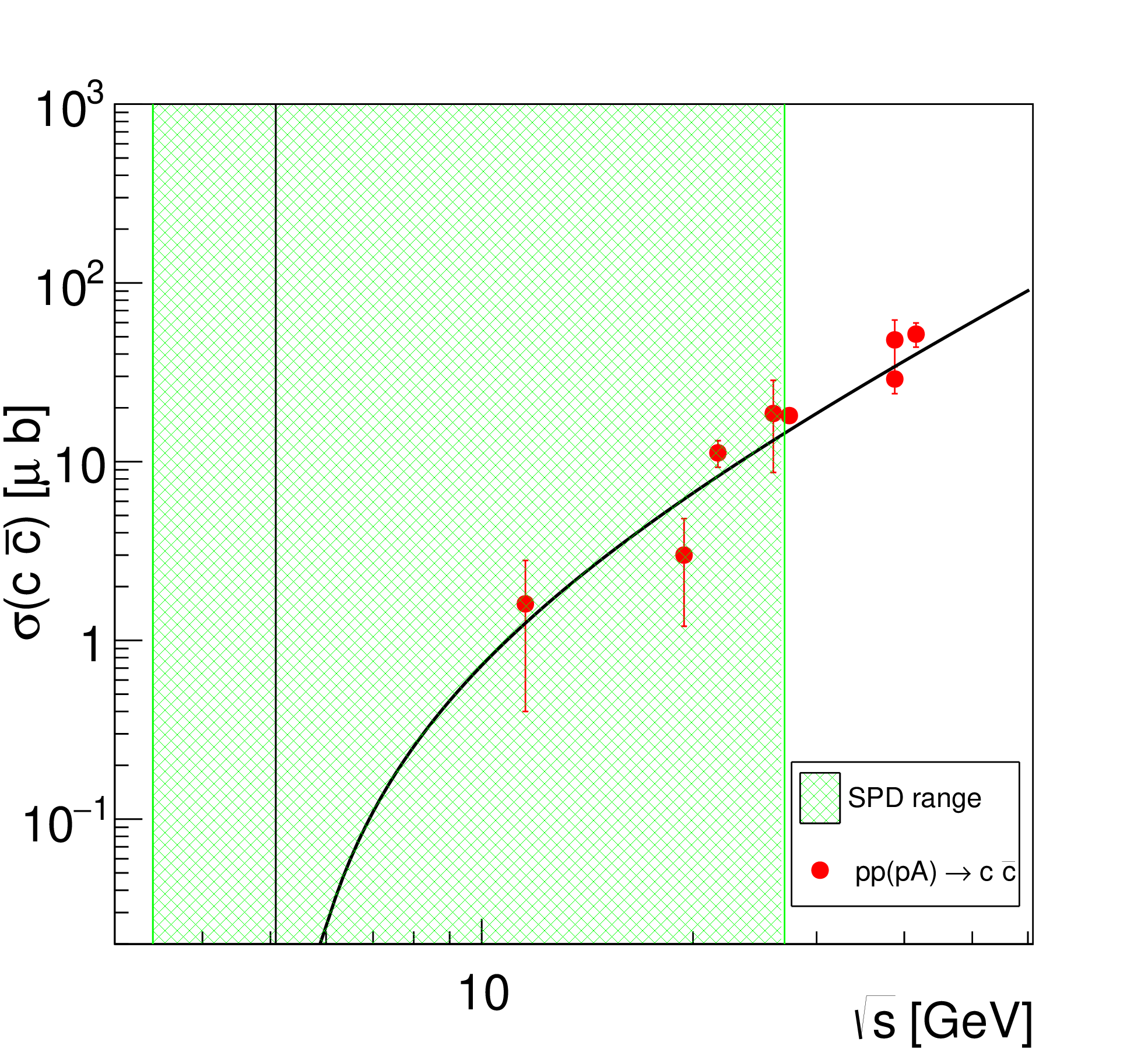}
  \caption{Total charm production in $pp$ and $pA$ collisions. Data are from  Ref. \protect\cite{Louren_o_2006}. The line is a threshold parametrization (see text).}
 \label{Fig:opencharm}
  \end{center}
\end{figure}

Open charm production, $N+N\to N+\bar{D}+\Lambda_C(\Sigma_C)$  gives information on scattering lengths, effective radius, hadronic form factors, and coupling constants and is also related to the dynamics of charm creation in $NN$, $NA$, $AA*$ collisions. Some predictions can be done from an analogy with strangeness production, relying on the equivalence of SU(3) and SU(4) symmetries, that is however, not totally reliable. Existing information and estimation indicate that near threshold cross section can be of the order of microbarns. The threshold cross section, normalized at the lowest existing value is plotted in Fig. \ref{Fig:opencharm}, where the insert highlights the threshold region. A dedicated simulation should be done to evaluate the counting rates, as the charmed particles should be reconstructed from the most suitable decay channels. 
The spin and isospin structure of the matrix element for the reactions $N+N\to \Lambda_C(\Sigma_C) +\bar D +N$  was derived for open charm in Ref. \cite{Rekalo:2002wg}. Detailed estimation of cross sections and the expressions of the polarization observables can be found there.

The  charm production near threshold cross section follows the behaviour: 
\be
\sigma [\mu b]=0.03 (Q [\mbox{GeV}])^2
\label{eq:open}
\ee
 that can be useful for simulation purposes. 
It is plotted in Fig. \ref {Fig:opencharm} over a collection of data from Ref. \cite{Louren_o_2006} reanalyzed from several experiments on charm production in $pp$ and $pA$ collisions at different facilities. 

We stress that these are difficult measurements, with low counting rates, but that even setting upper limits will be important, as no data at all are present in the threshold region.

\subsection{Backward meson production}
\label{Section:bmp} 
\begin{figure}
  \begin{center}
    \includegraphics[width=0.98\columnwidth]{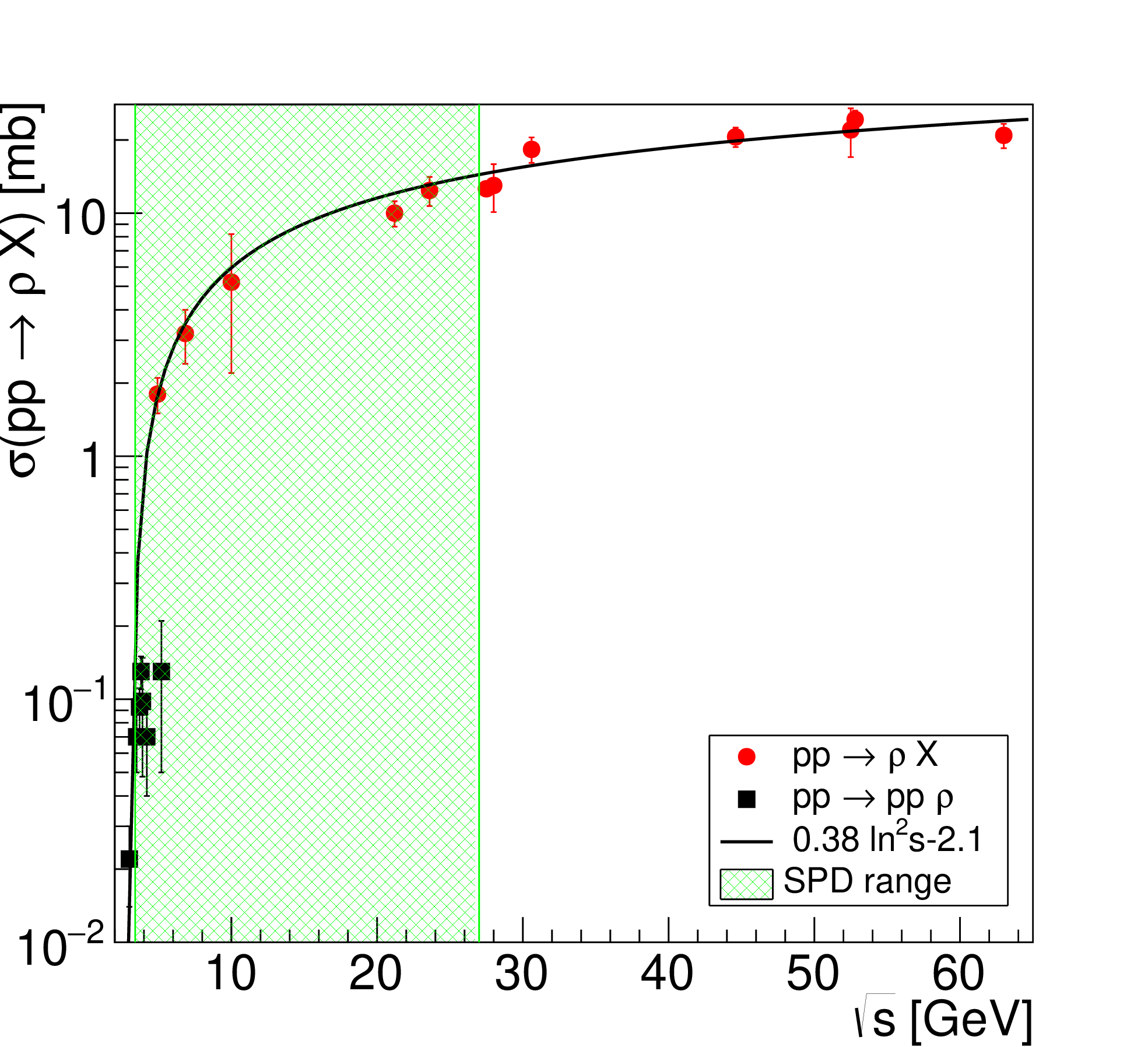}
  \caption{ $\rho$ production in $pp$ and $pA$ collisions. The red circles (black squares) are for inclusive (exclusive) $\rho$ production in different experiments.  The line is the  parametrization from Ref. \protect\cite{Albrow:1978uu}). The shaded region represents the SPD energy range.}
 \label{Fig:rhoinc}
  \end{center}
\end{figure}

Larger counting rates  are expected for light meson productions, since cross sections are of the order of mb. The $\rho^0$ meson production in elementary collisions and on nuclei has been discussed 
for example in Ref. \cite{Sibirtsev:1997ac} and references therein. The $\rho^0$ inclusive cross section has been measured at different accelerators since the 70s, mostly at CERN \cite{Blobel:1973wr}, and more recently by the HADES collaboration \cite{Rustamov:2010zz}. In Ref. \cite{Machavariani:2017erq}, the inclusive cross section for $\rho$ production in $pp$ collision is calculated in frame of a generalized vector meson dominance model, and the existing data up to $\sqrt{s}=65$ GeV are fairly reproduced and compared to other models. In Ref. \cite{Albrow:1978uu} the following parametrization was suggested 
\be 
\sigma (pp\to\rho^0 X) =  (0.38\pm 0.02)\ln^2 s -(2.1\pm 0.4).
\label{eq:Albrow}
\ee
This parametrization is shown together with the data for the inclusive cross section of 
$p+p\to \rho+X$ are in Fig. \ref{Fig:rhoinc}.
 One can see that it is of the order of the mb from the near threshold region, and therefore measurable at SPD already in the first phase of the experiment.

In Ref. \cite{Kuraev:2013izz} a specific kinematics,  the backward light meson production in $pp$ or $pA$ collisions, was discussed in similarity with the 'quasi real electron method', where a hard photon is produced on the collision of electrons on any target \cite{Baier:1973ms}.  Two important characteristics have been proved for the electron case:
-the collinear emission probability has a logarithmic enhancement - the cross section can be factorized in a term related to the probability of the meson emission  with a given energy at a given angle from the beam particle, and a term related to the interaction of the beam remnant after emission on the target. 

\begin{figure}
\begin{center}
\includegraphics[width=8cm]{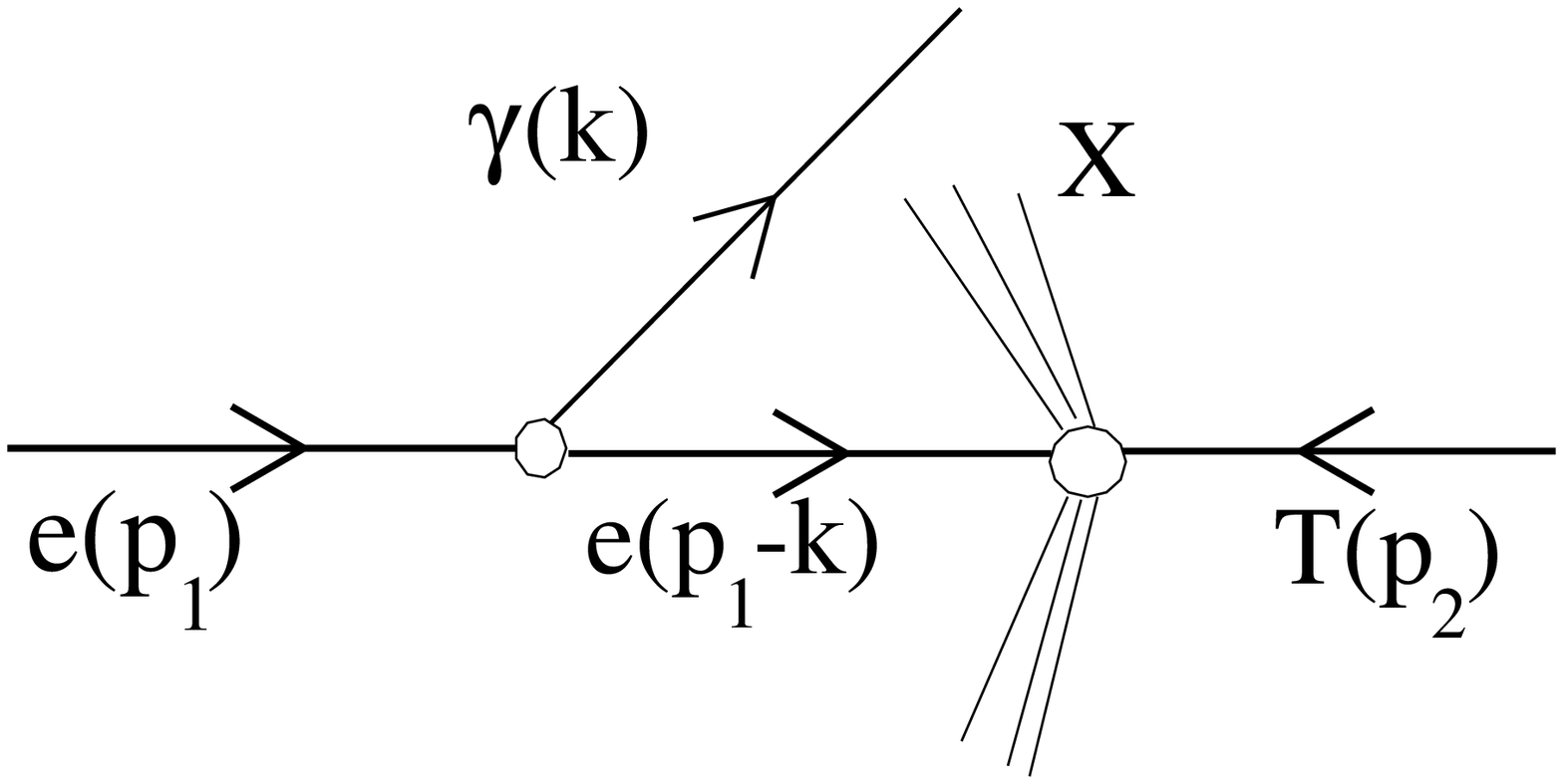}
\caption{Feynman diagram for collinear hard photon emission in $e T$ reactions (T stands for any target). The hadron equivalent is obtained by replacing the photon by a $\rho$-meson and the electron by a proton.}
\label{Fig:Fig2}
\end{center}
\end{figure}

The cross sections for the reactions of interest are:
\ba
d\sigma^{pT\to h_+X}(s,x)&=&\sigma^{nT\to X}(\bar{x}s)d W_{h_+}(x), \nn \\
d\sigma^{pT\to h_0X}(s,x)&=&\sigma^{pT\to X}(\bar{x}s)d W_{h_0}(x),
\ea
where $h$ is a hadron. The quantity $d W_{\rho}(x)$ can be inferred using the QED result:
\ba
\frac{dW_{\rho^i}(x)}{dx}&=&\frac{g^2}{4\pi^2}
\frac{1}{x}\sqrt{1-\frac{m_\rho^2}{x^2E^2}}\label{eq:eqrho} \\
&&
\left [\left (1-x+\frac{1}{2}x^2\right )L-(1-x)\right],\nn\\
&&\hspace{-2true cm} 1>x=\frac{E_\rho}{E}>\frac{m_\rho}{E}, 
L= \ln\left (1+\frac{E^2\theta_0^2}{M^2}\right ), \rho^i=\rho^+,\rho^-,\rho^0,
\nn
\ea
where $M$, $ m_\rho$, $E$, and $E_\rho$ are the masses and the energies of the initial proton and the emitted $\rho$-meson in the Laboratory system.

The integrated quantities $W_h$, $h={\rho,\pi}$ can, in general, exceed unity, violating unitarity. To restore unitarity, we have to take into account virtual corrections: the vertex for the emission of a single pion (charged or neutral) from the proton has to include 'radiative corrections', which account for emission and absorption of any number of virtual pions. For this aim we use the known expression for the probability of emission of $n$ "soft" photons in processes of charged particles hard interaction, $i.e.,$ the Poisson  formula for emission of $n$ soft photons $W^n=(a^n/n!)e^{-a}$ (where $a$ is the probability of emission of a single soft photon) \cite{AkhBer81}. 

The probability of emission of 'soft' neutral pions follows a Poisson distribution, which is not the case for the emission of charged pions. Fortunately, in our case, it is sufficient to consider the emission of one charged pion at lowest order (the process of one charged pion emission) plus any number of real and virtual pions with total charge zero. In such a configuration, this vertex has the form of the product of the Born probability of emission of a single pion multiplied by the Poisson-like factor:
\be
P_{\pi,\rho}=e^{-W_{\pi,\rho}},
\label{eq:probh}
\ee
which takes into account virtual corrections.

The final result is obtained using the replacement:
\be
\sigma(s) \to \sigma(s)\times {\cal R_{\pi}},~{\cal R_{\pi}}=P_\pi\sum_{k=0}^{k=n}\frac{W^k_\pi}{k!}, 
\label{eq:eqppi}
\ee
where ${\cal R}_\pi$ is the renormalization factor in order to account for the emission of $n$ real soft neutral pions escaping the detection.

\begin{figure}[t]
\begin{center}
\includegraphics[width=8cm]{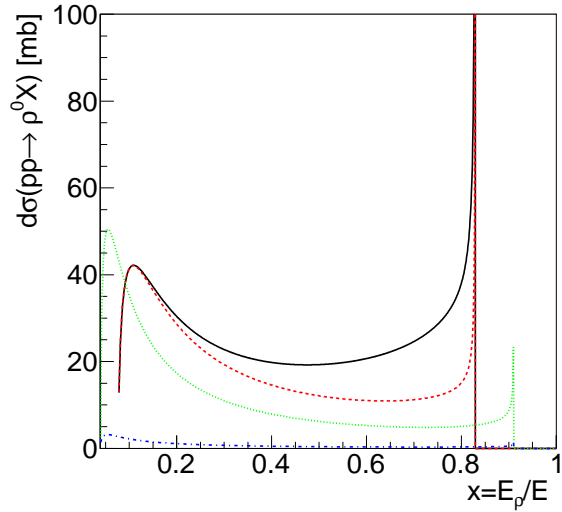}
\caption{Cross section $d\sigma(p,\bar p\to \rho^0 X)$ as function of the $\rho$ energy fraction for two values of the incident energy and of the $\rho$ emission angle: $E = 10$ GeV and $\theta_0=10^{\circ}$ (black, solid line),   $E = 10$ GeV and $\theta_0=20^{\circ}$ (red, dashed line), $E = 20$ GeV and $\theta_0=10^{\circ}$ (green, dotted line),   $E = 20$ GeV and $\theta_0=20^{\circ}$ (blue, dash-dotted line).}
\label{Fig:cs}
\end{center}
\end{figure}

Concerning the production of two charged pions, accompanied by a final state $X$, we can write:
\be
d\sigma^{p\bar p\to \rho^0 X}= 2\frac {d W_\rho(x)}{dx}\sigma^{p\bar p\to X}(\bar{x}s)\times P_{\rho},~
\label{eq:rhocs}
\ee
where the factor of two takes into account two kinematical situations, corresponding to the emission along each of the initial particles and $P_{\rho}$ is the survival factor (\ref{eq:probh}) which takes into account virtual radiative corrections. The cross section is shown in Fig. \ref{Fig:cs}
as a function of the $\rho$ energy fraction, for two values of the incident energy and of the  emission angle. 
The $x$ dependence shows a characteristic peak at $x=x_{max}$ that has the same nature as for the QED process 
$e^+ + e^-\to \mu^+ +\mu^- +\gamma$. As explained in Ref. \cite{Baier}, it is a threshold effect, corresponding
to the creation of a muon pair, where $x_{max}=1-4M_\mu^2/s$, $M_{\mu}$ is the muon mass.

\begin{figure}[t]
\begin{center} 
\includegraphics[width=8cm]{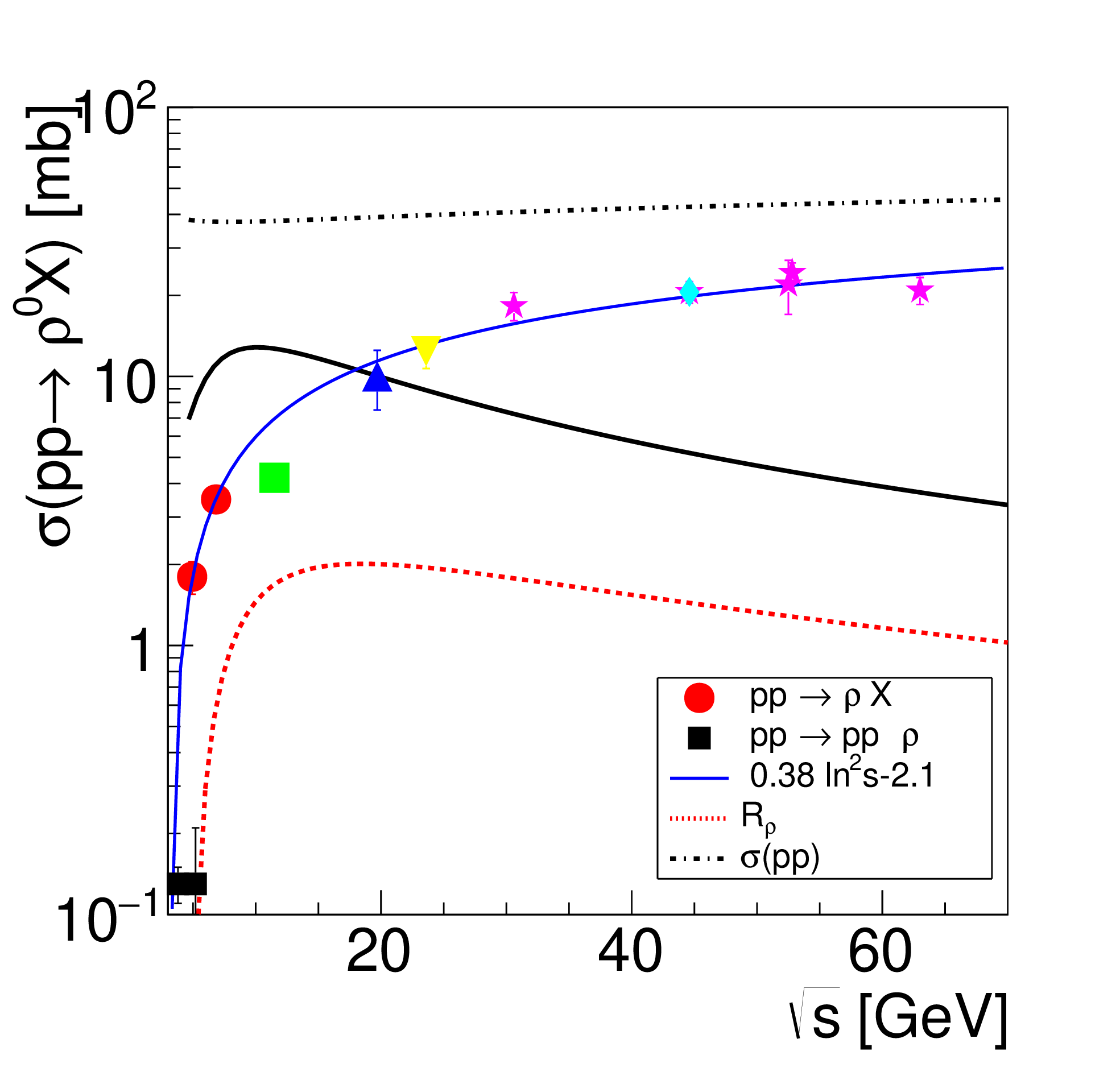}
\caption{Cross section for $\rho$-meson production in $p p $ collisions: inclusive (different symbols and colors from different experiments) and exclusive data from $pp\to pp\rho$ (black squares). The present calculation is shown as a black line. The red dashed line is the renormalization factor from Eq.  (\protect\ref{eq:eqppi}). The black dash-dotted line is the total $pp$ cross section.The first red point is the inclusive measurement from Ref. \cite{Rustamov:2010zz}. The blue line is the  parametrization from Ref. \protect\cite{Albrow:1978uu}. }
\label{Fig:rhofinal}
\end{center}
\end{figure}

The prediction of the model for backward  $\rho$-meson production in $p p $ collisions is shown in Fig. \ref{Fig:rhofinal}, as a black solid thick line. The red dashed line is the renormalization factor,  from Eq. (\ref{eq:eqppi}), integrated over $x$. The total $pp$ cross section is the black dash-dotted line, of the order of 40 mb, and it is quite flat in all the considered energy region. The blue line is the  parametrization from Ref. \protect\cite{Albrow:1978uu}) of the inclusive $\rho$ cross section.  The available data are also shown, as different symbols and colors for inclusive measurements and as black squares for exclusive $\rho$ production. Backward production can be of the order of several mb, therefore accessible at NICA-SPD also with the initial lower luminosity.

An original application is the possibility of creating neutron beams by tagging the incident proton beam with a negative meson emitted backwards. Charge exchange reaction takes place, and the beam remnant is a neutron impinging on the target beam. 

\subsection {Conclusions}
\label{Egle:conclusion} 

The understanding of charm production (open or hidden) should unify the different steps: parton-level hard process with production of $c\overline{c}$ pairs, after hadronization of
$c\overline{c}$ into $J/\psi$ or into charmed hadrons (mesons and baryons) including the final state interaction of the produced charmed hadrons with other particles. The relatively large transferred momenta involved in most processes of $J/\psi$ production in hadron-hadron collisions allow to treat the first step in framework of perturbative QCD. But the applicability of QCD is not so straightforward for the description of the $c$-quark hadronization. In this respect, precise data collected in the NICA-SPD energy range will bring important information, especially if covering a wide range above threshold. Light meson as $\rho$ meson production is definitely easier to be measured. Collecting precise, systematic data should help to refine the models and of great interest also for the collision on heavy targets. Backward kinematics could constitute an original  contribution to the field, offering an alternative possibility to produce neutron beams.

\section{\bf Exclusive hard processes with deuteron at NICA\protect\footnote{This section is written by   M. Strikman; E-mail: mxs43@psu.edu }}


\begin{abstract}
 
We argue that reaction $p^2H \to ppn$ at large momentum transfer to one of the nucleons of the deuteron provides a sensitive probe of space time evolution of hard  $pN$ scattering . The same process in a different kinematics allow to study short range correlations in the deuteron. Use of the polarized deuteron beams would provide a unique  opportunity to separate S- and D-wave contributions to the high momentum component of the deuteron wave function. A possibility to look for nonnucleonic components of the short range correlations is also outlined.
\end{abstract}
\vspace*{6pt}

PACS number(s) 13.75.Cs, 25.10.+s, 25.40Ep
\vskip 6mm
Our understanding of the dynamics of NN interactions at the energy range of
$\sqrt{s}\sim 5 \div 20\, GeV$ 
is still rather limited. In particular, it is not clear yet where transition occurs from nonperturbative to perturbative dynamics in few body  processes with a large momentum transfer ($-t$). This includes even the most basic  process of the large   $-t$ elastic nucleon - nucleon scattering at large $-t$. Among the puzzles are large spin effects  in large angle scattering of polarized  protons\cite{Crabb:1978km}  and a complicated  energy  dependence of the  nuclear transparency in large angle scattering of incident protons off the protons   embedded in nuclei \cite{Aclander:2004zm}. Also the recent observations of two nucleon short range / high momentum   correlations in nuclei mostly in electron - nucleus scattering (see review in \cite{Frankfurt:2008zv,Hen:2016kwk})  require confirmation and  testing universality of the SRCs using other projectiles - protons, photons, etc.

Questions involved in studies of the short-range / high momentum nuclear structure and understanding microscopic   nucleon structure and dynamics of large momentum transfer processes are delicately intertwined: 
understanding of hard dynamics of two body processes is also necessary for precision 
 studies of the short range nuclear structure.

Several strategies are possible   to address these questions. Here we will concentrate on reactions with the  deuteron  since the nonrelativistic deuteron  wave function is  well known and hence the measurements could be matched to detailed theoretical calculations. Also, the use of the deuteron  allows to choose special kinematic domains where $p^2H$ scattering is sensitive to  the short-range nuclear correlations. The collider kinematics presents a number of advantages as all particles in the reactions in question have large momenta and 
 hence can be easily detected.
 
 \subsection{Probing dynamics of nucleon - nucleon interaction in proton  - deuteron quasielastic scattering}
 
 The simplest reaction which would be interesting to study is the process
 $p^2H \to ppn$ where one of the nucleons has small transverse momentum and two others have approximately back to back large transverse momenta\cite{Frankfurt:1994nw, Frankfurt:1996uz}.  
 
 In the impulse approximation this process corresponds to elastic scattering of the  projectile proton off a quasifree nucleon of the target. There exist however 
  kinematical conditions where the dominant contributions are due to soft rescatterings of the initial and final nucleons, which accompany the hard $pp (pn)$ reaction. The eikonal approximation, which accounts for relativistic kinematics as dictated by the  Feynman diagrams, reveals the important role played by the initial and final state interactions in the angular and momentum dependences of the differential cross section in well defined kinematics. The condition for the applicability of the generalized eikonal approximation \cite{Frankfurt:1996xx} is  that 
 the c.m. scattering angle and invariant mass of the two nucleon system are large enough so that $-t, -u \ge \mbox{2 GeV}^2$.

 It was suggested in 
 \cite{Mueller, Brodsky}
 that nucleons in the elementary  reaction  interact in small size configurations with a small cross section - so called color transparency phenomenon. This effect is suppressed by the space - time evolution of nucleon wave packets \cite{Farrar:1988me,Frankfurt:1988nt}. However  effect of evolution  is very small for the deuteron where typical distances between nucleons in the rescattering amplitude  are is $\le$ 1.5 fm. Hence the discussed process allows {\it  to measure the wave packet size of a nucleon practically right in the interaction point}.

It was pointed out  that the hard dynamics in $pp$ and $pn$
elastic scattering may be rather different \cite{Granados:2009jh}.
Hence it would be instructive to compare the channels  where $pp$ and $pn$ 
are produced with  large $p_t$.

Experiments with polarized beams would greatly add to this program: a) study of the dependence of the cross section on the deuteron polarization allows to improve separation of kinematic domains where impulse approximation, double and triple scattering dominate, while the studies $\vec{p}\vec{d}\to  pNN$ processes will allow  both to study spin structure of $pp$ and $pn$ elastic  scattering   at large $t$ (the later is practically not known). Also, it would be possible to find out whether   the  
A.~D.~Krisch effect \cite{Crabb:1978km} (a strong difference between the cross sections of  elastic scattering of 
protons with parallel and antiparallel spins) involves collisions of protons  in configurations with  sizes depending  on the spin orientation.  

It would be possible also to study effects of coherence  in the channels where exchange by gluons in t-channel is not possible, for example $pd\to \Delta NN$. In particular, it would be possible to test the effect of chiral transparency suggested in \cite{Frankfurt:1996ai} - suppression of the pion field in the nucleons experiencing large $-t$ scattering.

\subsection{Probing microscopic deuteron structure}

As emphasized in the introduction, the dominant source of the SRC in nuclei are proton - neutron correlations with the same quantum numbers as the deuteron and with high momentum tail similar to that in the deuteron. Hence the deuteron serves as a kind of the hydrogen atom of the SRC physics. Only after it would be tested experimentally that approximations currently used for the description of the $p^2H$  reaction work well, it would be possible to perform high precision studies of SRC in heavier nuclei. 

 It was demonstrated in  Ref.\cite{Frankfurt:1994nw,Frankfurt:1996uz}  that 
    under specific kinematical conditions (in particular low transverse momenta of a slow nucleons in the deuteron rest frame))  the effect of initial and final state interactions can be accounted for by rescaling the cross section calculated within the plane wave impulse approximation.  
 In this kinematics it would be possible to check universality of the wave function - in particular its independence on the momentum transfer in the elementary reaction. Such factorization is expected to break down at sufficiently large $-t$ and  $-u$ where scattering involves interaction of nucleons in the small size configurations (the color transparency regime) since the small size configurations are suppressed in bound nucleons with suppression growing with the nucleon off shellness
\cite{Frankfurt:1988nt}.

Studies of the nonnucleonic configurations in   the deuteron as well as relativistic effects.
are of separate interest.
One option is 
a search  for non-nucleonic degrees of freedom like 6 quark, two$\Delta$ isobars via production reaction $p^2H\to \Delta^{++} +p +\Delta^- $
with 
 $\Delta^{++} $ and proton back to back and $\Delta^-$  in the deuteron fragmentation region for the light cone fraction 
 $\alpha_{\Delta}= 2p^{-}_{\Delta}/p^{-}_{d} \ge 1$ (slow in the deuteron rest frame;
 proton momentum is along $z$ direction).
 
 In a long run,  when polarized deuteron beams  would become available, 
 it would be possible to to  separate contribution of the S- and D- wave to the SRCs and compare different relativistic models of the deuteron -- light -cone vs virtual nucleon 
  approximations. The difference between  two predictions  is large already for the nucleon momenta   $\sim \mbox{300 MeV/c}$ in the deuteron rest frame.

\section{\bf Scaling behaviour of exclusive reactions with lightest nuclei and spin observables \protect\footnote{This section is written by 
V.P. Ladygin (E-mail: vladygin@jinr.ru) and  Yu.N.\, Uzikov}}
\label{uz-ladyg} 
\begin{abstract}
Differential cross sections of various binary reactions with the lightest nuclei
at  large fixed scattering angles 
are in qualitative agreement with the $s$- power-law dependence dictated by
the constituent counting rules.
We propose to measure differential cross section and deuteron analyzing powers of the $dp$-
elastic scattering at the  SPD NICA 
 to search for  transition region from meson-baryon to  quark-gluon  degrees of freedom in the
 deuteron structure.
 \end{abstract}
\vspace{5mm}

\noindent
PACS: 21.45.+v; 24.70.+s; 25.10.+s
\vspace{5mm}

 The structure of the lightest nuclei at short distances $r_{NN}<0.5$ fm or high relative momenta
 ($q> \hslash /r_{NN}\sim$ 0.4 GeV/$c$ constitues a fundamental problem in nuclear physics. One of 
  the most important questions is  related to search for onset of transiton region
  from meson-baryon to quark-gluon picture on nuclei.  A definite signature for transition to the
  valence quark region is given the constituent counting rules (CCR) \cite{Matveev:1973ra,Brodsky:1973kr}.  According the dimensional scaling the differential cross section
 of a binary reaction  at enough high incident energy can be parametized as
 $d\sigma/dt\sim s^{-(n-2)} f(t/s)$,
 where $n$ is the sum of costituent quarks in all participants, $s$ and $t$ are Mandelstam variables. 
 Many hard processes with free hadrons are consistent with CCR at energies of several GeV.
 The CCR propetrties of the reactions with the lightest nuclei were  observed
 in photodisintegration of the deuteron $\gamma d\to pn$ at $E_\gamma =1-4$ GeV and
 $^3He$ nucleus $^3He(\gamma,pp)n$, $\gamma ^3He\to dp$.  Earlier data on the reaction
 $dd\to ^3Hp$, \, $dd\to ^3Hen$  \cite{Bizard:1980aj} and $pd\to pd$, as was show in Ref.\cite{Uzikov:2005jd}
 also follow CCR  behavior $s^{-22}$ and $s^{-16} $, respectively, at surprising low energies,  0.5 GeV.  Recently the 
  CCR behaviour  of the reaction $pd\to pd$ was observed in
  \cite{Terekhin:2018gst,Terekhin:2019bsc} at
   higher energies. On the other hand, the reaction with pion production
   $pp\to d\pi^+$ does not follow CCR rule demostrating the differential
   cross section $\sim s^{-9}$ instead of $s^{-12}$. One possible way to explain this is  a partial restortaion of chiral symmetry 
   at enough high excitaion energy \cite{Uzikov:2016myj}.
   However, systematic study  of these properties of the  reactions with lightest nuclei are absent. So, important to know whether reaction 
   $pn\to d\rho^0$  follows the CCR  behavior and at what 
    minimal energy there is  the CCR onset. Assuming the model of the vector meson dominance and taking into account the observed   CCR behavior of the $\gamma d\to pn$ reaction, one may expect
    the $\sim s^{-12}$ dependence of the cross section of the  reaction  $pn\to d\rho^0$.
   Furthermore, possible  relation between CCR behavior of the upolarized cross  section and spin observables of the same reaction  are practically not known.  The SPD NICA  facility  provides a good opportunity  for this study using polarized beams  in $pp$, $dd$  and $pd$ collisions.


The results on the polarization observables for deuteron induced reactions at large transverse 
momenta are controversial.
Results on the tensor polarization $t_{20}$
for the $ed$- elastic scattering
\cite{Abbott:2000fg}
obtained at JLab can be 
reproduced quite well at $Q^2\le 1.7$ (GeV/c)$^2$ by the covariant
relativistic model
\cite{Carbonell:1998rj}
without contribution of
non-nucleonic degrees of freedom. 
The perturbative QCD (pQCD) predictions 
\cite{Brodsky:1992px}
are not reliable for 
these momentum transfers. On the other hand, while  
the electromagnetic form factors for the deuteron in the
soft-wall AdS/QCD model
\cite{Gutsche:2015xva}
are well in agreement with the
experimental data and the form factors display correct
$1/Q^{10}$ power scaling for large $Q^2$ which is consistent with
the CCR, the tensor analyzing power  $t_{20}$
\cite{Abbott:2000fg}
demonstrates 
the discrepancy with the soft-wall AdS/QCD  predictions 
\cite{Gutsche:2016lrz}.

The cross section of high energy two-body photodisintegration of the deuteron,  
$\gamma d\to pn$, at large angles in the cms 
\cite{Bochna:1998ca,Schulte:2001se}
has shown the scaling behavior 
up to 5.5 GeV  predicted by CCR
\cite{Matveev:1973ra,Brodsky:1973kr}.
Recent measurements of the proton 
polarization 
\cite{Wijesooriya:2001yu}
at  energies up to 2.4 GeV are also  consistent  with 
the pQCD hadron helicity conservation prediction \cite{Brodsky:1981kj}, but not the polarization
transfers obtained in the same experiment.

\begin{figure}[h]
\begin{minipage}[t]{0.47\textwidth}
 \centering
  \resizebox{7cm}{!}{\includegraphics{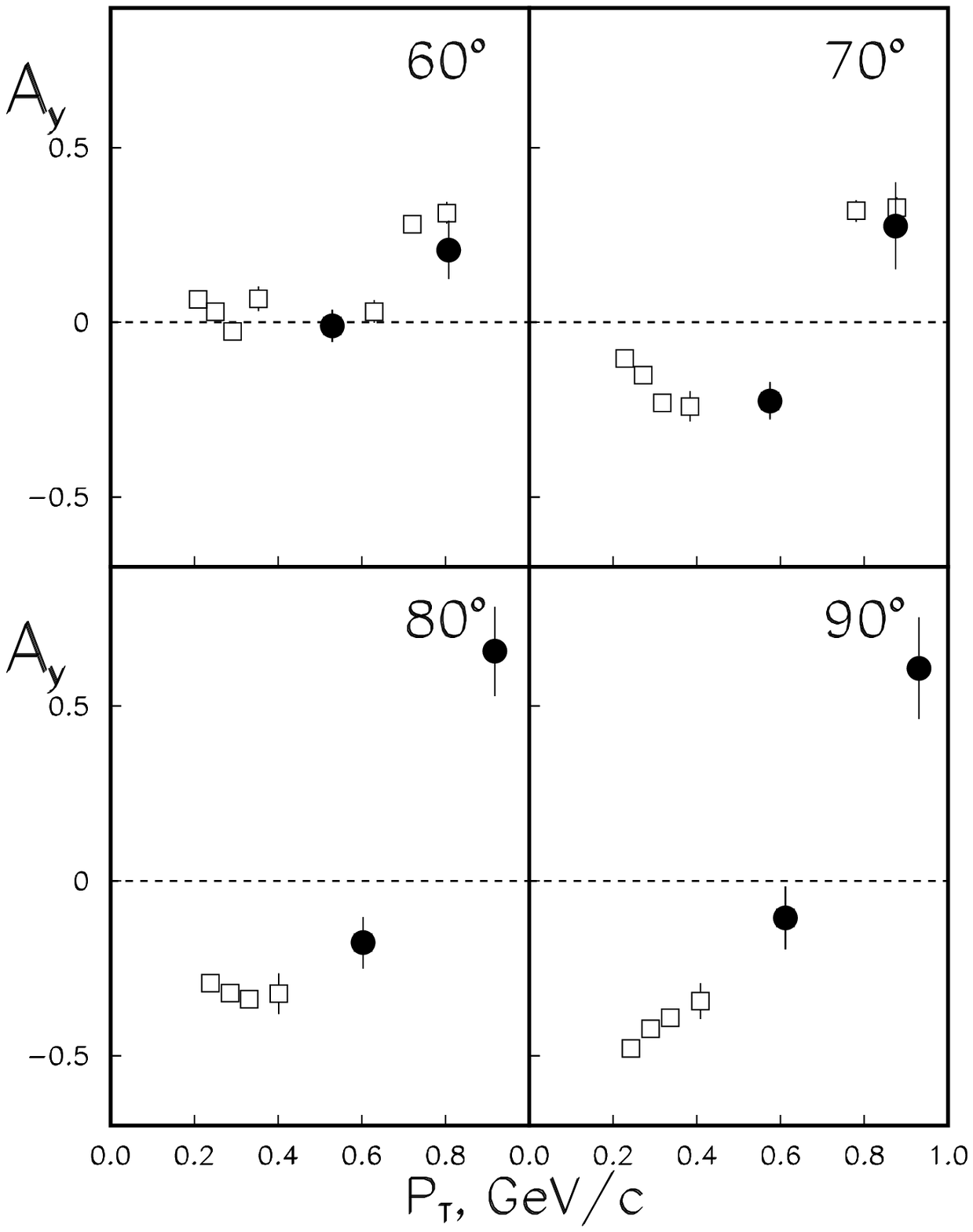}}
\end{minipage}\hfill
\begin{minipage}[t]{0.47\textwidth}
 \centering
  \resizebox{7cm}{!}{\includegraphics{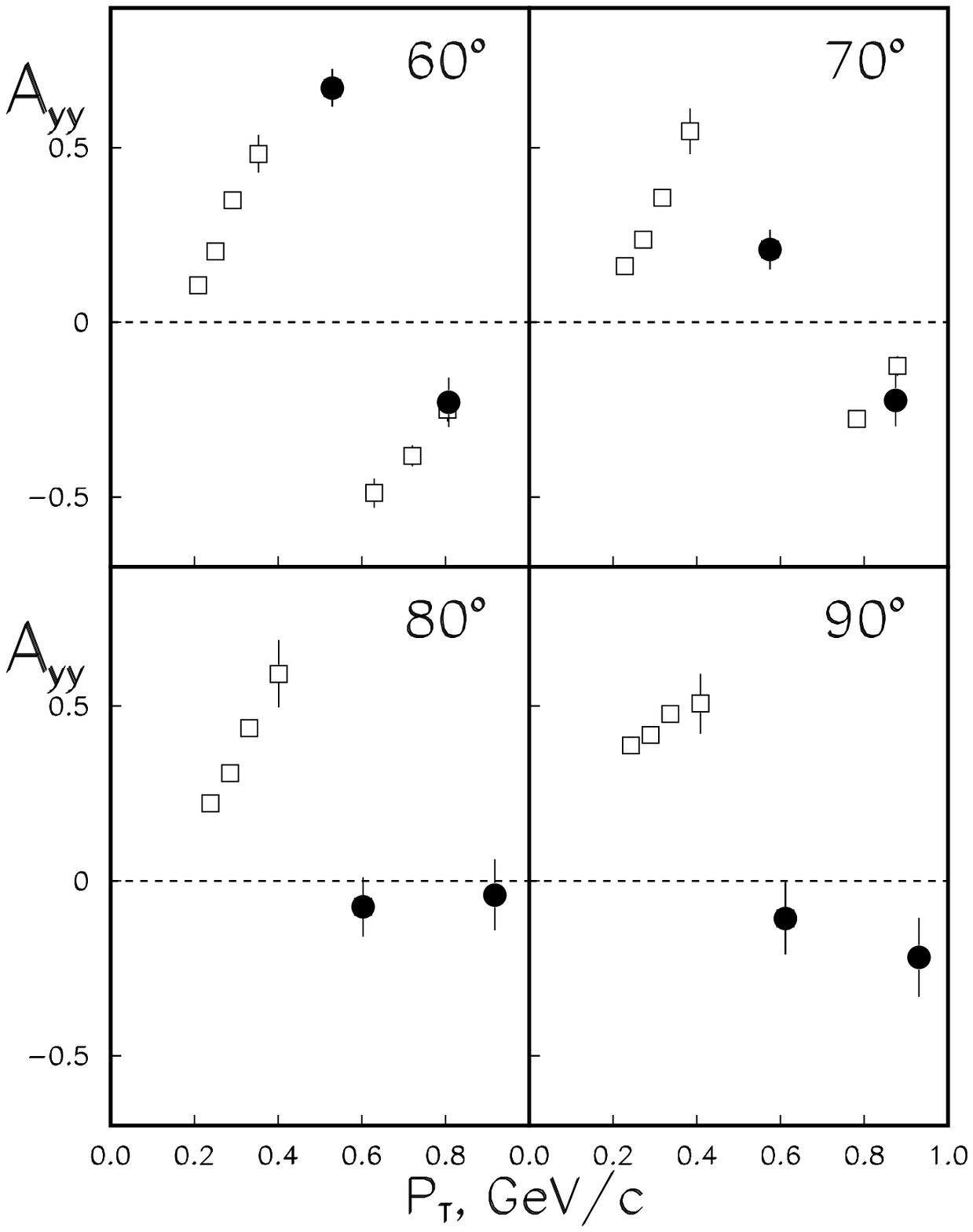}}
 \end{minipage}
\caption{Vector $A_{y}$ (left panel) and tensor $A_{yy}$ (right panel) analyzing powers
in $dp$- elastic scattering obtained at the
fixed angles in the cms: $60^\circ$, $70^\circ$, $80^\circ$ and $90^\circ$.}
\label{fig:ladygin_fig}
\end{figure}

The dependencies of the vector  $A_{y}$ and   tensor $A_{yy}$ analyzing powers 
in $dp$- elastic scattering
obtained at $60^\circ$, $70^\circ$, $80^\circ$ and $90^\circ$ in cms versus 
transverse momentum $p_T$ are shown in Fig.\ref{fig:ladygin_fig} in the left and
right panels, respectively. The open points are the world
data 
obtained at RIKEN, Saclay and ANL, while the black ones represent the results obtained at
Nuclotron 
\cite{Kurilkin:2012wk,Kurilkin:2011zz}
The analyzing powers  demonstrate the sign changes at $p_T\sim$ 650~MeV/$c$ and 
the asymptotic behavior at large transverse momenta. Note, that negative sign of
$A_{yy}$ is observed also in deuteron inclusive breakup 
at large $p_T$ 
\cite{Ladygin:2005wq},\cite{ayy}.


It would be interesting to extend the range of the measurements
to larger $p_T$, where the manifestation of non-nucleonic degrees of 
freedom is expected. New precise measurements with small statistical and systematic uncertainties
at the energies higher than  $\sqrt{s}\ge$3.3~GeV and at different scattering angles  are required to make a conclusion about the validity of CCR  \cite{Matveev:1973ra,Brodsky:1973kr} in $dp$ - elastic scattering. We propose to measure also different vector and tensor analyzing powers in $dp$-  elastic scattering at SPD energies. 


The tensor $A_{yy}$ and vector $A_{y}$ analyzing power in $dp$- elastic scattering
obtained at $60^\circ$, $70^\circ$, $80^\circ$ and $90^\circ$ in cms versus 
transverse momentum $p_T$ \cite{Kurilkin:2012wk,Kurilkin:2011zz} demonstrates the negative and positive asymptotics, respectively. Note, that negative sign of $A_{yy}$ is observed also in deuteron inclusive breakup 
at large $p_T$ \cite{Ladygin:2005wq},\cite{ayy}. It would be interesting to extend the range of the measurements
to larger $p_T$, where the manifestation of non-nucleonic degrees of 
freedom is expected. New precise measurements with small statistical and systematic uncertainties
at the energies higher than  $\sqrt{s}\ge$3.3~GeV and at different scattering angles  are required to make a conclusion about the validity of CCR \cite{Matveev:1973ra,Brodsky:1973kr} in $dp$ - elastic scattering. We propose to measure also different vector and tensor analyzing powers in $dp$-  elastic scattering at SPD energies. 

The measurements of $dp$- elastic scattering can be performed either with polarized deuterons and unpolarized protons, or with unpolarized deuterons and polarized deuterons. The $dp$- elastic
scattering events can be selected using cuts on the azimuthal and polar scattering  angles correlations. 
The vector $A_y$ and tensor $A_{yy}$ and $A_{xx}$ analyzing powers 
will be measured simultaneously in the case of the vertically polarized deuteron beam. 
The precision on the tensor   $\Delta A_{yy}\sim$0.09 and $ \Delta A_{xx}\sim$0.09 and on the 
vector $\Delta A_y\sim$0.03 analyzing powers can be achieved  for the scattering angle 
$\sim$90$^\circ\pm$5$^\circ$  
at $\sqrt{s}\sim$4.5~GeV ($p_T\sim$1.7~GeV/$c$) for 30 days of the beam time at the luminosity 
${\cal L}\approx$10$^{29}cm^{-2}\cdot s^{-1}$. We assume $\sim$75\% of the beam polarization from the 
ideal values of polarization for different spin modes.
The counting rate has been estimated using
$dp$- elastic scattering cross section
parameterization from Ref.\cite{Terekhin:2019bsc}.
The spin correlations can be obtained in quasi-free $dp$- elastic scattering using $dd$- collisions.

\section{\bf Multiquark correlations and exotic hadron state production
\protect\footnote{This section is written by V.T. Kim (kim\_vt@pnpi.nrcki.ru), A.A. Shavrin (shavrin.andrey.cp@gmail.com) and A.V. Zelenov (zelenov\_av@pnpi.nrcki.ru)}}



Multiquark correlations in the collisions of particles and nuclei at NICA energies play an important role in understanding of QCD. Processes involving multiquark degrees of freedom can shed light on various aspects from multiquark fluctons, diquarks, multiparton scattering  to exotic resonance production and fulfil a broad and rich physics program at SPD experiment.

\vspace*{6pt}

\noindent
PACS: 44.25.$+$f; 44.90.$+$c

\label{sec:multiquark_intro}
\subsection{Multiquark correlations and exotic state production at SPD NICA}

Multiquark correlations in the collisions of particles and nuclei play an important role in understanding of QCD. Multiquark correlation phenomena may be divided into four classes.

The first one can be related with parton distribution functions (PDFs) of the colliding nuclei. 
In the leading twist approximation, in the nuclear PDFs there is a contribution at large $x > 1$, which is related with objects fluctuations of nuclear matter~\cite{Blokhintsev:1957} known as multiquark fluctons~\cite{Efremov:1976xw,Lehman:1976xr,Burov:1976xd} (see for a review, e. g., Refs.~\cite{Efremov:1982tn,Burov:1984cb} and references therein) or few-nucleon short-range correlations 
(see, e. g., Ref.~\cite{Frankfurt:1981mk,Frankfurt:1988nt,Alvioli:2011aa} and references therein). Fluctons are compact multinucleon states with defrosen color quark-gluon degrees of freedom, see, e. g., ~\cite{Efremov:1976xw,Lehman:1976xr,Burov:1976xd,Matveev:1977ha,Matveev:1977xt,Blokhintsev:1978nf,Efremov:1978qy,Burov:1978zi,Smirnov:1978ux,Mulders:1980vx,
Obukhovsky:1980rh,Pirner:1980eu,Carlson:1983fs,Efremov:1987mx,Kaptar:1991vk}.

The second class is connected with higher twist contributions of two- and/or three- quark correlations in PDFs of hadrons and nuclei.

The third class is dealing with multiparton scattering subprocesses in hadronic and nuclear collisions. Multiparton scattering occurs
when two or more partons  from each colliding objects simultaneously scatter off each other.

The forth class phenomena can be related with production of exotic multiquark resonance states, e. g., pentaquark and tetraquark states. 

Below one can briefly outline possible studies, which can shed light on the all mentioned above classes of multiquark phenomena and fulfil a broad and rich physics program at SPD experiment. 

\label{sec:multiquark_fluctons}
\subsection{Multiquark correlations: fluctons in nuclei}

Fluctuations of nuclear matter were considered in Ref. \cite{Blokhintsev:1957}, which were discussed after discovering backward scattering off nuclei  \cite{Leksin:1956yh,Azhgirei:1957},  would form fluctons, compact multi-nucleon states or in other language, short-range few nucleon correlations.

Fluctons are directly connected with cumulative hadron production in the nuclear fragmentation region~\cite{Baldin:1971zga,Baldin:1974sh}. 
The flucton  approach~\cite{Efremov:1987mx}, which is based on hard QCD-factorisation and EMC-ratio constraints, predicts 
an extra nuclear quark sea, which has rather hard momentum distribution:   the extra nuclear sea $x$-slope is equal to the $x$-slope of the valence quarks at $x > 1$. It leads to "superscaling" for cumulative hadron production at $x > 1$ in the nuclear fragmentation region: the $x$-slope of all cumulative hadron distributions including "sea" ones~\cite{Efremov:1987mx,Kim:2018saa} are the same. The superscaling phenomenon  was experimentally confirmed by ITEP group~\cite{Boyarinov:1988ee,Boyarinov:1991up}.

Nuclear fluctons consist of the nucleons compressed in distances
comparable with nucleon size, so the flucton with five or six nucleons could be considered as a cold dense baryon matter since the effective nuclear density would be high as that in the core of neutron stars~\cite{Frankfurt:2008zv}. In such a dense nuclear medium there would be $CP$-violating effects~\cite{Andrianov:2016qgy}.  

In high-$p_T$ cumulative processes at the central region, other contributions  should be added to the contribution of the nuclear PDFs at $x > 1$, such as the contributions from the PDFs of the other colliding object and possible intranuclear rescattering effects~\cite{Efremov:1985cu,Braun:1994bf}. So, beyond the nuclear fragmentation region one should observe deviations from superscaling for cumulative production.

For many observables there are two popular approaches: fluctons and short-range nucleon correlations yield similar predictions, e. g., cumulative particle production, nuclear structure functions in deep inelastic processes with leptons, etc. However, there is the main difference with flucton and   short-range nucleon correlation approaches: extra sea quark degrees of freedom atributted to flucton sea in nuclear PDF at  $x > 1$.

The hard flucton antiquark sea can manifest in massive lepton pair \cite{Matveev:1969zz,Matveev:1969px,Drell:1970wh} and $J/\psi$ production in cumulative region~\cite{Zotov:1986ct,Harindranath:1986ke}. SPD can study $J/\psi$ production process in $pd$ and $dd$ collisions in $x > 1$, which should be highly sensitive for hard flucton antiquark sea of deutrons.

\label{sec:diquarks}
\subsection{Few-quark correlations: Diquarks}

\begin{figure}[t]
\begin{center}
\includegraphics[width=100mm]{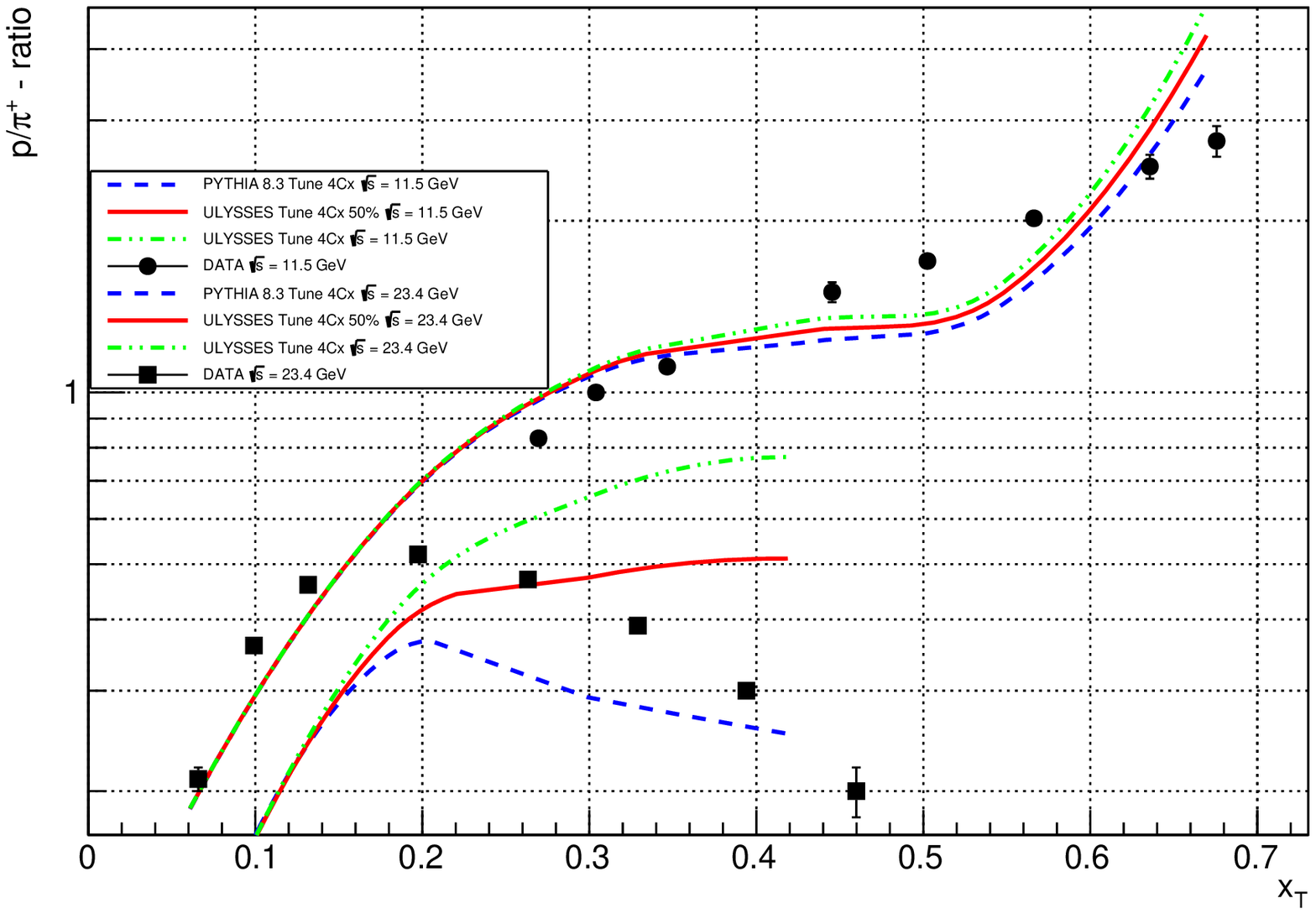} 
\end{center}
\vspace{-3mm}
\caption{$p$/$\pi^+$-ratio of inclusive invariant cross sections at $\theta_{cms}=90^0$ in $pp$-collisions confronted with the data from  IHEP, Protvino~\cite{Abramov:1985ct}  ($\bullet$) at $\sqrt{s} = 11.5$ GeV and from  FNAL, Batavia~\cite{Antreasyan:1978cw} ($\blacksquare$) at $\sqrt{s} = 23.4$ GeV. The preliminary predictions from MC event generator ULYSSES~\cite{Kim:2020a} with incorporated diquark subprocesses: the solid red line corresponds diquark state probablity in proton equals 0.5, the dash-dotted green line - 1.0. Prediction by the standard PYTHIA8.3 4Cx~\cite{Sjostrand:2014zea} without diquark subprocesses is shown by the blue dashed line.}
\label{fig01kim}
\end{figure}

Another type of multiquark correlations is few-quark short-range correlations: diquark and triquark states in baryons. Diquark states were discussed soon after suggesting quark model for hadrons, see for a review Refs.~\cite{Anselmino:1992vg,Barabanov:2020jvn}. This is an important source~\cite{Laperashvili:1982xj,Larsson:1984sd,Breakstone:1985ef,Kim:1987ec,Kim:1988gg} of large-$p_T$ baryon production~\cite{Antreasyan:1978cw,Abramov:1985ct,Breakstone:1985ef}. Being a higher-twist the diquark contribution can describe  
the strong scaling violation in deep inelastic scattering off nucleons and in large-$p_T$ baryon production in hard nucleon collisions at SPD energies~\cite{Kim:1987ec,Kim:1988gg,Kim:2020a}
(Fig. \ref{fig01kim})

Most of diquark studies were done within inclusive approach. ABCDHW Collaboration at CERN ISR have found more convincing indications ~\cite{Breakstone:1987ca,Geist:1989gw} on diquark manifestations in two-particle correlations.
SPD allows to study two-particle correlations between large-$p_T$ baryons (proton, $\Delta^{++}$, $\Lambda^0$, etc. ) and back-to-back hadrons  to reveal  of quark structure of proton in more detail.

Therefore, measuring by SPD experiment various two-particle correlations 
should reveal more detail diquark structure of proton.

\newpage
\label{sec:multiquark_scattering}
\subsection{Multiparton scattering}

There is an important aspect of hadronic collsions, which was studied since 70s ~\cite{Landshoff:1973wb,Landshoff:1974ew,Landshoff:1980pb,Goebel:1979mi,Paver:1982yp,Paver:1984ux,Mekhfi:1983az,Humpert:1984ay}. There are two basic approaches for description of multiparton scattering, depending on view whether it is considered as a higher twist effect or not~\cite{Berger:2009cm,Blok:2011bu,Ryskin:2012qx,Diehl:2011yj,dEnterria:2016ids}. In the all approaches the main object is momentum multiparton momentum-dependent distribution function, which is complicated in addition by the multiparton distribution in space (impact plane).
Therefore, measuring few-particle correlation at SPD one can study multiparton scattering processes~\cite{Kim:2020a}, which are related with 2D- and 3D- PDFs in momentum and coordinate spaces.    

\begin{figure}[!ht]
\centering
\includegraphics[width=0.60\textwidth]{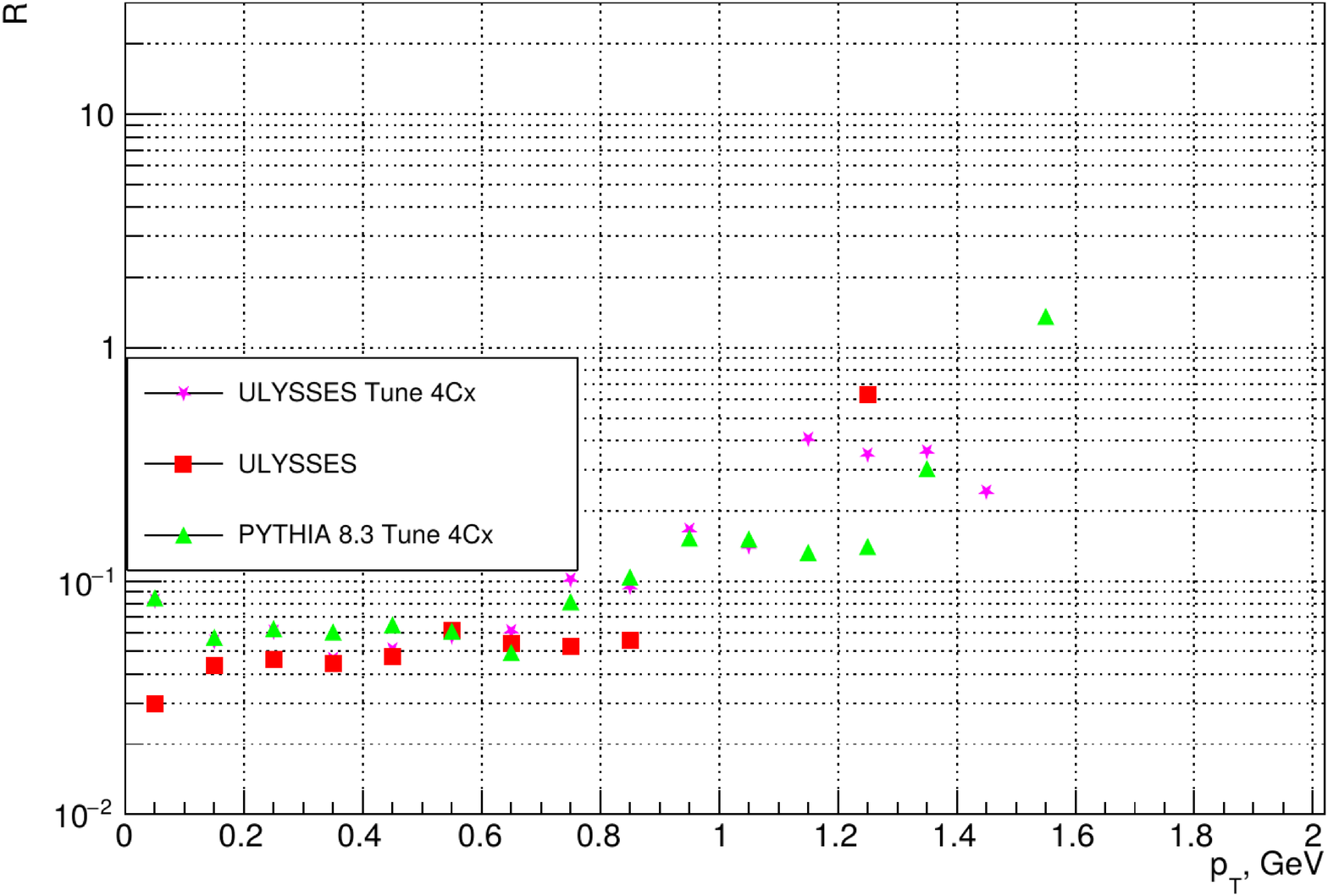}
   \\ (a) \\
\includegraphics[width=0.60\textwidth]{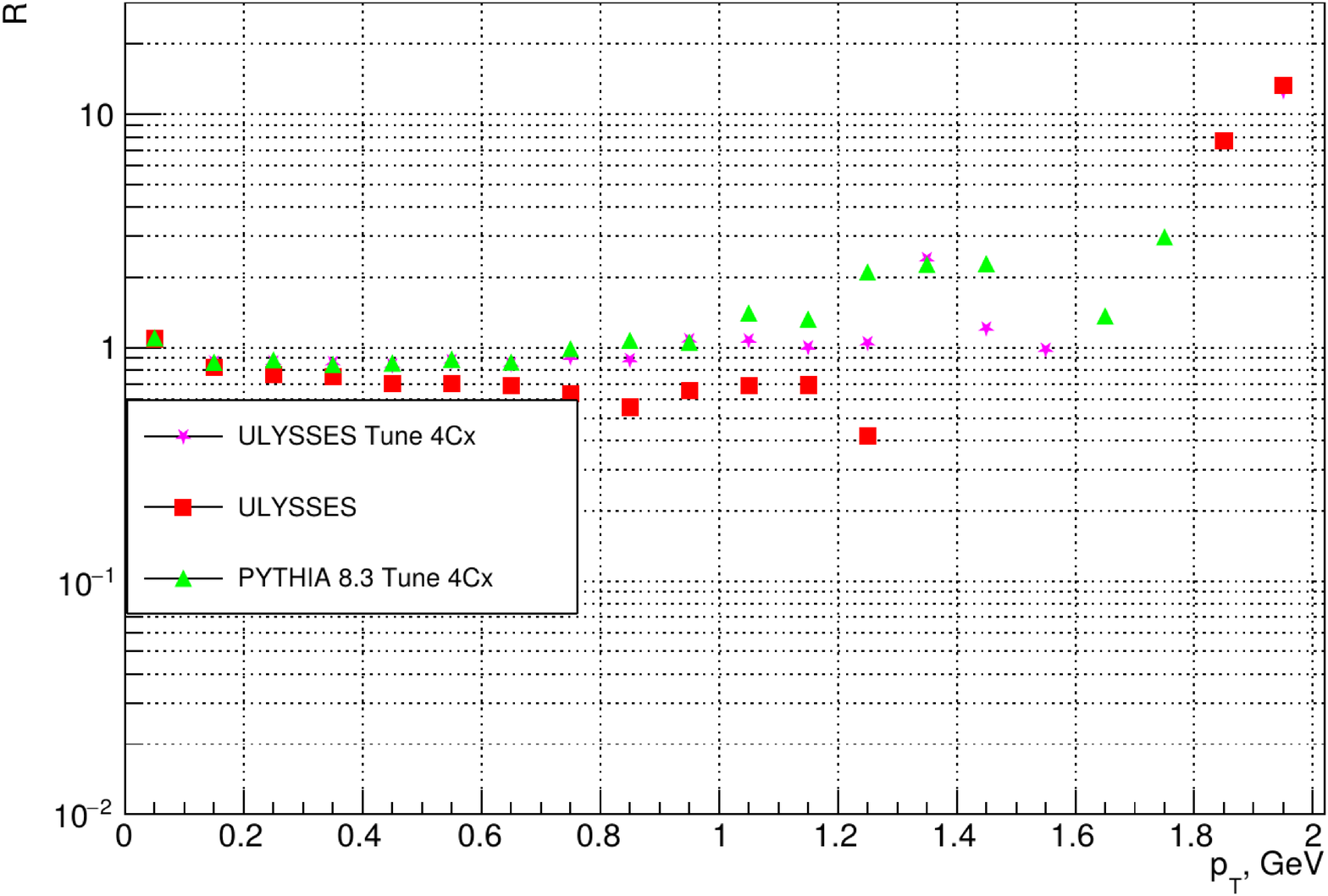}
   \\ (b)
\caption{Two-proton correlation function at $\theta_{cms}=90^0$ in $pp$-collisions at $\sqrt{s} = 23.4$ GeV as a function of $p_T$ ($\approx p_{T 1} \approx p_{T 2}$) for two regions of azimuthal angle difference: (a) for back-to-back region ($\Delta \varphi = |\varphi_1 - \varphi_2| \approx \pi$) and (b) out of it ($\Delta \varphi \not \approx \pi$). The preliminary predictions from MC event generator ULYSSES~\cite{Kim:2020rwn} with incorporated diquark subprocesses in different tunes (red squares and magenta stars) and PYTHIA8.3 4Cx~\cite{Sjostrand:2014zea} (green triangles) without diquark subprocesses.}
\label{fig02}
\end{figure}


SPD experiment can study multiparton distribuion in proton via multiparton scattering, which yield, in particular, certain azimuthal two-particle correlations. For instance, double-parton scattering should increase two-particle correlations
out of back-to-back scattering region, where single parton-parton scattering is dominant~\cite{Efremov:1987mx} yielding a ridge structure \cite{Dremin:2010yd}.
Predictions from MC event generators ULYSSES ~\cite{Kim:2020a} with incorporated diquark subprocesses and PYTHIA8.3 for two-particle correlation function of two protons 
$$R_{12}= \sigma^{in} \frac{\frac{E_1 E_2 d^6 \sigma}{d^3 p_1 d^3 p_2}} {\frac{E_1 d^3 \sigma}{d^3 p_1} \frac{E_2 d^3 \sigma}{d^3 p_2} }$$
is shown in 
Fig.~\ref{fig02} 
for back-to-back region and outside it at $\sqrt{s} = 23.4$ GeV.

Two-particle correlation function for out of back-to-back azimuthal region is directly connected with double-parton scattering cross section.

Multiparton scattering can be significant for production of multiquark systems and and nuclei~\cite{Efremov:1987eh,Kim:2020a}. 

\label{sec:multiquark_states}
\subsection{Multiquark exotic state production}

Multiparton scattering provide a unique opportunity to study production of various multiquark states and light nuclei~\cite{Efremov:1987eh,Kim:2020a} at SPD energies.

An excitement has been flourished when LEPS~\cite{Nakano:2003qx}, DIANA~\cite{Barmin:2003vv} and CLAS~\cite{Stepanyan:2003qr} Collaborations observed a resonance state, which was  correspondent to quantum flavour numbers of  $\Theta^+$-pentaquark, 
with the narrow width as predicted in Ref.~\cite{Diakonov:1997mm}.

However, in a meantime there
was a somewhat contradictory situation. Most of $\Theta^+$-pentaquark observation claims were ruled out while the others \cite{Barmin:2003vv,Asratyan:2003cb,Aleev:2004sa,Aleev:2008wd} were still not, see for a review of the experimental situation, e. g., Ref.~\cite{Danilov:2005kt,Danilov:2007bp}. There were also papers tried to explain why one experiments could observe the $\Theta^+$-pentaquark state while the others would not, see, e. g., Ref.~\cite{Azimov:2007hq}. A similar situation with the other multiquark states: H-dihyperon \cite{Jaffe:1976yi,Dorokhov:1986ew}, etc. \cite{Barabanov:2014lka,Barabanov:2016jjv}.

\begin{figure}[th]
\begin{center}
\includegraphics[width=55mm]{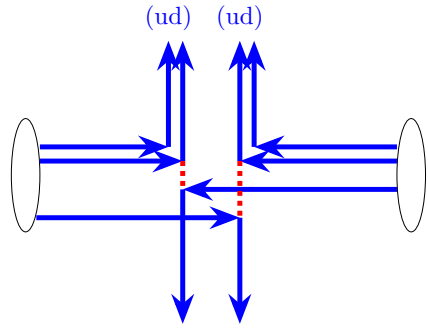} 
\end{center}
\caption{Production of multiquark systems (pentaquarks,H-dihyperons, etc.) and deutrons on large angles via double diquark-quark scattering~\cite{Efremov:1987eh,Kim:2020a} in nucleon collisions.}
\label{fig03}
\end{figure}

Multiquark state production cross section depends from its 
 structure and production mechanism. For instance, the atomic or molecular structures lead to
different mechanisms of production and to different branching ratios of the final states.

Production of multiquark systems with possible diquark 
structure~\cite{Jaffe:2003sg,Maiani:2004vq,Bashkanov:2013cla,West:2020rlk,Kim:2020rwn} on large
angles can be enhanced due to double diquark-quark scattering~\cite{Efremov:1987eh,Kim:2020a}. Inclusive and two-particle correlation studies with production of light nuclei and antinuclei should help establishing two production mechanism stages: multiparton scattering and fragmentation/fusion process ~\cite{Efremov:1987eh,Kim:2020a} 
(Fig. \ref{fig03}).

So far, most of the negative pentaquark and H-dihyperon searches were performed at small production angles. Therefore, SPD experiment provides an uniqe opportunity to search such multiquark states at unexplored kinematic domain.

\label{sec:Multiquark Correlations Summary}
\subsection{Summary}

To summarise this section: 
\begin{itemize}
\item
SPD with study of the inclusive 
particle production and few-particle correlations at different kinematic regions in $pp$-collisions at high luminosity should reveal such quark correlations as diquark states in proton;

\item
Multiparton distribution in proton can be studied via multiparton scattering, which yield certain azimuthal two-particle correlations;
In particular, double-parton scattering should increase two-particle correaltions
out of back-to-back scattering region;

\item SPD can carry out investigations of the novel production mechanisms as double quark-diquark scattering and the other multiparton scattering subprocesses, which would lead to enhanced production of exotic  multiquark resonance states and light nuclei at large angles;  

\item
$pd$- and $dd$-collisions at high luminosity can shed light on the difference between multiquark fluctons and short-range nucleon correlations by measuring cumulative production of antiprotons and $K^-$ mesons. Collision processes with light nuclei, if available, should enhance sensitivity to the  hard flucton antiquark sea.  
\end{itemize}

Therefore, SPD experiment at NICA has a unique opportunity, from one side, to test and extend our knowledge of various aspects of multiquark correlations: from diquark substructure of proton, double-parton scattering, the cold dense baryon matter, to the exotic multiquark resonance production, and, from the other side, it has a rather high potential for discovering new exotic multiquark states and novel mechanisms of production at high energy hadronic collisions.




\section{\bf {Study of inelastic d-d and p-d interactions for observation of
neutron-proton system under strong compression}
\protect\footnote{ This section is written by B.F.\,Kostenko, {E-mail: bkostenko@jinr.ru}}}
\begin{abstract}

In this experimental proposal, we consider a possibility of registering light dibaryons at the NICA SPD facility, experimental and theoretical indications of the existence of which were previously obtained at JINR. The main attention is paid to the description of the observable effects, as well as the requirements for measuring instruments necessary for their successful detection.
\end{abstract}

\vspace*{6pt}
\noindent

PACS: 24.85.+p, 13.88.+e, 13.85.Fb 

\label{sec:introa}
\subsection{Introduction}
At present, there are different theoretical indications that cold nucleonic matter under strong compression should undergo phase transitions, see, e.g., \cite{PhaseTran}. Experiments at NICA MPD will be devoted to the search for signs of these transitions, in which the manifestations of the transformations of the nucleon phase into quark-gluon one, as well as other transitions, is expected to be seen in collisions of heavy nuclei. Unfortunately, traditional nuclear physics, as well as astrophysical restrictions on the equation of state for cold dense matter, do not yet give direct indications of the need to take into account such processes \cite{Astro}. A similar situation has arisen in the description of small nucleon systems, which can be used as the basis for the microscopic theory of phase transitions in nuclear matter. Thus, it has long been suggested that various multibaryon states can exist in nature in the form of a configurational admixture together with ordinary nuclei \cite{Matv-Sorba}. At the same time, models continued to develop, suggesting that nucleons in nuclei can also retain their individuality even in very exotic high-momentum states \cite{GG}.
An attempt to find experimental manifestations of the multiquark component in the wave function of atomic nuclei was undertaken in the frame of the flucton model\cite{Efremov,BLT}. In parallel, the model of short-range correlations according to which, to explain the same processes, the nucleon models, taking into account the high-momentum component of the wave function of nuclei, were also quite suitable \cite{FrStr}.

Experiments on the NICA MPD facility should in the future answer some of the questions associated with the problem of phase transitions of nucleonic matter into other forms of nuclear substance at high densities, which can be achieved in collisions of heavy nuclei accelerated to relativistic velocities. The theory of these phenomena is extremely difficult, so that the interpretation of the obtained experimental data may turn out 
to be highly uncertain. In this regard, it is of particular interest to study the transitions of nucleonic matter to other states, which can occur in few-nucleon systems. Drawing an analogy with the theory of the formation of an electron-ion plasma, it can be argued that, in this respect, 
knowledge of the characteristics of few-nucleon systems under conditions of strong compression can contain information as indispensable as knowledge of the ionization characteristics of atoms for building a theory of ordinary plasma.

\eject

\label{sec:idea2}
\subsection{Search for new dibaryons at the NICA SPD facility}
There are a number of theoretical and experimental indications of the existence of light dibaryon excitations in 2-nucleon systems \cite{Kostenko,Troyan,Baldin}.
An analysis of the experimental data suggests the possible observation of events, the cross sections of which in a certain kinematic region are comparable to the cross sections of hard elastic nucleon-deuteron collisions.
For example, in \cite{Baldin} a peak was observed, which is interpreted to date as a quasi-free knockout by the target deuteron of a nucleon from the incident deuteron  in the reaction d $ + $ d $ \to $ d $ + $ X.
However, its large width can hardly be explained only by the internal motion of nucleons in the deuteron being destroyed.
At the same time, kinematic analysis shows \cite{Kostenko} that  scattering processes with the participation of virtual dibaryons preexisting in the incident deuteron in a virtual form before the collision can also contribute to the observed peak.
In experiments \cite{Troyan}, where the effective masses of two-nucleon systems formed after the emission of several (up to 5) pions from highly excited neutron-proton systems were measured, there are indications of the observation of narrow dibaryons, the masses of which can be described by a simple formula \cite{Kostenko} $M_n = M_d +10.08 n$ MeV, where
$M_d = 1875.6$ MeV is the value of deuteron mass\footnote{Hereinafter, the system of units is used, in which $ c = 1 $.}.
It can hardly be interpreted as a mere coincidence that the broadening of the peak caused by the ``quasi-free'' nucleon knockout in the experiment \cite{Baldin} can also be explained \cite{Kostenko} by scattering d $ + $ d $ \to $ d $ + $ d $ ^ * $ with the participation of the same dibaryons as in the experiment \cite{Troyan}. Of course, these arguments can only be considered as indirect evidence in favor of the existence of light dibaryons. 
On the other hand, it must be emphasized that the criticism of Troyan's experiment presented in \cite{Abramov}  also does not look convincing enough.
Firstly, in  \cite{Troyan} a larger number of events were studied than in \cite{Abramov}; secondly, in \cite{Abramov} , events with a poorer ``cooling'' of a highly excited dibaryon were used, i.e. the work is based on events with fewer secondary pions emitted by a highly excited dibaryonic system.
And finally, even in spite of all that, significant variations in the measured cross sections were observed in \cite{Abramov}, which can hardly be explained by simple statistical fluctuations.
The above facts suggest that light dibaryons could have been present in a sufficiently large number already in some of the earlier experiments, but were not recognized due to insufficient spectral resolution of the corresponding experimental facilities.
Drawing an analogy with the history of discoveries associated with the invention of the microscope, we can say that a medium that can contain a large number of interesting microscopic objects is  presumably known, but a device that allows them to be clearly seen has not yet been created.
Below we shall consider  the demands  that would have to be advanced to an experimental setup so that it could be used to obtain unambiguous answers to all the above questions. 

In experiments on colliding beams of deuterons with momenta of the same modulus, the energy $ q_0 $ transferred in the reaction d + d $ \to $ d + d $ ^ * $ from one of the deuterons to the dibaryon being produced is
\begin{equation}\label{q0CollBeams}
{q_0} = \frac{{M_*^2 - M_d^2}}{{4{E}}}, 
\end{equation}
where $ {E} = \sqrt {P ^ 2 + M_d ^ 2} $ is the deuteron energy in the laboratory coordinate system (l.c.s.).
For the $ n- $th excited level of the dibaryon 
$ {M_ *} = {M_d} + n \varepsilon $.
Here $ \varepsilon $ is the distance between the levels, which we will take equal to 10 MeV \cite{Kostenko}. The relation (\ref {q0CollBeams}) allows one to estimate the influence of the instability of the beam energy $ E $ on the accuracy of determining the value of $ \varepsilon $.
For this purpose, let us bring into correspondence the spread in energy $ \Delta E $ of deuterons in beams and the inaccuracy of setting the level number $ 0 < \delta <1 $, which will describe the deviation of the measured level energy from its exact value, referred to $ \varepsilon $.
Substitution of $ E \to E + \Delta E $ and $ n \to n + \delta n $ into the relation (\ref {q0CollBeams}) gives a quadratic equation for $ n $, which determines the highest level for which the uncertainty in the distance between adjacent levels is does not yet exceed the value of $ \delta $:
\begin{equation}\label{delta} 
\frac{{\Delta E}}{E} =  \frac{{2{M_d} + (2n + \delta )\varepsilon }}{{2{M_d}n + {n^2}\varepsilon }} \delta.
\end{equation}
For $ n \varepsilon \ll {M_d} $, this relation is reduced to a directly proportional relationship between the relative fluctuation of the beam energy and the relative error in measuring the distance between adjacent levels, caused by these fluctuations:
\[\delta  \approx \frac{{\Delta E}}{E}n.\]
It turns out to be rather weak (so far, the influence of the beam momentum spread on the accuracy of measuring the $ M _ * $ dibaryon mass is not taken into account).
For example, for $ \Delta E = 100 $ MeV, the error in determining the energy of the tenth level of a dibaryon is only 1 MeV for the momentum of colliding beams at the level of 10 GeV in l.c.s.

It is also useful to have the correspondence between the 4-momentum $ q $ transferred to the nascent dibaryon, which can be measured experimentally, and the relativistic Mandelstam  variable $ t \equiv q_0 ^ 2 - \mathbf {q} ^ 2. $
One can make sure after elementary but cumbersome calculations that for the experiments with colliding beams the following simple formula holds:
\begin{equation}\label{tCollBeams}
 t=2(E{q_0} - P{q_\parallel }).
\end{equation}
Here $ q_ \parallel $ is longitudinal, i.e. directed along the beams$'$ axis, the component of the momentum transfer, which is directly responsible for the excitations.
It can be seen from (\ref{tCollBeams}) that the processes of dibarion production from a kinematic
point of view are possible at both negative and positive, as well as zero values of
the varibale $t$, which seems to be extremely unlikely.
It is clear that the dynamic theory of the processe should  be based on the relativistic-invariant
amplitude using the variable $t$, since the fact of dibarion production does not depend on the frame
in which it is registered. In the absence of such a theory, the answer 
to the question of whether the dibaryons under consideration are produced,
for example, ar zero values of the variable $t$, can only be given by experiment.
Since the relativistic-invariant amplitude also depends on the Mandelstm variable 
$s$, the search for the region of localization of dibaryons should also be accompanied
by a variation in the momentum of deuterons in colliding beams.

Traditionally, formula (\ref{q0CollBeams}) is used to identify resonances, in which $ M_* $ is the effective mass of  decay products of a resonance.
An individual resonance can be seen if an error of its mass determining, which includes its own width, and also takes into account the measurement errors of the momenta of its decay products, is significantly less than distances to adjacent levels. It is clear that, to recognize resonances, one can also use the  relationship (\ref{tCollBeams}), in which resonances are seen as separate lines. However, to formulate the requirements for the experimental setup,
the most appropriate is a method of  resonance recognition directly based on the quantities to be measured. Indeed, resonances, as well as the degree of reliability of their recognition, can also be seen on graphs of the dependence $ q_\bot (q_\parallel) $ with a direct indication of the measurement errors of  longitudinal $ q_ \parallel $ and perpendicular $ q_\bot $ components of the momentum transfer. The exact analytical dependence $ {q_ \bot} ({q_\parallel}) $ for colliding beams with equal momenta has the form
\begin{equation}\label{q_t(q_||)}
{q_\bot }({q_\parallel }) = \sqrt {{{\left( {\frac{{4{E^2} + M_*^2 - M_d^2}}{{4E}}} \right)}^2} - M_*^2 - {{(P - {q_\parallel })}^2}}. 
\end{equation}
\begin{figure}
\begin{center}
\includegraphics[width=127mm]{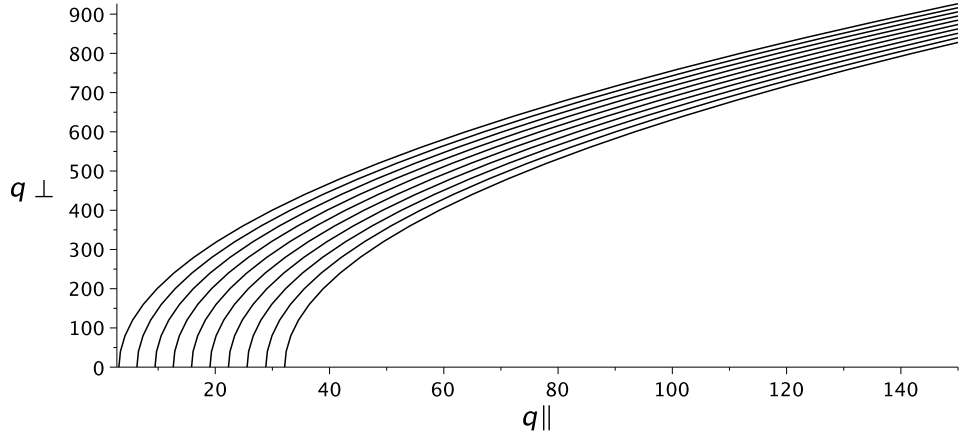}
\end{center}
\caption{\small 
Dependence of the transverse momenta of light dibaryons on their longitudinal momenta measured in MeV for an experiment with colliding beams with  $P=3$ GeV. The uppermost curve corresponds to the first excited level, and the lowest one corresponds to the tenth level. 
}\label{q_t_on_qL}
\end{figure}

This dependence is shown in Fig. \ref{q_t_on_qL} for the supposed theoretical
values of the masses of dibaryons and $P=3$ GeV. In a real experiment, values of the momenta 
of decay products of a dibaryon should be measured. For light dibaryons, these are the momenta
$\mathbf{p}_n$ and $\mathbf{p}_p$ of the secondary neutron and proton. The transverse
component of the momentum transferred to the dibaryon is equal to the transverse component
of the total momentum of the secondary nucleons 
$ q_\bot = p_{n \bot} + p_{p \bot} $, which is known with the measurement accuracy $ \Delta p_{n, p \bot} = \Delta p_{n \bot} + \Delta p_{p \bot} $. In addition, the nonzero fluctuation of the transverse momentum of the beam $ \Delta P_\bot $ makes a contribution to the measurement error of the transverse momentum transfer.
Thus, the total measurement error of the transverse component of the momentum transfer is calculated by the formula
$$\Delta q_\bot = \Delta p_{n\bot} + \Delta p_{p\bot} + \Delta P_\bot.$$ 
Similarly, taking into account that $ q_\parallel = P- (p_{n \parallel} + p_{p \parallel}) $, we have
$$ \Delta q_\parallel = \Delta p_{n \parallel} + \Delta p_ {p \parallel} + \Delta P_\parallel, $$
where $ \Delta P_\parallel $ is a value of the beam momentum fluctuations.

In paper \cite{Baldin}, in which the light dibaryons could be observed at $t = $ -0.5 GeV$^2$, interactions  of deuterons with momentum $ P = $ 8.9 GeV with a stationary deuteron target were used.
The colliding beam experiments at $ | P | = $ 3 GeV correspond to the deuteron momentum $ P = $ 11.3 GeV in experiments with the stationary target, which is close to the value in \cite{Baldin}.

Using the formulas (\ref{q0CollBeams}) and (\ref{tCollBeams}),
it can be verified that the kinematics of colliding beams corresponds to similar events at $ q_\parallel = $ 100 $ \pm $ 20 MeV.
In its turn, the formula (\ref{q_t(q_||)}) tells us that at $ q_\parallel  = $ 100 MeV the value of $ q_\bot $ will be  755.8 MeV for elastic d-d scattering, 743.2MeV for scattering with excitation of the first level, 730.3 MeV for scattering with excitation of the second level, ... 631.0 MeV -- for the tenth level. In this range, the distances between the levels grow as follows:  12.3, 12.6, 12.9, ... 14.9, 15.3 MeV. Thus, it can be seen that the measurement accuracy of the transverse momentum at a level of 3 MeV is quite enough to register all the dibaryons described above.

To estimate the required accuracy of measuring the longitudinal component of the momenta of the secondary particles, the dependence (\ref{q_t(q_||)}) should be reversed, expressing $ P-q_\parallel $ in terms of $ q_\bot $.
One can check that the distances in the horizontal direction
between the levels of dibaryons shown in Fig. \ref{q_t_on_qL}, in this case only slightly more than 3 MeV at full longitudinal
momentum $ P-q_\parallel $ of the decay products at a level of 3000 MeV. This dictates very high requirements  both for the measurement accuracy of the longitudinal components of  momentum of the secondary particles and the degree of monochromaticity of the beam  -- at the level of $10 ^{- 4}$. 
Nevertheless, even in this respect, the experiments with colliding beams have significant advantages in relation to experiments with the stationary target.
In the latter case, a similar experiment would require increasing relative accuracy by a factor of 7, because of the larger momentum $ P $ of the accelerated beam and also because of a different analytical dependence $ q_\bot (q_\parallel ) $.  

The requirements for the accuracy of measuring the longitudinal component of the momentum transfer very fast decrease with decreasing momentum of the colliding beams. So, if it is equal to 1 GeV, then the required relative accuracy becomes equal to 10$^{-2}$.
Since the exact range of values of the Mandelstam variable $s$, within which the production of 
light dibaryons is possible, we do not know, then it makes sense to set up an experiment
even at lower than in the experiment \cite{Baldin} values of $s$.

Fig.\ref{q_t_on_qL} suggests the possibility of searching for any dibaryons, including heavier
ones, the existence of which quark models  predict. For this, it is enough to use the fact 
of a distinct concentration events near some line $ {q_ \bot }({q_\parallel }) $, described by formula (\ref{q_t(q_||)}), which should be very similar to those shown in Fig.\ref{q_t_on_qL}.



\section{\bf Proposal for the study of lightest neutral hypernuclei\\ with strangeness $-1$ and $-2$
\protect\footnote{This section is written by J.-M. Richard,
Q. Wang and Q. Zhao;\, E-mail:{zhaoq@ihep.ac.cn}.}}

\begin{abstract}
  \vspace{0.2cm}

  Based on our recent study of the lightest neutral hypernuclei with strangeness $-1$ and $-2$, we propose to look for the neutral hypernucleus \isotope[4][\Lambda\Lambda]{n} in deuteron-deuteron collisions which can be accessed by SPD NICA in the future. Some advantages and opportunities for hypernuclei and exotic hadrons in the double $K^+$ production channels at NICA are addressed.
\end{abstract}
\vspace*{6pt}

\noindent
PACS: 21.80.+a, 21.30.Fe, 21.30.-x, 21.45.-v, 25.45.De


There have been significant progresses in the study of hypernuclei in both experiment and theory during the past decades. The discovery of a number of hypernuclei with $S=-1$ and $S=-2$ has greatly enriched our knowledge about the hyperon-nucleon and hyperon-hyperon interactions~\cite{Rijken:2013wxa,Gal:2011tb}, and the search for stable hypernuclei still serves as a direct probe for many interesting physics with the presence of hyperons in nuclear matter.

So far, there is no evidence for a stable $(n,\Lambda)$ nor $(p,\Lambda)$ bound state for baryon number $A=2$, except for a resonance peak of about 2.06\,GeV seen in $d+\pi^-$ \cite{Rappold:2013jta}. The situation becomes more complicated in the 3-body or 4-body systems made of nucleons and hyperons. For $A=3$ in the sector of isospin $I=1$, $\isotope[3][\Lambda]{He}=(p,p,\Lambda)$ and $\isotope[3][\Lambda]{H}=(n,p,\Lambda)$ are found bound. The latter has a 3-body binding energy of  $E_3=-2.45$ MeV, which is just below the threshold of a deuteron plus an isolated $\Lambda$, i.e. $E_2=-2.20\,$ MeV. In contrast, the existence of $\isotope[3][\Lambda]{n}=(n,n,\Lambda)$ still needs confirmation~\cite{Rappold:2013jta}.

For $A=4$ with $I=0$, the nonstrange tetraneutron, $\isotope[4]{n}=(n,n,n,n)$, is naively suggested to exist by the stability  of \isotope[8]{He} isotope, but so far has received only controversial experimental indications~\cite{Marques:2001wh}. Calculations based on realistic neutron-neutron potentials \cite{Tang:1965zza,Bertulani:2002px,Pieper:2003dc,Timofeyuk:2003ya,Lazauskas:2005ig} or potentials made artificially deeper to produce a dineutron  \isotope[2]{n} do not support its stability due to the Pauli principle.  To avoid the Pauli blocking effects and benefit from the convening role played by the hyperon, it is thus interesting to consider the stability of the doubly-strange tetra-baryon system, $T=\isotope[4][\Lambda\Lambda]{n}=(n,n,\Lambda,\Lambda)$ with $S=-2$ and $I=1$. In particular, the ground state should favor the spin-singlet assignments for both the $nn$ and the $\Lambda\Lambda$ pairs in order to satisfy the antisymmetrisation.

In the sector of $A=4$ and $S=-2$, namely the ``double $\Lambda$'' hypernuclei, it includes the states, $\isotope[4][\Lambda\Lambda]{He}$, $\isotope[4][\Lambda\Lambda]{H}$, and $\isotope[4][\Lambda\Lambda]{n}$. Note that the ``Nagara'' event \cite{Takahashi:2001nm}, has set a limit on the $\Lambda\Lambda$ effective attraction in $\isotope[6][\Lambda\Lambda]{He}$ with a precise measurement of its binding energy. This may provide some constraints on lighter $S=-2$ systems~\cite{Garcilazo:2012qv,Gal:2013bw,Garcilazo:2013zwa}, and in particular hypernuclei with $A=4$.

The $\isotope[4][\Lambda\Lambda]{H}$ was studied in
Ref.~\cite{Filikhin:2002wp} by Filikhin and Gal, and it was found to be unbound within the models adopted. However, with a more sophisticated method for solving the four-body problem, the calculation by Nemura, Akaishi and Myint \cite{Nemura:2002hv} found a small amount of binding below the threshold of dissociation into $\isotope[3][\Lambda]{H}+\Lambda$. As stressed in Refs.~\cite{Filikhin:2002wp,Nemura:2002hv,Garcilazo:2012qv}, the interaction of $\Lambda\Lambda\leftrightarrow N\Xi\leftrightarrow \Lambda\Lambda$, which is sizeable in free $\Lambda\Lambda$ scatterings, will be suppressed in a dense nucleus due to the antisymmetrisation between the nucleons in the core and the nucleon in $N\Xi$. Such a Pauli suppression effect was invoked to explain the relatively weak binding of $\isotope[6][\Lambda\Lambda]{He}$. However, since it requires a $(n\Lambda\Lambda)$ correlation it should not play a significant role in the limit of weak binding.

It is no doubt that the question of stability of the $A=4$ double-$\Lambda$ hypernuclei would be crucial for our understanding of the role played by hyperons in nuclear matter. While it is still controversial for model calculations of such a four-body problem in the regime of weak binding, we would propose that some general properties arising from the weakly binding systems involving the 2-body and 3-body bound-state energies may provide a guidance for a possible stability of \isotope[4][\Lambda\Lambda]{n}. Meanwhile, we propose a sensitive reaction process for the search for \isotope[4][\Lambda\Lambda]{n} in deuteron-deuteron scatterings, i.e. $d+d\to K^+ + K^+ +\isotope[4][\Lambda\Lambda]{n}$, which is accessible at NICA. After all, it would rely on the experimental study to decide the dedicate dynamics for such an exotic system.

As follows, we will first demonstrate that \isotope[4][\Lambda\Lambda]{n} may exist given the satisfaction of Thomas condition, where \isotope[4][\Lambda\Lambda]{n} is likely to be a ``Borromean'' system. We then propose and discuss its production mechanism in $d+d\to K^+ + K^+ +\isotope[4][\Lambda\Lambda]{n}$. Advantages of using the double $K^+$ production channels to probe exotic hypernuclei and exotic hadrons are also addressed. A brief summary is given in the end.

\label{binding-cond}
\subsection {Binding conditions for 3 and 4-body systems with strangeness $-1$ and $-2$}

The Thomas condition is referred to the relation between the range of nuclear forces and the ratio of 3-body to 2-body bound-state energies which was discovered by Thomas in 1935~\cite{Thomas:1935zz}. Namely, it shows that the ratio of 3-body to 2-body bound-state energies, $E_3/E_2$, becomes very large if the range of the interaction decreases.
In such a case one may have a deep 3-body binding of $E_3/E_2\to\infty$ for a given (short) range if the coupling $g$ approaches (from above) the minimal value $g_2$ required by the 2-body binding.  Here, the coupling $g$ is defined by the  potential energy $g\,\sum v(r_{ij})$, where $v$ accounts for the attractive parts of the potential, and $r_{ij}$ denotes the inter-particle separation. Implication of the Thomas condition is that the minimal coupling $g_3$ to bind three particles is smaller than $g_2$. Therefore, it allows a coupling value of $g_3<g<g_2$, which will lead to a ``Borromean'' 3-body binding system. Namely, the 3-body system become bound but its two-body subsystems are unbound.

The rigorous boundaries on the allowed domain of coupling constants for the Borromean system have been studied in the literature~\cite{Richard:1994mc}, and the Borromean window as a function of the potential shape~\cite{Moszkowski:2000ni} or of the low-energy parameters of the pair interactions~\cite{Thogersen:2008zz} have been investigated. In Ref.~\cite{Richard:2014pwa} it was shown that such a Borromean binding for 3-body (such as $nn\Lambda$) and 4-body (such as $nn\Lambda\Lambda$ systems may exist based on some general features implemented for the 2-body interactions.

While the detailed discussions on the Thomas conditions can be found in Refs.~\cite{Richard:2014pwa,Richard:2014txa}, we only outline the key points which are relevant to the study of the $nn\Lambda\Lambda$ system. First, we adopt some simple potentials for the 2-body interaction:
\begin{align}
   & -g\,\exp(-\mu\,r)\qquad \text{(exponential)}~, \label{eq:exp}                        \\
   & -g\,\exp(-\mu\,r)/r\quad\text{(Yukawa)}~, \label{eq:yuk}                             \\
   & g \exp[-2\,\mu\, (r-R)]-2\,g\,\exp[-\mu\,(r-R)] \  \text{(Morse)}~, \label{eq:Morse}
\end{align}
where for the Morse potential, we use $R=0.6$ for the illustration purpose.

\begin{table}[ht]
  \label{tab:tab1}
  \caption{Values (in fm) adopted for the scattering length and effective range parameter in the two models.}
  \begin{center}
    \begin{tabular}{cccccc}
      \hline\hline
      \multirow{2}{*}{Pair} & \multicolumn{2}{c}{ESC08} &                & \multicolumn{2}{c}{CEFT}                          \\
      \cline{2-3}\cline{5-6}
                            & a                         & $r_\text{eff}$ &                          & a     & $r_\text{eff}$ \\
      \hline
      $nn$                  & -16.51                    & 2.85           &                          & -18.9 & 2.75           \\
      $(n\Lambda)_{s=0}$    & -2.7                      & 2.97           &                          & -2.9  & 2.65           \\
      $(n\Lambda)_{s=1}$    & -1.65                     & 3.63           &                          & -1.51 & 2.64           \\
      $\Lambda\Lambda$      & -0.88                     & 4.34           &                          & -1.54 & 0.31           \\
      \hline\hline
    \end{tabular}
  \end{center}
  \label{tab:multicol}
\end{table}

We then solve the 2-body problem to reproduce the deuteron binding energy, and $nn$, $n\Lambda$ and $\Lambda\Lambda$ scattering lengths and effective ranges extracted by the Nijmegen-RIKEN model~\cite{Rijken:2013wxa,Rijken:2010zzb,Rijken:2013wxa} and the chiral effective field theory (CEFT)~\cite{Polinder:2007mp,Haidenbauer:2013oca}. The values for the scattering lengths and effective ranges from these two models are listed in Table~\ref{tab:tab1}. It shows that these two models have similar results for the $nn$ and $n\Lambda$ interactions, but they produce quite different values for the $\Lambda\Lambda$ interaction.
In particular, the effective range of $\Lambda\Lambda$ is remarkably small, and for a given scattering length, this eases the occurrence of Borromean binding involving a pair of $\Lambda$. However, in a more advanced CEFT study~\cite{Haidenbauer:2015zqb}, the same group found a larger value for $r_{\rm eff}$, and this modifies the conclusions for the 3-body and 4-body systems at the edge of binding.  Hopefully, some new experimental results, e.g., from final-state correlations in heavy-ion collisions~\cite{Acharya:2019yvb} would allow for a better tuning of the models.
The main results can be outlined as follows:

\begin{itemize}
  \item The model reproduces the observed binding of the \isotope[2]{H}\ and \isotope[3][\Lambda]{H}\ systems. For \isotope[3][\Lambda]{H}\, both spin $s=1/2$ and $s=3/2$ are found bound since there is no much difference between the spin-triplet and the spin-averaged nucleon-hyperon interactions.

  \item We find that \isotope[3][\Lambda]{H} with isospin $I=1$ and spin $s=1/2$ is marginally unbound. However, it may become bound if some masses increase by about 10\%. Namely, the unequal masses between nucleon and hyperon will bring more binding.

  \item We also find that $\isotope[3][\Lambda]{n}$ is marginally unbound. Our results for $\isotope[3][\Lambda]{n}$ agree with the conclusions of the recent studies~\cite{Gal:2014efa,Hiyama:2014cua,Garcilazo:2014lva}.

  \item The state \isotope[4][\Lambda\Lambda]{H} with isospin $I=0$ is found weakly bound (about 3\,MeV) in the Nijmegen-RIKEN model, and slightly more (about 9\,MeV) in the CEFT one.  The state \isotope[4][\Lambda\Lambda]{H} with isospin $I=1$ and \isotope[4][\Lambda\Lambda]{n} deviate from binding by a very small amount with the Nijmegen-RIKEN parameters, but become bound by about 1\,MeV with the CEFT parameters.
\end{itemize}

The above analysis is based on some general properties arising from few-body systems. In particular, the satisfaction of the Thomas condition is crucial for the stability of \isotope[4][\Lambda\Lambda]{n} as a Borromean system. However, it should be noted that the detailed dynamics for the nucleon-hyperon and hyperon-hyperon interactions will decide whether \isotope[4][\Lambda\Lambda]{n} could exist at all. At this moment, there are still significant discrepancies between some of the most popular models. For instance, as shown in Table~\ref{tab:tab1}, the $\Lambda\Lambda$ scattering length and effective range determined by the Nijmegen-RIKEN model and CEFT model turn out to be signficantly different. It indicates that experimental constraints on the $\Lambda\Lambda$ interaction is desired.

Another point needs to be addressed is that so far the 3-body forces have not been considered. Whether or not they contain an attractive component would be crucial for the stability of close-to-binding systems. In case that some spin-dependence of the 3-body forces can play a significant role, it would keep the spin $s=1/2$ state of \isotope[3][\Lambda]{H} bound and move the $s=3/2$ in the continuum. It is also possible that the 3-body and $n$-body forces with $n>3$ contain a short-range repulsive component. This is due to the Pauli exclusion of the constituent quarks when several hyperons (or several hyperons and nucleons) overlap within a small distance. The repulsive component seems to be necessary in large systems containing strangeness~\cite{Lonardoni:2013rm}. While the calculations of hyperon-nucleon and hyperon-hyperon forces should be pushed to higher order within theoretical models, experimental search for these strangeness $-2$ hypernuclei would provide crucial constraints on the model parameters.

\label{prod-mech}
\subsection {Production mechanism for $\isotope[4][\Lambda\Lambda]{n}$ and advantages of double $K^+$ productions }

We turn to the possible experimental search for $\isotope[4][\Lambda\Lambda]{n}$ and propose a production mechanism which can be accessed at SPD NICA. In Ref.~\cite{Richard:2014pwa} we have shown that the deuteron-deuteron collisions around the energy region above $E_{cm}\simeq 5.2$ GeV is favored to produce $\isotope[4][\Lambda\Lambda]{n}$ with the total cross section of about 2.5 nb. Here, based on the same analysis we try to clarify some key points and make a rough estimate of its production rate at the kinematics of SPD NICA.

As mentioned earlier, the quantum numbers of the ground state $\isotope[4][\Lambda\Lambda]{n}$ will favor $J^P=0^+$, where the neutron pair and $\Lambda$ pair have spin 0, namely, their spins are anti-parallel, respectively. Meanwhile, the total isospin is $I=1$. Thus, the total wavefunction of the ground state is anti-symmetric under the interchange of the two nucleons or the two $\Lambda$. In principle, one has to construct a dynamic wavefunction for the $(n,n,\Lambda,\Lambda)$ system, which is a nontrivial work and strongly model-dependent due to the unknown $\Lambda\Lambda$ interactions. But for the purpose of making an estimate of the production rate, we can simply introduce a momentum distribution for the $n\Lambda$ clusters for the $(n,n,\Lambda,\Lambda)$ system~\cite{Richard:2014pwa}.

The most ideal reaction for producing $\isotope[4][\Lambda\Lambda]{n}$ should be $d+d\to K^+ + K^+ +T$ which is an extremely clean process since the background processes involving the $K^+K^-$ productions become irrelevant. It makes the measurement of the missing mass spectrum recoiling against the $K^+K^+$ pairs sensitive to the existence of any pole structure in the $nn\Lambda\Lambda$ system. The transition matrix element can be expressed as
\begin{eqnarray}
  {\cal M}&=&\int \psi_T^*(\boldsymbol{p}_1',\boldsymbol{p}_2',\boldsymbol{p}_3',\boldsymbol{p}_4';\boldsymbol{P}_T')\psi^*_{K_1}(\boldsymbol{P}_{K_1})\psi^*_{K_2}(\boldsymbol{P}_{K_2})\hat{{\cal O}}(\boldsymbol{p}_1',\boldsymbol{p}_3',\boldsymbol{p}_1, \boldsymbol{p}_3, \boldsymbol{P}_{K_1}, \boldsymbol{P}_{K_2})\nonumber\\
  &\times & \psi_{d_1}(\boldsymbol{p}_1,\boldsymbol{p}_2; \boldsymbol{P}_{d_1})  \psi_{d_2}(\boldsymbol{p}_3,\boldsymbol{p}_4; \boldsymbol{P}_{d_2}) \delta(\boldsymbol{P}_T'+\boldsymbol{P}_{K_1}+\boldsymbol{P}_{K_2}-\boldsymbol{P}_{d_1}-\boldsymbol{P}_{d_2})\nonumber\\
  &\times & \delta(\boldsymbol{p}_1+\boldsymbol{p}_2-\boldsymbol{P}_{d_1})\delta(\boldsymbol{p}_3+\boldsymbol{p}_4 -\boldsymbol{P}_{d_2})\delta(\boldsymbol{p}_1'+\boldsymbol{p}_2'+\boldsymbol{p}_3'+\boldsymbol{p}_4'-\boldsymbol{P}_T')\nonumber\\
  &\times & \delta(\boldsymbol{p}_2-\boldsymbol{p}_2')\delta(\boldsymbol{p}_4-\boldsymbol{p}_4')d\boldsymbol{p}_1 d\boldsymbol{p}_2 d\boldsymbol{p}_3 d\boldsymbol{p}_4 d\boldsymbol{p}_1' d\boldsymbol{p}_2' d\boldsymbol{p}_3' d\boldsymbol{p}_4' \ ,
\end{eqnarray}
where the kinematic variables are defined in Fig.~\ref{fig:reac}. Note that Fig.~\ref{fig:reac} illustrates one of the leading transitions favored by the \isotope[4][\Lambda\Lambda]{n} production in the central deuteron-deuteron collisions.

\begin{figure}[ht]
  \begin{center}
    \includegraphics[width=80mm]{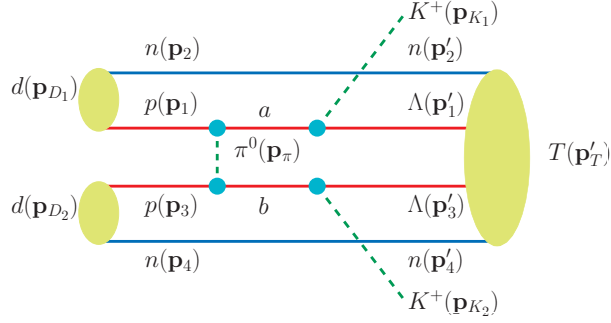}
    \vspace{-3mm}
    \caption{(Color online) Mechanism for $T$ production in $dd$ collisions.}
  \end{center}
  \labelf{fig:reac}
  \vspace{-5mm}
\end{figure}

In Fig.~\ref{fig:cross} the total cross section is estimated by the leading transition process of Fig.~\ref{fig:reac}.
As the leading-order test the relative internal momenta between the proton and neutron inside the incident deuterons has been neglected. Namely, the Fermi motion inside the initial deuterons is neglected. Also, the final-state interactions among the final-state baryons are overlooked, which means that the neutrons are treated as spectators and their contributions to the amplitude will be via the convolution of the final-baryon momentum distributions. The $S_{11}(1535)$ resonance is included in the transition amplitude which is found to be relatively small, mainly due to the smaller couplings to $\pi N$ and $K\Lambda$.  Since other processes with intermediate $N^*$ excitations may contribute, our estimate including only the Born term and $S_{11}(1535)$ excitations can be regarded as a conservative estimate of the production cross section for $\isotope[4][\Lambda\Lambda]{n}$.

\begin{figure}[ht]
  \begin{center}
    \includegraphics[width=80mm]{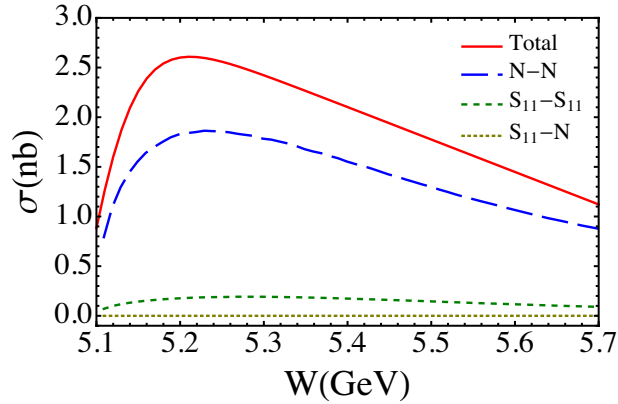}
    \vspace{-3mm}
    \caption{(Color online) Total cross section for $d+d\to K^+ + K^+ +(n,n,\Lambda,\Lambda)$). From upper to lower the curves stand for the total cross sections of the full calculations, exclusive process from the nucleon Born terms, exclusive process from the double $S_{11}(1535)$ excitations, and exclusive process from the one Born transition and one $S_{11}(1535)$ excitation.}
  \end{center}
  \labelf{fig:cross}
  \vspace{-5mm}
\end{figure}

In Fig.~\ref{fig:missingmass} the $K^+K^+$ missing mass spectra for the recoiled $(n,n,\Lambda,\Lambda)$ at different energies are shown above the production threshold. The peak position is located at the four baryon $nn\Lambda\Lambda$ threshold since only a momentum distribution for \isotope[4][\Lambda\Lambda]{n} is considered. However, our estimate is sufficient to demonstrate the behavior of the correlated system recoiled by the $K^+$ pair. For uncorrelated $K^+K^+$ events, i.e. the final-state $(n,n,\Lambda,\Lambda)$ are not bound, there would no peak in the missing mass spectrum.

\begin{figure}[htb]
  \begin{center}
    \includegraphics[width=80mm]{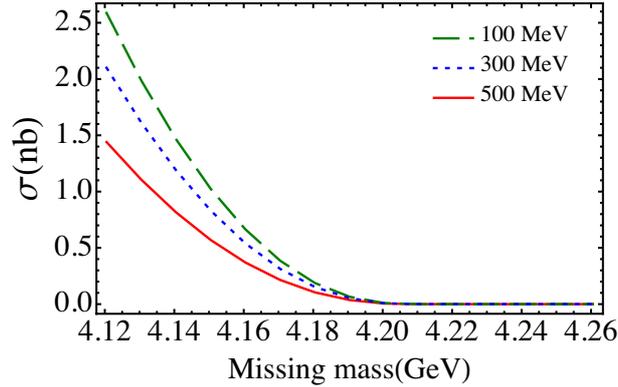}
    \vspace{-3mm}
    \caption{(Color online) The missing mass spectra for the recoiled $(n,n,\Lambda,\Lambda)$ at different energies above the production threshold in  $d+d\to K^+ + K^+ +(n,n,\Lambda,\Lambda)$}
  \end{center}
  \labelf{fig:missingmass}
  \vspace{-5mm}
\end{figure}

The c.m. energy at SPD NICA starts at $E_{cm}=6.7$ GeV with a luminosity of $L=10^{27}\mbox{cm}^{-2}\mbox{s}^{-1}$. We estimate that the total cross section at $E_{cm}=6.7$ GeV will drop about one order of magnitude compared with the peak value of about 2 nb. Thus, the events expected in one-year runtime are
\begin{equation}
  N=\sigma_{total}\times L\times t= 0.2\mbox{nb}\times 10^{27}\mbox{cm}^{-2}\mbox{s}^{-1}\times 1 \ \mbox{year}\simeq 6.3 \ ,
\end{equation}
which is a small event counting. It could be even smaller taking into account the detection efficiency which generally will reduce the event counting by one order of magnitude. However, if the luminosity can reach $10^{29}$, which is an approximate average between the lower limit of $10^{27}\mbox{cm}^{-2}\mbox{s}^{-1}$ and the upper limit of $2\times 10^{30}\mbox{cm}^{-2}\mbox{s}^{-1}$, the event counting can be significantly increased:
\begin{equation}
  N_{m}=\sigma_{total}\times L\times t= 0.2\mbox{nb}\times 10^{29}\mbox{cm}^{-2}\mbox{s}^{-1}\times 1 \mbox{year}\simeq 630 \ ,
\end{equation}
which is sufficient for establishing the state. For the highest luminosity, one would expect about 12000 events in one-year runtime. Even though the detection efficiency will reduce the events, there will be tens to hundreds of events to  count.

It shows that the double $K^+$ production channel has great advantages for the study of hypernuclei and exotic hadrons. Apart from the proposed process, $d+d\to K^+ + K^+ + (n,n,\Lambda,\Lambda)$, it is also interesting to look at the proton-proton collisions, $p+p\to K^++ K^+ +\Lambda +\Lambda$, where the missing mass spectrum of $K^+K^+$ also provides a clean and direct way to search for the di-baryon $\Lambda\Lambda$, or to study the $\Lambda\Lambda$ interactions.

For the proton-deuteron collisions, the double $K^+$ channel is $p+d\to K^++ K^+ +n +\Lambda +\Lambda$. The recoiled part of  the double $K^+$ is $n\Lambda\Lambda$. It is also unknown whether such an exotic system, i.e. $H$ di-baryon, can exist or not. A direct measurement of such a system would provide rich information about both $\Lambda\Lambda$ and $n\Lambda$ interactions. Nevertheless, notice that the final states have access to the $nK^+$ invariant mass spectrum. The exclusive measurement of this process can also tell whether the light pentaquark state $\Theta^+(1540)$ exists or not.

\subsection{Summary}
The stability of $\isotope[3][\Lambda]{n}$ and, more likely, of $T=\isotope[4][\Lambda\Lambda]{n}$ is within the uncertainties of our knowledge of the baryon-baryon interaction. Although many effects need to be considered in order to refine the predictions, we propose that direct experimental evidence would be extremely useful for model constraints and a better understanding of the hyperon dynamics within nuclear matter. We also emphasize the advantages of the double $K^+$ production channels as a probe for topical exotic hypernuclei and exotic hadron studies. The SPD NICA facility would be the ideal place for such an experimental effort in the future.



\section{\bf Problems of soft $pp$ interactions
\protect\footnote{This section is written by A.\,Galoyan  and V.\,Uzhinsky.
}}

\begin{abstract}


\vspace{0.2cm}

Experiments are proposed directed on solution of three main problems of physics of soft
$pp$ interactions: understanding/description of baryon spectra in $pp$ collisions, evolution
of $<P^2_T>$ -- $x_F$ correlations with energy growth and two-particle $P_T$ correlations.

\end{abstract}
\vspace*{6pt}

\noindent
PACS: 24.10.Lx, 13.85.Ni, 14.20.-c  

Description of proton spectra in hadron-nucleon interactions is a long standing problem of high
energy physics. By tradition, three reggeon phenomenology is used for its solution at $x_F \rightarrow$ 
1 (see Ref.~\cite{KaidalovPhysRep}). Especially, it is assumed that the three pomeron graph (PPP) is 
responsible for the high mass diffraction dissociation. The non-vacuum reggeon -- two pomeron's graph
(RPP) is connected with so-called low mass diffraction dissociation. The three non-vacuum reggeon's
graph (RRR) gives a contribution at $x_F \geq$ 0.8. A.B.~Kaidalov and O.I. Piskunova 
\cite{KaidalovPiskunova} proposed a method of the spectra description in the central region in 
the framework of the Quark-Gluon-String  model. Up to now various Monte Carlo models cannot describe 
the spectra sufficiently well \cite{Uzhi2014}. All of them are used the LUND fragmentation scheme
\cite{LUND} for a treatment of decays of quark-diquark strings. An example of the description is presented in
Fig.~1 where Pythia 6.4 \cite{Pythia6_4} and Geant4 FTF model's predictions are shown.

The Geant4 FTF model \cite{Geant4FTF} considers RRR graphs. As seen, the Pythia model gives abnormal 
humps at $y\sim$ 1.7 and 2.3 at $P_{lab}=$ 158 and 400 GeV/c, correspondently. Thus, detailed 
experimental data at low energies, where the diffractions are dominant processes, are highly 
desired for a development of the theoretical models.

The most impressive experimental data on $pp$ interactions were presented in the last decade by 
the NA61/SHINE collaboration for $\sqrt{s_{NN}}=$ 6.2, 7.6, 8.8, 12.3 and 17.3 GeV \cite{NA61pp}. 
It is pitty, but the data are not sufficiently detailed. Before there were data at $P_{lab}=$ 12
and 24 GeV/c \cite{PP12_24}. They were based on low statistics according to a modern point of view.

On the whole, we can say that the FTF model of Geant4 toolkit describes quite well the multiplicities and
kinematical spectra of produced particles in proton-proton interactions in a wide laboratory energy range 
from 1 GeV up to 1000 GeV.

\begin{figure}[bth]
	\begin{center}
		\resizebox{5in}{2in}{\includegraphics{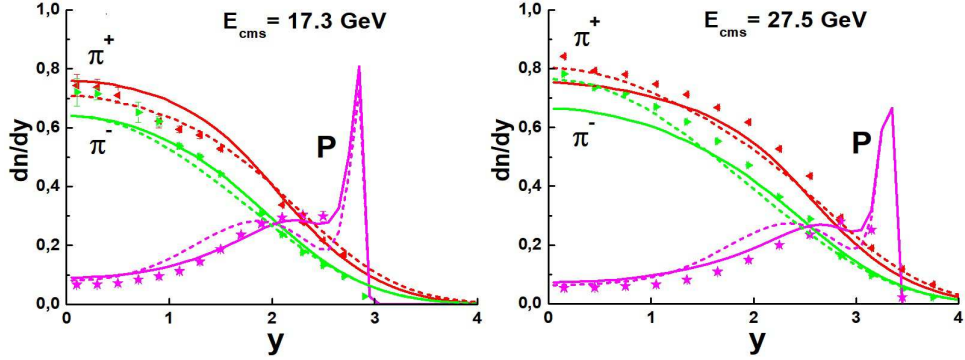}}
		\vspace{-3mm}
		\caption{Distributions of $\pi^\pm$ and protons on rapidity in the
			center-of-mass system of $pp$ interactions at $P_{lab}$= 158 and 400 GeV/c. Points are experimental
			data \protect{\cite{NA61pp,LEBC-EHS400}} without systematical errors. Solid and dashed curves are FTF and Pythia 6.4 
			model calculations.
		}
	\end{center}
	\label{Fig1}
	\vspace{-5mm}
\end{figure}

Another problem, recognized by us, was observed in $<$$P^2_T$$>$ -- $x_F$ correlations in $pp$ interactions.  
The NA61/SHINE collaboration did not provide corresponding experimental data for the correlations,
though, they can be extracted in principle. The NA49 collaboration presented the needed data \cite{NA49pi,NA49pn,NA49K} 
only at the page -- \newline  http://spshadrons.web.cern.ch/ppdata.html. 
The data are very different from analogous data by
the LEBC-EHS collaboration \cite{LEBC-EHS400} at $\sqrt{s_{NN}}=$ 27.5 GeV. The models cannot describe 
the last data quite well.
\begin{figure}[bth]
	\begin{center}
		\resizebox{5in}{2in}{\includegraphics{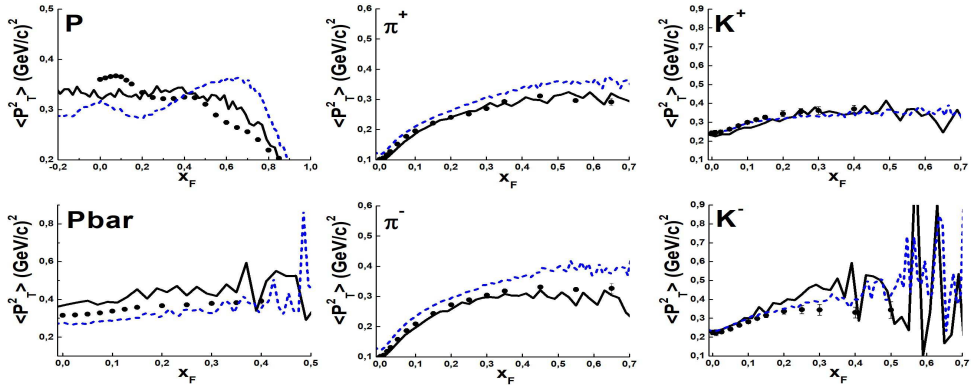}} 
		\vspace{-3mm}
		\caption{$<P^2_T>$ -- $x_F$ correlations in $pp$ interactions at $\sqrt{s_{NN}}$=17.3 GeV. Points are
			the NA49 data \protect{\cite{NA49pi,NA49pn,NA49K}}. Solid and dashed lines are FTF and Pythia 6.4 model calculations.
		}
	\end{center}
	\label{Fig2}
	\vspace{-5mm}
\end{figure}

As seen in Fig.~2, the models describe the general behaviour of the data except the correlations for protons. Predictions
of the models are close to each other and to the data for $\pi^+$, $\pi^-$, $K^+$ and $K^-$ mesons at
$\sqrt{s_{NN}}=$ 17.3 GeV. The FTF model overestimates the correlations for protons at $x_F \geq$ 0.2, and
does not reproduce the hump at $x_F \sim$ 0. The Pythia model gives the hump, but it is essential lower than
the experimental data. In other regions of $x_F$, the Pythia model does not describe the shape of the data.
At higher energy, the situation becomes worse for the proton correlation. This shows that both models
have a problem with baryon production in $pp$ interactions.

It is interesting to look at evolution of the correlations with energy. In Fig.~3 we present $<P^2_T>$ -- $x_F$ correlations
at various energies. As seen, there is a smooth evolution, though there is a change of process contributions with energy in
the FTF model. Quark exchange processes and one-vertex diffractions dominate at low energies. At higher energies, the diffraction
is stay on the same lavel, but an yield of non-diffractive processes is increased.
\begin{figure}[bth]
	\begin{center}
		\resizebox{4in}{4in}{\includegraphics{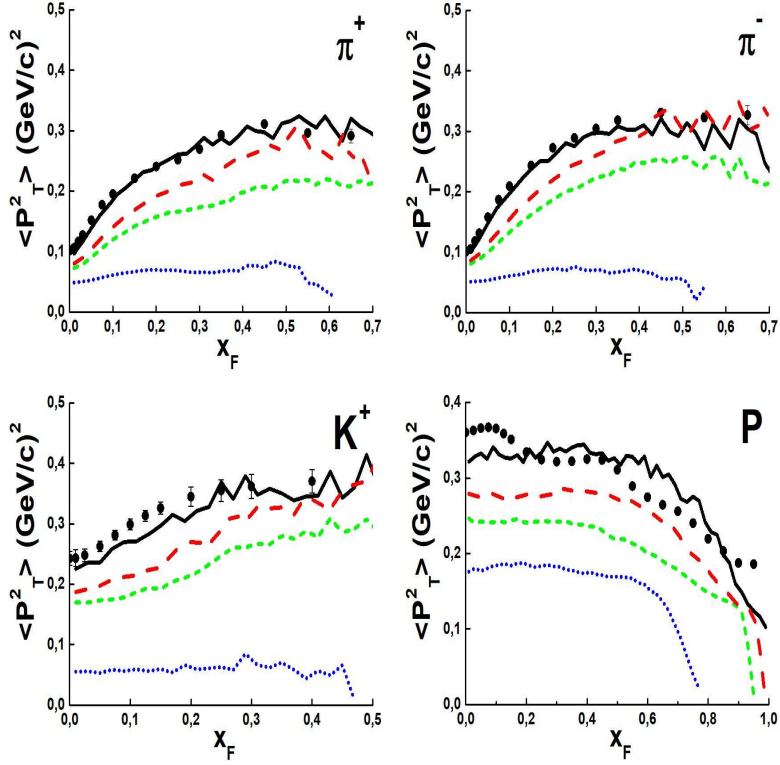}} 
		\vspace{-3mm}
		\caption{$<P^2_T>$ -- $x_F$ correlations in $pp$ interactions at $\sqrt{s_{NN}}$=3.2, 6.3, 8.8 and 17.3 GeV
			(lines from bottom to top). Points are the NA49 data \protect{\cite{NA49pi,NA49pn,NA49K}}. Lines are FTF 
			model calculations.
		}
	\end{center}
     \label{Fig2a}
	\vspace{-5mm}
\end{figure}

In order to clarify a nature of the correlations, we propose to study two-particle $P_T$ correlations
in soft interactions. The correlation can be also studied by current experiments at RHIC and the LHC.

As known, two-particle correlations were intensively used in the past at study of particles and
jets in high energy physics (see, for example, \cite{Phenix200,ATLAS_2T,ATLAS_pPb,ATLAS_5T,ALICE}). 
At the present time, there 
are well developed methods of jet recognition \cite{anti-kt1,anti-kt2} and analysis of thier properties.
More complicated situation takes place at sufficiently lower energies, especially for 
nucleus-nucleus interactions, when hard jets cannot be produced. Though, some methods of high 
energies can be adjusted for lower energies. Instead of jets with high $P_T$, one can choose
as a trigger a particle of any type in an event. An associated particle can be choosen in the 
same manner in the event, where the trigger paticle is produced.

In general, two-particle $P_T$ correlation function can be determined as:
\begin{equation}
C(\vec P^{tr}_T,\vec P^{as}_T)=\frac{1}{N_{tr}} \ \frac{d\ N(tr,as)}{d^2P^{tr}_T\ d^2 P^{as}_T},
\label{Eq6}
\end{equation}
where $\vec P^{tr}_T$ is a transverse momentum of a trigger particle. $\vec P^{as}_T$ is a
transverse momentum of an associated particle. $N_{tr}$ is a number of the trigger 
particles. $N(tr,as)$ is a number of pairs -- trigger particles and associated particle
having pre-determined values of $\vec P^{tr}_T$ and $\vec P^{as}_T$.

The function $C$ is a function of 4 independent variables. Though, accounting azimuthal symmetry
of interactions of unpolarized particles there must be only 3 independent variables. We propose to
use as the variables the module of transverse momentum of the trigger particle ($|\vec P^{tr}_T|$), 
and 2 projections of the vector $\vec P^{as}_T$ on the direction of the vector $\vec P^{tr}_T$,
and on the direction perpendicular to $\vec P^{tr}_T$ (see Fig.~4). In Fig.~4
we choose $\Lambda$ as a trigger particle, and $K$-meson or $\pi$-meson as associated particles.  

\begin{minipage}{5.5cm}
	\resizebox{2in}{2in}{\includegraphics{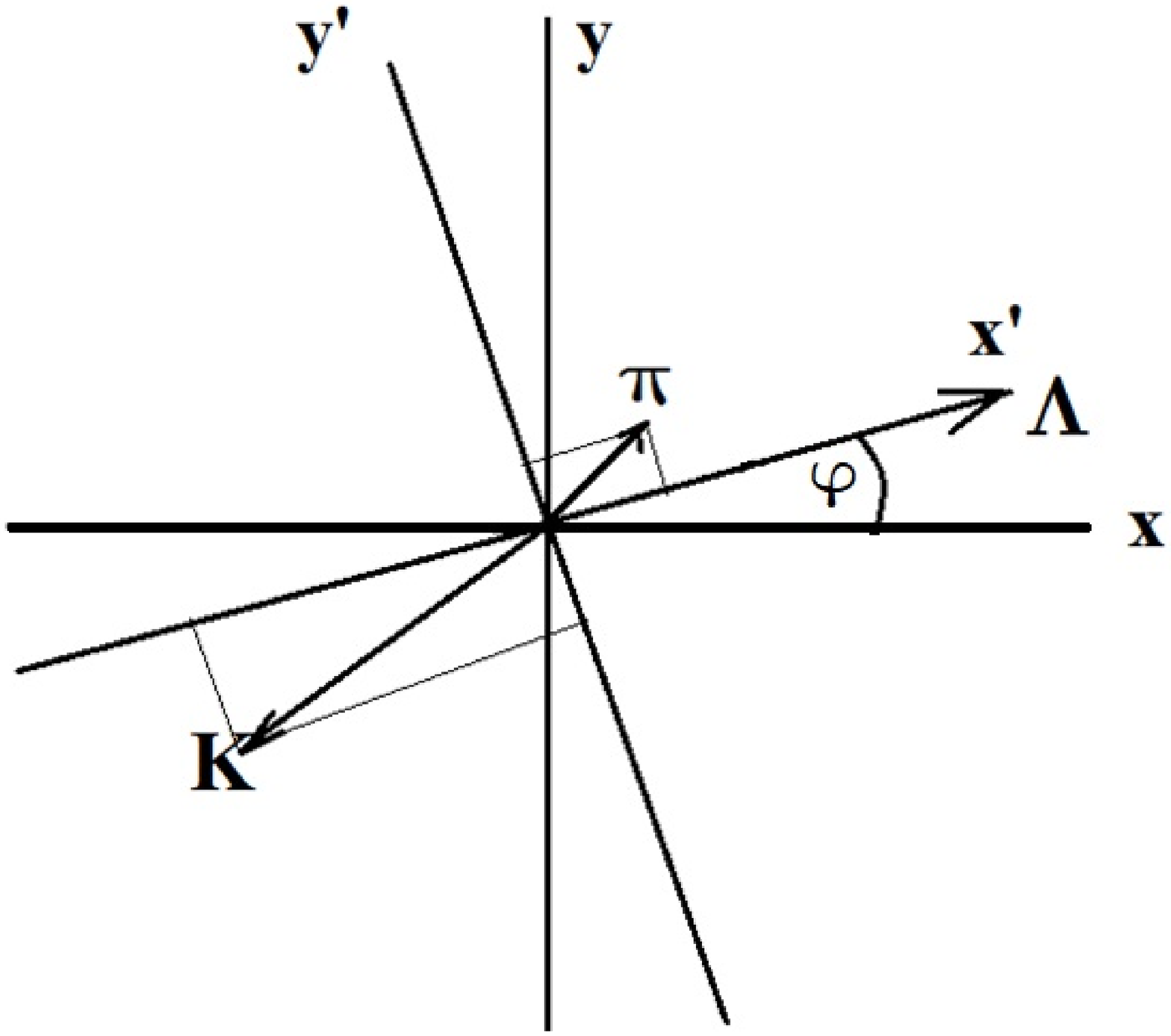}}
	\centerline{Fig. 4} 
\addtocounter{figure}{1}
	\label{Fig16}
\end{minipage}
\begin{minipage}{7.4cm} 
	\begin{equation}
	P^{as}_{T,L}\ =\ \vec P^{as}_T \cdot \vec P^{tr}_T/|\vec P^{tr}_T|,
	\label{Eq7}
	\end{equation}
	\begin{equation}
	P^{as}_{T,T}\ =\ |\vec P^{as}_T \otimes \vec P^{tr}_T|/|\vec P^{tr}_T|
	\label{Eq8}
	\end{equation}
	
	Instead of calculations of scalar and vector products, one can use the following method:
	Determine the azimuthal angle of flying of the trigger particle -- $\phi = \arctan{P^{tr}_{T,y}/P^{tr}_{T,x}}$.
	Make the Euler transform of the coordinate system, and find new components of $\vec P^{as}_T$. 
	\begin{eqnarray}
	P^{as}_{T,x'}=\ P^{as}_{T,x}\cdot \cos{\phi}\ +\ P^{as}_{T,y}\cdot \sin{\phi}, \label{Eq9}\\
	P^{as}_{T,y'}= -P^{as}_{T,x}\cdot \sin{\phi}\ +\ P^{as}_{T,y}\cdot \cos{\phi}. \label{Eq10label} 
	\end{eqnarray}
\end{minipage}


It is obvious, that the new components of $\vec P^{tr}_T$ will be $P^{tr}_{T,x'}=|\vec P^{tr}_{T}|$
and $P^{tr}_{T,y'}=0$. In the following, we will omit subscripts "$T$", and apostrophes of $x$ and $y$ 
for the new components of vectors.

Trigger and associated particles can be chosen in various rapidity/pseudo-rapidity windows. Types
of the particles can or cannot coincide. It is useful to consider also the correlation functions
integrated on one or two variables.

Let us consider a connection of the functions with transverese momentum generation mechanism
(see Fig.~5).
Let us take a baryon ($\Lambda$) as a trigger particle. The transverse
momentum of the baryon ($\vec P^{\Lambda}$) is a sum of the transverese momentum of diquark 
($\vec  P_{qq}$) after previouse fragmentation steps and the transverse momentum of $s$-quark
($P_s$) produced from the vacuum: $\vec P^{\Lambda} =\vec  P_{qq}\ +\ \vec P_s$. The transverse momentum
of an associated hadron $\vec P^{as}=-\vec P_s\ + \vec P_q$. $\vec P_q$ is a transverse
momentum of a quark produced at the next fragmentation step from the vacuum.

\begin{minipage}{4cm} 5.5
	\resizebox{1.5in}{1in}{\includegraphics{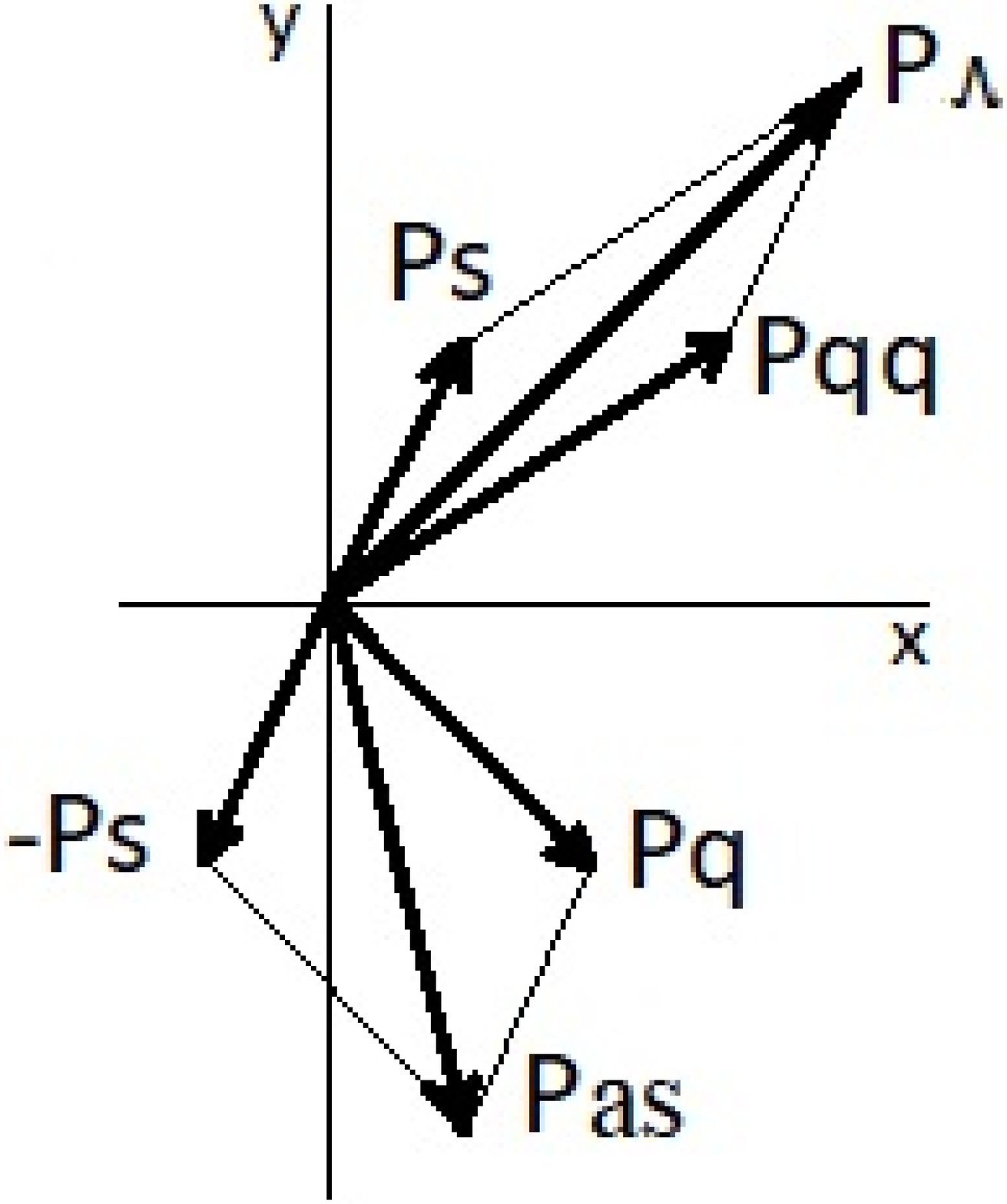}} 
	\centerline{Fig. 5} 
	\addtocounter{figure}{1}
	\label{Fig4}
\end{minipage}
\begin{minipage}{8.5cm} 
According to the Schwinger's model \cite{Schwinger} implemented in the LUND fragmentation model \cite{LUND}
distributions of $\vec P_s$ and $\vec P_q$ are Gaussian ones. Assume for simplicity, that
distribution of $\vec  P_{qq}$ has also Gaussian shape. 

~~~~~~~~In the case, a probability density of the process is given by the expression:
\begin{equation}
W=
\left[\frac{1}{\pi\ \sigma_1}\ e^{-\vec P_{qq}^2/\sigma_1}\right] 
\left[\frac{1}{\pi\ \sigma_2}\ e^{-\vec P_{s}^2 /\sigma_2}\right] 
\left[\frac{1}{\pi\ \sigma_2}\ e^{-\vec P_{q}^2 /\sigma_2}\right]
\nonumber
\label{Eq11}
\end{equation} 
\end{minipage}

$$
\propto \exp(-\vec (P^{tr})^2/\sigma_1\ -\ \vec (P^{as})^2/\sigma_2\ +\ 
\frac{\sigma_1\ \sigma_2}{2 \sigma_1\ +\ \sigma_2}(\vec P^{tr}/\sigma_1 - \vec P^{as}/\sigma_2)^2).
$$

\noindent Choosing $x$-axis direction along the vector $\vec P^{tr}$, we have:  
\begin{equation}
W(\vec P^{tr},\vec P^{as})\ \propto\ \exp\left(-\frac{\sigma_1+\sigma_2}{\sigma_2 (2 \sigma_1+\sigma_2)}
\left[P^{as}_x+\frac{\sigma_2}{\sigma_1+\sigma_2} P^{tr}\right]^2 - 
\frac{\sigma_1+\sigma_2}{\sigma_2 (2 \sigma_1+\sigma_2)} (P^{as}_y)^2  \right)
\label{Eq13}
\end{equation}

Using Exp.~\ref{Eq13}, one can obtain for average values of projections of vector $\vec P^{as}$
on $x$ and $y$ axes the following expressions:  
\begin{equation}
<P^{as}_x>=-\frac{\sigma_2}{\sigma_1+\sigma_2} P^{tr},
\label{Eq14}
\end{equation}  
\begin{equation}
<(P^{as}_y)^2>=\frac{1}{2} \frac{2 \sigma_1+\sigma_2}{\sigma_1+\sigma_2}\cdot \sigma_2.
\label{Eq15}
\end{equation}

It is useful to consider the following cases:

$1)\ \ \sigma_1=0,\ \ \ <(P^{as}_y)^2>=\sigma_2/2,\ \ \ \ <P_{as,x}>=-P_{tr}$

$2)\ \ \sigma_1=\sigma_2,\  <(P^{as}_y)^2>=\frac{3}{4}\sigma_2,\ \ \ <P^{as}_x>=-\frac{1}{2}P^{tr}$

$3)\ \ \sigma_2\simeq 0,\ \ \ <(P^{as}_y)^2>\simeq 0,\ \ \ \ \ \ <P_{as,x}>\simeq 0$

In the first case, we assume that diquarks have no transverse momentum. The second case corresponds
to the assumptions that average momenta of quarks and diquarks are equal. The last one shows the
obvious results, when transverse momenta of sea quarks are equal zeros. 

Let us test the method on events generated by Pythia 6.4 and Geant4 FTF models 
at $\sqrt{s_{NN}}=$ 10 GeV. We chose for the testing $\Lambda$-hyperons as trigger particles, 
and considered anti-$\Lambda$ hyperons, $K$-mesons and $\pi$-mesons as associated particles.
We expected that there will be an essential difference between the model preductions due to the
difference in the baryon production mechanisms. As seen in Fig.~6, it is so. We present
in Fig.~6 $<P^{as}_x>$ as functions of triggered $\Lambda$ momentum for various associated
particles. 
\begin{figure}[bth]
\begin{center}
\resizebox{5in}{4.5in}{\includegraphics{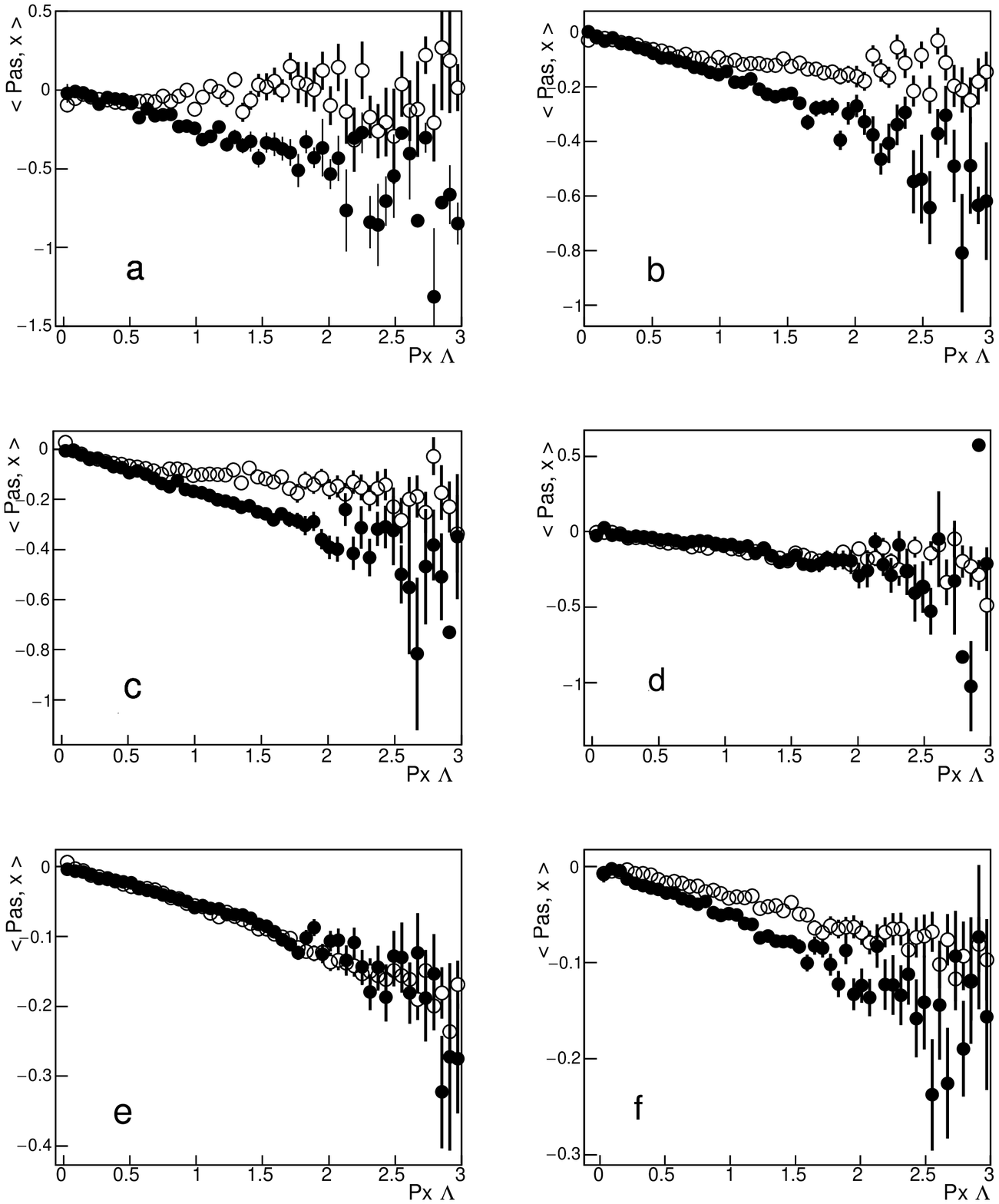}} 
	\vspace{-3mm}
	\caption{$<P^{as}_x>$ as functions of triggered $\Lambda$ momentum for associated particles -- 
		$\bar \Lambda$, $K^0$, $K^+$, $K^-$, $\pi^+$ and $\pi^-$, figures {\it a, b, c, d, e} and {\it f},
		respectively, in $pp$ interactions at $\sqrt{s_{NN}}=$ 25 GeV. Open and filled symbols are Pythia 
		and FTF calculation results, respectively.}
\end{center}
\label{Fig6}
\vspace{-5mm}
\end{figure}

As seen, first of all, the correlations are approximately linear functions of  $P^{tr}$, as it was expected.  
A large deviation between the model results takes place for $\Lambda$ correlations with $\bar \Lambda$, 
$K^0$, $K^+$ and $\pi^-$ mesons. The models give close predictions for $\Lambda$ correlations with  
$K^-$ and $\pi^+$ mesons. In general, the FTF model predicts strongest correlations between
baryons and anti-baryons. Other strong correlations are between triggered $\pi$ mesons and baryons. The correlations
are weaker in the Pythia model. There is practically no correlations between $\Lambda$ and $\bar \Lambda$
hyperons in the Pythia model.  

According to the FTF model, there is an evolution of the correlations from strong to less strong ones
in the energy range $\sqrt{s_{NN}}=$ 3 -- 15 GeV. It is connected with dying out of the quark exchange 
processes in the FTF model. At higher energies the correlations become "frozen". A next step in 
the function's evolution can be at copious gluon production.

Correlations of $<(P^{as}_y)^2$ with $P^{tr}$ give a possibility of a direct check of the Schwinger's mechanism.
The correlations are presented in Fig.~7.

As seen, the correlations are practically constant. The model predictions are close to each other for $\pi^\pm$ and
$K^\pm$ mesons. There is an essential difference between the predictions for $\bar \Lambda$ hyperons and $K^0$ 
mesons. 

\begin{figure}[bth]
\begin{center}
\resizebox{5in}{4.5in}{\includegraphics{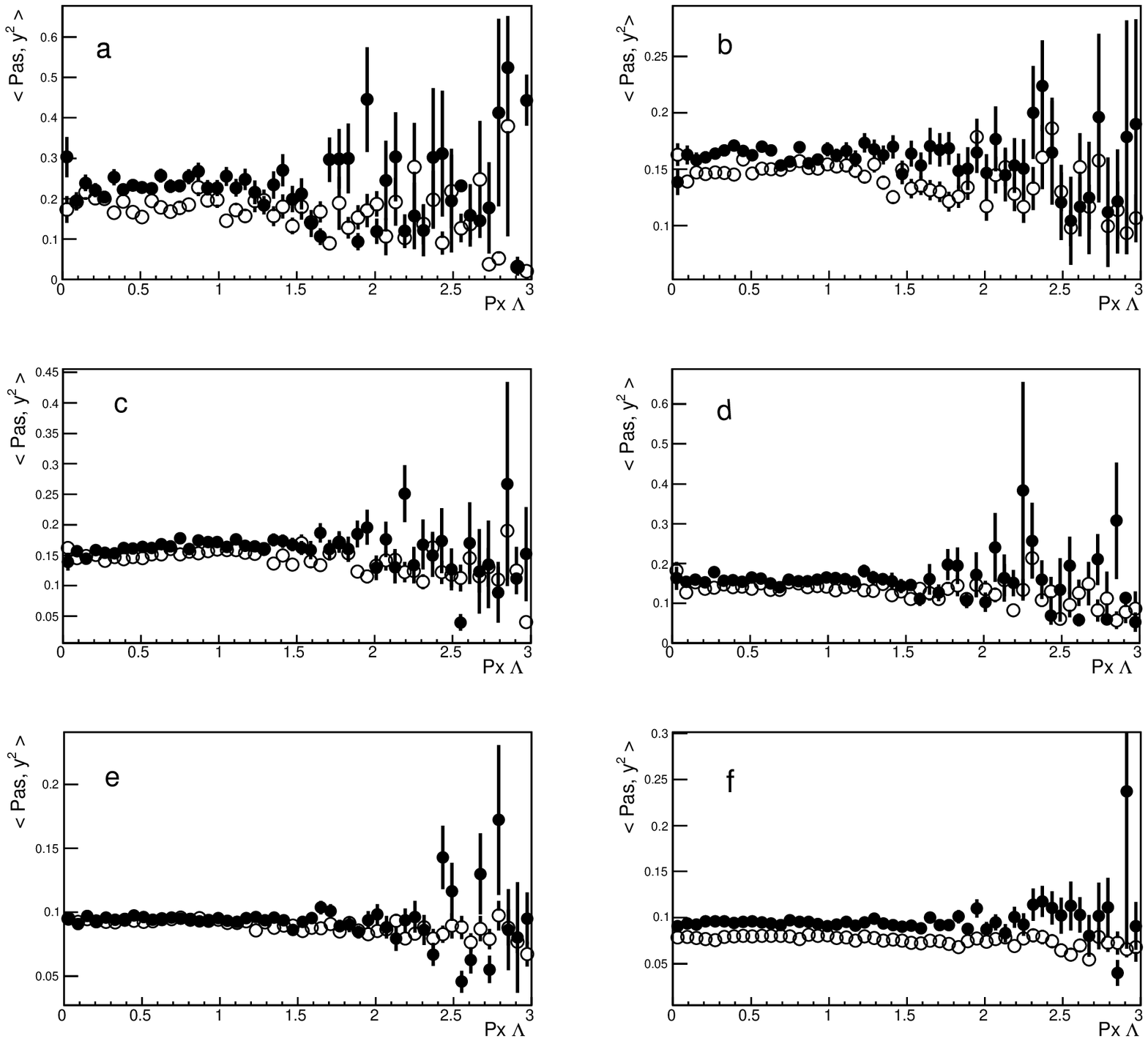}}    
	\vspace{-3mm}
	\caption{$<(P^{as}_y)^2>$ as functions of triggered $\Lambda$ momentum for associated particles -- 
		$\bar \Lambda$, $K^0$, $K^+$, $K^-$, $\pi^+$ and $\pi^-$, figures {\it a, b, c, d, e} and {\it f},
		respectively, in $pp$ interactions at $\sqrt{s_{NN}}=$ 25 GeV. Open and filled symbols are Pythia 
		and FTF calculation results, respectively.}
\end{center}
\label{Fig7}
\end{figure}

The correlations evolve in the energy range $\sqrt{s_{NN}}=$ 3 -- 15 GeV in the FTF model. We believe all
of the considered correlations can be studied at the SPD experiment. 




\section{\bf Puzzles of soft photons pp, pA and AA interactions \protect\footnote{This section is written by E.\,Kokoulina ({E-mail:kokoulina@jinr.ru}) and  
V.A.\,Nikitin ({E-mail: nikitin@jinr.ru}).}}

\begin{abstract}
\vspace{0.2cm}

Over three decades there has been no comprehensive understanding of the mechanism of soft photons (energy smaller 50 MeV) formation. Experimental data indicate an excess of their yield in hadron and nuclear interactions in comparison with calculations performed theoretically. In JINR, in connection with the building of a new accelerator complex NICA, it has become possible to carry out such studies in pp, pA and AA interactions at energies up to 25 $A\,$GeV. We prepared the extensive physical program for soft photons that covers the wide region of investigations in high energy physics. To carry out this program our group develops the conception of an electromagnetic calorimeter on type ``shashlik'' based on gadolinium-gallium garnet (GaGG) crystals, which have significantly lower the threshold for registration of photons. The first tests of electromagnetic calorimeters manufactured at JINR on the basis of the GaGG and a composite of tungsten and copper confirm that choice.
\end{abstract}
\vspace*{6pt}

\noindent

PACS: 12.39.{-}x, 12.40.{-}y


Over more than three decades, a question about the mysterious origin and properties of soft photons (SP), has been lacking a comprehensive understanding. A few experimental  groups found out  an increased yield of photons in the region of low transverse momentum, $p_T$, smaller than 50 MeV/$\it c$ \cite{Chliapnikov:1984ed,HELIOS:1989,NA22:1991,Banerjee:1992ut,WA91:1997,WA102:2002,Z0:2006,PI0:2010,MU:2008}. These photons are not the decay products of short-living particles (including resonances). 

According to QCD, at high energies, $q\overline q $, $qg$ and $gg$ interactions lead to  the emission of photons called direct photons. In spite of the fact that the majority of photons  in high energy interactions comes from decays of secondary hadrons (neutral mesons and others), the direct photons present a unique opportunity to study the soft gluon component of a nucleon and a stage of hadronisation.  Direct photons interact with the surrounding medium only electromagnetically as opposed to the strong interaction of hadrons. So SP keep more information about the medium at all stages of the interaction. This is especially  valuable, along with the information about the secondary hadrons.

The  first convincing experiment on the detection SP has been carried out at the Big European Bubble Chamber (BEBC) \cite{Chliapnikov:1984ed}. The study of SP production in $K^+p$ interactions at 70 GeV/{\it c} beams of $K^+$-mesons after the subtraction of all known hadron decays indicated their excess compared to what was expected from QED inner bremsstrahlung. 

The next series of experiments have been conducted at higher energies \cite{HELIOS:1989,NA22:1991,Banerjee:1992ut,WA91:1997,WA102:2002}. The excess of the SP yield was fourfold or even  higher, eightfold. The last European experiment has been realised at the SPS accelerator, CERN by the DELPHI Collaboration \cite{Z0:2006}. The yield of SP as a function of the neutral pion number turned out to be completely unexpected \cite{PI0:2010}. The excess of the inner bremsstrahlung rates over the predicted ones as a function of the neutral multiplicity in the quark jet turned out to be seventeen-fold for the largest recorded number of pions $N_{neu}$ = 6. The muon bremsstrahlung photons in the reaction {\it e}$^+${\it e}$^-$ $\to $ Z$^0$ $\to \mu ^+\mu ^- $  has demonstrated a good agreement of the observed photon rate with predictions from QED for the muon inner bremsstrahlung \cite{MU:2008}. 

Our BGO calorimeter has been installed at the NIS-GIBS setup at an angle of 16$^\circ $ relative to the beam direction, registering the energy release of gamma quanta. The Monte Carlo simulation has also been carried out in the conditions of the last assembly and with the beam energy 3.5 $A\,$GeV. 
Spectra of photon energy release in deuterium-carbon and
lithium-carbon (see \cite{spa:2020}) interactions have been obtained. In the region
of energy release below 50 MeV, a noticeable
excess over the Monte-Carlo simulation 
is observed. 

There are a few phenomenological models which have been worked out to describe the SP spectra \cite{Van_Hove:1989,Barshay:1989,Wong:2020, GDM:2004}. The most attractive among them is the model of the cold quark-gluon plasma of L. Van Hove \cite{Van_Hove:1989}. But until now, a complete comprehensive understanding of the nature of the anomalous yield of SP has not been achieved.

\subsection{The scientific program of SP study} 
\label{section:context}
In accordance of the gluon dominance model (GDM) design \cite{GDM:2004,GDM1:2016} multiparticle production is described as the convolution of two stages. The first stage is described in QCD as a quark-gluon cascade. For the second one, the phenomenological scheme of hadronization is applied. GDM evidences that the sources of secondary particles are gluons that we call active, and an abundance of soft gluons can be the sources of SP. They are picked up by newly born quarks with a subsequent dropping of energy by emission of SP:  $g + q \to \gamma + q$ or $q+\bar q \to \gamma $. At that valence quarks are staying in the leading particles. Our estimations of the emission region of SP (its linear size) for $pp$ $\to $ hadrons + photons at U-70  in case of  a near-equilibrium state at using the black body radiation  exceeds the typical hadronisation region (1 fm) and reaches a value of about 4--6 fm \cite{GDM1:2016}. We consider the soft gluon component of protons is relevant for the understanding of the spin structure of nucleons. 

The formation of the pionic (Bose-Einstein) condensate in the region of high total multiplicity  (N $>>$ ${<N>}$, $N=N_{ch}+N_{0}$, ${<N>}$ -- the average multiplicity of charged and neutral pions) can be related with the exceeding yield of SP \cite{Barshay:1989}. The growth of the scaled variance $\omega ^0~=~({<N_0^2>}-{<N_0>}^2)/{<N_0>}$ will be the signal of pionic condensate formation  \cite{Gor:2007}. 

The increasing yield of photons at low energy gives us the opportunity to calculate two particle correlations of direct photons. Similar results were obtained by the WA98 Collaboration \cite{WA98:2004}. The deviation from theoretical predictions is at the level of a fraction of a percent.

The next item of our scientific program appeared thanks to the RHIC experiment, namely with using the variable flow or {\it v}$_2$. There is an interesting prediction in \cite{Kodama:2016} about the growth of {\it v}$_2$ in the region of small $p_T$ of photons. This dependence can be the evidence of a coherent SP emission. Our program proposes to test this behaviour.

The scientist from the US Cheuk-Yin Wong \cite{Wong:2020} develops an intriguing model of open string QED mesons to explain and describe the $p_T$-spectrum of SP. Wong believes that  since $q$ and $\bar q$ cannot be isolated, the intrinsic motion of the $q \bar q$ system in its lowest-energy state lies predominantly in 1+1 dimensions as in the open string with $q$ and $\bar q$ at its two ends. 
Extrapolating into the $q\bar q$ QED sector in which $q$ and $\bar q$ interact with the QED interaction, he finds an open string QED meson state at 17.9 $\pm $ 1.5 MeV and QED meson state at 36.4 $\pm $ 3.8 MeV. These predicted masses of the isoscalar and isovector QED mesons are close to the masses of the hypothetical X17 \cite{X17:2016} and E38 \cite{E38:2019} particles observed recently, making them good candidates for these particles. This hypothesis has generated a great deal interest \cite{NA64:2018}. 

The decay of these particles can show up as an excess of $e^+e^-$ and $\gamma \gamma$ pairs in the SP phenomenon \cite{Wong:2020}. 
Wang shows that  an astrophysical object consisting of a large assembly of QED mesons such as the X17 particle with a mass $m_X $ = 17 MeV will be an electron-positron and gamma-ray emitters. If the temperature of such an assembly is low, it can form a Bose-Einstein condensate. Such assemblies of QED mesons present themselves as good candidates as $e^+e^-$ emitters, gamma-ray emitters, or a part of the primordial cold dark matter. 

\subsection{The preparation to experimental SP study} 
\label{section:prep}
To carry out our scientific program we have to manufacture a space saver electromagnetic calorimeter (ECal) with the capacity of low energy gamma
quanta detection . As it is known the production of homogeneous crystalline ECal's is expensive. Usually physicists prefer making heterogeneous assemblies \cite{Grup:2011}. They are cheaper and at that possess satisfactory properties. Our activity was  aimed at testing of two types of ECal's. The ``spaghetti'' type was the first prototype that we produced and irradiated with photon beams in Germany \cite{spa:2020}. 

We substituted 
light scintillator material for a very dense (heavy) crystal having a high specific
light yield \cite{Grup:2011}. It allowed us to build a device that is capable of maintaining considerable compactness (space-saving of about 30$\%$). To produce the ``spaghetti'' type we have chosen a mono-crystal of gadolinium-gallium garnet, Gd$_3$Al$_2$Ga$_3$O$_{12}$:Ce (GaGG), as a scintillator and
tungsten+copper composite by way of the absorber. We expect the decay time to be $\sim $ 90 ns; 
the light yield to be $\sim $ 45,000--55,000 ph/MeV; an estimated price of 25--35 $\$ $/cm$^3$ of volume, and a good radiation resistance. 

GaGG crystal is a fast-acting scintillator, its light yield is 4 times greater than that of BGO crystals. 
And we can  acquire these crystals from the well-known domestic firm ``Fomos-Materials''.
The Monte-Carlo simulation has shown that the necessary energy resolution for SP in the case of the ECal ``spaghetti'' type
turns out lower than when at using the ``shashlik'' type. See figures in \cite{spa:2020}) for an example of a typical view of  the ECal ``shashlik'' type. 

We plan to assemble the ECal ``shashlik''. It is assumed that it will consist or 16 or more of GaGG plates (100$\times $100$\times $3)mm$^3$, 15 plates of 2-mm-absorber (W:Cu composite, 1:19), and its total length will be
about 138 mm. GEANT4 simulation shows that the energy resolution of the  ``shashlik''  is considerably better than ``spaghetti'' type. The dimensions of both assemblies were optimised by the Monte-Carlo simulation for obtaining the required characteristics. The design of the ``shashlik'' allows expanding it up to any length. We also can exclude the W/Cu absorber from the ``shashlik'' and then to assemble the homogeneous device for the detection of MeV photons if the energy resolution of the ECal will not be of a good enough quality.

We would like to thank all participants  of SP study activity for a long years.and also the leadership of our laboratory for significant support.

\newcommand{\ltsim}{\protect\raisebox{-0.5ex}{$\:\stackrel{\textstyle <}
	{\sim}\:$}}
\newcommand{\gtsim}{\protect\raisebox{-0.5ex}{$\:\stackrel{\textstyle >}
	{\sim}\:$}}
\newcommand{\ltsimscript}{\protect\raisebox{-0.5ex}{$\stackrel{\scriptstyle <}
	{\sim}$}}
\newcommand{\gtsimscript}{\protect\raisebox{-0.5ex}{$\stackrel{\scriptstyle >}
	{\sim}$}}

\originalTeX

\section{\bf {Hadron formation effects in heavy ion collisions}
\protect\footnote{This section is written  by A.\,B.\,Larionov; {E-mail:
larionov@theor.jinr.ru}}}

\begin{abstract}

The space-time picture of hadron formation in high energy processes with nuclear targets is still poorly known.
It is suggested to test different models of hadron formation by using  collisions of heavy ions.
Results of microscopic transport calculations of proton and charged pion rapidity and transverse
momentum distributions in C+C and Ca+Ca collisions at $\sqrt{s}_{NN}=11$ GeV are presented. 
\end{abstract}
\vspace*{6pt}

\noindent
PACS: 25.75.-q; 25.75.Bh; 25.75.Dw

\label{intro}

Hadrons produced in a hard exclusive particle-nucleon collisions emerge first in a form of prehadrons having
a reduced transverse size $\sim 1/\sqrt{Q^2}$ where $Q^2$ is a hard scale. 
For exclusive meson electroproduction $e N \to e^\prime N^\prime M$ ($M=\pi,\rho$), the hard scale is
the photon virtuality (cf. Ref. \cite{Larson:2006ge}). 
In the case of large-angle hadronic scattering $h N \to h^\prime N^\prime$ with $-t \simeq -u \simeq s/2$,
the hard scale is given by the momentum transfer, i.e. $Q^2=-t$, (cf. Ref. \cite{Larionov:2019xdn} and refs. therein).
The prehadrons, called also the point-like configurations (PLCs), interact with nucleons with reduced strength
which is known as the color transparency (CT) phenomenon, see Ref. \cite{Dutta:2012ii} for the most recent review of CT.

In inclusive production channels of the $NN$ collisions, there is no fixed hard scale, even at large $\sqrt{s}$.
Thus, formation of PLCs is questionable here. However, the hard scale will fluctuate event-by-event and one can expect
that in average the transverse size of the outgoing hadrons will be still reduced. This motivates inclusion of the
hadron formation effects in microscopic transport models for high-energy heavy ion collisions, such as UrQMD \cite{Bass:1998ca},
HSD \cite{Cassing:1999es}, GiBUU \cite{Buss:2011mx}. \footnote{Based on the CT concept, also the incoming nucleons are transversely
'squeezed' if they are in the initial state of a hard collision. This effect is totally neglected in transport models.}

Hadron formation studies in the current fragmentation region \cite{Gallmeister:2007an} based on the Deep Inelastic Scattering (DIS) data
from HERMES and EMC concluded the linear increase of the prehadron-nucleon cross section with time (or length traveled by a prehadron),
in agreement with the quantum diffusion model (QDM) \cite{Farrar:1988me}.
In the QDM, the prehadron is converted to the 'normal' hadron after passing the distance called formation (or coherence) length
\begin{equation}
    l_h  \simeq \frac{2p_h}{|M_h^2-M_{h^\prime}^2|} \sim 0.4-0.6 \mbox{(fm/GeV)} \cdot  p_h\mbox{(GeV)}~,    \label{l_h}
\end{equation}
where $p_h$ is the momentum of hadron $h$, $M_h$ is the mass of the hadron $h$, and $M_{h^\prime}$ is roughly the mass of the closest
radially excited state $h^\prime$.
Thus, based on the QDM one can estimate that particles with momenta above $\sim 10$ GeV leave the nucleus almost without interactions.
On the other hand, the analyses of E665 data on the low-energy ($E < 10$ MeV) neutron production in the 470 GeV muon DIS on Pb target
\cite{Strikman:1998cc,Larionov:2018igy} favor a surprisingly low cutoff value of the momentum $\sim 1$ GeV above which the
prehadrons do not interact in the nucleus. This can not be explained by the QDM or any other existing model of hadron formation and
calls for the new experiments on hadron formation in the target fragmentation region, e.g. in virtual photon - nucleus
collisions at the future Electron-Ion Collider (EIC) or in ultraperipheral heavy ion collisions at the LHC and RHIC \cite{Larionov:2018igy}.
The physics of target fragmentation is one of perspectives for the experimental program of the EIC  \cite{CFNSAdhoc}. 

Another opportunity to study hadron formation effects is offered by heavy ion collisions
at NICA-SPD regime ($\sqrt{s}_{NN} \sim 10$ GeV). Here, the stopped nucleon in the $NN$ center-of-mass system (c.m.s.)
has $x_F = m_N/\sqrt{s}_{NN} \sim 0.1$. Therefore, slow nucleons in the $NN$ c.m.s. represent a mixture of the current ($x_F \ll 1$)
and target ($x_F \ltsimscript 1$) fragmentation products. This paper addresses the sensitivity of heavy ion collision observables to the different
treatments of hadron formation. The calculations are performed within the Giessen Boltzmann-Uehling-Uhlenbeck (GiBUU) model. 

In sec. \ref{model}, the GiBUU model is briefly described with a focus on hadron formation.
Sec. \ref{results} contains the results of numerical calculations of the  the rapidity- and $p_t$ spectra
of protons and pions in C+C and Ca+Ca collisions at $\sqrt{s}_{NN}=11$ GeV.
The summary and estimation of rates at NICA-SPD are given in sec. \ref{summary}.

\subsection{The model}
\label{model}

The GiBUU model (see detailed description in ref. \cite{Buss:2011mx}) solves the system of kinetic equations for the baryons ($N$, $N^*$, $\Delta$, $\Lambda$, $\Sigma, \ldots$),
corresponding antibaryons ($\bar N$, $\bar N^*$, $\bar\Delta$, $\bar\Lambda$, $\bar\Sigma, \ldots$), and mesons ($\pi$, $K, \ldots$).
The kinetic equations are coupled via collision integrals and mean field potentials. The latter are determined self-consistently,
i.e. they depend on the actual particle distributions in the six-dimensional phase space of position and momentum.
The distribution function is projected onto the set of point-like test particles. Kinetic equation is then solved if the test particle positions and momenta propagate in time
according to the Hamiltonian equations of motion between collisions. \footnote{In the present calculations we disregard mean field potentials. Thus, the test particles propagate along
  straight-line trajectories.}
The two test particles experience a two-body collision during the time interval $[t-\Delta t/2;t+\Delta t/2]$ if they approach their minimal distance $d$
in position space during this time interval and if $d < \sqrt{\sigma_{\rm tot}/\pi}$.
Here, $\Delta t$ is the time step in the computational frame that is chosen to be the nucleus-nucleus c.m.s. and 
$\sigma_{\rm tot}$ is the total interaction cross section of colliding particles. 

The model also includes three-body collision processes $\pi N N \to N N$ and $\Delta N N \to N N N$ which are simulated by the direct calculation of the three-body absorption rate.     
Unstable resonances experience decays
with a probability $P=\Gamma \Delta t$ where $\Gamma$ is the total decay width of a resonance. The final states of the two- and three-body collisions and resonance decays are sampled by the Monte-Carlo method
according to the partial channel cross sections, widths and angular distributions. Pauli blocking is taken into account for the final state nucleons.

For low-energetic elementary binary collisions, the resonance model complemented with empirical background cross sections (e.g. $s$-wave direct pion production $N N \to N N \pi) $ is implemented.
For such collisions, it is assumed that all produced particles are formed immediately and interact with usual hadronic cross sections. 

High-energy elementary binary collisions are simulated by the PYTHIA and FRITIOF models.
(The latter is applied for antibaryon-baryon collisions only.)
The corresponding transition values of $\sqrt{s}$ for meson-baryon, baryon-baryon, and antibaryon-baryon collisions
are $2.2, 4.0$ and 2.38 GeV. Note that in the narrow $\sqrt{s}$ region $\pm (0.1-0.2)$ GeV centered at the transition value,
the low- and high-energy events are mixed to ensure a smooth behaviour of the cross sections between the two energy regimes.

The particles produced in a high-energy  binary collision are supposed to be in a prehadronic state.
Their effective interaction cross section $\sigma_{\rm eff}$ with nucleon becomes time-dependent and deviates from the
ordinary interaction cross section $\sigma_0$.
We applied the three different models for the ratio of the effective and ordinary interaction cross sections:\\
(i) Based on JETSET-production-formation points (see ref. \cite{Gallmeister:2005ad}, used as default in GiBUU) favored by analysis of hadron
    attenuation at HERMES and EMC \cite{Gallmeister:2007an}:
\begin{equation}
  \sigma_{\rm eff}(t)/\sigma_0=X_0+(1-X_0)\frac{t-t_{\rm prod}}{t_{\rm form}-t_{\rm prod}}~,          \label{sigma_eff_GiBUU}
\end{equation}
where $X_0=r_{\rm lead} a/Q^2$, $a=1$ GeV$^2$, $r_{\rm lead}$ is the ratio of the number of leading quarks to the total number of quarks in the prehadron.
$Q^2$ is defined for the hard $2 \to 2$ subprocess by the PYTHIA variable VINT(52). By default, it gives $Q^2 \sim p_\perp^2$ where $p_\perp$
is the transverse momentum of the hard scattering (see PYTHIA 6.4 manual \cite{Sjostrand:2006za} for detail).\\   
(ii) QDM \cite{Farrar:1988me}:
\begin{equation}
  \sigma_{\rm eff}(t)/\sigma_0=X_0+(1-X_0)\frac{c(t-t_{\rm hard})}{l_h}~.          \label{sigma_eff_QDM}
\end{equation}
Here we set $X_0=0$ for simplicity. $t_{\rm hard}$ is the time moment of the collision in GiBUU calculation.\\
(iii) Momentum cutoff:
\begin{equation}
  \sigma_{\rm eff}/\sigma_0=\Theta(p_{\rm cut}-p_h)~, ~~~p_{\rm cut} \sim 1-2~\mbox{GeV/c}.                \label{sigma_eff_cut}
\end{equation}

Some comments are in order with regard to the Lorentz covariance. In principle, all three above models are explicitly non-covariant.
Thus, one needs to specify the frame in which they are applied. In the case of models (i) and (ii) the non-covariance,
however, should largely cancel-out in the ratios of time- or length differences. The model (iii) is most problematic: it is clear
that the result will certainly depend on the frame in which the particle momentum is calculated. Note that originally Eq.(\ref{sigma_eff_cut})
has been suggested for the target nucleus rest frame \cite{Larionov:2018igy} where DIS is calculated in GiBUU.
In contrast, in the present work, the nucleus-nucleus c.m.s. is used as the calculational frame. The problem related to the choice of
the calculational frame is beyond the scope of this work and should be addressed in future studies.

\newpage

\subsection{Numerical results}
\label{results}

The calculations were performed for the systems C+C and Ca+Ca at $\sqrt{s}_{NN}=11$ GeV.
The maximum impact parameter was set to $8$ fm for C+C and to $11$ fm for Ca+Ca 
that corresponds to the minimum bias trigger.  
The time evolution of the colliding system was calculated until the maximum time 30 fm/c
with the time step $\Delta t=0.2$ fm/c. The total accumulated statistics is $2\cdot10^6$ ($1\cdot10^6$)
nucleus-nucleus collision events for C+C (Ca+Ca).

\begin{figure}[t]
 \begin{center} 
 \includegraphics[width=127mm]{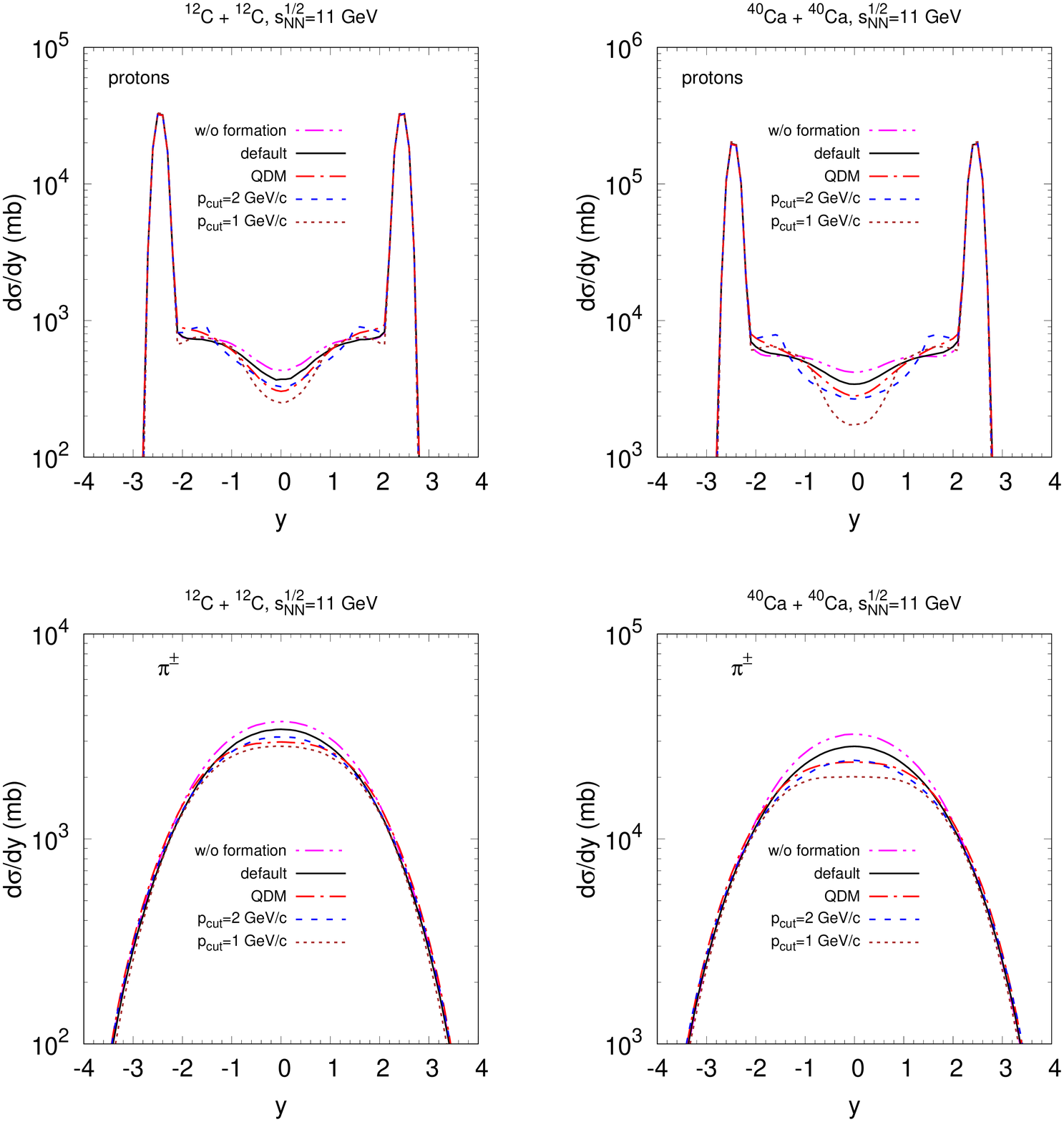}
 \vspace{-3mm}
 \caption{Proton and charged pion rapidity distributions in C+C and Ca+Ca at $\sqrt{s}_{NN}=11$ GeV 
   as indicated. Dot-dot-dashed (magenta) line -- calculation without formation (immediately formed hadrons),
   Solid (black) line -- default formation (i). Dot-dashed (red) line -- QDM model (ii).  
   Dashed (blue) line --  $p_{\rm cut}= 2$ GeV/c (iii).
   Dotted (brown) line --  $p_{\rm cut}= 1$ GeV/c (iii).
   Here (i), (ii) and (iii) refer to the models defined in the text,
   see Eqs.(\ref{sigma_eff_GiBUU}),(\ref{sigma_eff_QDM}),(\ref{sigma_eff_cut}), respectively.}
 \end{center}
 \labelf{fig:dNdy}
 \vspace{-5mm}
\end{figure}
Fig.~\ref{fig:dNdy} shows the rapidity spectra of protons and pions.
Note that we do not distinguish free protons from those bound in nuclear clusters.
This explains the maxima in the proton rapidity spectra at the target and projectile rapidities.   
Calculation without hadron formation gives
the largest production at midrapidity ($y=0$). Hadron formation leads to the reduced FSI of the outgoing hadrons.
Hence, the produced protons lose less longitudinal momentum which results in the depletion of
the proton spectrum at midrapidity. Since less FSI implies less inelastic production by secondary protons,
also the pion spectrum gets depleted at midrapidity in calculations with hadron formation. As expected,
the depletion is stronger for heavier system and for calculations with more restrictions on FSI
($p_{\rm cut}=1$ GeV/c).

\begin{figure}[t]
 \begin{center} 
 \includegraphics[width=127mm]{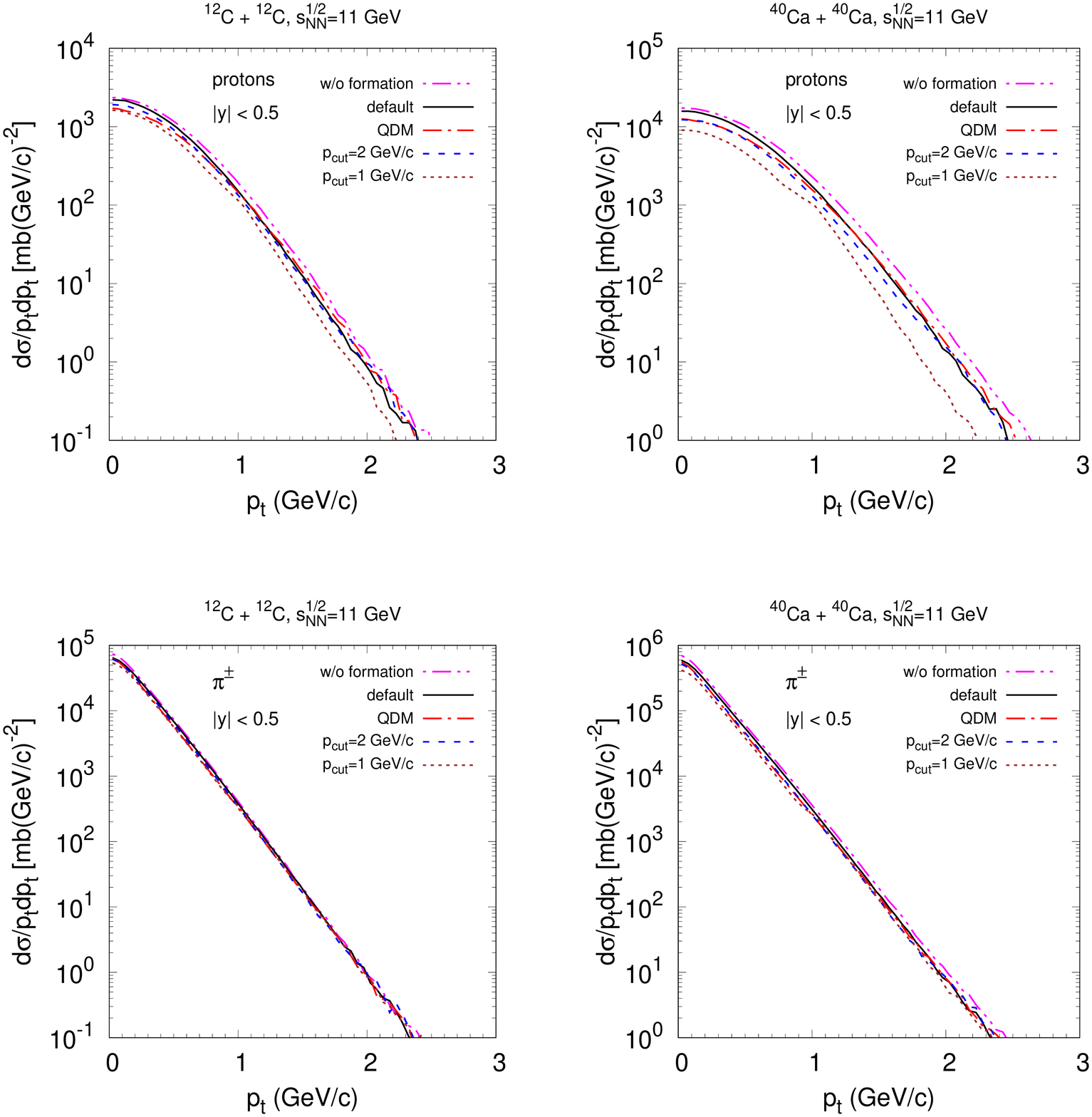}
 \vspace{-3mm}
 \caption{Proton and charged pion  transverse momentum distributions in C+C and Ca+Ca at $\sqrt{s}_{NN}=11$ GeV 
   as indicated. C.m.s. rapidity cut $|y| < 0.5$ is applied. Line notation is the same as in Fig.~\ref{fig:dNdy}.}
 \end{center}
 \labelf{fig:dNdpt}
 \vspace{-5mm}
\end{figure}
Fig.~\ref{fig:dNdpt} shows the proton and charged pion transverse momentum spectra at midrapidity.
The proton $p_t$ spectrum becomes softer due to hadron formation. Indeed, protons effectively gain transverse momentum
in the rescattering processes while hadron formation suppresses rescattering. Pion $p_t$ spectrum is influenced by
hadron formation only very weakly. In the collisions of heavier nuclei \cite{Larionov:2020lzk} the hadron formation effect becomes
more pronounced, especially for central events.

\subsection{Summary and conclusions}
\label{summary}

The purpose of this work was to clarify whether heavy ion collisions can be used to test various models of hadron formation.
To this end, the GiBUU transport model was used that includes the three optional models of hadron formation.
The first two models, (i) and (ii) [Eqs.(\ref{sigma_eff_GiBUU}),(\ref{sigma_eff_QDM})] are based on the CT mechanism
and differ only in the details of the hadron formation length. Both these models were successful in the description
of hadron attenuation in DIS reactions. The third model (iii) [Eq.(\ref{sigma_eff_cut}] assumes that the prehadrons
with momenta above 1-2 GeV/c do not interact with nucleons which is in stark contrast
with widely accepted momentum dependence of the hadron formation length, Eq.(\ref{l_h}). This model was originally adopted
to describe the low-energy neutron production induced by the muon DIS on nuclear targets \cite{Larionov:2018igy}.

The GiBUU model calculations were performed for inclusive proton and charged pion production in C+C and Ca+Ca at $\sqrt{s}_{NN}=11$ GeV. 
In the C+C system, the secondary rescattering effects are minimal and the reaction process is governed 
by first-chance $NN$ collisions. The Ca+Ca system is more suitable for the study of FSI effects.
Proton and charged pion rapidity distributions and proton $p_t$ distributions are strongly sensitive to the hadron formation
effects. More restrictions on FSI of prehadronic states leads to less production at midrapidity and to softer
proton $p_t$ spectra.

In the C+C system the proton and $\pi^\pm$ production cross sections in the c.m.s. rapidity window $|y| < 0.5$ are
about 0.4 b and 3 b, respectively. In the Ca+Ca system the corresponding cross sections are about
one order of magnitude higher.
Thus, the measurements of the proton and charged pion spectra seems easily possible
at the first stage of NICA-SPD operation ($L=10^{25}$ cm$^{-2}$ s$^{-1}$) with the production rate of about
1-10 events per second (without efficiency and acceptance corrections). 
The accuracy needed to separate different models of hadron formation should be better than $\sim 10\%$ for
the production cross sections at midrapidity which is reachable within few hours of beam time. 
  



\section{\bf {
Measurement of characteristics of the processes of pair production of polarized tau leptons in the SPD experiment. 
\protect\footnote{ This section is written by A.\,Aleshko,
E.\,Boos ({E-mail: boos@theory.sinp.msu.ru}),
V.\,Bunichev ({E-mail: bunichev@theory.sinp.msu.ru})}}}

\begin{abstract}


\vspace{0.2cm}

It is proposed to use the Drell-Yan process with pair production of $\tau$ leptons to measure the parameters of polarized parton distribution functions of a proton at the NICA collider in the SPD experiment. To determine the polarization of tau leptons, we propose to use decays of $\tau$ leptons into a single charged pi-meson and neutrino. To parameterize the polarization state of $\tau$ leptons,
it is proposed to use the energy of single $\pi$-mesons.

\end{abstract}
\vspace*{6pt}

\noindent
PACS: 24.10.Lx, 13.85.Ni, 14.20.-c 
\vspace*{6pt}

The probability of the process with a particular parton is described by the function, which is dependent on the fraction of the total momentum of the proton carried by that parton and the momentum transfer scale. These functions are known as parton distribution functions (PDF) and contain information about the internal structure of corresponding nucleons \cite{ss1,ss2}. Hence, for a consistent description of processes with polarized nucleons in a wide energy range and the most complete description of the proton structure a generalized parton model was developed \cite{gpm1,gpm2,gpm3}, which consist of 8 specialized parton distribution functions. These functions are: the distribution of the parton density in unpolarized nucleon (Density); distribution of the longitudinal polarization of quarks in the longitudinally polarized nucleon (Helicity); distribution of transverse polarization of quarks in transversely polarized nucleon (Transversity); the correlation between the transverse polarization of a nucleon and the transverse momentum of unpolarized quarks (Sivers); the correlation between the transverse polarization of a nucleon and the longitudinal polarization of quarks (Worm-gear-T); the distribution of the transverse momenta of quarks in unpolarized nucleon (Boer-Mulders); the correlation between longitudinal polarization of nucleon and the transverse momenta of quarks (Worm-gear-L); the distribution of the transverse momenta of quarks in transversely polarized nucleon (Pretzelosity).

At the current level of our knowledge of the nucleon structure, parton distribution functions cannot be obtained analytically and have to be measured experimentally. The deep inelastic scattering (DIS,SIDIS) and the Drell-Yan (DY) processes are typically used for the measurements of PDFs \cite{exp1,exp2,exp3,dis1,dis2, dy3, dy1, dy2, dy3, dy4, dy5, dy6, dy7, dy8}.

In the Drell-Yan processes, the annihilation of a quark and antiquark, from
colliding nucleons, through an intermediate photon or Z-boson leads to
the production of a lepton-antilepton pair \cite{dy0,dy3}. The matrix element of
the Drell – Yang process consists of a convolution of two tensors \cite{dyf1,dyf2,dy3}. The first one is responsible for the hadron part with initial partons and
includes a dependence on the polarized parton distribution functions.
The second tensor corresponds to the lepton part of the process. Such a
simple structure and the fact that Drell-Yan process is one of the
cleanest hard hadron-hadron scattering processes makes it a very
convenient tool for studying PDFs.

The conventional way for studying parton distribution functions via Drell-Yan process is to use processes with the production of electron-positron (and muon anti-muon) pairs \cite{dy3,dy6,dy8}, because in this case the mass of leptons can be neglected and one can consider them as having some definite helicity. Within this approach angular parameters, which define the spatial orientation of the momenta of leptons relative to the momenta of colliding nucleons, are used as parameters describing the polarization state of the lepton pair. Then one can measure azimuthal asymmetries and extract particular structure functions from them. 

In the case of $\tau$ production, this approach will work only at very high energies. At lower energies, however, the mass of tau cannot be neglected. Massive particles do not have definite helicity. Therefore, for $\tau$ leptons, the
angular parameters as spin-sensitive variables can be used only at collision energy much greater than the mass of the $\tau$ lepton and are
not suitable for energies comparable to its mass. Consequently, we need a more suitable parameterization of the polarization state of $\tau$ lepton. 

\begin{figure}[t]
	\centering
	\includegraphics[width=0.49\textwidth,clip]{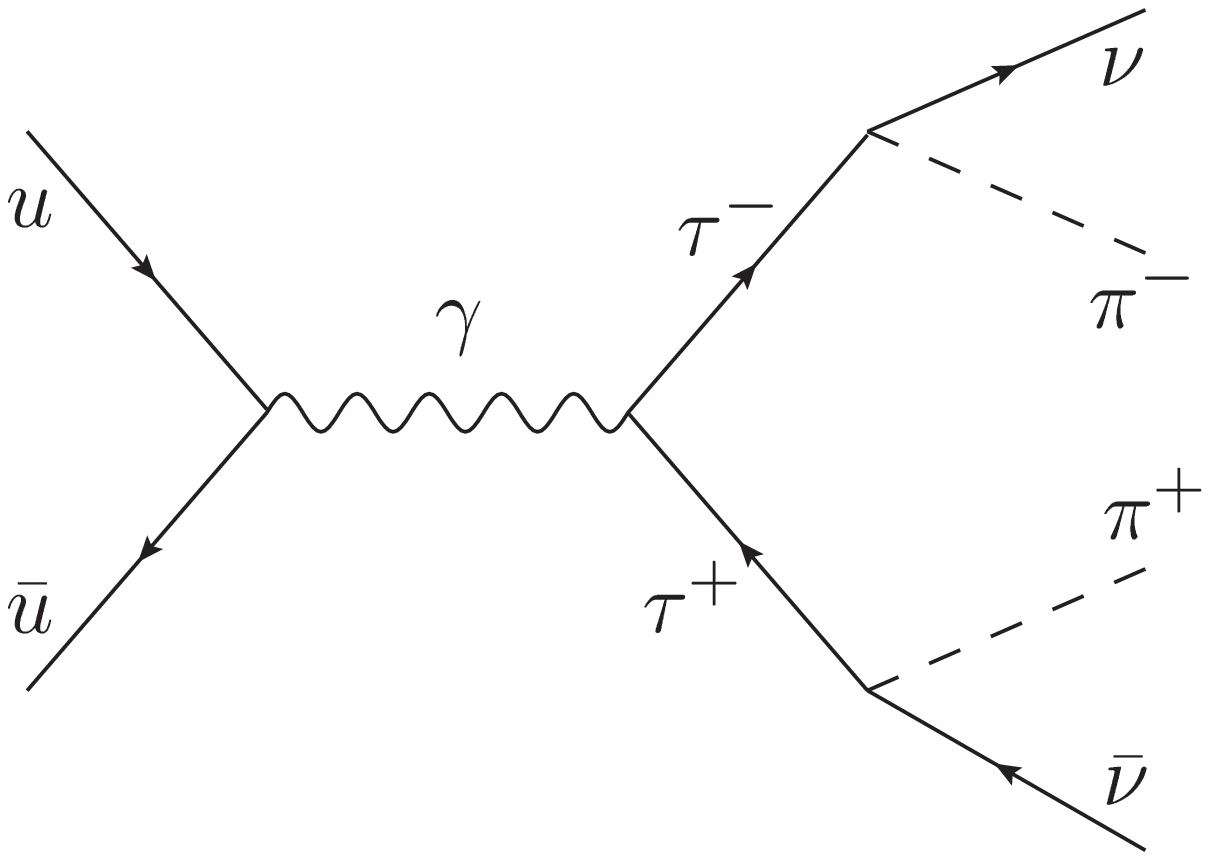}
	\caption{Drell-Yan process with the pair-production of tau leptons and their
		subsequent decay into single charged $\pi$-meson and neutrino.}
	\label{taup1}      
\end{figure}

Our idea to tackle this problem is to utilize the unique decays properties of $\tau$-lepton. Consider the hadronic decay of $\tau$ to a single charged $\pi$-meson and neutrino. Due to the weak nature of this decay and the fact that neutrinos are always left-handed, the energy spectra of $\pi$-meson is strictly correlated with the polarization state of decaying $\tau$ lepton \cite{ch,tp1,tp2,tp3}. For instance, in the case of mostly right-handed $\tau$ decay, due to the angular momentum conservation law, most of the tau’s momentum is transferred to the $\pi$-meson. On the contrary, in the case of mostly left-handed $\tau$, the significant part of its momentum is transferred to the neutrino. Thereby, the energy of single $\pi$-meson is a convenient characteristic of the polarization state of its parent $\tau$. 
So, to determine the polarization of $\tau$ leptons, we suggest to use hadron decays of $\tau$ leptons to a single charged $\pi$-meson and neutrino. The diagram of the corresponding process is shown in the Figure \ref{taup1}. We propose to use the energy of single $\pi$ meson as variable that parametrize the polarization state of $\tau$ leptons. A key feature of our method is that we do not summarize the $\tau$ lepton polarizations, but keep information about the polarizations of both $\tau$ leptons through the energies of single $\pi$ mesons from $\tau$ decays. Due to the fact that a pair of $\tau$ leptons is produced at the NICA collider near the reaction threshold, depending on the parton momenta, the $\tau$ leptons can be produced almost at rest or have a significant momentum, which is reflected in the decay products of the $\tau$ leptons and makes the process with $\tau$ leptons are very sensitive to the state of the spins and momenta of the initial partons and nucleons.

One of the closest opportunities for testing this approach is the future NICA SPD experiment \cite{spd}. Since at the NICA collider a pair of $\tau$ is produced near the reaction threshold, the leptons can be at rest or possess a significant momentum, which affects the $\tau$ decay products and makes this process very sensitive to the polarization states and momenta of the initial hadrons.

To confirm the efficiency of our method, we carried out a detailed numerical simulation. 
First, using the CompHEP \cite{comphep} package, we computed the tree level cross-section of the Drell-Yan $\tau$ pair production process. Also, we used the \(MadGraph5\_aMC@NLO\) package \cite{mg5} to estimate the cross section of the same process  at the next to the leading order in perturbation theory. In both cases we have used the NNPDF3.1 PDF set \cite{nnpdf_un}. As it has been mentioned, the important advantage of the NICA facility is its energy range \(\sqrt{s} = 10-26\) GeV, which lies close to the $\tau$ pair production threshold. However, one of the problems arising in calculating the cross-sections at such low energies is the  correct choice of the QCD factorization scale. At present, certain ambiguity exists as to how to choose this scale properly.  
When dealing with Drell-Yan process it is often suggested to choose the scale to be of the order of the invariant mass of the lepton pair \(Q\), since it is the most obvious characteristic scale of the process. However, the important point given in the aforementioned work is that the choice of the factorization scale \(\mu_{F} = Q\) is not always lead to satisfactory results. The problem especially evident at lower energies, which is the case for NICA. In \cite{scale} authors suggest using scale \(\mu_{F} \sim 0.5 Q\). Other popular choices of the factorization scale include the transverse energy \(E_{T}\) and transverse mass \(m_{T}\) of the event. The results obtained using different types of scales at energy \( \sqrt{s} = 24 \) GeV shown in Table \ref{tautab1}.

\begin{table}
\begin{center}
\begin{tabular}{ |c|c|c| } 
\hline
 {Order} & \multicolumn{2}{c}{LO} \\
\hline
 Scale & cross-section, \(\sigma\) pb & MC error, \(\Delta \sigma\) pb \\ 
 \cline{2-3}
 \(Q \)     & 70.0 & 0.3  \\
 \(Q/2 \)   & 84.8 & 0.4  \\ 
 \(E_{T} \) & 75.7 & 0.4  \\
 \(m_{T} \) & 70.7 & 0.3  \\ 
 \hline
 \hline
 {Order} & \multicolumn{2}{c}{QCD NLO} \\
 \hline
 Scale & cross-section, \(\sigma \) pb & MC error, \(\Delta \sigma\) pb \\
  \cline{2-3}
 \(Q \)     & 83.5 & 0.7  \\
 \(Q/2 \)   & 83.8 & 0.7  \\ 
 \(E_{T} \) & 108.5 & 0.6 \\
 \(m_{T} \) & 105.7 & 0.7 \\ 
 \hline
\end{tabular}
\end{center}
\caption{Cross-sections of the Drell-Yan tau pair production at energy \( \sqrt{s} = 24 \) GeV calculated with different choices of factorization scale \(\mu_{F}\).}
\label{tautab1}
\end{table}
To assess the approximate birthrate of corresponding events in NICA conditions, we utilize the known branching ratio of corresponding $\tau$ decay channel: \(Br(\tau^{-} \to \pi^{-} \nu_{\tau}) =  (10.83 \pm 0.06 )\) \% according to \cite{pdg}. Table \ref{tautab2} contains the estimation of the number of events at two energies, which correspond to the highest proposed luminosities.

\begin{table}
\begin{center}
\begin{tabular}{ |c|c|c| } 

 \multicolumn{3}{c}{\( \sqrt{s} = 24 \text{GeV} (L = 1.0 \cdot 10^{32}\ \text{cm}^{-2} \text{s}^{-1}) \)} \\
\hline
 Lower cut on \(M_{l^{+}l^{-}}\), GeV & 3.56 & 4  \\ 
 \hline
 
 \(\sigma_{pp \to \tau^{+}\tau^{-}} \cdot Br_{\tau^{-} \to \pi^{-} \nu_{\tau}} \cdot Br_{\tau^{+} \to \pi^{+} \bar{\nu_{\tau}}}\), pb & 1.00 & 0.71 \\
 \hline
 Approximate number of events per 7000h & 2500 & 1800 \\ 
 \hline
 \hline

\multicolumn{3}{c}{\( \sqrt{s} = 26 \text{GeV} (L = 1.2 \cdot 10^{32}\ \text{cm}^{-2} \text{s}^{-1}) \)} \\
\hline
 Lower cut on \(M_{l^{+}l^{-}}\), GeV & 3.56 & 4  \\ 
 \hline
 
 \(\sigma_{pp \to \tau^{+}\tau^{-}} \cdot Br_{\tau^{-} \to \pi^{-} \nu_{\tau}} \cdot Br_{\tau^{+} \to \pi^{+} \bar{\nu_{\tau}}}\), pb & 1.22 & 0.88 \\
  \hline
 Approximate number of events per 7000h & 3100 & 2200 \\ 
 \hline 
 
\end{tabular}
\end{center}
\caption{Estimation of the number of events \(pp \to \pi^{-} \nu_{\tau}\pi^{+} \bar{\nu_{\tau}}\) per year (\(\sim7000\)h) of data taking, assuming 100\% efficiency of the detector. }
\label{tautab2}
\end{table}

The next important step is to demonstate the manifestation of the effect discussed. In order to do so, we have carried out numerical simulation of the full process \(pp \to \pi^{-} \nu_{\tau}\pi^{+} \bar{\nu_{\tau}}\) with polarized protons and different polarization of initial partons. Events were generated in CompHEP package. For the simulation of the polarized processes, we have used polarized PDF set NNPDFpol provided by NNPDF collaboration \cite{nnpdf}. 
Currently, only the longitudinally polarized proton version of the PDF is available. 
Generated events were transferred to Pythia 8 package \cite{pyth} for parton showering and hadronization. Finally, events were treated by Delphes package \cite{del} for detector simulation and reconstruction.

\begin{figure}[t]
\centering
\includegraphics[width=0.49\textwidth,clip]{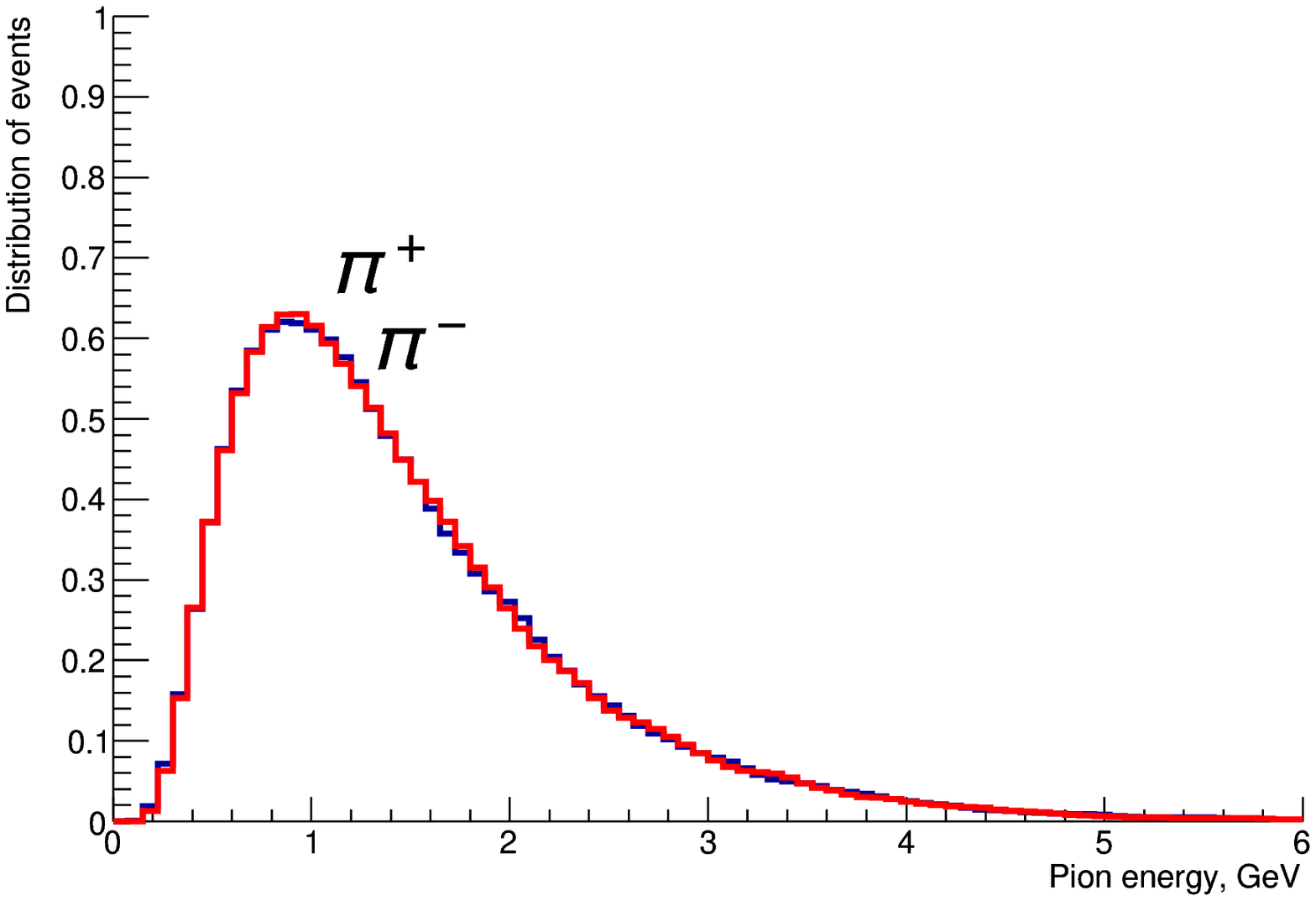}
\caption{Energy spectra of $\pi$-mesons in unpolarized case.}
\label{taup2}      
\end{figure}

\begin{figure}[t]
	\centering
	\includegraphics[width=0.49\textwidth,clip]{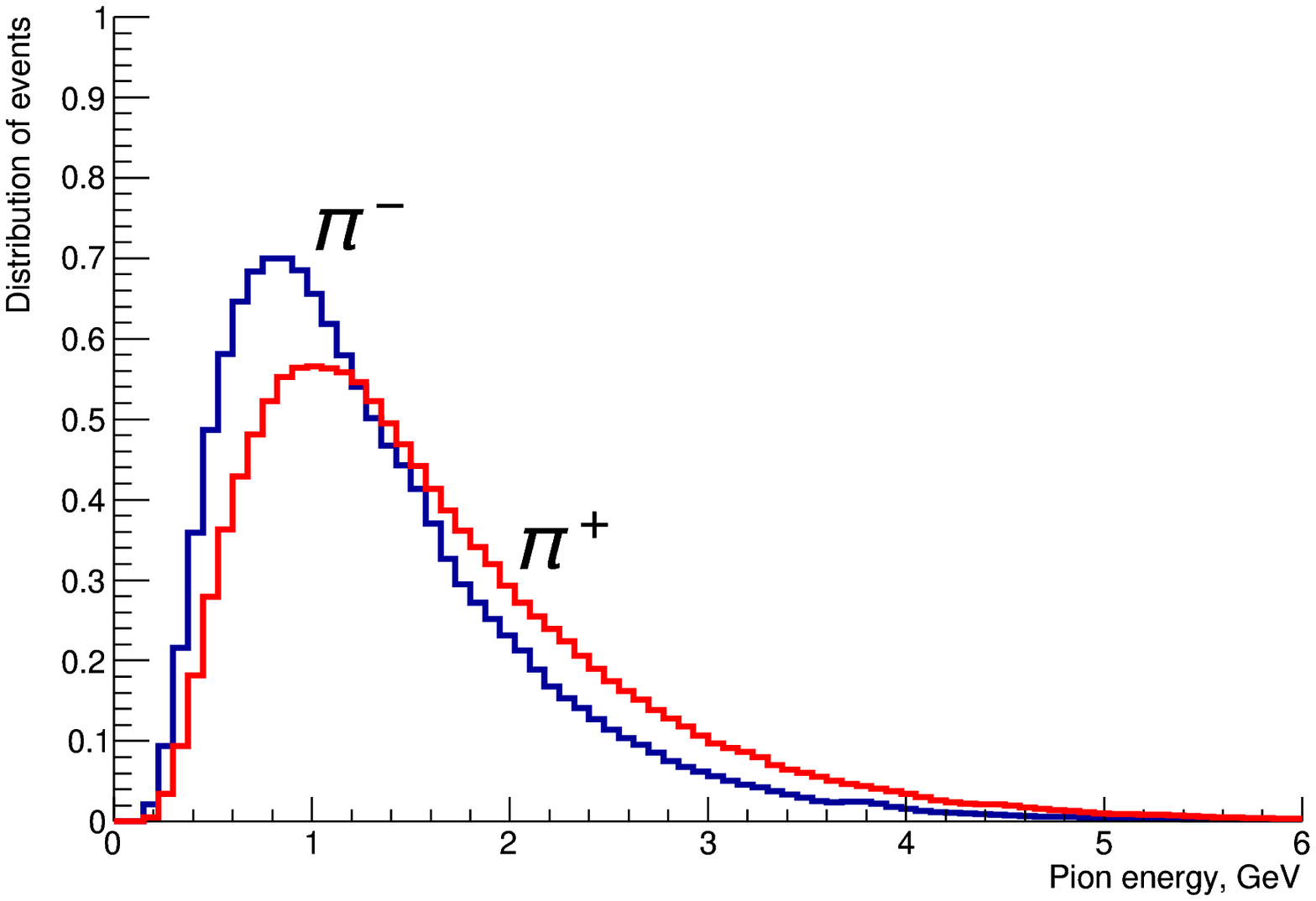}
	\includegraphics[width=0.49\textwidth,clip]{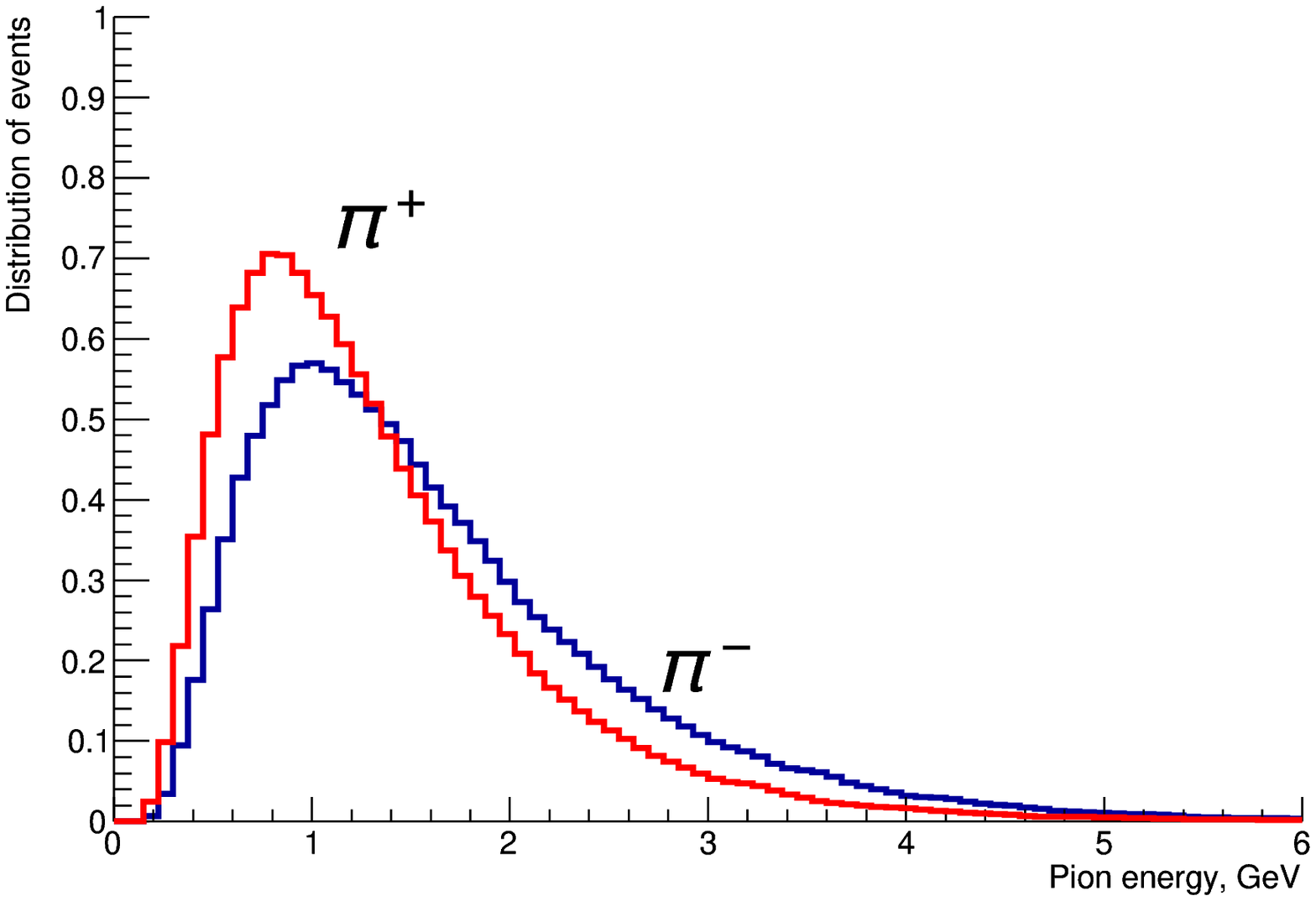}
	\caption{Energy spectra of $\pi$-mesons produced in processes with different polarization of the initial partons: left picture corresponds to the case of left-handed quark and right-handed anti-quark, while right one corresponds to the case of right-handed quark and left-handed anti-quark
	}
	\label{taup3}      
\end{figure}

The energy spectra of pions produced in the unpolarized case can be seen in Figure \ref{taup2}. Both $\pi^+$ and $\pi^-$ mesons manifest a similar behaviour.
The picture becomes interesting in the case of polarized initial states. The Figure \ref{taup3} shows the energy spectra of pions produced in the processes with different polarization of initial partons. The first figure corresponds to the case of left-handed quark interacting with right-handed anti-quark while the second one corresponds to the case of right-handed quark interacting with left-handed anti-quark. As we can see, the energy spectra of the pions are correlating with the polarization states of quarks in the way described in the previous section. Moreover, the clear asymmetry is seen between the energy spectra of $\pi^+$ and $\pi^-$ mesons, which can also be utilised as a characteristic of the polarization state of the initial quarks.

The first numerical evaluations carried out show the potential of the new method.
In the next step, we will carry out simulations with different initial combinations of longitudinally and transversely polarized protons. Next, we will compile a detailed table of correspondence between the pair distributions of $\pi$ mesons and various combinations of the initial polarizations of protons and their constituent quarks. 

Following this approach one can determine the polarization of interacting partons by measuring energies of $\pi$-mesons produced in corresponding decays of $\tau$-leptons. Considering the fact that pions are rather easy detectable \cite{pd}, the proposed approach is potentially very convenient and powerful tool for studying nuclons PDF via tau production in polarised Drell-Yan process.






\section{\bf On Measuring Antiproton-Production Cross Sections
for Dark Matter Search
\protect
\footnote{This section is written by R.\,El-Kholy; {E-mail: relkholy@sci.cu.edu.eg}.}}

\begin{abstract}
 Firm interpretation of the recent results from the AMS-02 and PAMELA spectrometers, regarding the antiproton yield in $p$-$p$ and $p$-$d$ collisions, has been hindered by uncertainties in production cross section, angular, and momentum spectra of the produced antiprotons. The proposed measurements of antiproton yield at the planned SPD experiment at the NICA collider could significantly contribute to enhancing the situation in favor of the search for dark matter WIMPs.
 
\end{abstract}
\vspace*{6pt}
\noindent

PACS: 95.35.+d, 95.55.Vj, 96.50.sb

\vspace{0.5cm}

The dark matter (DM) problem is a long-standing puzzle in modern cosmology. Even though DM is known to make more than 26\% of the total energy-matter content of the Universe\cite{Ade2014}, we still have no confirmed conclusions about its identity. Evidence of DM is abundant and diverse, but mainly gravitational. The effects of DM have been observed in the rotation curves of galaxies, the mass discrepancy of galaxy clusters, and the lens-less gravitational lensing, among others\cite{Majumdar2014}. Perhaps most of our knowledge about DM properties was a product of the \emph{Bullet Cluster} event, where two galaxy clusters collided with one another. Astronomical observations and lensing analyses of the event assured astronomers that DM particles are only weakly interacting\cite{Clowe2006}. Once merely a tentative hypothesis, this has become widely accepted. Consequently, among many DM candidates, Weakly Interacting Massive Particles (WIMPs) became the most favored\cite{Buchmueller2017}.

There are several search approaches applied in the case of DM. Each of them has an underlying paradigm. The main three approaches are collider searches, direct detection, and indirect detection. Collider searches are based on the hypothesis that DM particles can be pair-produced in collisions of Standard-Model (SM) particles, either directly or via a mediator. Direct detection experiments try to measure the recoil energy of DM particles colliding with SM nuclei. Indirect detection seeks to prove the hypothesis that DM particles decay and pair-annihilate producing SM particles as final products\cite{Buchmueller2017}. Thus, astrophysical searches try to accurately measure rare fluxes in Cosmic Rays (CRs), in order to detect a secondary anomalous signature of DM. To identify an anomalous signal, it is crucial to first subtract other ordinary astrophysical sources. Secondary CR fluxes are produced in collisions of primary CRs on Inter-Stellar Medium (ISM). In particular, secondary antiprotons are produced in proton-nucleus, nucleus-proton, and nucleus-nucleus collisions. 

\begin{figure}[ht]
	\includegraphics[width=\linewidth]{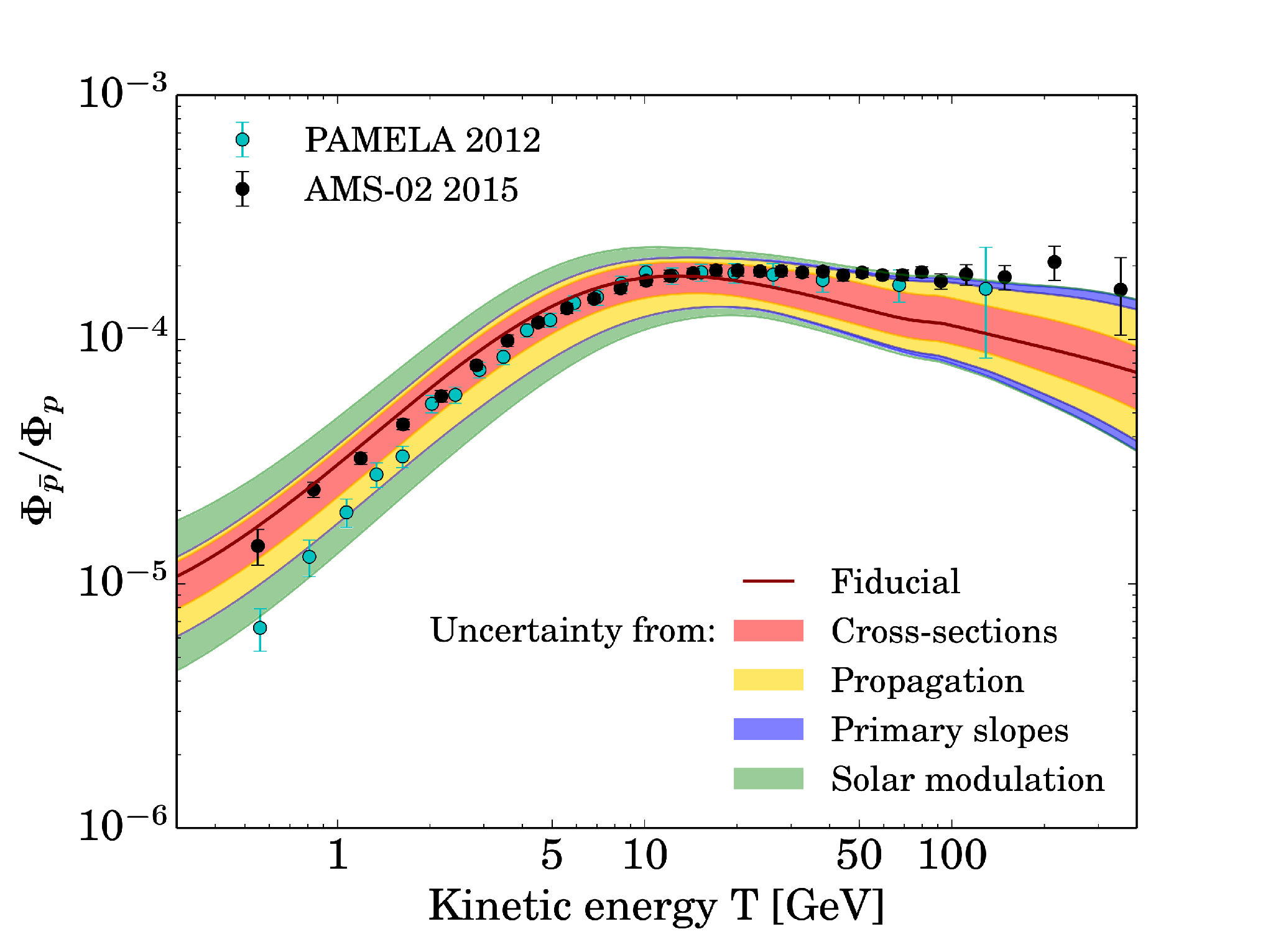}
	\caption{The combined total uncertainty on the predicted secondary $\bar{p}/p$ ratio superimposed to the PAMELA and the AMS-02 data. Each envelope accounts for uncertainties enclosed in it; such that the green band is the total band of all four sources of uncertainty\cite{Giesen2015}.}
	\label{fig:uncertainties}
\end{figure}

During the last two decades, two satellite-borne experiments, namely, PAMELA\cite{Adriani2014} and AMS-02\cite{Aguilar2013}, have measured the secondary antiproton flux and antiproton-to-proton ratio with unprecedented accuracy, covering a wide energy range. The AMS-02 experiment has measured antiproton-to-proton ratio with an accuracy higher than 95\% for kinetic energies from 1 to 450 GeV\cite{Aguilar2016}. However, we still cannot come to firm conclusions about any exotic signal because the AMS-02 measurements are still surrounded with several sources of uncertainties as shown in Fig.\ref{fig:uncertainties}. These sources stem from uncertainties on: (i) the primary spectra slopes at high energies, (ii) the solar modulation at lower energies, (iii) the propagation parameters in the galactic environment, and (iv) the antiproton-production cross sections. The first two can be relatively reduced through new AMS-02 measurements. The third can be minimized via parameter variation. However, the last of these sources is the most significant and ranges between 20\% and 50\%, depending on the energy\cite{Giesen2015}.

In addition to the scarcity of cross-section measurements of antiproton-production, most available datasets date back to before 1980, and did not account for hyperon decay to antiprotons\cite{Mauro2014}. Thus raising the need for new measurements of antiproton-production cross sections. The kinematic phase-space coverage required to catch up to the AMS-02 measurements has been outlined\cite{Donato2017}. Here, we affirm the capability of the Spin Physics Detector (SPD), planned at the NICA collider currently under construction at JINR, to make a sizable contribution to the desired measurements\cite{Guskov2019,Alexakhin2020}.

\label{sec:pbar-production}
\subsection{{Antiproton Production Cross Sections}}
The contribution of each interaction channel of primary CRs with ISM to the production of secondary antiprotons depends on the abundances of different nuclei in both CRs and ISM. Based on the abundances outlined in\cite{Guskov2019}, the most dominant channel of antiproton production would be proton-proton collisions, followed by proton-helium, and proton-deuteron collisions, respectively. Other nucleus-nucleus collisions have a negligible contribution. In addition to different production channels, there are also different production mechanisms. The dominant mechanism is the direct production. However, antiprotons are also produced via decay of anti-baryons; namely, antineutrons, and the $\bar{\Lambda}$ and $\bar{\Sigma}^{-}$ hyperons. The overall antiproton-production cross section can be expressed as\cite{Winkler2017}
\begin{equation}
	\sigma_{\bar{p}} = \sigma_{\bar{p}}^{0} (2+\Delta_{\text{IS}}+2\Delta_{\Lambda}),
\end{equation}
where $\Delta_{\text{IS}} = \sigma_{\bar{n}}^{0}/\sigma_{\bar{p}}^{0} - 1$ is the isospin enhancement factor of antineutron direct production, and $\Delta_{\Lambda} = \sigma^{\Lambda}_{\bar{p}}/\sigma_{\bar{p}}^{0}$ is the hyperon factor, assuming $\sigma_{\bar{n}}^{\Lambda} = \sigma_{\bar{p}}^{\Lambda}$.

\begin{figure}[ht]
	\includegraphics[width=\linewidth]{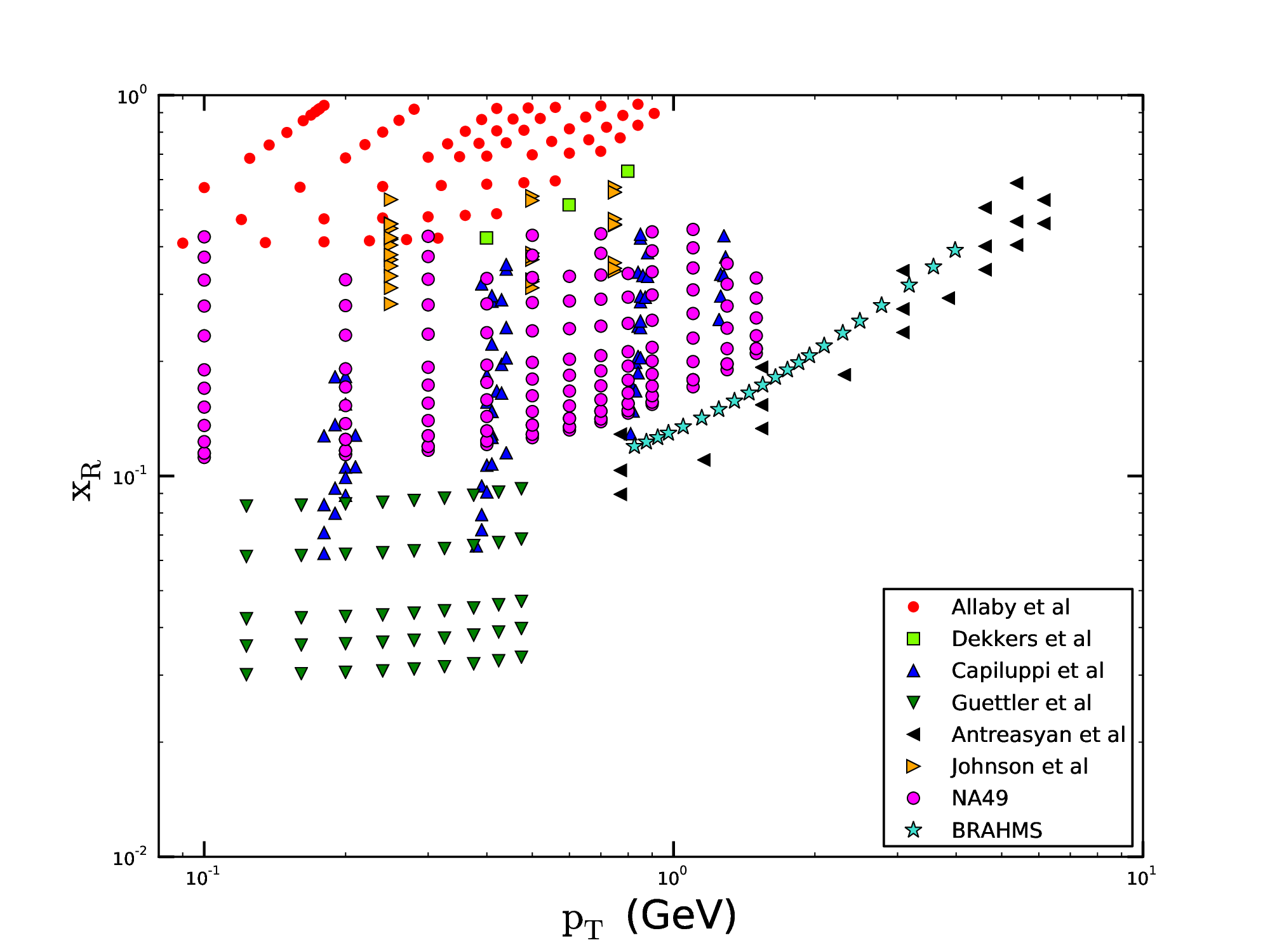}
	\caption{The available data on $d^{3}\sigma_{pp\rightarrow\bar{p}}/dp^{3}$ in the $p_{T}-x_{R}$ space\cite{Mauro2014}.}
	\label{fig:datasets}
\end{figure}

The currently-available data on proton-proton collisions is extremely scarce. It is graphically summarized in Fig.\ref{fig:datasets}, in terms of transverse momentum, $p_{T}$, and the radial-scaling variable, $x_{R}$, which is given by\cite{Mauro2014}
\begin{equation}
	x_{R}=\dfrac{E_{\bar{p}}^{*}}{E_{\bar{p}.\text{max}}^{*}},
\end{equation}
where $E_{\bar{p}}^{*}$ is the antiproton energy, and $E_{\bar{p}.\text{max}}^{*}$ is the maximal energy it can acquire, both in the CM frame. The maximal antiproton energy can be obtained by\cite{Mauro2014}
\begin{equation}
	E_{\bar{p}.\text{max}}^{*}=\dfrac{s-8m_{p}^{2}}{2\sqrt{s}}.
\end{equation}
Moreover, most of these datasets date back to before 1980, and thus do not include a hyperon-decay feed-down. As for datasets on other production channels, they are essentially non-existent. The first ever proton-helium dataset has been released by the LHCb collaboration in 2018\cite{Aaij2018}.

In addition to measuring the cross sections of direct antiproton production, it is necessary to accurately estimate the hyperon factor, $\Delta_{\Lambda}$, as well. The hyperon factor can be expressed as\cite{Winkler2017}
\begin{equation}
\label{eq:hyperon-factor}
\Delta_{\Lambda} = \frac{\bar{\Lambda}}{\bar{p}} \times \text{BF}_{\bar{\Lambda}\to \bar{p} \pi^+}
+  \frac{\bar{\Sigma}^-}{\bar{p}} \times \text{BF}_{\bar{\Sigma}^-\to \bar{p} \pi^0},
\end{equation}
where $\bar{\Lambda}/\bar{p}$ and ${\bar{\Sigma}}^-/\bar{p}$ are the hyperon to promptly-produced antiproton ratios, and $\text{BF}_{\bar{\Lambda}\to \bar{p} \pi^+}$ $= 0.639\pm0.005$ and $\text{BF}_{\bar{\Sigma}^-\to \bar{p} \pi^0}=0.5157\pm0.0003$ are the branching fractions of the corresponding decays. While there are some existing data on the ratio $\bar{\Lambda}/\bar{p}$, there is not any data on the ratio $\bar{\Sigma}^{-}/\bar{p}$; and it is only estimated based on the symmetry argument that $\bar{\Sigma}^{-}/\bar{\Lambda}=0.33$\cite{Kappl2014}. Thus, expression \eqref{eq:hyperon-factor} can be rewritten as
\begin{equation}
\Delta_{\Lambda}=(0.81\pm0.04)\times\bar{\Lambda}/\bar{p}.
\end{equation}
Even existing data on the ratio $\bar{\Lambda}/\bar{p}$ has a high energy-dependent uncertainty ranging from 12\% to 18\%\cite{Winkler2017}.

\begin{figure}[ht]
	\includegraphics[width=\linewidth]{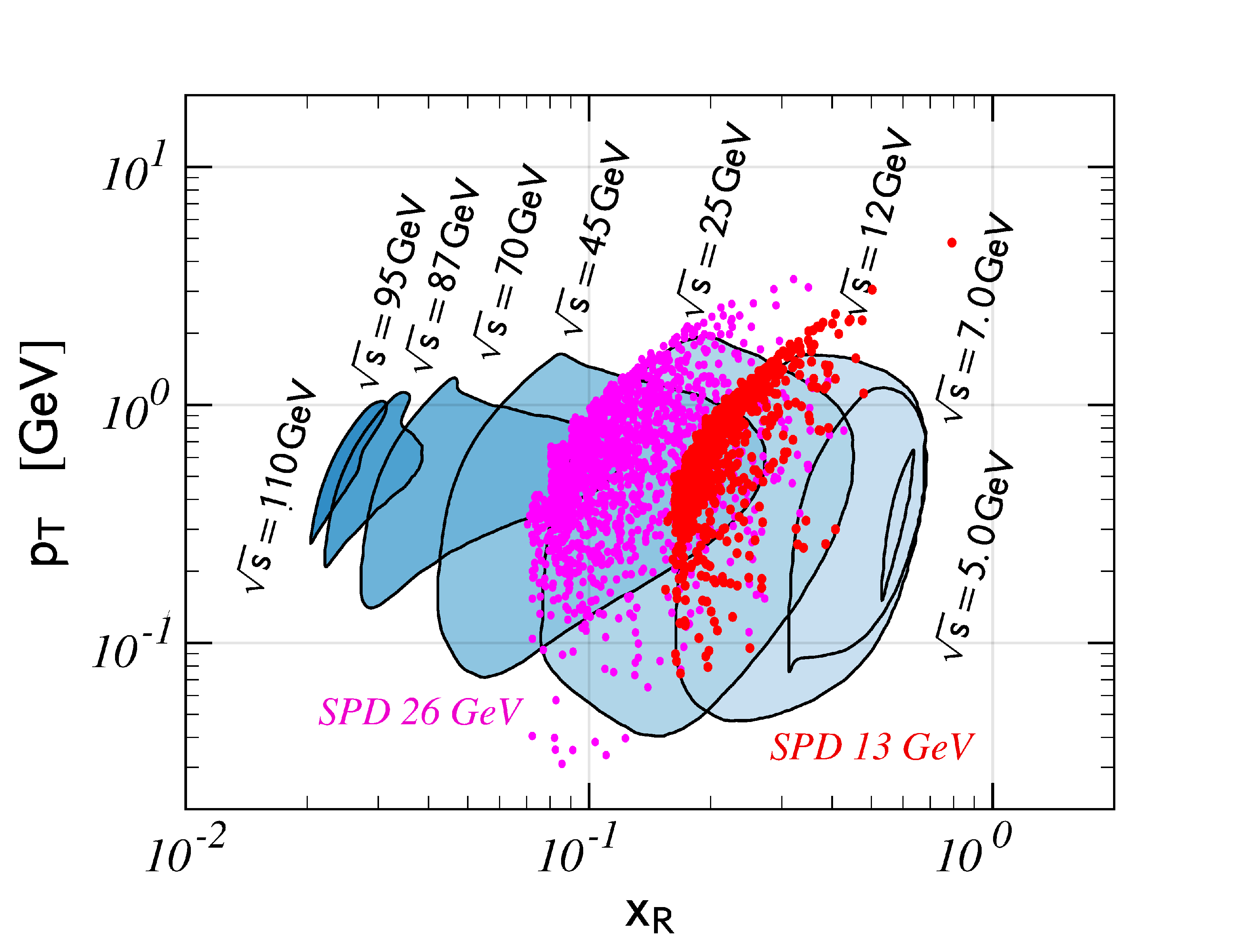}
	\caption{The kinematic range accessible to the SPD\cite{Alexakhin2020} superimposed on the required measurement range to match the AMS-02 accuracy\cite{Donato2017}.}
	\label{fig:spd-coverage}
\end{figure}

From all of the above, the need for new measurements, of the antiproton-production cross sections and hyperon-to-antiproton production ratios, becomes clear. The phase space that requires coverage in order to bridge the gap between antiproton data and the AMS-02 measurements has been outlined. Fig.\ref{fig:spd-coverage} shows it in terms of transverse momentum and the radial-scaling variable, where the areas within the contours need to be covered with uncertainty no more than 3\%, while the areas outside the contours require coverage with uncertainty no more than 30\%\cite{Donato2017}.

\break
\label{sec:nica-spd}
\subsection{{NICA SPD Contribution}}
The SPD detector is planned to operate at the NICA collider currently under construction at JINR\cite{Savin2015}. Polarized proton and deuteron beams will be used. The $pp$ collisions are expected to reach $\sqrt{s}=27$ GeV, and a luminosity of $5\times10^{30}$ cm$^{-2}$s$^{-1}$ should be achieved\cite{Meshkov2019}. Preliminary MC results using PYTHIA8\cite{Sjoestrand2015} show that the antiproton-production cross section in $pp$ collisions multiplied by the average antiproton-multiplicity in this energy range is on the level of a few millibarns\cite{Guskov2019}, corresponding to a production rate $>10^{5}$ s$^{-1}$. Thus, the measurements will not be hindered by statistical uncertainty. The $4\pi$ angular acceptance of the SPD will also allow access to a wider kinematic range, in comparison to fixed-target experiments. Fig.\ref{fig:spd-coverage} shows the kinematic range accessible to the SPD based on a preliminary MC study\cite{Alexakhin2020}, superimposed on the required measurement range to match the AMS-02 accuracy.

\begin{figure}[ht]
	\includegraphics[width=\linewidth]{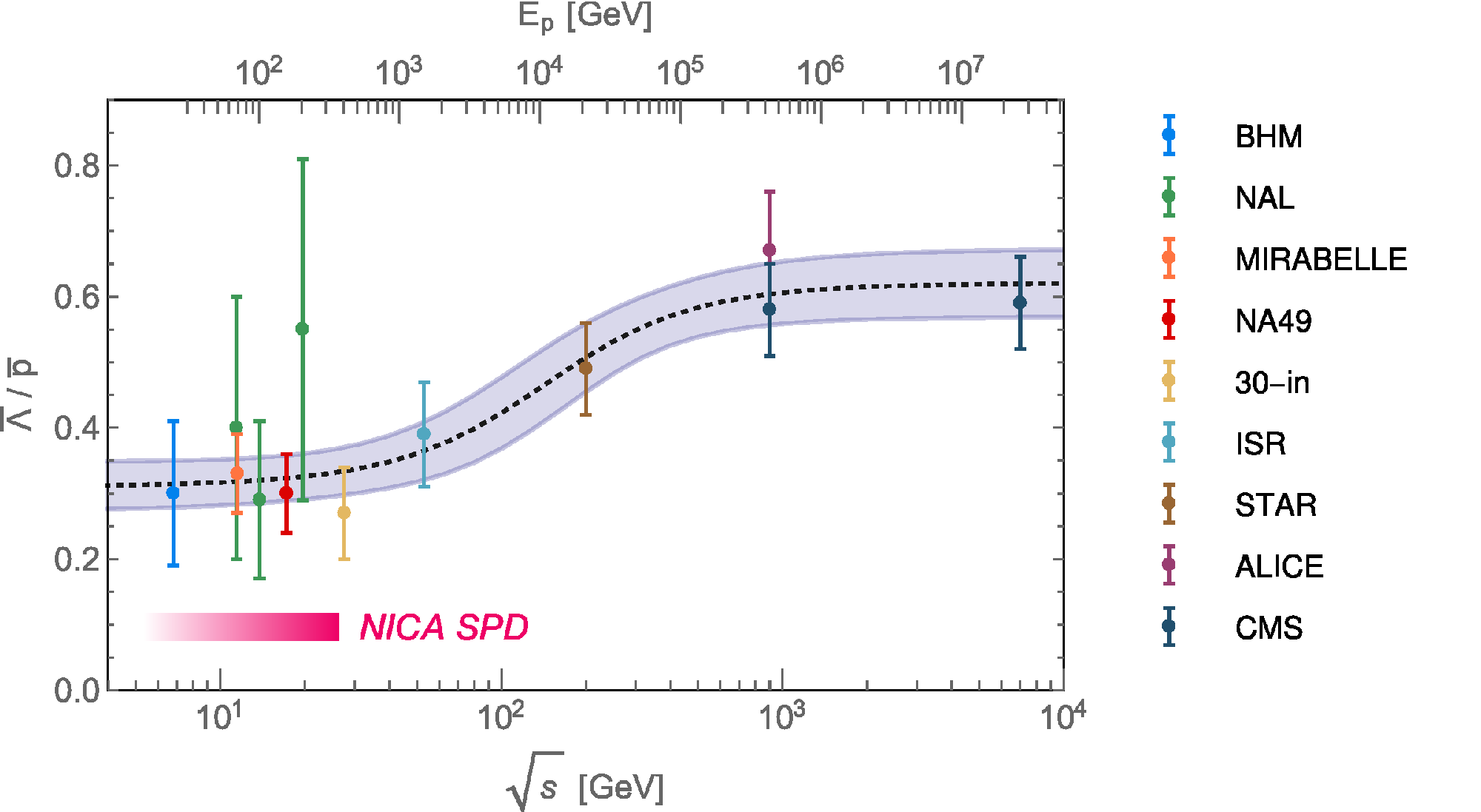}
	\caption{$\bar{\Lambda}/\bar{p}$ ratio in proton-proton collisions as measured by several experiments\cite{Winkler2017}. The range of $\sqrt{s}$ accessible at the NICA SPD is indicated at the lower left part\cite{Alexakhin2020}.}
	\label{fig:hyperons-range}
\end{figure}

The time-of-flight system of the SPD detector will enable particle identification. In addition, reconstruction of secondary-vertices will allow the study of secondary hyperon decays\cite{Alexakhin2020}. Thus, the SPD can also contribute to measurement of the $\bar{\Lambda}/\bar{p}$ ratio, where Fig.\ref{fig:hyperons-range} shows the energy range accessible to the SPD.

\label{sec:summary}
\subsection{{Summary}}
The Spin Physics Detector can make a sizable contribution to the search of physics beyond the Standard Model. The SPD can measure energy and angular distributions of antiprotons produced in proton-proton and proton-deuteron collisions both directly and from the decays of $\bar{\Lambda}$ and $\bar{\Sigma}^{-}$ hyperons in the kinematic range starting from the threshold, for the astrophysical searches for dark matter. The collider mode and the geometry of the SPD detector provide a unique possibility to study the production of antiprotons at high transverse momenta which is unavailable for fixed-target experiments. The possibility for NICA to operate with beams of light nuclei, such as $^{3}He$ and $^{4}He$, could extend this program.



	\section{\bf{Tests of fundamental discrete symmetries at NICA facility: addendum to the spin physics programme 
	\protect\footnote{ This section is presented  by I.A.\,Koop, A.I.\,Milstein,
	N.N.\,Nikolaev ({E-mail: nikolaev@itp.ac.ru}),
    A.S.\,Popov, S.G.\,Salnikov, P.Yu.\,Shatunov,Yu.M.\,Shatunov.}}}
	
	\begin{abstract}
		We present new ideas on tests of fundamental symmetries in polarization experiments at the NICA facility. Specifically, we explore the possibilities of high precision tests of the Standard Model by parity violation and searches of Beyond the Standard Model semistrong breaking of time reversal invariance in double polarized proton-deuteron scattering, taking advantage of high intensity beams of polarized protons and deuterons available at NICA. In both cases, we propose to use the new technique of polarized beam with  precessing horizontal polarizations, and polarized deuterons are the favored choice. The external target in the extracted beam is optional for the parity violation experiment, which requires furnishing Nuclotron and/or new Booster with very modest new instrumentation. One should not overlook this potential for substantial broadening of horizons of spin physics at the NICA facility.
		 
	\end{abstract}	
\vspace*{6pt}

\noindent
PACS: 13.40.Em, 11.30.Er, 29.20.Dh, 29.27.Hj


\subsection*{Introduction}


 The SPD project at NICA aims primarily at polarization experiments in the collider mode. In the broader context, the main virtue of NICA facility is high current beams of polarized protons and deuterons  \cite{Agapov:2016fbg, Savin2014SPD}.  We report here new ideas developed  in the framework of the  Russian Fund for Basic Research Grant No. 18-02-40092 MEGA. Basically, we propose addendum to the previously discussed spin physics programme of the NICA  facility. For decades to come, NICA  will be a holder of a unique potential to conduct fixed target experimental tests of fundamental symmetries within or beyond the Standard Model (SM), and one should not overlook this outstanding opportunity.

Our knowledge about high energy  parity violation (PV) in the pure non-leptonic sector of SM is as yet scarce. The most accurate result on PV asymmetry in $pp$ scattering at 45\ MeV, $ A_{PV} = - (1.5 \pm 0.22) {\cdot} 10 ^ { -7} $, is based on the data collected for several years \cite{Kistryn1987tq}. A result of PV experiment at ZGS wtih 5.1\,GeV protons interacting with the water target, $ A_{PV}= - (26.5 \pm 6.0 \pm 3.6) {\cdot} 10 ^ {- 7} $,     was also collected for several years \cite {Lockyer:1984pg}. Theorists failed to acomodate the latter, anomalously large, asymmetry in the SM (see review \cite{Gardner:2017xyl}). Only modest upper bounds were set at Fermilab in $pp$ and $p\bar{p}$ interactions at  200\, GeV: $A_{PV} < 10^{-5}$ \cite {Grosnick1996sy}. 

Of our prime concern is a feasibility of precision measurements of PV asymmetry at the NICA facility \cite{Koop:2020yzn,MilsteinPP,MilsteinPD}. 
Our key suggestion is using beams with polarization idly rotating in the ring plane. Such a polarization has for the first time been successfully applied in 1986 at the Budker INP, with participation of members of  the present team, to a  record precision comparison of magnetic anomalies of $e^+$ and $e^-$ \cite{Vasserman:1987hc}. We also mention the 2002 idea to accelerate  oscillating polarization beams in Nuclotron \cite{Sitnik2002}. Subsequently, the JEDI collaboration, including members of the present team, had extended this approach to studies of fine aspects of  spin dynamics of polarized deuterons stored at the COSY accelerator \cite{Bagdasarian:2014ega,Eversmann:2015jnk,Guidoboni1000s,GuidoboniChromaticity,Hempelmann:2017zgg}. The principal point is that time-stamped  precessing polarization has matured, and can be viewed as useful as the static ones. Namely, oscillating spin asymmetries in cross section generated by the in-plane-precessing polarization, can readily be isolated by Fourier analysis. We find it imperative to dwell on the aspects of precessing polarization approach which are not that familiar to the NICA community. 

In this report we mostly focus on the preferred external fixed target PV experiments at the new Booster or Nuclotron. The already published collaboration results \cite{Koop:2020yzn,MilsteinPP,MilsteinPD} are  mentioned only briefly. Execution of the PV experiment would require furnishing the Nuclotron or Booster ring with only very modest extra instrumentation: the radiofrequency spin flipper and internal target polarimeter to fix the time stamp of the spin phase. Extraction of the signal of PV violation from the difference of attenuations of extracted beams of opposite helicity will not require any sophisticated external detectors -- the anticipated non-invasive measurement of the total charge of the bunch incident on the condensed matter target and of the transmitted bunch will be performed by a system of Rogowski coils. 

 Polarized deuterons are favored because the NICA energy range is free of the deuteron spin resonances. An outstanding record of experimentation with polarized deuterons at the Nuclotron is noteworthy \cite{Fimushkin2008,Vokal:2009zz}. Our starting point was the PV studies, but the project outgrew its original boundary: it was understood that the  precessing polarization deuterons can give an access to tests of still another fundamental symmetry --- the time reversal invariance \cite{Lenisa:2019cgb,Nikolaev:2020wsj}. Specifically, one can search for the semistrong T- and CP-violating, P-conserving and flavor-conserving interaction, suggested by  Okun\,\cite{Okun:1965tu}, Prentki \& Veltman\,\cite{Prentki:1965tt}, and Lee \& Wolfenstein\,\cite{Lee:1965hi}. An intriguing open issue is whether this manifestly beyond SM semistrong CP-violation can resolve the puzzle of the anomalously large baryon asymmetry of the Universe, where SM fails by many orders in magnitude \cite{Bernreuther:2002uj}. On top of that, we mention also a possibility of spin crisis experiments with  oscillating deuteron polarization at the electron-ion collider eIC \cite{Nikolaev:2020wsj}. 



\subsection{Precessing spin asymmetries in the total $pd$ cross section} 


We illustrate the polarization effects on the example of total $pd$ cross section:
\begin{equation}
\begin{split}
\label{sigmatotal}
\sigma_\text{tot} = & \,\sigma_0 + \sigma_\text{TT} \left[ \left( {\bf P}^\text{d} \cdot {\bf P}^\text{p} \right) - \left( {\bf P}^\text{d}\cdot {\bf k} \right) \left( {\bf P}^\text{p}\cdot {\bf k} \right) \right]+ \sigma_\text{LL} \left( {\bf P}^{\rm d} \cdot {\bf k} \right) \left( {\bf P}^\text{p}\cdot {\bf k} \right)
+\sigma_\text{T} Q_{mn}k_m k_n \\
& + \sigma_\text{PV}^\text{p} \left( {\bf P}^\text{p} \cdot {\bf k} \right) + \sigma_\text{PV}^\text{d} \left( {\bf P}^\text{d} \cdot {\bf k} \right) +\sigma_\text{PV}^\text{T} \left( {\bf P}^\text{p} \cdot {\bf k} \right) Q_{mn}k_m k_n \\
& + \sigma_\text{TVPV} \left( {\bf k} \cdot \left[ {\bf P}^\text{d} \times {\bf P}^\text{p} \right] \right) + \sigma_\text{TVPC} k_m Q_{mn} \epsilon_{nlr}P_l^\text{p} k_r\, .
\end{split}
\end{equation}
Here ${\bf P}^{\rm d}$ and ${\bf P}^{\rm p}$ are polarizations of the deuteron and proton, $Q_{mn}$ is the tensor polarization of the deuteron, and ${\bf k}$ is the collision axis (the $z$-axis). The $y$-axis is normal to the ring plane,  $x$-axis is the radial one. In the tensor polarization dependent terms  $Q_{mn}k_mk_n =Q_{zz}$, and $
k_m Q_{mn}\epsilon_{nlr}P_l^{\rm p} k_r
= Q_{xz}P_y^{\rm p}-Q_{yz}P_x^{\rm p}\,$  is odd under flip of the proton polarization. The cross sections $\sigma_0$, $\sigma_\text{TT}$, $\sigma_\text{LL}$, and $\sigma_\text{T}$ correspond to the ordinary P-even and T-even interactions, $\sigma_\text{PV}^\text{p}$, $\sigma_\text{PV}^\text{d}$, and $\sigma_\text{PV}^\text{T}$ give PV signals , and $\sigma_\text{TVPV}$ is the  TVPV component, while  $\sigma_\text{TVPC}$ is the null observable for the TVPC  semistrong interaction \cite{Temerbayev:2015foa,Uzikov:2015aua,Uzikov:2016lsc}.The last three polarization effects have not been studied before.




In our approach, the equilibrium vertical spin of the stored beam will be subjected to parametric resonance driven by the radiofrequency (RF) flipper (solenoid): 
\begin{eqnarray}
\vec{P}(n)  = P_y(0)[&\cos (\epsilon n) \vec{e}_y 
+ \sin(\epsilon n)  [ \cos (\theta_s n) \vec{e}_x - \sin (\theta_s n) \vec{e}_z ]\,. \label{eq:Vector}
\end{eqnarray}
Here $n$ is the turn number,  $\theta_s =2\pi \nu_s$, where $\nu_s=G\gamma$ is the spin tune, $G$ is the magnetic anomaly of the particle, $\gamma$ is its relativistic $\gamma$-factor, $\epsilon = 2\pi \nu_R$ and $\nu_R$ is the spin resonance tune, related to the amplitude of spin rotation in a single pass through the spin flipper. In this equation,  $\cos (\epsilon n)$ and $\sin(\epsilon n)$ have conspicuous interpretation as envelopes of the vertical and in-plane idly precessing polarizations,  respectively.
The flipper is turned off when $\epsilon n = \pi/2$ is reached and then spin will keep idly precessing in the storage ring plane. 

Oscillating $P_z=-P_y(0)\sin (\theta_s n)$ gives rise to the oscillating PV signal in total cross section, see Eq.\ \ref{sigmatotal}. The internal target polarimetry of oscillating $P_x$ is used by JEDI collaboration to monitor the spin precession frequency \cite{Bagdasarian:2014ega,Eversmann:2015jnk}  and simultaneously to provide the time stamp for $P_z$ . In our proposal this time stamp will be used to trigger the single-turn dump of beam of the desired $P_z$ onto the external target. In JEDI experiments \cite{Guidoboni1000s,GuidoboniChromaticity} the spin coherence time of idly precessing deuterons has been maximized up to 1400 s by fine tuning the sextupole lenses, as was suggested already in 1987 by the present collaborations members (Koop and Shatunov \cite{Koop1988}). 

We mention briefly only major points about flipper driven evolution of the tensor polarization of deuterons starting with  $Q_{yx}(0) = Q_{yz}(0) =  Q_{xz}(0) =0$, and $Q_{xx}(0) =  Q_{zz} (0)= -\frac{1}{2}Q_{yy}(0)$ \cite{Nikolaev:2020wsj}:
\begin{equation}
\begin{split}
Q_{yy}(n) & = \phantom{-}\frac{1}{2}Q_{yy}(0)  \cdot \left[ -1 + 3 \cos^2( \epsilon n )\right] \,,\\ 
Q_{xx}(n) & = \phantom{-}\frac{1}{2}Q_{yy}(0)  \cdot \left[ -1 + 3 \sin^2( \epsilon n) \cdot  \cos^2(\theta_s n) \right] \,, \\ 
Q_{zz}(n) & = \phantom{-}\frac{1}{2}Q_{yy}(0)  \cdot \left[ -1 +3 \sin^2 (\epsilon n) \cdot \sin^2(\theta_s n) \right]\,, \\
Q_{yx}(n) & = \phantom{-}\frac{3}{2}Q_{yy}(0)  \cdot \sin (\epsilon n) \cdot \cos (\epsilon n) \cdot \cos(\theta_s n)\,, \\ 
Q_{yz}(n) & = -\frac{3}{2}Q_{yy}(0) \cdot \sin ( \epsilon n) \cdot \cos (\epsilon n) \cdot \sin(\theta_s n)\,, \\
Q_{xz}(n) & = -\frac{3}{4}Q_{yy}(0) \cdot \sin^{2} (\epsilon n) \cdot \sin (2\theta_s n)\,. 
\end{split}
\label{eq:Qyy-Qzz}
\end{equation}
At the working point $\epsilon n = \pi/2$ we have $Q_{yx}=Q_{yz}=0$,~ $Q_{yy}= const$ and $Q_{xx,zz} \propto 1\pm 3\cos 2\theta_s n$. There is a conspicuous difference of Fourier harmonics of the vector and tensor polarizations, and the P-conserving effects of tensor polarizations are easily separated from the PV signal by Fourier analysis. On the other hand, the tensor asymmetry is well understood, and the signal of oscillating component of $Q_{zz}$ can serve as an important cross check of our technique. We emphasize that in search for the single-spin PV asymmetry one only needs unpolarized targets.  
 
Of particular interest is the off-diagonal  $Q_{xz}$. It enters the TVPC asymmetry which probes the semistrong CP-violating interaction \cite{Uzikov:2015aua,Temerbayev:2015foa}.  The TVPC asymmetry $\propto Q_{xz}$ has the unique Fourier signature, is $P_y^p$-odd, and is free of systematic background \cite{Nikolaev:2020wsj}. However, it is the double-polarization observable, and search for semistrong CP-violation with stored polarized deuterons requires the internal polarized hydrogen target \cite{Lenisa:2019cgb}.  


\subsection{PV asymmetry: expectations from Standard Model}


The theoretical results of the team were  reported in three publications \cite{MilsteinPP,MilsteinPD,Nikolaev:2020wsj}.  The salient feature of the tree-level PV weak Hamiltonian is a strong suppression of the PV $pp$ amplitude, because numerically $ | 4\sin^2\theta_W-1 | \ll 1$. However, the effective PV neutral current can be generated from charged current $np$ interaction by radiative corrections from charge exchange strong interaction with  encouraging magnitude, although the uncertainties are inevitably substantial \cite{MilsteinPP}. The initial and final state strong interactions, aka  absorption corrections, endow  the initially real valued tree-level PV amplitude with the imaginary part, which can be evaluated in the eikonal approximation. The  PV contribution to the total cross section is nearly exhausted by PV in elastic scattering, what  entails a strong suppression of PV in inelastic scattering.  The corollary is that in comparison to the total cross section, the PV asymmetry in elastic scattering will be enhanced by large factor of  $\sigma_{tot}/ \delta\sigma_{el} \sim 3$.  The expectations from the SM for PV in proton-nucleon interactions at NICA are
\cite{MilsteinPP}: ${\cal A}_{tot}^{pn}\sim 10^{- 7}, \,\, {\cal A}_{tot}^{pp} \sim 0.4 \cdot 10^{- 7}  $.

New features of $pd$ interactions are Glauber screening and quasielastic scattering, aka diffractive breakup of the deuteron \cite{Glauber:1955qq,Franco:1965wi}. The technicalities of calculations of P-odd asymmetries are found in the publication \cite{MilsteinPD}. The
enhancement of the P-odd asymmetry in elastic $pN$ scattering  vs. total cross section will persist in both
elastic and quasi-elastic $pd$ scattering. A substantial difference between the scattering of polarized deuterons on unpolarized protons and the scattering of polarized protons on unpolarized deuterons has its origin in the interference of the P-odd $pp$ and $pn$ amplitudes.

Remarkably, to the same accuracy as in the case of $pp$ scattering, the P-odd component of the total $pd$ cross section is exhausted by the sum of P-odd components of the total elastic and quasielastic cross sections. Now the strong suppression of the parity violation holds for truly inelastic $pd$ collisions with production of new particles (mesons).

We first report expectations for PV cross sections and  asymmetries ${\cal A}=\sigma_W/\sigma_s\,$ in the
scattering of polarized deuterons with $\lambda_d=1$ on unpolarized protons:
\begin{align}\label{delsigd}
&\sigma_{s,\,tot}^{pd} = 96\, \mbox{ mb},\, \quad
\sigma_{W,\,tot}^{pd}=2.1\,\mbox{nb},\quad {\cal A}_{tot}^{pd}=2*10^{-8}\,,
\nonumber\\
&\sigma_{s,\,el}^{pd} = 20\,\mbox{ mb},\quad
\sigma_{W,\,el}^{pd}=0.7\,\mbox{nb}\,,\quad
{\cal A}_{el}^{pd}=3.5*10^{-8}\, , \nonumber\\
&\sigma_{s,\,qel}^{pd} = 22.4\mbox { mb},
\quad\sigma_{W,\,qel}^{pd}=1.4\,\mbox{nb}\,,\quad {\cal A}_{qel}^{pd}=6*10^{-8}\,.
\end{align}
For the interaction of a polarized protons with $\lambda_p=1$ with unpolarized
deuterons, we have
\begin{align}\label{delsigp}
& \sigma_{W,\,tot}^{pd}=-0.8\,\mbox{nb}\,,\quad {\cal
	A}_{tot}^{pd}=-0.9*10^{-8}\,,\nonumber\\
& \sigma_{W,\,el}^{pd}=-0.6\,\mbox{nb}\,,\quad {\cal A}_{el}^{pd}=-3*10^{-8}\,,
\nonumber\\
& \sigma_{W,\,qel}^{pd}=-0.2\,\mbox{nb}\,,\quad {\cal A}_{qel}^{pd}=-10^{-8}\,.
\end{align}

\subsection{The experimental strategies}

{\bf Generic considerations of the external target option.} The analysis of the optimal strategy for the PV experiment is in the formative stage. To illustrate the challenges one faces, we focus here on the option of extracted polarized deuteron beams  interacting with the external target, similar to that  used in the ZGS experiment with accelerated protons \cite{Lockyer:1984pg}.  A source of polarized deuterons with unique parameters \cite{Fimushkin2008} has been commissioned at the Joint Institute for Nuclear Research decades  ago, and successful sessions of their acceleration in the Nuclotron \cite{Vokal:2009zz} have been carried out.  According to \cite {Savin2014SPD}, the Nuclotron is able to accelerate in one cycle up to $ 1.6 {\cdot} 10 ^ {11} $ polarized protons and deuterons.

The principal target is a reach to PV asymmetries at the level $\sim 10^{-7}$ or better. On generic grounds that requires about $10^{15}$ events. The physical observable will be a difference of attenuations of the positive and negative helicity beams in thick dense target. To maximize the statistics, one needs large number of cycles. The typical cycle will consist of (1) injection of vertically polarized particles, (2) acceleration to the required energy, (3) rotation of  polarization from the vertical to horizontal one by RF flipper, (4) the polarimetry of the in-plane precessing spin and the determination of the spin-tune and the spin phase, (5) single-turn extraction of the beam of desired helicity onto the target, (6) comparison of beam currents upstream and downstream of the target.

We skip a discussion of the routine stages (1) and (2). 

{\bf The spin coherence time}. The stages (3) and (4) require more attention.  The vertical polarization is preserved by the vertical guiding field in the ring. The in-plane idly precessing spins  decohere with time. Therefore, the stages (3) and (4) together must be shorter than the spin coherence time. Ever since the experiment \cite{Vasserman:1987hc},  the RF flippers are being routinely used in the spin experiments. More detailed discussion of the proposed fast flipper will be presented below, for the purposes of the present discussion it suffices to know that the vertical spin can be rotated to the horizontal one faster than in 1 s.  As the reference point we cite the JEDI result, that with beams of $10^9$ deuterons of momentum 0.97 GeV/c in COSY, the spin precession phase can be  measured to the accuracy of $\sim 0.2$ in 2 s \cite{Eversmann:2015jnk,Hempelmann:2017zgg}. Steady operation with spin coherence time exceeding 1000 s has been achieved  \cite{Guidoboni1000s,GuidoboniChromaticity}. The educated guess is that, at the same energy of deuterons, the spin coherence time will be sufficient for less that 3 s of idle precession in Nuclotron or the new Booster rings even without cooling the beam and a single cycle can be as short as $\sim 5$ sec.
The radial polarization cycles bring the effective cycle length to $\sim 10$ s. Making allowance for the contingency factor of 2, we end up with $\sim 130 000$ effective cycles per month.  By the rule of thumb, in one month data taking with thick target of one absorption length, the total number of interactions can reach  $\sim 10^{16}$. A parasitic data taking, when Nuclotron and/or Booster are idling during  operation of NICA in the collider mode, makes possible a further gain in the statistics.

{\bf Polarization of the ejected beam.}  One needs the single-turn extraction of the stored bunch. At the discussed energy, the spin tune of deuterons $\nu_s=G\gamma = -0.160977$. After 50 particle revolutions in the ring, the spin will make 8.048 in-plane rotations, after 99 revolutions the spin will make 15.966 rotations, and 23.986 in-pane rotations after 149 revolutions  etc. This shows that with time stamp it  will take not much longer than few hundred turns of the beam, {\sl i.e.,}, few decimal fractions of a millisecond, to decide when to extract the beam polarized in any desired orientation. A good option is a sequence of two cycles with alternating $P_z$ to measure the PV asymmetry,  and two more cycles to  crosscheck the equality of attenuations of beams with alternating radial $P_x$. Tensor asymmetry of the total cross section is large and PV cycles can easily be interspersed with the control sequence of cycles to extract the second harmonics signal from the precessing tensor polarization. 

{\bf Polarimetry issues.} Internal cylindrical scintillation polarimeter made of four, top--bottom and right--left sectors, will provide time resolution to dynamically measure the oscillating transverse polarization of the beam from the oscillating up-down asymmetry  \cite{Bagdasarian:2014ega,Eversmann:2015jnk}. The periphery of the beam can be brought to collisions with the carbon target either by stochastic heating of the beam or generating the bump by beam steerers. The polarimetry will consume only a small fraction of the beam before it is ejected into the target channel. A cycle-to-cycle stability of orbits will be controlled by beam position monitors along the ring circumference,  the magnetic field will be controlled by  NMR sensors in a special dipole magnet powered serially with the ring dipoles. Specific to the approach is a high precision cycle-to-cycle comparison of spin tunes, which amounts to a comparison of energies.  
The supplementary polarimetry of the beam after the target will provide important cross check of orientation of the polarization vector of the beam incident on the target.

{\bf Flipper implementation issues.}
For deuterons with momentum $p = 0.97$\,GeV/$c$ (kinematic parameters $\gamma = 1.125$, $\beta = 0.46$), it is rational to apply the longitudinal magnetic field oscillating at a relatively low frequency $f = f_c \cdot \gamma |G| = 88.3$\,kHz. 
The ceramic vacuum chamber must have conductive longitudinal stripes on the inner surface so that the beam image currents can freely propagate along these metalized tracks. The outer side of the ceramic chamber will serve as a skeleton for winding the solenoid turns. The approximate technical parameters of the flipper \cite{Koop:2020yzn} are shown in Table~\ref{tab1}. The necessary power to the RF generator, 5\,kW, can be provided, for instance, by the generator triode GI-50, capable of delivering up to 40\,kW of power in a continuous mode, with the help of modern semiconductor amplifiers. We leave these issues for a future technical study.

\begin{table}[htb]
	\centering
	\caption{The main flipper parameters for the deuteron momentum $0.97$\,GeV/$c$ and the amplitude of its circular harmonic $w = 2.5 \cdot 10^{-5}$ (field integral $B l = 1.2 \cdot 10^{-3}$\,T$\cdot$ m)}\label{tab1}
	\begin{tabular}{l l l}
		Solenoid length & 1.0 & m \\
		Magnetic field amplitude & 0.0012 & T \\
		Spiral winding diameter & 150 & mm \\
		Aperture of the ceramic vacuum chamber & 120 & mm \\
		Case diameter & 400 & mm \\
		Solenoid turns & 80 & \\
		Winding inductance & 150 & $\mu$H \\
		Characteristic impedance of the circuit & 75 & Ohm \\
		Active loss resistance & 0.2 & Ohm \\
		Quality factor of the oscillating circuit & 375 & \\
		Winding current & 150 & A \\
		Inductive voltage & 11 & kV \\
		Active loss power & 4.5 & kW \\
	\end{tabular}
\end{table}
{\bf The accuracy issues in the external target mode.} With  $N_1$ particles impinging on the target and $N_2$ particles left behind the target, the total beam loss cross section $\sigma_{\textnormal{tot}}$ per target nucleus is derived from the exponential attenuation law, $
N_2 = N_1 \cdot \exp(-\sigma_{\textnormal{tot}} \rho)$, where $\rho$ is target density: $
\sigma_{\textnormal{tot}} = \rho^{-1} \ln(N_1/N_2), \quad \delta\sigma_{\textnormal{tot}}= \quad  \rho^{-1}(\delta N_1/N_1 -\delta N_2/N_2).$
We estimate the dispersion of the measured number of particles $N$ following the $\sqrt{N}$ law, so that $\langle \delta N_1 ^2\rangle = N_1$. For the transmitted beam, allowance for the dispersion of the transmission coefficient $p$ gives the corrected formula $\langle \delta N_2 ^2\rangle = N_2 +p(1-p)N_1$. 
The best rms accuracy of measuring  loss cross-section is achieved at $p=e^{-2}$:
\begin{equation}
\frac{\delta(\sigma_{\textnormal{tot}})}{\sigma_{\textnormal{tot}}}= \sqrt{\frac{2}{p\ln^2 p}}\cdot \frac{1}{\sqrt{N_1}} \Rightarrow \frac{e}{\sqrt{2}}\cdot \frac{1}{\sqrt{N_1}} = \frac{1.92}{\sqrt{N_1}}\, .\end{equation}

Above we argued that $\sim 1.6\cdot 10^5$ cycles per month are feasible. Consequently, in order to achieve the statistical accuracy of $10^{-7}$ in measuring the loss cross-section in the one-month run, it is necessary to ensure the accuracy of the asymmetry measurement in the single cycle at the level of $A_1 \approx 4\cdot 10^{-5}$. The number of particles in the bunch impinging on the target must be no less than $\approx 2.3 \cdot 10^{9}$. This leaves a certain room for the further improvement of sensitivity to PV asymmetry increasing the number of particles in the bunch. Furthermore, in the parasitic mode the data taking can be stretched beyond one month.

One can view two options to measure the number of particles in the beam. The first one is to resort to ionization chambers or secondary emission sensors with multiplication of secondary particles. With $n$ secondary particles  produced per one primary particle in the final state, total number of secondary particles will be  $N \cdot n$, what entails the relative rms fluctuation in the determination of the number of particles in the beam 
\[
\frac{\delta N}{N} = \frac{\delta(N n)}{N n} = \sqrt{\frac{1}{N n}}.
\]

An alternative option is a non-destructive measurement of the total charge of the bunch before and after interaction in the target - a comparison of charges before and after amounts to desired comparison of particle numbers $N_{1,2}$. Such an approach will take advantage of a bunched beam required for the time stamp of the precessing polarization. Namely, the Rogowski coils with  high permeability amorphous iron core are known to  be good transformers of the current from the primary circuit, {\sl i.e.,}  the beam current, to the secondary circuit with a very high degree of identity. This is largely due to the very large ratio of the magnetizing inductance of the core to the leakage inductance. 

The primary signal from the Rogowski coil is a voltage proportional to the time derivative of the beam current: $U = L \dot{I}$. This signal is applied  to the infinitely large resistance of the amplifier buffer stage. Next, it is subjected to analog integrations on operational amplifiers (OA),  composed of an RC chain. The first integration will give at the output a signal $U_1(t) = L \cdot I(t)/R_1 C_1$. After the second integration, we get at the output an almost constant voltage $U_2$ and the accumulated charge $q_2$ on the capacitor $C_2$ will equal
\[
U_2 = \frac{q \cdot L}{R_1 R_2 C_1 C_2}, \qquad q_2 = \frac{q \cdot L}{R_1 R_2 C_1}.
\]
With a large ratio $L/(R_1 R_2 C_1)$, one can get a significant gain in the accumulated charge on the capacitor of the second integrator. Note that the values of the time constants of the RC chains do not in any way affect the linearity of signal conversion by the integrator on the OA, in contrast to its passive analogue, where the signal is integrated imperfectly, with damping determined by the time constant $\tau = R C$.
Leaving aside the question of the magnitude of the noise in the signal processing circuit for the current coils, we can state that it is promising to use the above approach to measure the transmission coefficient of a beam through a dense target in the transport channel from the Nuclotron or the new Booster.

One can further increase the overall statistical accuracy of measuring the beam transmission coefficient installing 3--5 identical devices both in front of, and after the target. Besides better statistics, that will allow for the mutual control of the received data from all sensors. In principle, the above considerations of performance of the Rogowski coils as beam current transformers can be studied in the test stand experiment simulating  the particle bunches by the current pulses.

\subsection{Summary and outlook}

High intensity beams of polarized deuterons available at NICA facility make feasible a high precision PV tests of the Standard Model. We consider the external fixed dense target experiments  at either Nuclotron or new Booster the most promising ones. At the core of our proposal is a new technique of polarization precessing in the ring plane, which enables one to eject onto the target beams of any desired spin orientation.  We  anticipate a non-invasive measurement of the total charge of the bunch incident on the condensed matter target and of the transmitted jne as well by a system of Rogowski coils, so that the PV experiment  will not require any sophisticated external detectors. The only two new devices,  the RF spin flipper and internal target polarimeter, can be made sufficiently compact to fit into the Nuclotron and/or the new Booster ring. The Booster may be preferred for the less crowded ring lattice. A possibility of conducting the PV asymmetry experiment in the parasitic mode needs more scrutiny.
There are still open questions, but by statistics considerations, the PV asymmetries smaller than $10^{-7}$ are within the reach of the proposed scheme.
 
For the lack of space, we omitted a number of items, including the spin resonance issues in operation with polarized protons, possible PV experiment with the internal dense target, selection of elastic events at high energies etc. We sketched only briefly a search for the semistrong CP violation which requires the internal polarized proton ABS target. The theoretical analysis of PV in polarized deuteron-nucleus interactions is in progress.

At electron-ion colliders, one can not produce longitudinal polarization of deuterons resorting to Siberian snakes, because of the impractically large required field integrals. The ideas of operation at the integer spin tune, developed at JINR \cite{Derbenev:2016chn,Filatov:2020ygb,Filatov:2020xsy}, have been further extended at BNL   \,\cite{MorozovRHIC,Huang:2020uui}. In view of simplicity of the approach, a fresh look at the possibility of oscillating in-plane polarization of ultra-relativistic deuterons is worthwhile. A solution has to be found to increase the horizontal spin coherence time of $\sim {1400}$\ s, achieved so far at COSY\,\cite{Guidoboni1000s}, by more than one order of magnitude to match the expected storage time of $\sim {10}$\ h at eIC\,\cite{Montag:2019PoS}.

%

\section*{\bf Acknowledgments}

V.~Baskov, O.~Dalkarov, A.~L'vov and V.~Polyanskiy acknowledge a
    support by the Russian Fund for Basic Research, Grant No. 18-02-40061.
V.A. Ladygin  acknowledges the support by the Russian Fund for Basic Research, Grant No.19-02-00079a.
A.\, Larionov acknowledges the support by the Frankfurt Center for Scientific Computing 
and financial support by the German Federal Ministry of Education and
Research (BMBF), Grant No. 05P18RGFCA.\,
I.A.\,Koop, A.I.\,Milstein,
	N.N.\,Nikolaev,
    A.S.\,Popov, S.G.\,Salnikov, P.Yu.\,Shatunov,Yu.M.\,Shatunov
acknowledge a support 
by the Russian Fund for Basic Research, Grant No. 18-02-40092 MEGA.

 \bibliographystyle{pepan}

\bibliography{Introd,uzikova,lvov,Selugina,SPD11,egle-arxiv,strikman,Uz-Ladyg,kimvt,
Kostenkoeng,Zhao,Uzhinsk,Kokoulina,Larionov,bunichev,Elkholy,NIKOLAEV-SPD}


\end{document}